\documentclass[a4paper,fleqn,usenatbib]{mnras}

\usepackage{amssymb}
\usepackage{amsmath}
\usepackage{graphicx}
\usepackage{natbib}
\usepackage{hyperref}
\usepackage{pdflscape}
\usepackage{longtable}
\usepackage{multicol}
\usepackage{multirow}
\usepackage{xspace}

\usepackage{url}
\usepackage{graphicx}
\usepackage{color}
\usepackage{txfonts}

\DeclareRobustCommand{\ION}[2]{%
\relax\ifmmode
\ifx\testbx\f@series
{\mathbf{#1\,\mathsc{#2}}}\else
{\mathrm{#1\,\mathsc{#2}}}\fi
\else\textup{#1\,{\mdseries\textsc{#2}}}%
\fi}

\newcommand{\nii}{[\ION{N}{ii}]}

\newcommand{\oii}{[\ION{O}{ii}]}
\newcommand{\oiii}{[\ION{O}{iii}]}
\newcommand{\sii}{[\ION{S}{ii}]}

\newcommand{\Ha}{$\rm{H}\alpha$}
\newcommand{\Hb}{$\rm{H}\beta$}
\newcommand{\Hg}{$\rm{H}\gamma$}
\newcommand{\Hd}{$\rm{H}\delta$}

\newcommand{\EWa}{EW(\Ha)}

\newcommand{\HII}{\ion{H}{ii}}

\newcommand{\Ssfr}{$\Sigma_{\rm SFR}$\,}
\newcommand{\Sst}{$\Sigma_{\rm *}$\,}
\newcommand{\Sgas}{$\Sigma_{\rm mol}$\,}

\newcommand{\kms}{km\,s$^{-1}$}

\newcommand{\fCAL}{10$^{-16}$\,erg\,s$^{-1}$\,cm$^{-2}$}
\newcommand{\fluxSB}{erg\,s$^{-1}$\,cm$^{-2}$$\,\AA^{-1}$\,arcsec$^{-2}$}

\newcommand{\ageLW}{\ifmmode \mathcal{A}_{\star,L} \else $\mathcal{A}_{\star,L}$\fi\xspace}
\newcommand{\age}{\ifmmode \mathcal{A}_\star \else $\mathcal{A}_\star$\fi\xspace}
\newcommand{\met}{\ifmmode \mathcal{Z}_\star \else $\mathcal{Z}_\star$\fi\xspace}
\newcommand{\metLW}{\ifmmode \mathcal{Z}_{\star,L} \else $\mathcal{Z}_{\star,L}$\fi\xspace}
\newcommand{\ageMW}{\ifmmode \mathcal{A}_{\star,M} \else $\mathcal{A}_{\star,M}$\fi\xspace}
\newcommand{\metMW}{\ifmmode \mathcal{Z}_{\star,M} \else $\mathcal{Z}_{\star,M}$\fi\xspace}

\title[]{The Calar Alto Legacy Integral Field Area Survey: extended and remastered data release}

\author[S.F.S\'anchez et al.]{S.F. S\'anchez$^{1}$,
  L. Galbany$^{2,3}$, 
  C.J.Walcher$^{4}$,
  R.Garc\'\i a-Benito$^{5}$,
  J.K. Barrera-Ballesteros$^{1}$
\\
$^{1}$Instituto de Astronom\'ia, Universidad Nacional Aut\'onoma de  M\'exico, A.~P. 70-264, C.P. 04510, M\'exico, D.F., Mexico. \\
$^{2}$Institute of Space Sciences (ICE, CSIC), Campus UAB, Carrer de Can Magrans, s/n, E-08193 Barcelona, Spain.\\
$^{3}$Institut d’Estudis Espacials de Catalunya (IEEC), E-08034 Barcelona, Spain.\\
$^{4}$Leibniz-Institut f\"ur Astrophysik Potsdam (AIP), An der Sternwarte 16, 14482 Potsdam, Germany.\\
$^{5}$Instituto de Astrof\'isica de Andaluc\'ia (IAA/CSIC), Glorieta de la Astronom\'{\i}a s/n Aptdo. 3004, E-18080 Granada, Spain.\\
}

\date{Accepted XXX. Received YYY; in original form ZZZ}

\pubyear{2020}

\begin{document}
\label{firstpage}
\pagerange{\pageref{firstpage}--\pageref{lastpage}}
\maketitle

\begin{abstract}
 This paper describes the extended data release of the Calar Alto Legacy Integral Field Area (CALIFA) survey (eDR). It comprises science-grade quality data for 895 galaxies obtained with the PMAS/PPak instrument at the 3.5 m telescope at the Calar Alto Observatory along the last 12 years, using the V500 setup (3700-7500\AA, 6\AA/FWHM) and the CALIFA observing strategy. It includes galaxies of any morphological type, star-formation stage, a wide range of stellar masses ($\sim$10$^7$-10$^{12}$M$_\odot$), at an average redshift of $\sim$0.015 (90\% within 0.005$<z<$0.05).  Primarily selected based on the projected size  and apparent magnitude, we demonstrate that it can be volume corrected resulting in a statistically limited but representative sample of the population of galaxies in the nearby Universe. All the data were homogeneous re-reduced, introducing a set of modifications to the previous reduction. The most relevant is the development and implementation of  a new cube-reconstruction algorithm that provides with an (almost) seeing-limited spatial resolution (FWHM$_{\rm PSF}\sim$1.0$\arcsec$).To illustrate the usability and quality of the data, we extracted two aperture spectra for each galaxy (central 1.5$\arcsec$ and fully integrated), and analyze them using {\sc pyFIT3D}. We obtain a set of  observational and physical properties of both the stellar populations and the ionized gas, that have been compared for the two apertures, exploring their distributions as a function of the stellar masses and morphologies of the galaxies, comparing with recent results in the literature.  
\end{abstract}


\section{Introduction}
\label{sec:intro}

The exploration of galaxy properties and the understanding of their
evolution along cosmological times has been significantly improved in
the last few decades by the combination of large cosmological surveys
\citep[e.g.][]{york2000,gamma}, detailed N-body and hydrodynamical
simulations \citep[e.g., IllustrisTNG][]{springel18}, and in particular by the massive
exploitation of novel techniques like integral field spectroscopy
\citep[IFS, e.g.][]{cappellari16,ARAA}. Recent IFS galaxy surveys
(IFS-GS) in the nearby universe ($z\sim$0.01-0.03), like CALIFA
\citep{califa}, MaNGA \citep{manga} or SAMI \citep{sami}, have uncovered
new spatial resolved relations that rules the star-formation and
chemical enrichment in galaxies
\citep[e.g.][]{rosales-ortega:2012,sanchez14,mariana16,lin19,sanchez21},
defined which is their dynamical stage
\citep[e.g.][]{cappellari16,zhu18b}, and uncovered the patterns that
define their mass assembly and metal enrichment
\citep[e.g.][]{eperez13,rosa16,ibarra16,camps20}, among many other
results. Those explorations were possible thanks to the unique
combination of (i) the use of statistically representative and
significantly large samples of galaxies (from $\sim$1000 to
$\sim$10,000 objects), (ii) a wide spectroscopic coverage (from
$\sim$3600\AA~up to $\sim$10,000\AA) with intermediate resolutions
($R\sim$1000-2000), (iii) the spatial coverage of a substantial
fraction of their optical extension (between 1.5 to 2.5 effective
radius, R$_e$), and (iv) a spatial resolution of $\sim$1 kpc.

More recently, the advent of new IFS instruments \citep[e.g.,
MUSE][]{muse} and techniques \citep[e.g., SITELLE][]{grand12}, that
allow us to obtain natural seeing-limited spatial-resolved spectroscopic
information covering a wide field-of-view (FoV, above $\sim$1
arcmin$^2$), has allowed to perform systematic explorations at a
sub-kiloparsec scale. This has lead to a new wave of IFS surveys and
compilations, like GASP \citep{GASP}, AMUSING++ \citep{carlos20} and
in particular PHANGS-MUSE \citep{emse22}, that have improved our
understanding of different processes, in particular those related with
the details of star-formation and quenching mechanisms and the nature
of diffuse ionized gas
\citep[e.g.][]{rous18,george19,vulcani19,belfiore22,pan22,pessa22}. These
explorations highlight the importance of a improved spatial resolution
when exploring galaxy properties using IFS. However, despite of their
outstanding results, the explorations performed using MUSE or SITELLE
present intrinsic problems due to the limitations in the spectral
coverage compared to those of previous IFS galaxy surveys: for
instance, the spectral range covered by MUSE does not allow to sample
important stellar spectral features like the 4000-break or emission
lines like \oii$\lambda$3727. Therefore, it is important to
determine if it is possible to improve the spatial resolution of
already existing IFS GS data.

The seek for a improved spatial resolution and a complete sampling of
the covered FoV using IFS data lead to the development of innovative
observational techniques and cube reconstruction techniques in
parallel with the development of the IFS-GS. Most of the surveys
listed before (CALIFA, SAMI and MaNGA) adopted Integral Field Units
(IFUs) that integrate fiber bundles as the basic systems to sample the
observed objects in different discrete apertures \citep[e.g., PPAK
mode of the PMAS spectrograph][]{kelz06,roth05}. By construction the
spatial resolution provided by those IFUs is limited by the fiber-size
(convolved by the natural seeing). Furthermore, they present an
incomplete coverage of the FoV (hexagonal or circular in general),
sampling $\sim$60-65\% of it, depending of the fiber packing. To
overcome those two limitations it was proposed an observational scheme
that includes a minimum of three dithered exposures
\citep{sanchez06a,marmol-queralto11}, to cover the entire FoV, and a flux-conservative variation of the Shepard's
interpolation method \citep{shepard1968} to re-sample the discrete
observations into a regular sampled datacube \citep{califa}.

This scheme, with little variations, was adapted and adopted by the
main IFS-GS described before, establishing a more or less standard
procedure. One of the main requisites of this procedure is to
reconstruct the image preserving the flux, minimizing the co-variance
between adjacent data, and generating a PSF with the minimum possible
structure. Due to that it was preferred to other schemes, like
drizzling \citep[e.g.,][]{dr2}. However, this procedure does not
guarantee that it is recovered the best possible spatial resolution of
the data, that in general is more limited by the average distance
between sampled points \citep{shannon1948}, and the natural-seeing, rather by the size of
the original fiber (what defines and limits the minimum co-variance
between adjacent spaxels). For this reason, the spatial resolution
achieved was just slightly better than the original fiber size convolved 
by the natural-seeing PSF, ranging between 1.8$\arcsec$ (SAMI) and
2.4$\arcsec$ (CALIFA, MaNGA), despite the fact that the average
distance between adjacent fibers is of the order of two times smaller in 
the case of SAMI than in the case of CALIFA.

In this article we present a remastered version of the extended CALIFA
dataset \citep{lacerda20,espi20}, in which we applied a new image
reconstruction algorithm. We re-reduce just the data corresponding to
the low-resolution setup (V500) and not those observed using the
high-resolution one (V1200), as those ones have not suffered any
significant increase in the number of objects since the last
data-release \citep{sanchez16}.  The new data present a clear improved
spatial resolution, achieving a PSF size near to the limit imposed by
the natural seeing, preserving the photometry and the image
quality. Similar improvements in the spatial resolution have been
achieved by recent studies by adopting an {\it a posteriori}
deconvolution scheme, wavelength-by-wavelength, to the reconstructed
datacubes for the MaNGA survey \citep{chung21}. Despite of the clear
improvements of that procedure, it essentially inheritate all the
issues of the original observing procedure and adopted interpolation
scheme { ,as already noticed and explored by \citet{liu20}}. In
particular, those introduced by (i) the large size and aperture of the
interpolation kernel, (ii) the possible mismatch of the foreseen
dithering scheme and the actually adopted along the observations, and
(iii) the inevitable problems in deriving and correcting for the
differential atmospheric refraction (DAR) for different dithered
pointings \citep[e.g.][]{sanchez06a,dr2,law15,sanchez16}. Our
procedure involves a new interpolation scheme in which the spatial
resolution is improved and the covariance minimized by implementing a
narrower interpolation kernel, taking into account the observational
adopted dither and the DAR at the moment of the image reconstruction,
and correcting by the irregular shape of the PSF (introduced by the
dithering and discrete sampling of the fiber bundle). Furthermore we
introduce additional modifications to the existing data reduction
pipeline \citep{sanchez22} that homogenize the dataset in terms of
spectrophotometric calibration, mask and clean the foreground stars
and improves the treatment of the vignetted and
low-transmission/broken fibers.

Finally, we perform an {\it a posteriori} selection considering the
quality of the data. In addition to the prior selection based mostly
on the spatial coverage of the observed galaxies and their apparent
magnitude, the selection function defines a sample of galaxies which
main properties are described. We illustrate the quality and usability
of the new reduced dataset for this final galaxy sample by exploring
the main properties of the stellar populations and ionized gas at two
different apertures: (i) the central region and (ii) a fully
integrated aperture covering the entire FoV of the instrument. We present
the main distributions of those properties as a function of the stellar 
masses and morphologies of the galaxies, comparing them with recent results in the literature.

The structure of this articles is as follows: (i) Sec. \ref{sec:data} describes the data adopted in this study; (ii) the modifications introduced in the reduction are described in  Sec. \ref{sec:dr}, making a particular emphasis in new image reconstruction algorithm (Sec. \ref{sec:img_rec}) and its implementation in the cube reconstruction  (Sec. \ref{sec:cube_rec}); (iii) the identification of the galaxies and the derivation of their structural parameters is described in Sec. \ref{sec:gal_seg}; (iv) the analysis of the quality of the data is introduced in 
\ref{sec:qc}, which leads to the selection of the final sample of galaxies described in  Sec. \ref{sec:sample}; (v) the demonstration that this sample behaves primarily as diameter and magnitude limited sample, and the estimation of the volume correction is introduced in Sec. \ref{sec:Vcor}; (vi) the analysis performed on the new dataset is explained in Sec. \ref{sec:analysis}, summarizing the adopted spectral fitting procedure (Sec. \ref{sec:spec_fit}), and describing the main properties obtained for the stellar populations (Sec. \ref{sec:LW}) and the ionized gas (Sec. \ref{sec:SFR}, \ref{sec:Mgas} and \ref{sec:OH}); (vii) the results from this analysis are included in Sec. \ref{sec:results}, describing the distribution of the stellar population (Sec. \ref{sec:stars}) and ionized gas (Sec. \ref{sec:ion}) as a function of the stellar masses and morphologies; (viii) the analysis is also used to characterize the spectrophotometric calibration, explore the behaviour of the residuals from the spectral fitting (Sec \ref{sec:res_spec}), how they compare with the errors estimated by the reduction and how the covariance affects their propagation (Sec. \ref{sec:covar}); (ix) a summary of the main results is included in Sec. \ref{sec:covar}. How to access to the delivered dataset and the results from the analysis is described in the CALIFA webpage \footnote{official: \url{http://califa.caha.es/}, mirror:\url{http://ifs.astroscu.unam.mx/CALIFA_WEB/public_html/}}.

Throughout this article we assume the standard $\Lambda$ Cold 
Dark Matter cosmology with the parameters: $H_0$=71 km/s/Mpc, $\Omega_M$=0.27, $\Omega_\Lambda$=0.73.




\section{Data}
\label{sec:data}

\begin{figure}
 \minipage{0.99\textwidth}
 \includegraphics[width=8.5cm]{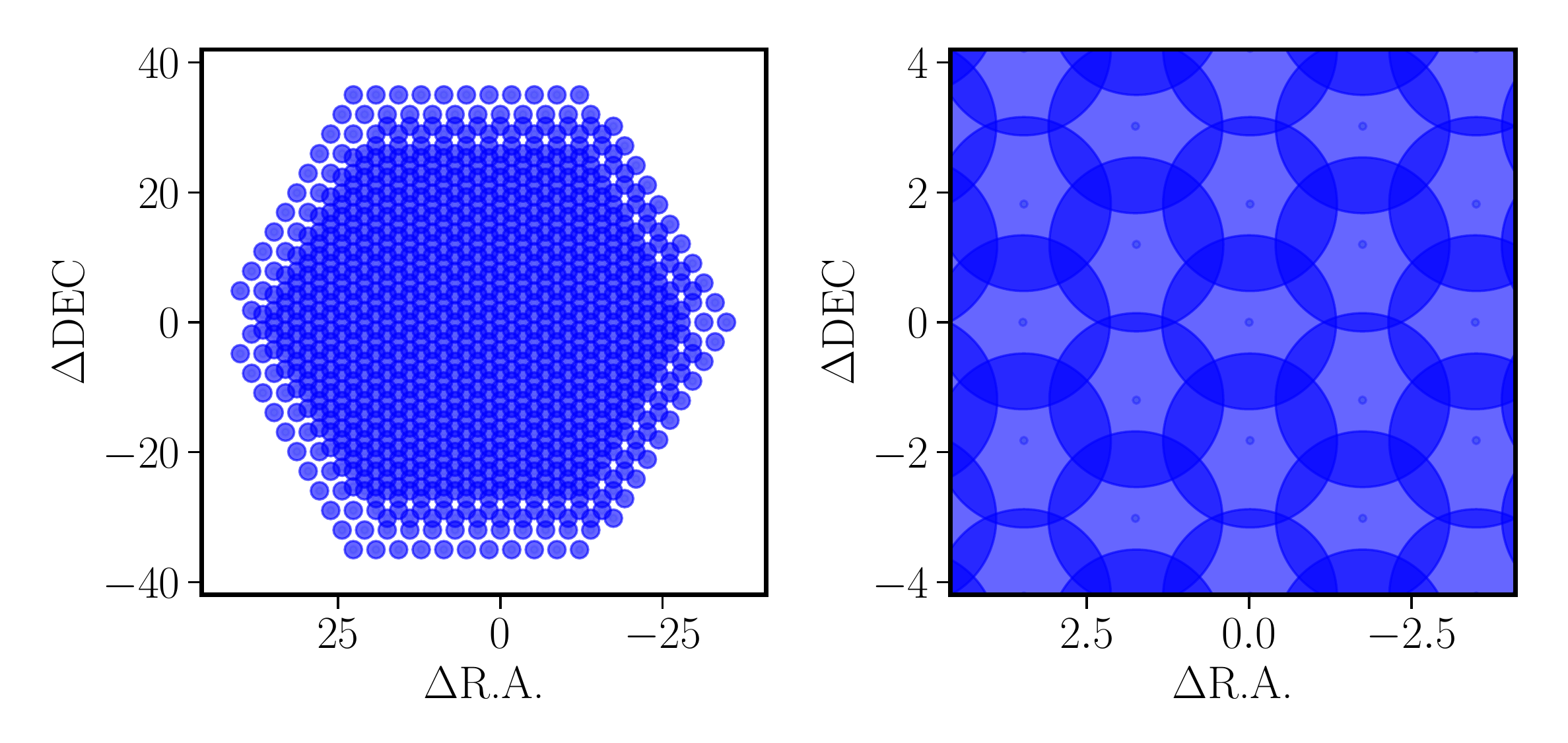}
 \endminipage
 \caption{Example of the three dithering scheme adopted by the CALIFA survey. Left panel shows the distribution of the 993 fibers using the nominal offsets between the three individual pointings, that comprises 331 fibers each one. Right panel shows a zoom of the central region for the same distribution. In both panels the center of each fiber is represented by a blue point, and its corresponding aperture with a light-blue solid-circle. The typical distance between adjacent fibers ($\sim$1.5$\arcsec$), the area covered by only one fiber ($\sim$2 arcsec$^2$), and the overlapping region between adjacent fibers is evident in this figure.}
 \label{fig:dither}
\end{figure}



\subsection{Dataset}
\label{sec:dataset}

We collect all the data acquired by the 3.5m telescope at the Calar
Alto observatory using the PMAS \citep{roth05} IFS spectrograph in the
PPAK mode \citep{kelz06}, using the same observing strategy as the one
adopted by the CALIFA survey. This is, using the V500 grating,
configured at the same goniometer angle, with the same integrating
time (900s per pointing\footnote{for a handful targets the exposure time was slightly larger, $\sim$1000-1200s}), and the same dithering scheme as the one
adopted by the CALIFA observations \citep[see details
in][]{califa,sanchez16}. This configuration corresponds to a
low-resolution setup (R$\sim$850, instrumental
FWHM$\sim$6.5\AA), that samples a wavelength range between 3745 and
7500 \AA, covering an hexagonal FoV of 74$\arcsec$ $\times$
64$\arcsec$.

The science fiber-bundle of the PPAK IFU comprises 331 fibers with a
2.7$\arcsec$/diameter. Thus, the adopted dithering scheme provides
with 993 individual sampling points of the FoV, with an average
distance between adjacent fibers of $\sim$1.5$\arcsec$. This involves
an overlapping between the apertures of adjacent fibers, and therefore
an inevitable co-variance between the information when applying any
image reconstruction scheme using spaxels smaller than the original
apertures. This reconstruction is required to fully correct some
observational features, such as the DAR, as we will explain below. An
additional consequence of this observing strategy is the differential
depth achieved across the FoV, even for a target with an uniform light
distribution, as there are regions sampled by three, two or one fiber,
respectively. Figure \ref{fig:dither} illustrates the final
distribution of the PPAK science fibers adopting the nominal offsets
defined by the CALIFA observing strategy, highlighting the overlaps
between adjacent apertures. This is relevant for the modifications
introduced in the data reduction.

We should recall that the original CALIFA survey observed a sub-sample
of the objects using a V1200 setup, that provides with a higher
spectral resolution covering only the blue regime of the optical
spectral range. So far we have restricted the current analysis to the
V500 setup only, comprising a total of 1088 individual observed
galaxies.  The final galaxy sample would be extracted from this
compilation based on (i) the quality of the data and (ii) the
properties of the galaxies themselves, as we will describe later-on.

\subsection{Data Reduction}
\label{sec:dr}

Data reduction (DR) was performed using a modified version of the
CALIFA DR pipeline, formerly used in the third public data release
\citep[i.e., version 2.2][]{sanchez16}. This new DR-pipeline, version
2.3, comprises exactly the same reduction steps and procedures of the
former version up to the very last steps that involve the cube
reconstruction procedures. It follows the prescriptions described in
\citep{sanchez06a}, using a modified version {\tt py3D} (PI:
B. Husemann) adapted to {\tt python3} as the basic reduction package.

\begin{figure*}
 \minipage{0.99\textwidth}
 \includegraphics[width=18cm,clip,trim=0 20 0 10]{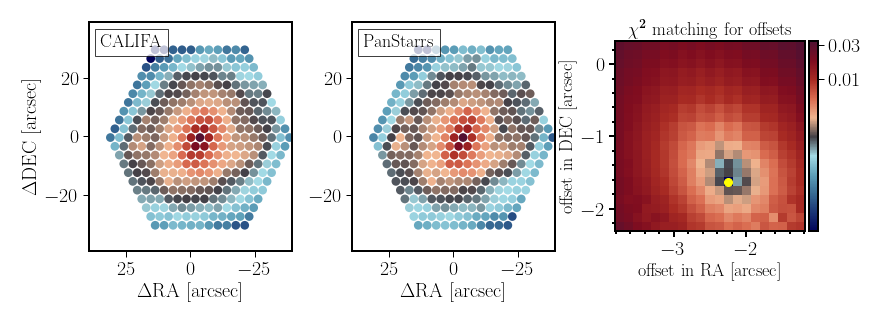}
 \endminipage
 \caption{Illustration of the registration procedure for each individual dithering. {\it Left-Panel:} Distribution of flux intensities in the $r-$band synthesized from the 331 spectra obtained for the galaxy NGC 5947, corresponding to the 1st pointing of the three dithered exposures. {\it Middle-Panel:} Similar distribution derived by coadding the flux intensities in the corresponding PanStarrs $r-$band image for each aperture of each fiber, with the absolute location of the whole fiber-bundle shifted to match the flux intensities with those shown in the left-panel (i.e., the shift that minimizes the $\chi^2$ value between both distributions). {\it Right-Panel:} $\chi^2$-map derived from the exploration of different offsets,  with the location of the minimum value highlighted with a yellow solid-circle}
 \label{fig:d_astro}
\end{figure*}

The procedures in common with version 2.2 of the DR pipeline involve
(i) pre-processing the raw data to unify the data read by the
different CCD amplifiers (gluing in an homogeneous orientation,
removing the bias and normalizing by the GAIN), and cleaning
cosmic-rays, (ii) tracing the location across the CCD of the
science and calibration spectra corresponding to each fiber, deriving
simultaneously the FWHMs of the projected light-distribution in both
the cross-dispersion and dispersion axis, (iii) an optimal extraction
of the spectra using the tracing and FWHM derived in the previous
step, correcting the data for the presence of stray-light, (iv) a
wavelength calibration and regularization of the data following a
linear wavelength solution (with 2\AA~per spectral pixel), (v) an
homogenization of the spectral resolution along the wavelength range
to its nominal value of FWHM = 6.5\AA, (vi) a wavelength-dependence
fiber-to-fiber transmission correction, (vii) a flux calibration using
the standard transmission curve derived for the considered
instrumental along the survey, (viii) a separation of the science,
calibration and sky fibers, previous to the estimation and subtraction
of the night-sky spectrum for each dither pointing, and finally (ix)
an integration of the three dithering exposures into a single
raw-stacked spectra (RSS) file, with each spectra/fiber associated to
a particular (relative) location in the sky via a position-table (PT).
As the variance and the masked regions (broken fibers, cosmic-rays,
CCD defects and vignetted regimes) are propagated through the
reduction, the final product of this set of procedures is a set of
three RSS frames (flux, error and mask), each one comprising 993 spectra. All those procedures been
extensively explained in detail in previous articles
\citep[][]{sanchez06a,califa,dr1,dr2,sanchez16}.


We describe in more detail the modifications and new procedures introduced in version 2.3 of the DR pipeline, aimed to address different issues present on the data.\\

\subsubsection{ Preliminar astrometry}
\label{sec:pre_ast}

All eCALIFA observations were first reduced using version 2.2 of the
pipeline. This provides with a reconstructed datacube that adopts a
Shepard's interpolation scheme, with a sampling spaxel of 0.5$\arcsec$
\citep{sanchez16}. We synthesize the $g-$ and $r-$band images from
these datacubes, by convolving them with the corresponding filter
responses. Then, adopting the nominal coordinates of the targets, we
download two images of 2$\arcmin$ and 10$\arcmin$ size for the same
bands from the PanStarrs (PS; \citealt{PS1,PS1_data}) data archive corresponding to
the PS1 public data release
\footnote{\url{https://outerspace.stsci.edu/display/PANSTARRS/}}. Using
this dataset we perform a coordinates matching of the eCALIFA and PS
$r-$band images, selecting by hand the center of each object and
performing a baricenter estimation in a 10$\arcsec\times$10$\arcsec$
box. Then, we perform a refined image registration by (1) matching the
spatial resolution of the PS images ($\sim$1.5$\arcsec$) to that of
the eCALIFA v2.2 cubes ($\sim$2.5$\arcsec$), (2) reproject and
resample the PS images to that of the eCALIFA ones, using the {\sc
  reproject\_interp} routine implemented in the {\sc reproject} python
module
\footnote{\url{https://reproject.readthedocs.io/en/stable/index.html}},
and finally (3) register the eCALIFA images to the PS one, by running
the {\sc chi2\_shift} routine implemented in the {\sc
  image-registration} {\sc python} package
\footnote{\url{https://pypi.org/project/image-registration/}}. This
procedure is repeated for both the $g-$ and $r-$band images, and the
average WCS recovered is adopted as the preliminary astrometry of the
eCALIFA datacubes.


\subsubsection{Dither Registration}
\label{sec:dither}

The previous procedure was adopted assuming that the offset between
each dithering pointing follows the nominal/foreseen one. However, it
was already noticed along the CALIFA observations that this is not
always the case \citep[e.g.][]{dr2}. For this reason in previous
versions of the pipeline it was adopted a procedure to (i) obtain an
astrometric solution for each individual dither pointing, (ii)
determine the real offsets between the three dithering exposures, and
(iii) correct for the possible changes in the transparency of the
atmosphere between them. This procedure uses a broad-band image as a
reference. In previous versions of the reduction, it was adopted the
SDSS $r-$band image \citep{dr2}. This choice was valid for the
galaxies in the original CALIFA sample, since they were all extracted
from the SDSS catalog. However, our extended sample comprises galaxies
that have not been necessarily observed by SDSS survey. For this
reason we adopted the $r-$band images provided by the PS survey,
already described before.

The procedure is indeed very similar to the {\sc chi2\_shift}
algorithm indicated before, but implemented for discrete dataset
(i.e., not regular grided images). First, the
integrated flux is synthesized through the $r-$band filter using the RSS data of each
pointing, obtaining 331 photometric measurements (one for each science
fiber). Then, it is estimated the same flux intensities from the
reference image using the fiber apertures at the expected location on
the sky based on the preliminar astrometry described
before. Afterwards, we compare the two photometric datasets, obtaining
the corresponding $\chi^2$ distribution. The procedure was repeated by
shifting the assumed coordinates of the target within a box of
20$\arcsec$ width in both RA and DEC, following a regular grid with a
step of 1.5$\arcsec$ (in both directions). This generates a $\chi^2$
map, which minimum value is considered as the first guess of the
$\Delta RA$ and $\Delta DEC$ required to match the coordinates of the
target to that of the reference image. Finally the whole procedure is
repeated one more time by limiting the search around this new
coordinate using a smaller box 2.5$\arcsec$ width and a step of
0.125$\arcsec$. Once the offsets that minimize the $\chi^2$ are found, the
photometric values are re-evaluated deriving the average ratio between
the flux intensities measured through the fibers and those derived
from the reference image.

Figure \ref{fig:d_astro} illustrates this procedure, showing the
$r-$band photometric values extracted from one dither exposure on the
galaxy NGC\ 5947, the corresponding values extracted from the PS image
at the location that minimizes the $\chi^2$, and finally the $\chi^2$
map derived from the second iteration described before. This procedure
provide with (1) the best absolute astrometric solution for each
pointing, (2) the real offset between each dithering, and (3) a flux
re-calibration of each dither-pointing that anchors the photometry to
that of the reference image, limiting any possible relative
photometric fluctuation between each individual pointing (due to
variations in the atmospheric conditions along the observing process).

\subsubsection{Broken fibers}
\label{sec:broken}

Along the full decade in which the data
was taken the PPAK fiber-bundle has suffered an inevitable aging
process. Due to that a set of broken fibers has appeared. For the
first observations, covering the first four years, no fiber present a
degradation of the transmission. Afterwards, three fibers present a
continuous degradation. This number increased to five fibers in the
last two years of the collected period, with a negligible transmission
in the last observations. In previous versions of the DR these fibers
were identified by hand and masked. In this new version we introduce a
procedure that automatically detects those fibers and replaces the
spectra by an inverse-distance weighted average-spectra of the 5th
nearest (non broken) fibers.

\subsubsection{Vignetted regions}
\label{sec:vign}

As extensively described in
different articles \citep[e.g.][]{califa,sanchez16}, a 30\%\ of the
fibers are affected by different degrees of vignetting. This effect
degrades the transmission down to a 30-50\% in the extremes of the
wavelength range, with a fully unvignetted regime between
4240-7140\AA, in the worst cases. Due to the particular correspondence
between the location of fibers in the science bundle and the entrance
of the spectrograph (and final location in the CCD), those spectra
correspond to fibers located on an annular ring at $\sim$15$\arcsec$
from the center of the FoV. However, not all spectra in this location
are affected by vignetting. Thus, there are fully unvignetted spectra
spatially adjacent to any set of vignetted ones. Using this property
we apply a procedure similar to the one adopted to correct the effects
of broken-fibers, replacing the vignetted regime of the spectra of
those fibers affected by this effect by a flux-scaled version of the
inverse-distance weighted average-spectra of the 5th nearest (fully
unvignetted) fibers. We selected as vignetted those regions in which
the transmission is a 70\%\ of the average transmission within a
considered fiber. Despite of the benefits of both procedures, we keep
a record of those modified regions in the propagated mask frame.

\subsubsection{Image reconstruction}
\label{sec:img_rec}

{ One of the main goals in the new reduction is to increase the
  spatial resolution of the delivered datacubes. As indicated in the
  introduction, the intrinsic spatial resolution is limited more by
  the distance between sampling elements (1.5$\arcsec$, defined by the
  dithering pattern) than by the size of those elements (2.7$\arcsec$,
  defined by the aperture of the fibers). However, in practice, the
  spatial resolution is strongly affected by the method adopted to
  reconstruct the cubes/images and the parameters adopted in the
  interpolation kernel.  For instance, the main goal of the method
  adopted in previous versions of the CALIFA pipeline to reconstruct
  the cubes was to provide with a final smooth PSF, without evident
  sub-structures from the dither pattern, preserving the
  spectrophotometry \citep{califa}. To provide with the best possible
  spatial resolution was not among its main priorities. Indeed, other
  reconstructions procedures that produce better spatial resolutions
  has been explored before.  For instance, \citet{dr2} experimented
  with the use of the drizzle method \citep{fruchter:2002} in the
  CALIFA data, which produces a sharper PSF, but with a considerable
  amount of sub-structures (i.e., a PSF with secondary peaks). More
  recently, \citet{liu20} adopted a covariance-regularized
  reconstruction procedure to reconstruct the cubes from the MaNGA
  dataset. This method slightly improves the spatial resolution and
  limit considerably the covariance between adjacent spaxels. Finally,
  \citet{chung21} performed a spatial deconvolution of the original
  MaNGA cubes, without introducing a new image reconstruction
  procedure, by adopting the SDSS images a guidance for the
  deconvolution process. The method we describe in this section
  introduces a modification in the image reconstruction algorithm,
  following \citet{dr2} and \citet{liu20}, with the ultimate goal of
  improving the spatial resolution. We finally implement a
  deconvolution process just to mitigate the effects of a complex PSF.
  This deconvolution improves slightly the spatial resolution, but it
  is not the main reason of the achieved/improved spatial resolution. In this
  regard our method departs considerably from the one introduced in
  \citet{chung21}.}

As indicated
before, the image reconstruction in previous version of the pipeline
adopted a modified version of the Shepard's interpolation method,
being broadly adopted in the reduction of fiber-based IFS datasets
that adopts a dither scheme \citep[with certain
variations][]{law15,green18}. This method
assign to a certain spaxel\footnote{understood as the final sampling
  element of the outcome datacube} a flux that results from the
weighted-sum of all fluxes from all fibers within a certain distance:
\begin{equation}\label{eq:f_ij}
 f_{i,j} = N_{i,j} \sum_{k} f_A w_{k,i,j} F_k\ {\rm where}\  d_{k|(i,j)}<d_{\rm lim}
\end{equation}
where $k$ is the index of the fibers contributing to the flux in the spaxel (i,j), denoted as $f_{i,j}$. Only those fibers within a certain distance to the considered spaxel, $d_{\rm lim}$, are considered to contribute to its flux. $f_A$ is a scaling parameter corresponding to the ratio between the area covered by each original fiber and the one covered by the final spaxel. In previous versions of the pipeline this factor was 5.64, as the PPAK fibers have a diameter of 2.7$\arcsec$ and
the final spaxel a size of 1$\arcsec\times$1$\arcsec$. $F_k$ is the flux within a certain fiber, and $w_{k,i,j}$ is the weight (or interpolation kernel):
\begin{equation}\label{eq:w_ijk}
 w_{i,j,k} = \exp{-0.5 \left(\frac{d_{k|(i,j)}}{\sigma}\right)^{\alpha}}
\end{equation}
where $d_{k|(i,j)}$ is the distance between the fiber $k$ and the considered spaxel, $\sigma$ is a parameter controlling the width of the kernel, and $\alpha$ is a parameter controlling its shape. $N_{i,j}$ is just the normalization factor, i.e., the inverse of sum of all $w_{i,j,k}$ contributing to the flux in the spaxel (i,j). 

\begin{figure*}
 \minipage{0.99\textwidth}
 \includegraphics[width=18cm]{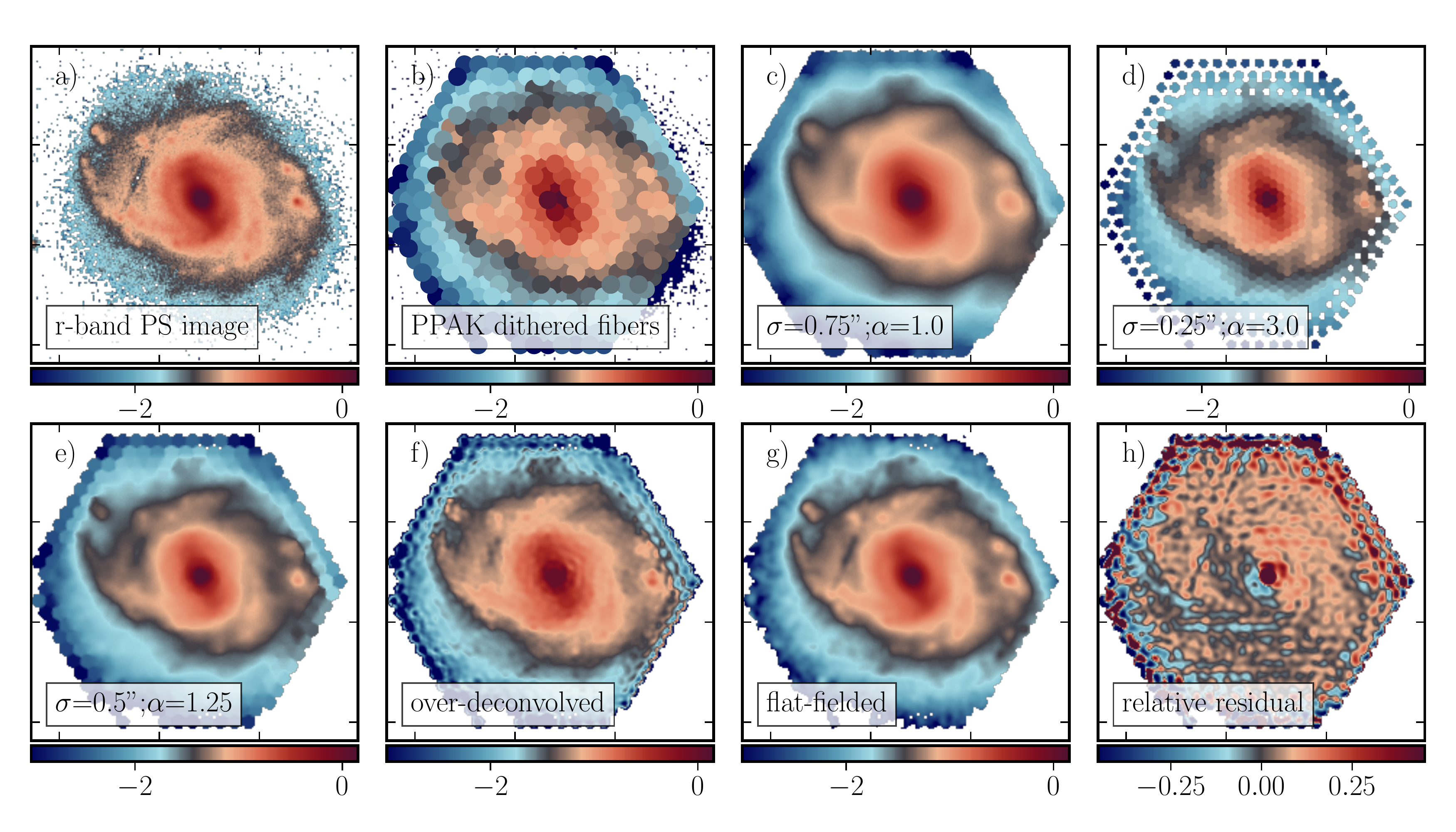}
 \endminipage
 \caption{Illustration of the new image reconstruction procedure. Each panel shows: (a) PS $r-$band image of the archetype galaxy NGC 5947; (b) discrete distribution of the $r-$band flux intensities extracted for the 993 individual spectra corresponding to the three dithering pointings observed with the 331 PPAK science fibers (color solid-circles). Note the overlapping between adjacent fibers due to the adopted dithering scheme; (c) $r-$band reconstructed image using the distribution of flux intensities shown in panel $b$ using the interpolation scheme and parameters adopted in the previous version of the data-reduction (i.e., Eq. \ref{eq:f_ij}, with  $\sigma$=0.75$\arcsec$ and  $\alpha$=1.0); (d) similar reconstructed image using a sharper interpolation kernel (i.e., Eq. \ref{eq:f_ij}, with $\sigma$=0.25$\arcsec$ and $\alpha$=3.0); (e)  similar reconstructed image using the parameters finally adopted in current version of the data reduction (i.e., Eq. \ref{eq:f_ij}, with $\sigma$=0.5$\arcsec$ and $\alpha$=1.25); (f) image obtained by deconvolving the image shown in panel $e$ using a large number of iterations in the deconvolution algorithm as indicated in the text. Note the patchy distribution with unreal sub-structures generated by the deconvolution algorithm; (g) final image generated by the full reduction algorithm, including the final flat-fielding procedure introduced in \citep{dr2}; (h) residual image obtained by subtracting the final reconstructed image shown in panel $g$ to the PS $r-$band image shown in panel $a$ (matching the spatial resolutions), relative to this later image. The difference between both images is below a $\sim$10\% for most of the FoV, with larger differences found in the very central and very outer regions ($\sim$30\%).}
 \label{fig:img_rec}
\end{figure*}

The adopted parameters for the image reconstruction changed along the
different implementations of the data reduction. In early versions of
the DR, we adopted the following values: $\sigma=1.0\arcsec$,
$\alpha$=1.0 and $d_{lim}=5\arcsec$ \citep[ver. 1.2 and
1.3][]{califa,dr1}. Different explorations lead to a tuning of those
parameters, that produce an sharper image and a lower co-variance
between adjacent spaxels: $\sigma=0.75\arcsec$, $\alpha$=1.0 and
$d_{lim}=3.5\arcsec$ \citep[ver. 1.5 and 2.2][]{dr2,sanchez16}. The
FWHM of the reconstructed cubes was estimated using the same algorithm
described above, when describing the preliminar astrometry: the
$r-$band image extracted from the IFS data was compared, by a
$\chi^2$, with a reference $r-$band image, convolved with a sequence
of Gaussian functions of different width. The best-matched image
provides with an estimation of the FWHM of the PSF. In average it was
found a value of $\sim$2.4$\arcsec$/FWHM for the DR3 CALIFA dataset
\citep{sanchez16}.

To optimize the procedure it is required to explore which are the set
of parameters that reproduce better the original image (i.e., the one
provided by the natural seeing). Since we do not have a direct image
in a wavelength covered by the spectroscopic data taken simultaneously
to the observations, once again, we adopted as a guiding images the PS
$g-$ and $r-$band images. On these images we simulate the observations
performed using the CALIFA setup, following the same procedure
described before 
i.e., we
extracted the flux corresponding to each fiber, following the nominal
dither pattern). Then, we use this simulated observation to explore
different image reconstruction schemes. The procedure is illustrated
using NGC\ 5947 as an archetype CALIFA galaxy.  Figure
\ref{fig:img_rec} shows how the original image of this galaxy ($a$
panel) compared with both the flux distribution through the three
dithered exposures ($b$ panel) and the reconstructed image using the
parameters adopted in the previous version of the DR ($c$ panel). The
resolution degradation is clearly appreciated { when using the same parameters for the interpolation Kernel (eq. \ref{eq:w_ijk}) adopted in the previous reductions of the CALIFA dataset.}

We explore a wide range of values for the considered parameters
(0.15$<\sigma<$1.5; $\Delta_\sigma=0.15$, 0.75$<\alpha<$4.0;
$\Delta_\alpha=$0.25), finding that the FWHM of the reconstructed
image increases with $\sigma$ (i.e., the kernel width), and decreases
with $\alpha$ (i.e., the kernel sharpness). Thus, in principal the
smaller $\sigma$ and the largest $\alpha$ may produce the best
reconstructed image { in terms of its PSF FWHM (i.e., a better resolution).
Indeed, for certain combination of parameters the recovered FWHM is well below the
  one obtained using the nominal parameters adopted in the previous reductions of
  the CALIFA dataset (i.e., FWHM$<$2.5$\arcsec$).} However, { resolution is not the only parameter to take in place when weighting the quality of the reconstructed image}. A narrow and sharp
interpolation kernel produces a reconstructed image with a lot of
non-realistic substructures, as illustrated in Fig. \ref{fig:img_rec}
$d$ panel ($\sigma=$0.25$\arcsec$ and $\alpha=$3.0). Those artifacts
can be characterized by the standard deviation of the residual image
derived from subtracting the best-matched image (i.e., the reference
image convolved with a Gaussian kernel to match the resolutions) to
the reconstructed image. As indicated before, this standard deviation
increases as sharper is the interpolation kernel. Thus, the optimal
parameters for the image reconstruction procedure result from a
compromise between minimizing both the FWHM and the standard deviation
of the residuals. This is achieved when $\sigma=$0.5$\arcsec$ and
$\alpha=$1.25. Fig \ref{fig:img_rec}, panel $e$, shows the
reconstructed image when adopting both values for the interpolation
kernel. The improvement in the resolution with respect to the image
obtained when using the parameters adopted in the previous version of
the reduction (panel $c$) is evident ($FWHM_c\sim$2.5$\arcsec$
vs. $FWHM_e\sim$1.8$\arcsec$). However, the resolution is still worse
than that of the original image (panel $a$, $FWHM_a\sim$1.0$\arcsec$),
and lower than the ultimate goal based on the sampling of the
dithering scheme ($\sim$1.5$\arcsec$). Furthermore the PSF present a
complex structure, due to the adopted reconstruction scheme.


In order to improve furthermore the resolution { and mitigate the presence of secondary peaks/substructures in the reconstructed PSF when using a sharp interpolation kernel (Fig. \ref{fig:img_rec}, panels (d) and (e) } we followed an approach
somehow similar to the one adopted by \citep{chung21}, i.e., we
perform a deconvolution of the reconstructed image with a PSF. { The choose
  of a realistic PSF is very important when applying a deconvolution procedure. Therefore, a particular care was taken during this process. First, we assume that the
  natural seeing PSF in the focal plane of the 3.5m telescope at Calar Alto is well represented by a Gaussian function, following \citet{sanchez:2008}. Introducing more complex shapes, like a Moffat function, does not increases significantly the characterization of the shape of the PSF \citep[e.g.][]{dr2}. Considering the average seeing in the observatory \citep[$\sim$0.9$\arcsec$][]{sanchez07a}, and the turbulence introduced by the dome \citep[$\sim$10\% in the best case][]{sanchez:2008}, a Gaussian function with a width of $\sigma=$0.5$\arcsec$ (FWHM$\sim$1.3$\arcsec$) is indeed a good representation of the typical PSF at the entrance of the IFU. However, this if not by far a good representation of the PSF for the reconstructed images, which shape and structure are vastly dominated by the fiber-size, fiber-bundle pattern, dithering scheme and adopted image reconstruction procedure \citep[e.g.][]{sanchez06a}. Thus, to generate a realistic PSF we }
followed the same procedure adopted to generate the reconstructed
image, i.e., we extract the flux through the dithered-fiber apertures
and we adopted the same image reconstruction procedure and same
parameters. Then we deconvolve the image using the Richardson-Lucy
algorithm
\citep{Richardson:72}\footnote{\url{https://en.wikipedia.org/wiki/Richardson-Lucy_deconvolution}}
implemented in the {\sc restoration} module of the {\sc scikit} python
module. We should note that this procedure involves a further
optimization, as this algorithm requires two parameters. One defines a
threshold below which the values are set to zero to reduce the noise
(and ghost generation) in the deconvolved image. The other is the
number of iterations for which the algorithm is performed. A larger
number of iterations generates sharper images. However it produces,
again, a PSF with unreal substructures and ghosts due to noise
aggregation.  This is appreciated in Fig \ref{fig:img_rec}, panel $f$,
where it is shown the result of a deconvolution adopting a very low
threshold level corresponding to 1\% of the 1$\sigma$ noise level, and a large number of 50 iterations of the deconvolution algorithm. { This is an example of an over-deconvolution of the data,
  that introduces unreal sub-structures in the PSF, degrading the image quality.}

We followed a procedure similar to the one adopted to define the
optimal parameters of the image reconstruction kernel to estimate the
optimal parameters to adopt in the deconvolution procedure.  Thus, we
run the algorithm covering an ample range of thresholds (from 10 to
0.01$\sigma$) and number of iterations (from 1 to 100), evaluating
final FWHM of the reconstructed image and the standard deviation of
the residual ($\sigma_{res}$) when comparing with the original image.
To evaluate both the FWHM and $\sigma_{res}$ we convolve the original
image with a set of Gaussian functions of variable width and compare
them with the reconstructed images using a $\chi^2$ criterion. The
image that minimize this $\chi^2$ provides with the FWHM of the final
PSF (i.e., the quadratic propagation of the FWHM of the PSF of the
original image and the width of the convolved Gaussian function), and
the $\sigma_{res}$ (the standard deviation of the residual of the
difference between the reconstructed and the convolved images).
We find that a threshold corresponding to $\sim$2$\sigma$
the noise level, and just 5  iterations of the deconvolution algorithm provides
the best compromise between minimizing both the FWHM and $\sigma_{res}$.
As described before larger (lower) number of iterations decreases (increases) the FWHM, but $\sigma_{res}$ increases (decreases). { The final FWHM estimated for panel (f), is 1.65$\arcsec$, which is significant improvement with respect to the value prior to deconvolving the image ($\sim$8\% FWHM reduction). However, we remind the reader that most of the improvement in the resolution was a result of the selection of a new set of parameters for the interpolation kernel, as indicated before ($\sim$28\% FWHM reduction)}


As a final refinement, following \citet{dr2}, we generate a flat-field as the ratio between the convolved PS image { (i.e., PSF-matched)} and the reconstructed image from the deconvolved spectral data, what removes any final defect. { The application of this flat-field solve some minor photometric differences between the three dithering pointings and anchor the absolute photometry to the reference one (i.e. PS), being already discussed in detail in \citet{dr2} and \citet{sanchez16}.}
Fig. \ref{fig:img_rec} includes the final reconstructed image once applied the full procedure (panel $g$) and the residual
with respect to the original image (panel $h$).
The comparison between the reconstructed images corresponding to the
procedure adopted in the previous data-reduction
(Fig. \ref{fig:img_rec}, panel $c$), and the new procedure
(Fig. \ref{fig:img_rec}, panel $g$), illustrates the clear improvement of the new procedure.
Quantitatively, the reconstructed PSF has a FWHM of
$\sim$1.65$\arcsec$, for the archetype image that illustrates the
procedure, that has an original (natural seeing) PSF FWHM of
$\sim$1.4$\arcsec$. { We should note that this FWHM is not affected by the flat-fielding process in any way.} This is a significant improvement compared
to the FWHM of the PSF provided by the previous version of the image
reconstruction procedure ($\sim$2.4$\arcsec$). The major drawback of the adopted procedure, as indicated before, is that it relies on the quality of the reference image. Therefore, any defect of problem on that image is propagated to the final reconstructed one { due to the flat-fielding procedure}.

%
\begin{table}
\begin{center}
\caption{Description of the DRSCUBE file.}
\begin{tabular}{cllr}\hline\hline
HDU	&  EXTENSION & Dimensions$^1$ & Format \\
\hline
  0  & PRIMARY        &  (159, 151, 1877)    & float64   \\
  1  & ERROR           &   (159, 151, 1877)   & float32  \\
  2  & ERRWEIGHT   &   (159, 151, 1877)   & float32  \\
  3  & BADPIX          &   (159, 151, 1877)   & uint8      \\
  4  & FLAT              &   (159, 151)             & float64   \\
  5  & GAIA\_MASK  &   (159, 151)             & int64   \\
\hline
\end{tabular}\label{tab:drscube} 
\end{center}
(1) The actual spatial dimensions may change galaxy by galaxy, depending
on the real dither scheme adopted during the observations.
\end{table}

\subsubsection{Cube reconstruction}
\label{sec:cube_rec}

The new image reconstruction procedure described before is fully implemented in the
data reduction in the following way: (1) for each wavelength along the
sampled range it is generated a reconstructed image using the
corresponding monochromatic flux intensities sampled by dithered
fibers, taking into account the offsets introduced by the DAR, and
using the interpolation kernel defined in Eq. \ref{eq:w_ijk} and the optimal values for the $\sigma$ and $\alpha$ parameters obtained
from the experiment described before; (2) the resulting image is
deconvolved using a PSF generated using the same reconstruction
algorithm; (3) along this process and error
propagation is performed adopting a Monte-Carlo iteration with 100
loops, perturbing the flux intensities measured through the fibers and
repeating steps (1) and (2).  The mean image from this MC iteration
and the corresponding standard deviation are stored, being adopted as
the reconstructed image and error at the considered wavelength; (4)
once this procedure is iterated over all the wavelength range a datacube is
obtained. From this datacube we generate the $g-$ and
$r-$band images, by convolving the corresponding filter curves with
the spectra at each spaxel.  Those images are used to derive the final
flat-field and FWHM, following the procedure described in the previous subsections (i.e., they are compared based on a $\chi^2$ with a set of
Gaussian convolved PS $g-$ and $r-$band images). { The procedure to estimate the FWHM was introduced and described in detail in \citet{dr2}, and adopted in further IFS studies \citep[e.g.][]{law15,renbin16,sanchez16}. This methodology is robust, as it was demonstrated in \citet{dr2} and \citet{sanchez16} by comparing its results with those obtained by measuring the FWHM directly on cube-reconstructed images corresponding to calibration stars.}
We  adopted
the $r-$band derived flat-field for all the wavelengths, as we find
not significant  difference in using this one, the $g-$band one or a combination
of both. We should highlight that the flat-field is purely monochromatic,
and it is not fine tuned for each different wavelength. { We note again that this flat-field was already introduced as part of the reduction of the CALIFA data in \citet{dr2}, being applied in the previous version of the data-reduction \citep{sanchez16}.}.

\subsubsection{Foreground stars masking}
\label{sec:fg_stars}

Following the procedure
described in \citet{sanchez22}, we search for the possible foreground
field stars in the FoV of the analyzed datacubes. We use the Gaia DR3
catalog\footnote{https://www.cosmos.esa.int/web/gaia/dr3}
\citep{gaia1,gaia3}, that comprises the most complete and accurate
survey of stars covering the full sky, at least regarding its
astrometric solution. We selected only those sources with a measured
parallax at least five times higher the reported uncertainty, to
ensure that they are Galactic stars. Then, we generate a mask covering
a circular aperture of 2.5$\arcsec$ around each star with the spatial size
format and astrometric solution of the reduced datacube. This mask
will be used later on when analyzing the data. We found a foreground star in $\sim$25\% of the datacubes.

\subsubsection{Format of the reduced data}
\label{sec:drscube}

For a more simple distribution, the final outcome of the reduction is stored in a single file for each
observed object. We adopted a format
similar to the one implemented in the CALIFA DR3 \citep{sanchez16},
and other IFS Galaxy Survey released \citep[e.g., MaNGA
format][]{DR17}. It consists on a multi-extension FITSFILE, in which
each extension comprises a particular outcome of the reduction.  Table
\ref{tab:drscube} lists each of the extensions in the file, their format and dimensions.  The first four extensions consists of a set of
datacubes. The first one ({\sc PRIMARY}) comprise the flux intensity
spectra corresponding to each spaxel, with the corresponding error
stored in the second extension ({\sc ERROR}). The required weights to
include in error propagation due the different coverage of the
dither pattern (i.e., due to the fact that each spaxel is covered by a
different number of fibers), as described in Sec. \ref{sec:data}, is
included in the third extension ({\sc ERRWEIGHT}). The fourth extension ({\sc BADPIX})
store the bad-pixel mask. The two final extensions store two images, the flat-fielding
described in Sec. \ref{sec:cube_rec}, and the mask of the foreground stars described
in Sec. \ref{sec:fg_stars}.

These files are named after the original object name included in the
PMAS software during the observations, adopting the nomenclature {\sc
  CUBENAME.V500.drscube.fits.gz}, where {\sc CUBENAME} corresponds to
the {\sc OBJECT} header in the original FITSFILE. We distribute these
datacubes in the following web page:
\url{http://ifs.astroscu.unam.mx/CALIFA/V500/v2.3/reduced/}, comprising a total of 1116 datacubes.

%
\begin{table}
\begin{center}
\caption{Description of the QC file.}
\begin{tabular}{cll}\hline\hline
column &  value &  meaning\\
  \hline
Name        &  cubename   & Prefix of the reduced datacube\\
  \hline
  QC\_flag  &  0  & OK:   Everything seems to be ok in the data.\\
       &  1  & BAD: The data are bad for a reason indicated\\
       &    & below. They should not be used.\\
       &  2  & WARNING: The data present some problems,\\
       &      & but they can be used for most science cases.\\
    \hline
  Reason   &  0  & No QC issue found.\\
       &  1  & The targets galaxy is considerably larger than the \\
       &      & FoV of the instrument.\\
       &  2  & Severe problems with the spectra, very low-S/N, \\
       &      & problems with the sky-subtraction.\\
       &  3  & Non severe problems with the spectra, affecting \\
       &      & just a small region within the FoV.\\
       &  4  & The galaxy is too small compared to the FoV, it is\\
       &      & essentially unresolved. \\
       &  5  & Repeated observation of the same galaxy. \\
       &  6  & FoV crowded with field stars affecting a significant\\
       &      & fraction of the FoV.\\
       &  7  & A single very bright field star is affecting a \\
       &      & significant fraction of the FoV.\\
       &  8  & Evident problem in the absolute spectro-\\
       &      & photometric calibration.\\
       &  9  & Problem with the image reconstruction \\
       &      & procedure or defects in the PS $g-$band image. \\
       &10  & Spectroscopic and photometric derived stellar \\
       &      & masses do not match.\\
       &11  & The stellar population analysis does not provide\\
       &      & a reliable result.\\
    \hline
 Multiple    & 0   & Only one galaxy detected in the FoV.\\ 
                 & 1    & Multiple galaxies detected in the FoV.\\
\hline
\end{tabular}\label{tab:qc} 
\end{center}
\end{table}

\subsection{Galaxies segmentation and structural properties}
\label{sec:gal_seg}

In some cases more than one galaxy is observed within the FoV of the
instrument. As one of the goals of this exploration is to provide with
a final galaxy sample, including its individual observational (and
physical) properties, we have performed an identification of the
objects in the field, segregating them, to treat them separately in
further analysis. For doing so, we make use of the routines included in
the {\sc photutils} python package.  We obtain from the PS survey a
square 10$\arcmin$ size $g-$band image centred in location of each
observed cube. On this image we detect all the objects detected about
2$\sigma$ the noise level, and based on an analysis of the ellipticity
and the size, we select all the stars in the field. This allow us to:
(i) estimate the FWHM of the $g-$band image, a required input of the
reduction as indicated before; (ii) detect small objects or faint
stars not included in the Gaia survey, that should be masked from the
data too; (iii) estimate the local background and noise level around
the science target in this image. Then, we select a central square
image of 1.5$\arcmin$ size, mask the field stars, remove the
background and perform an isophotal analysis of each galaxy detected
within the FoV of the original IFS data.  This isophotal analysis
comprises the derivation of the surface-brightness, ellipticity and
position angle profiles, together with the estimation of the
baricenter and effective radius. We selected as representative position angle and
ellipticity the one corresponding to a $g$ surface brightness of 22 mag arcsec$^{-2}$.
Then, the effective radius is found by deriving the
cumulative flux in successive radial apertures along the semi-major in elliptical apertures
following the previous estimated position angle and ellipticity.
Finally a new segmentation map is
generated with the same spatial size and astrometric solution of the
reduced datacube.

When more than one galaxy is detected in the FoV of a single datacube
the mask generated by the previous procedure is used to segregate
them. This way, a set of copies of the reduced datacube is created,
each one corresponding to each of the different detected galaxies,
masking the rest of them. The new files adopt as prefix the parent
{\sc CUBENAME} (Sec. \ref{sec:drscube}), with a running index
corresponding to each galaxy (i.e., {\sc CUBENAME\_I}, where {\sc I}
runs from 0 to the number of galaxies).  An index corresponding to the
ID in the original segregation mask is stored in the header in the
{\sc LABEL} keyword. Finally, the baricenter of each object and the
coordinates in the sky are stored in the {\sc XC, YC, RA} and {\sc
  DEC} header keywords. On the other hand, if only one galaxy is
detected in the FoV, it is created just a copy of the data cubes including
the corresponding header keywords.

Finally, the stars detected within the field-of-view by this analysis
and those included in the {\sc GAIA\_MASK} extension of the reduced
data are masked from the datacubes, interpolating the flux intensities within
each of those masks. This interpolation is just adopted for future explorations
of extensive quantities (e.g., the total stellar mass). For spatial resolved analysis
we recommend to mask those regions and exclude them.

We found 22 datacubes with multiple galaxies detected within the FoV, in most of the cases two galaxies (only in 4 cases we found 3 galaxies).
For completeness we distribute those (i) galaxy-segmented and (ii)
field-stars mask interpolated FITSFILES, using the same format
described in the previous section (Sec. \ref{sec:drscube}), through the webpage
\url{http://ifs.astroscu.unam.mx/CALIFA/V500/v2.3/reduced_masked/}. 

\begin{figure*}
 \minipage{0.99\textwidth}
 \includegraphics[width=8cm]{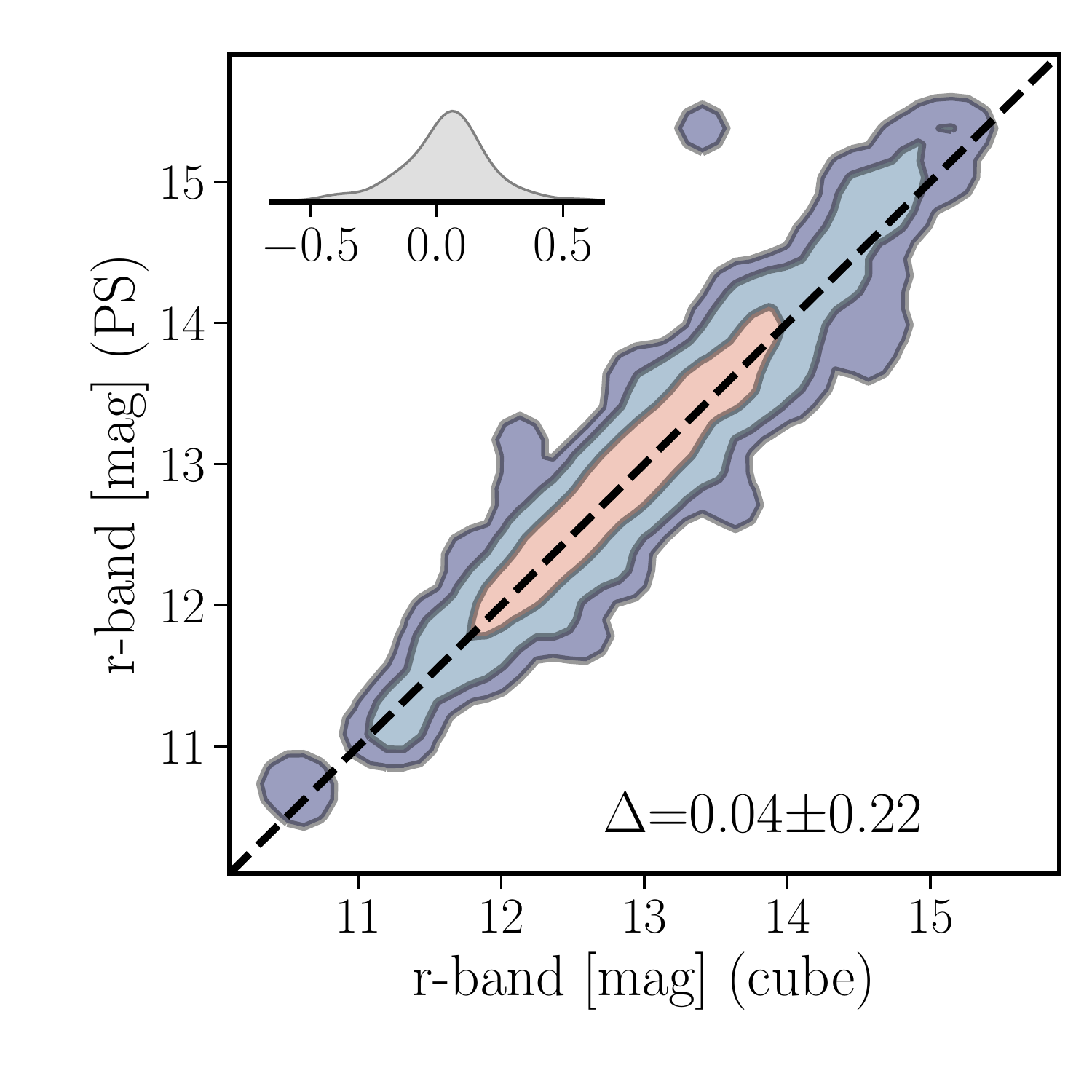}\includegraphics[width=8cm]{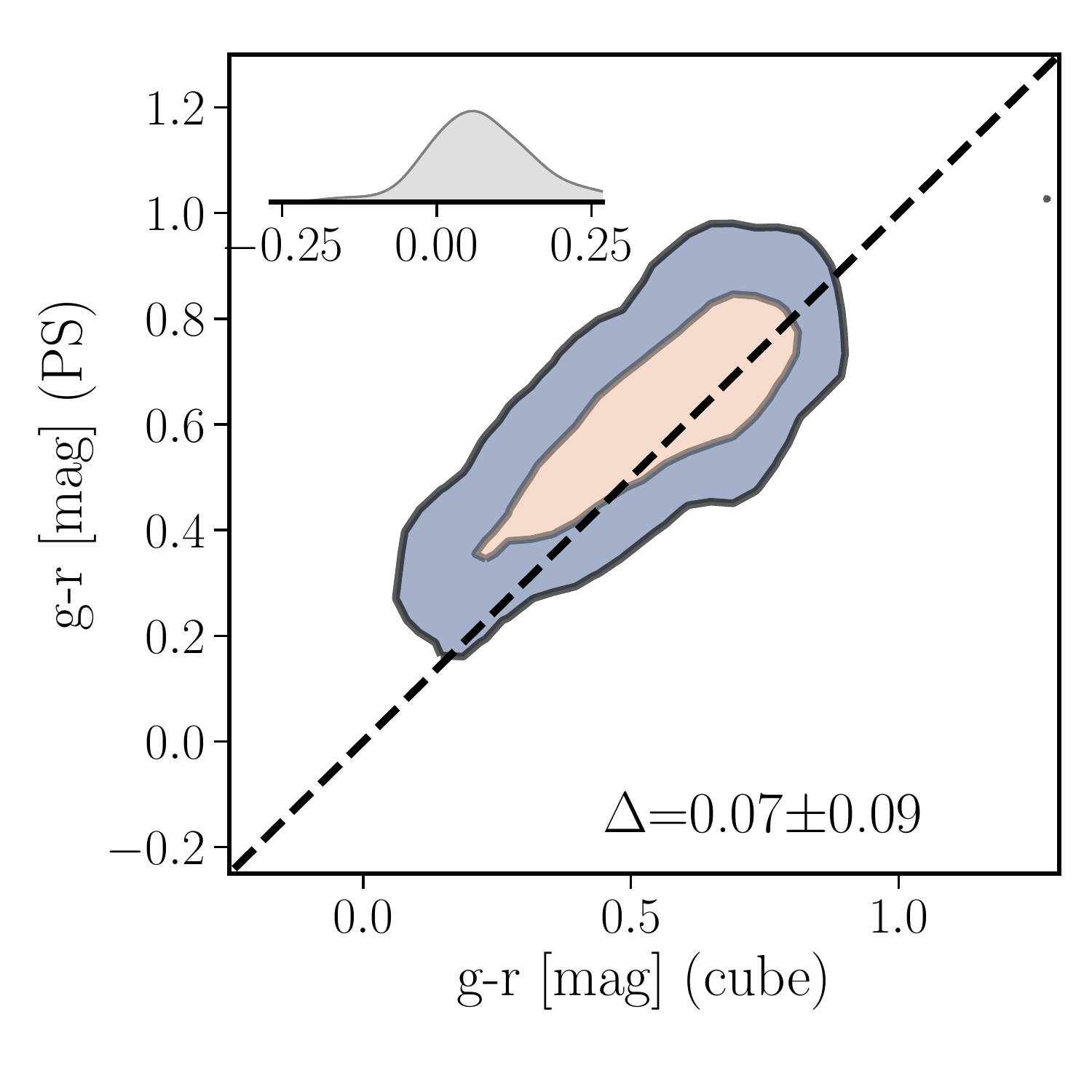}
 \endminipage
 \caption{Comparison between the $r$-band magnitudes (left panel) and $g-r$ colors (right panel) derived for the final sample of good quality datacubes using the segmentation maps described in Sec. \ref{sec:gal_seg} for the PS images and the images synthesized from the IFS datacubes. Each successive contour corresponds to a density distribution containing a 99\%, 95\% and 65\% of the total sample. The inset shows the kernel density distribution of the difference between both quantities, with the average (and standard deviation) value indicated as a legend.}
 \label{fig:qc_phot}
\end{figure*}

\subsection{Data Quality}
\label{sec:qc}

A quality control procedure is implemented to determine which data are
fully useful, flagging out the poor quality ones and indicating with a
warning those data that are still useful for most of the science cases
but may be taken with care for particular explorations. The quality of
data is weighted based on a set of simple visual, qualitative and
quantitative analysis: (i) first, we flag as bad those datacubes that
do not sample the center of the galaxies, covering just a small
portion of its optical extension (i.e., FoV$<<$Re), and
datacubes corresponding to galaxies too small compared to the original
fiber size (Re$\lesssim$5.0$\arcsec$), i.e., unresolved
targets; (ii) a visual inspection of the central aperture
(1.5$\arcsec$/diameter) and the FoV-integrated spectra and the
synthesized $u$-, $g-$ and $r-$band images allows to identify evident
bad quality data, that usually corresponds to very low-S/N data and/or
problems with the observations that are also evident in the QC-flags
included in version 2.2 of the data reduction \citep[strong
straylight, electric background, problems with the tracing and
extraction of the spectra, problems with the sky subtraction, as
described in ][]{sanchez16}. When those problems are observed, but
they affect just a few locations within the FoV or a limited spectral
range, and the central and integrated spectra does not show evident defects,
then the datacubes are flagged with a warning; (iii) repeated
observations on the same field are compared, the best quality one is
selected and the worst one is flagged as bad, to avoid duplication in
the final galaxy sample; (iv) the presence of a large number of
field-stars or a single bright star, contaminating a substantial
fraction of the FoV, is considered a reason to flag a datacube as
either bad or with a warning (depending on the relative importance of
the contamination); (v) when there is a clear absolute or relative
(blue-to-red) photometric mismatch between the datacubes and the PS
observations is used to flag the quality of the data; (vi) a visual inspection of the
reconstructed $g-$band image and its comparison the corresponding PS
image (e.g., Fig. \ref{fig:img_rec}, panels $g$ and $h$), is used to
identify clear problems (e.g., errors in the estimated dithering
pattern, image reconstruction algorithm, astrometric issues); finally
we perform two additional tests following \citep{sanchez22}, (vii) we
compare the stellar masses derived using the photometric and
spectrophotometric information, and (viii) we fit the central and
integrated spectra using {\sc pyFIT3D} to identify possible
issues/problems not evident based on a pure visual inspection. We will
provide more details on these two particular analysis in the
forthcoming sections.

It is possible that one cube is affected by more than one quality issues. In this
case the table reflects that issue with the strongest/worst impact in the quality of the data.

Of the total number of 1116 datacubes, 895 has been classified as good quality. Of them,
392 present some warning (i.e., a minor issue as described before), and 490 passed all
the explored quality criteria. A total of 234 cubes were rejected. This may be considered
as considerable large fraction ($\sim$20\% of the observed cubes). However, the vast
majority are rejected not due to its intrinsic bad quality, but due to the fact that
they correspond to observations covering just a small fraction of the targeted object,
or because the target object is too small for a proper spatial resolved analysis.
We provide with some characteristic of the good quality dataset.

\subsubsection{Spectrophotometric accuracy and precision}

We gauge the accuracy and precision of our spectrophotometric calibration by comparing the $r$-band magnitudes and $g-r$ colors derived from the datacubes with those extracted from the PS images. For this calculation we use the segmentation maps described in Sec. \ref{sec:gal_seg}, co-adding the fluxes within each segmentation corresponding to each detected galaxy in the PS images corresponding to the $g$- and $r$-band, and the same image bands synthetized from the reduced datacubes. Figure \ref{fig:qc_phot} shows the comparison between $r$-band magnitudes and $g-r$ colors for both datasets. There is a remarkable good agreement, with a distribution almost following the one-to-one relation, and with an average offset of $-$0.06 mag for the magnitude and $-$0.07 mag for the color. These offsets corresponds to just a 2-3\% mismatch in the absolute spectrophotometric and blue-to-red calibration between our reduced datacubes and the PS photometry. The difference between the $r$-band magnitudes present a standard deviation of 0.22 mag, that corresponds to a precision in the absolute photometric calibration of  $\sim$9\%. On the other hand, we estimated the blue-to-red spectrophotometric precision, based on the dispersion in the difference between the $g-r$ colors ($\sim$0.09 mag), in $\sim$4\%. Those numbers are similar to the ones reported in previous DR of the CALIFA survey \citep[e.g.][]{sanchez16} and of the same order and those found by other IFS surveys \citep[e.g.][]{DR17}. { As discussed on those articles there are multiple reasons for these differences, but the dominant one is the slight differences in the nominal transmission curve of the filters adopted to estimate the photometric values from the databucubes \citep{gunn98} and the real transmission curves of the PanStarrs observations \citep{stub10}. Those curves result from the convolution of the nominal curves with the optical system conformed by the telescope and instrument, that are difficult to reproduce.}

\subsubsection{Depth and signal-to-noise}

\begin{figure}
 \minipage{0.99\textwidth}
 \includegraphics[width=8cm]{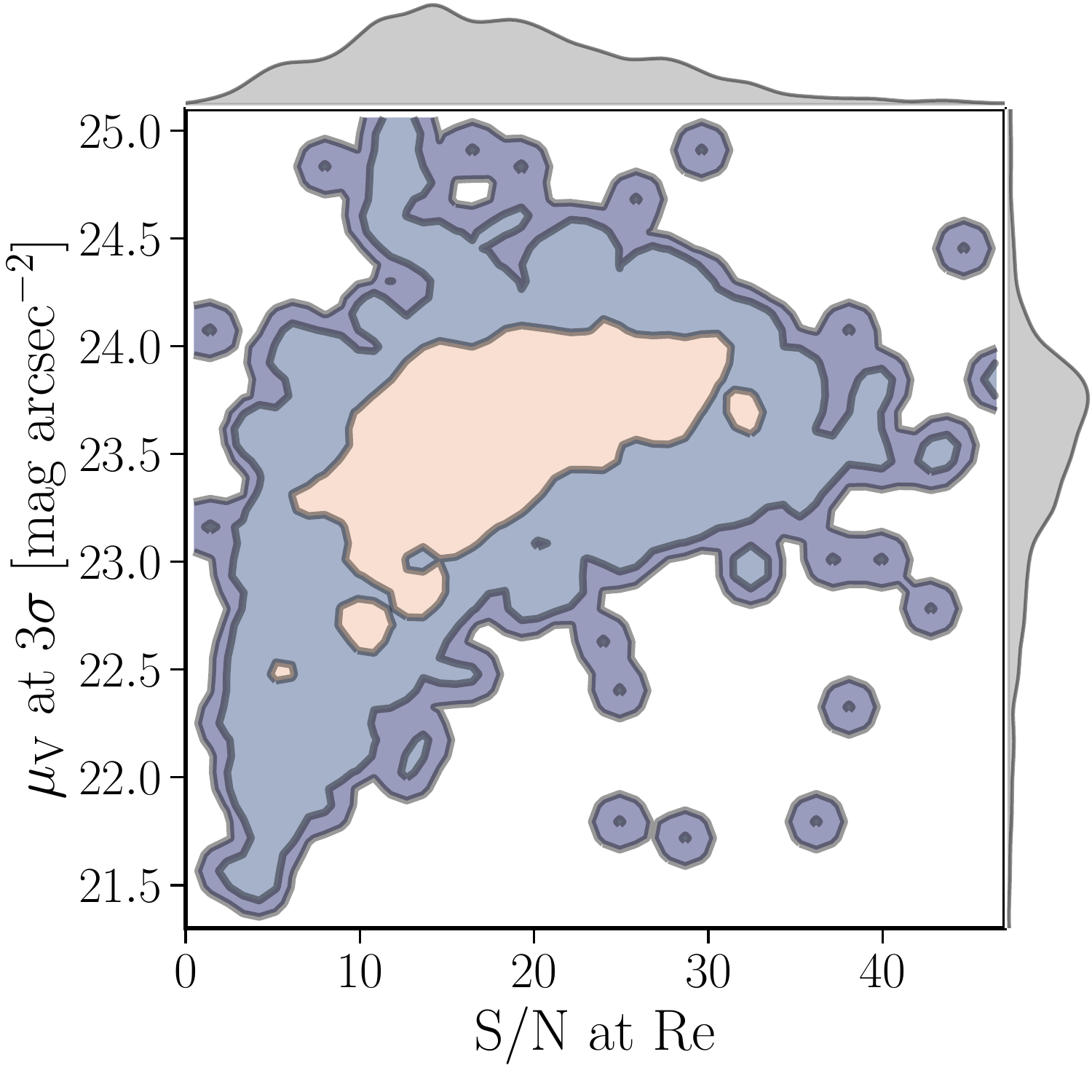}
 \endminipage
 \caption{Surface-brightness at 3$\sigma$ detection limit as a function of the S/N at the effective radius in the final good quality datacubes. Each successive contour corresponds to a density distribution containing a 99\%, 95\% and 65\% of the total sample. The kernel density distribution of both quantities is included in the top and right subpanels.}
 \label{fig:qc_SN}
\end{figure}

The depth of the datacubes is gauged by estimating (i) the $V$-band surface brightness magnitude (and flux intensity) at the 3$\sigma$ level for those datacubes fulfilling our good quality criteria in which this limit is reached, and (ii) the typical S/N at the effective radius. Figure \ref{fig:qc_SN} shows the distribution of both quantities, showing a wide range of values, from S/N from 10 to 40, and with $\mu_{\rm V}$ from 21.7 to 24.5 mag arcsec$^{-2}$, with a typical value of S/N$\sim$20 and $\mu_{\rm V}\sim$23.6 mag arcsec$^{-2}$. This later value corresponds to a flux surface intensity of $\sim$1 10$^{-18}$ \fluxSB . There is a clear trend between both quantities, with datacubes reaching a lower surface-brightness magnitude as higher is the S/N at Re, following a non-linear relation reaching a somehow asymptotic value at high-S/N. These distributions, trends, and typical values are indeed very similar to the values already reported in previous CALIFA data-releases \citep[e.g.][]{sanchez16}.

\subsubsection{Seeing and spatial resolution}

\begin{figure}
 \minipage{0.99\textwidth}
 \includegraphics[width=8cm]{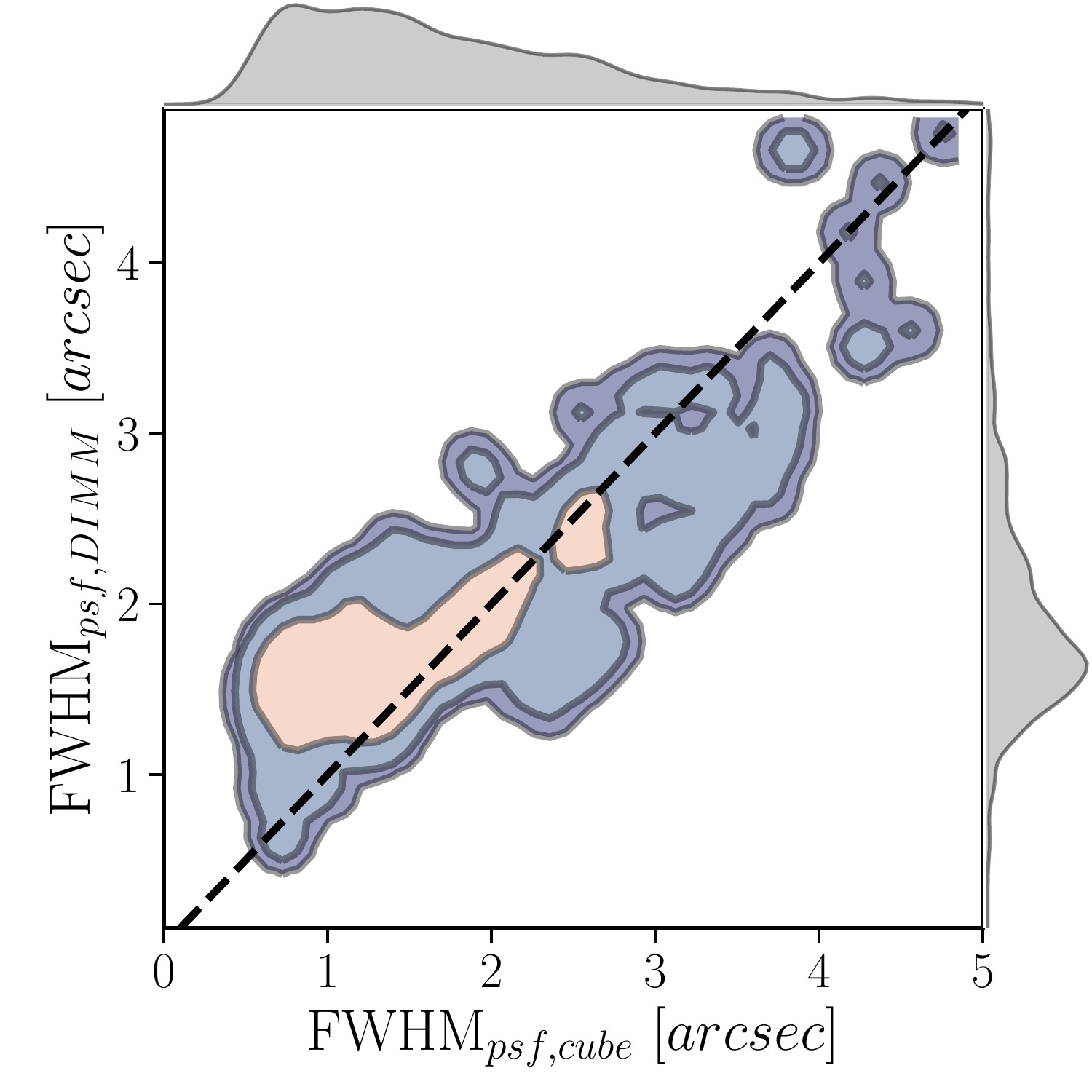}
 \endminipage
 \caption{Comparison between the FWHM of the natural seeing measured by the Calar Alto DIMM within a one hour range the observation and the FWHM of the PSF estimated on each cube as part of the new reduction scheme, as described in the text. Each successive contour corresponds to a density distribution containing a 99\%, 95\% and 65\% of the total sample.  The kernel density distribution of both quantities is included in the top and right subpanels.}
 \label{fig:qc_seeing}
\end{figure}

One of the main goals of the current re-reduction of the data is to achieve a better spatial resolution. Along the reduction the FWHM of the PSF is estimated (Sec. \ref{sec:img_rec}). This estimation is not very precise, due to the nature of the computation. 
Indeed in many cases the FWHM 
estimated by the reduction procedure is poorly constrained (i.e., a wide range of values provide similar likelihoods)\footnote{Note that this does not affect the reduction itself significantly, as the estimated FWHM is applied to derivethe last flat-fielding described in Sec. \ref{sec:img_rec}, that introduces minor changes}. However, for most of the galaxies is the only estimation we have. This value can be compared with the natural seeing FWHM provided by the Calar Alto seeing monitor (RoboDIMM). For doing so we downloaded all the available values provided by this monitor and cross-match the observing date/time for the three different pointings of each datacube with the date/time included in the seeing monitor catalog. We allow for a range of 20 minutes difference between both times, considering the exposure time and overheads of each observation. We finally obtain the average seeing FWHM reported by the monitor within in this range. The RoboDIMM is not always active when observations are taking place, being active of about 2/3 of the observing periods. We end-up with a subset of 372 estimations of the FWHM for both the datacube PSF and the seeing monitor. Figure \ref{fig:qc_seeing} shows the comparison between both quantities, showing a clear correspondence between both of them, nearly following a one-to-one relation. This is very encouraging as it shows that our estimation of the FWHM PSF is not off by a significant quantity. { However, the trend present an offset at low values, with the FWHM PSF showing lower values than the natural seeing one below 1.0 $\arcsec$. We consider that this is not realistic and most probably our procedure to estimate the FWHM of the PSF in the datacubes is not reliable for values near to the spaxel size (0.5$\arcsec$)  due to the limitations of sampling in the estimation of the FWHM. An additional problem is that it is not feasible to estimate the FWHM in the cubes below the PSF size of the reference PS image that we use to compare with (i.e., $\sim$1.1$\arcsec$ in average). To estimate the accuracy of our estimations of the FWHMs based on the PSF-matching described in Sec. \ref{sec:img_rec}, we followed  \citet{dr2} and \citet{sanchez16}, and explored the FWHM in the field-stars. For doing so we selected the foreground stars included in the Gaia DR3 adopted for the masking described in Sec. \ref{sec:fg_stars}. We applied an additional cut, selecting only those stars brighter than $g<$18 mag, and at a distance larger than 15$\arcsec$ from the center of the galaxies to minimize the possible contamination from this source. We end up with 91 field stars. We extracted a post-stamp image of $\sim$10$\arcsec$$\times$10$\arcsec$ size from the new $g-$band image reconstructed from the new (v2.3) and old (v2.2) datacubes, deriving their FWHM by fitting a Gaussian function. The average FWHM derived from field-stars is 1.43$\pm$0.40$\arcsec$ (2.62$\arcsec$$\pm$0.58$\arcsec$) that agrees with the value derived based on the PSF-matching procedure, 1.51$\arcsec$$\pm$0.32$\arcsec$ (2.64$\arcsec$$\pm$0.56$\arcsec$) for the v2.3 (v2.2) reduced data. Once determined the accuracy of our procedure it is evident that Fig. \ref{fig:qc_seeing}, illustrates clearly} the improvement in the spatial resolution of the data introduced by the new reduction. Even considering that we may be underestimating the FWHM of the PSF at low values, we can claim that the new datacubes have a typical spatial resolution of $\sim$1.0-1.5$\arcsec$ in most of the cases.

\begin{figure*}
 \minipage{0.99\textwidth}
 \includegraphics[width=18cm]{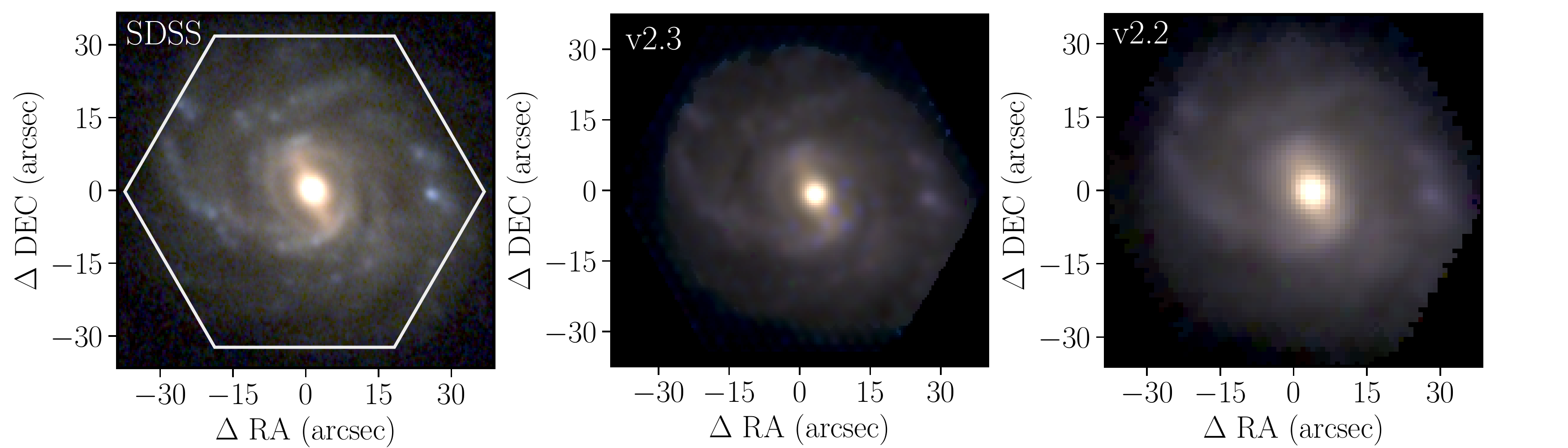}
 \includegraphics[width=18cm]{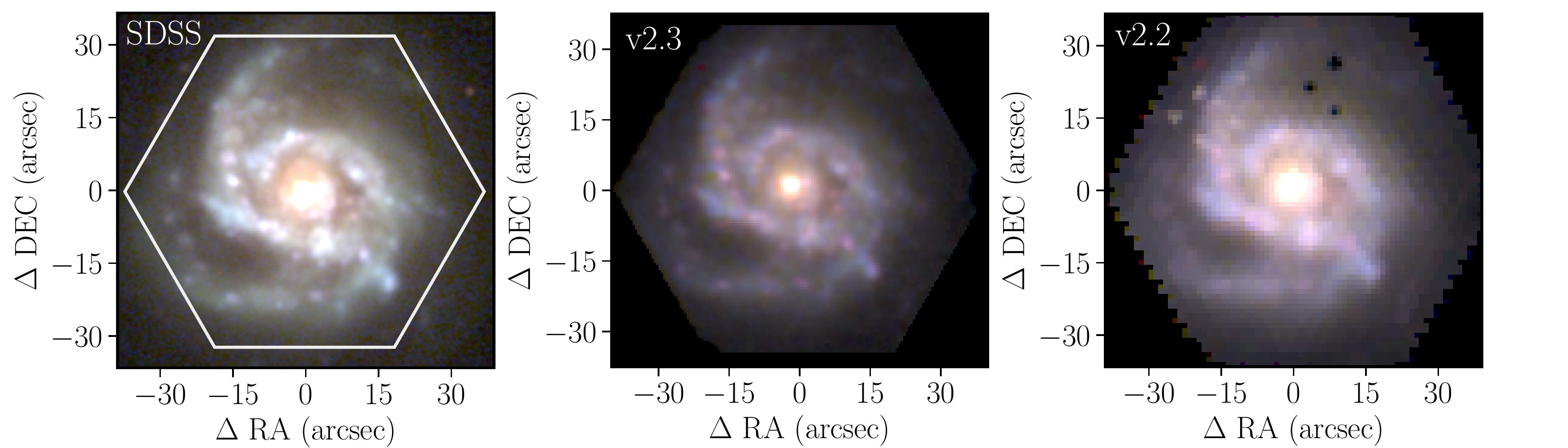}
  \includegraphics[width=18cm]{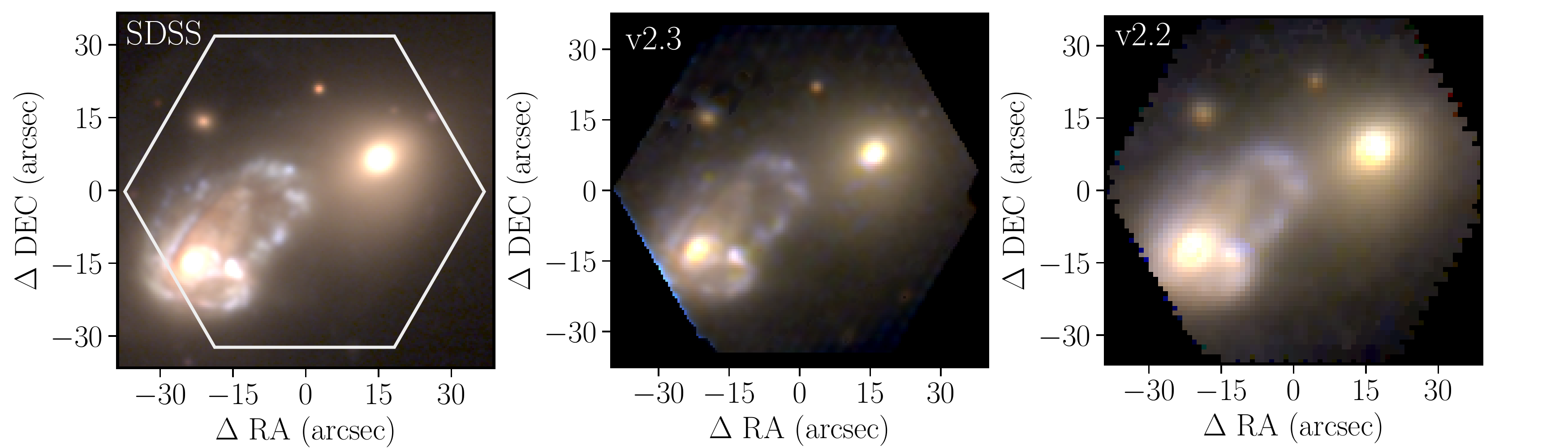}
 \endminipage
 \caption{Selected examples to illustrate the improvements in the image quality introduced by the new reduction procedure for three different galaxies: (i) the archetype galaxy shown along this study, NGC 5947 (top panels); (ii) a face-on spiral image with clear issues in the cube reconstruction in the previous reduction of the data, NGC 5936 (middle panels); and (iii) a merging system, ARP 118, observed during a period of extremely good seeing, $\sim$0.5$\arcsec$ (bottom panels). From left to right each panel shows the true color image generated using the $u-$, $g-$ and $r-$band images extracted from the SDSS survey (left panel), synthetized from the datacubes corresponding to the new reduction (central panel) and produced by the previous version of the data reduction (right panel). Note that there are some color differences due to the truncation of the $u-$ (and $r-$-band) synthetized image introduced by the wavelength range covered by the IFS data. Despite of them the improvement in the image quality is clearly appreciated in the central panels. Similar plots for the full dataset are included in the distribution webpage: \url{http://ifs.astroscu.unam.mx/CALIFA/V500/v2.3/}}
 \label{fig:rgb}
\end{figure*}

A more qualitative method to gauge the improvement in the image
quality and in particular in the resolution achieved by the new
reduction is to compare the images obtained by photometric observations with those synthesized from datacubes provided by the new
and the previous data reductions. Figure \ref{fig:rgb} shows this
comparison for three selected datacubes, corresponding to galaxies NGC
5947, NGC 5936 and ARP 118. It comprises the true color images created
using the $u-$, $g-$ and $r-$band images retrieved from the SDSS
survey and synthesized from the databuces of the current and the
previous reduction. A simple visual inspection of the three images demonstrates that
the resolution achieved by the new reduction is of the same order of the
one achieved by a photometric observation (i.e., natural seeing), with an evident improvement
with respect to the previous version. We should note that, as shown in Fig. \ref{fig:qc_seeing},
the final resolution depends on the seeing during the observation, and therefore
not all datacubes present the same image quality. Beside the improvement
in the resolution, other improvements introduced by the new reduction, including
the re-evaluation of the real location of the dithered pointings and the masking
and interpolation of the broken fibers, are also appreciated in Fig. \ref{fig:rgb} (e.g., in the
case of images corresponding to NGC 5936).

\section{Galaxy Sample}
\label{sec:sample}

We describe in this section the main properties of the compiled sample,
illustrating that indeed it can be used as representative sample
of the galaxies in its redshift footprint, despite of its heterogeneous nature.
The original CALIFA sample was primarily selected based on the
diameter of the galaxies in the sky \citep{walcher14}. This selection
was adopted to match their optical extension to the FoV of the PPAK
science bundle ($\sim$70$\arcsec$). Additional restrictions were
introduced in that selection to focus the sample in the nearby
Universe ($z\sim$0.015), but far enough to be out of the local
cosmological bubble dominated by the cosmic variance ($>$20 Mpc).
Finally a cut in the apparent magnitude of the targets was introduced
to explicitly exclude dwarf galaxies that would in other case dominate
the number statistics. This essentially defined the CALIFA mother
sample \citep[MS][]{califa,walcher14}.  These criteria were relaxed to
introduce galaxy types underrepresented in the MS, such as cluster
members, large ellipticals, dwarf galaxies, companions of galaxies in
interacting systems covered by the survey, and supernova host galaxies
\citep[compiled by the PISCO, PMAS/PPak Integral-field Supernova Hosts
Compilation survey][]{pisco}. All those galaxies comprise the CALIFA
extended sample (ES), described in 3rd data release
\citep{sanchez16}. In this previous data release it was distributed
646 datacubes corresponding to individual galaxies observed using the
V500 configuration, 529 of them extracted from the MS and 117 to the
ES. The current eCALIFA compilation comprises 895 individual galaxies, 519
corresponding to the original CALIFA MS and 376 more to the ES.

\begin{figure}
 \minipage{0.99\textwidth}
 \includegraphics[width=9.5cm]{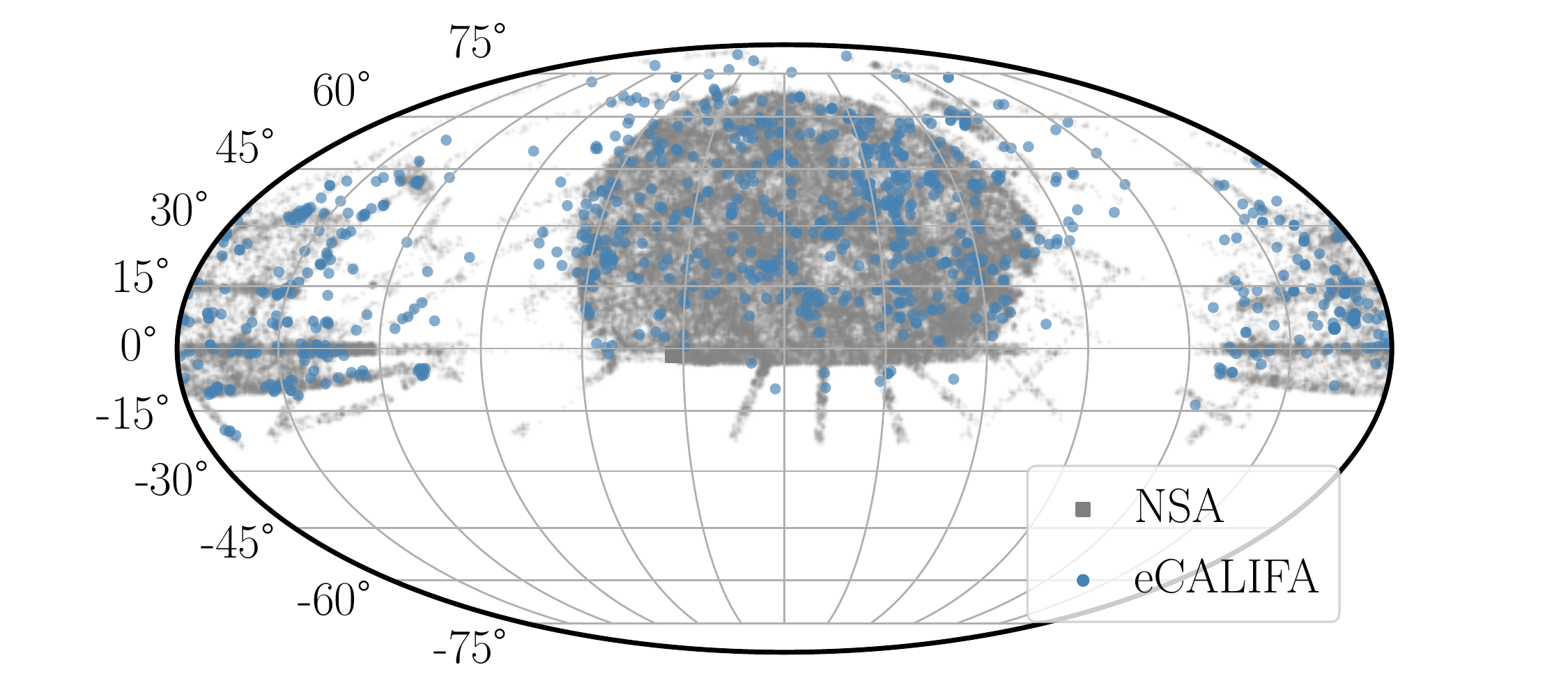}
 \endminipage
 \caption{Distribution of the eCALIFA galaxies in the sky (blue-solid circles) together with the distribution fo the SDSS-NSA galaxies (grey-solid squares) at a similar redshift range. The galaxies from the current IFS compilation are essentially distributed in the north hemisphere, with a large fraction in them in the same sky foot-print as the SDSS galaxies. Comparing both distributions we appreciate that the eCALIFA galaxies sample objects in different densities.}
 \label{fig:sky}
\end{figure}

Figure \ref{fig:sky} shows the distribution of galaxies from the
eCALIFA sample in the sky compared to that of the SDSS galaxies
extracted from the NASA Sloan Atlas
\citep[SDSS-NSA][]{blanton+2017}. The SDSS-NSA is one of the largest
and more complete catalog of galaxy parameters in the local universe,
comprising $\sim$150,000 objects. Being a volume/apparent magnitude
selected sample it can be volume corrected and it is frequently
used to explore the general properties of galaxy population at low$-z$
\citep[e.g.][]{sanchez18b}. Indeed, the full redshift range covered by the eCALIFA
sample (0.0005$<z<$0.08) match pretty well with the one covered by the
SDSS-NSA catalog (99.8\% of the galaxies included in this redshift
range). We should note that the redshift distribution of the eCALIFA sample
is not homogeneous, with a better coverage below $z<0.03$,
where 94\%\ of the objects is found. Fig \ref{fig:sky} shows that the
imprint of the eCALIFA sample in the sky is slightly wider than that
of the SDSS-NSA, being both restricted essentially to the north
hemisphere. For the region in common it is evident that the final
sample cover a wide range of galaxy densities in the sky, traced by
the darkest regions in the SDSS-NSA distribution.

Like in the case of the MS sample, the eCALIFA sample was primarily
selected by diameter, traced by the effective radius. We selected
objects which most of its optical extension is covered by the FoV of
the IFS data and they are well resolved by the observations. Indeed, 90\%\ of the sample is restricted to an Re between
4$\arcsec$ (Re$>$fiber-size) and 24$\arcsec$ (FoV$>$2 Re).  In
addition, we do not cover very faint objects. There is no galaxy
fainter than $r>18.5$ mag in the sample, and $\sim$95\%\ of the
objects are restricted to a $r-$band magnitude range between 11 and 15
mag. If this compilation is representative of the population in the
nearby universe it must have similar properties of that of a sample
selected using similar cuts in diameter and magnitude.  To explore
this we select a sub-sample of the SDSS-NSA catalog imposing an
effective radius (4$\arcsec<$Re$<$24$\arcsec$) and $r-$band magnitude
cut ($r<$18 mag), comprising $\sim$45,000 galaxies (NSA$|$Re$_{lim}$
sample hereafter).

\begin{figure}
 \minipage{0.99\textwidth}
 \includegraphics[width=8.5cm]{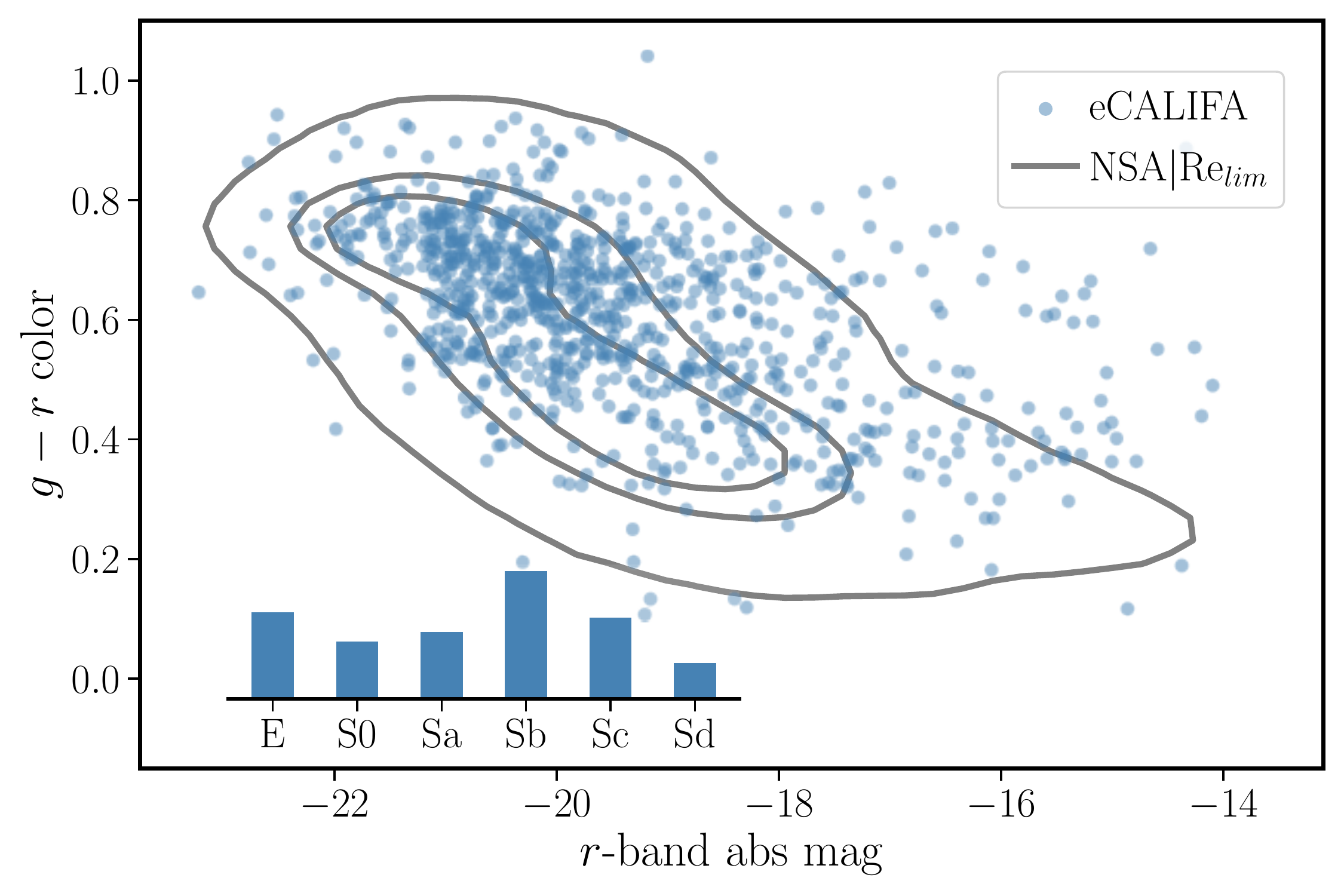}
 \endminipage
 \caption{Distribution in the $g-r$ vs. $r-$band absolute magnitude diagram of the eCALIFA galaxies (blue-solid circles), compared to the same distribution for sub-sample of SDSS-NSA galaxies selected using the same diameter, magnitude and redshift range as the eCALIFA compilation (grey contours). Each successive contour encircles a 95\%, 65\% and 40\% of the SDSS-NSA galaxies. The bottom-left inset shows the morphological distribution of the eCALIFA galaxies.}
 \label{fig:CMD}
\end{figure}

Figure \ref{fig:CMD} shows the distribution of  $g-r$ color as a function of the $r$-band
absolute magnitude (i.e.,  the color-magnitude diagram, CMD) for the eCALIFA sample
compared to the same distribution for the NSA$|$Re$_{lim}$ sub-sample. Like in the case of the
CALIFA sample the new compilation has a pretty good coverage of the
CMD, with a reasonable sampling of galaxies in the red-sequence
(mostly early-type and massive), the blue cloud (mostly late-type and
less massive), and the green-valley (intermediate morphological type
and mass ones). There is a rough agreement between the distributions
traced by both samples, despite the fact that a KS-test suggest that
there is significant difference between them. Actually, a 92\%\ of the
eCALIFA galaxies are indeed encircled by the lowest density contour
tracing the distribution of the NSA$|$Re$_{lim}$ sub-sample
(comprising to 95\%\ of these objects).  Some evident differences are
(i) the tail towards redder galaxies in the fainter regime of the
distribution, all
out of the 95\%\ contour. These galaxies belongs to a low-luminosity sub-set of SNIa hosts included in the PISCO sample; and (ii) the less defined green-valley
appreciated in the eCALIFA sample. This later difference may indicate
an excess of intermediate type galaxies, something that has been
already noticed for the CALIFA sample \citep[e.g.][]{lacerda20}. Indeed, the morphological
distribution, shown in the inset of Fig. \ref{fig:CMD} illustrate both
the wide coverage of morphological types of the sample, and the clear
peak in Sb (early spiral) galaxies.

\begin{figure*}
 \minipage{0.99\textwidth}
 \includegraphics[width=18cm]{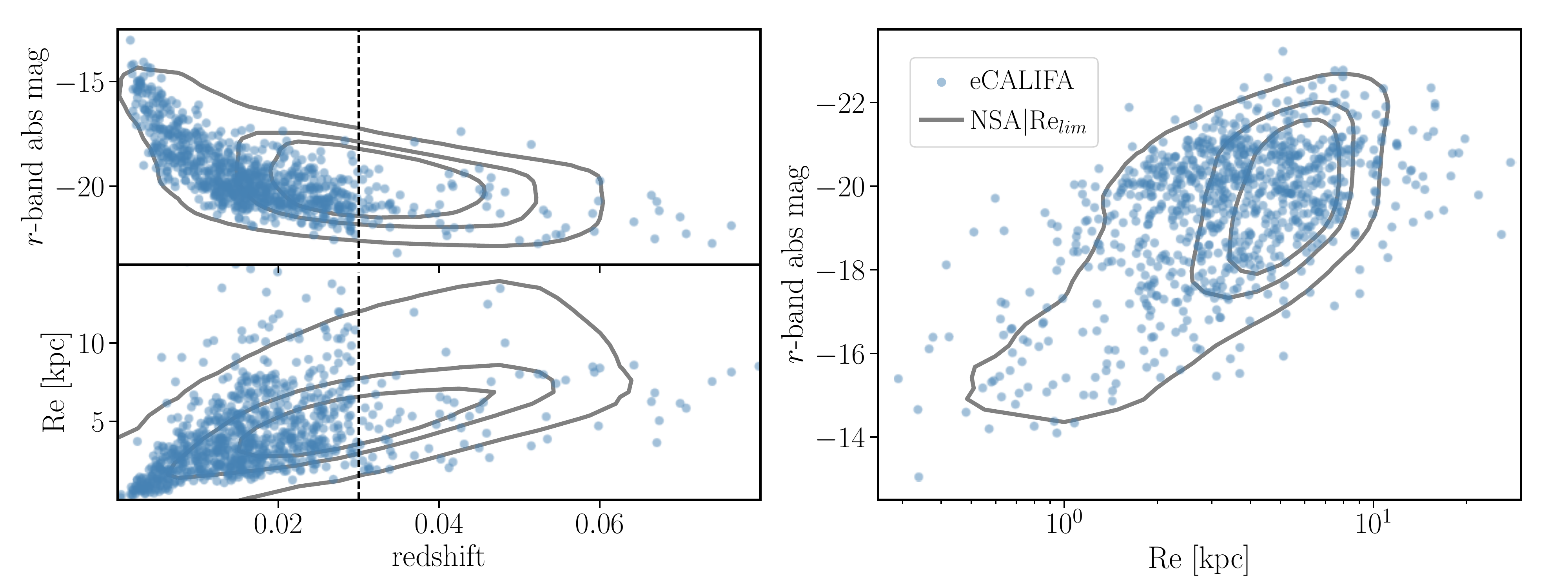}
 \endminipage
 \caption{Distribution of $r-$band absolute magnitude (top-left panel) and effective radius in kpc (bottom-left panel) as a function of the redshift, and one as a function of the other (right panel), for the galaxies in the eCALIFA compilation (solid-blue circles). For comparison purposes grey-contours show, in each diagram, the distribution of the SDSS-NSA galaxies using the same diameter, magnitude and redshift range as the eCALIFA galaxies.  Each successive contour encircles a 95\%, 65\% and 40\% of the SDSS-NSA galaxies. Dashed line in the left panels indicate the $z=0.03$ limit above which we consider that the eCALIFA sample lose completeness.}
 \label{fig:s_prop}
\end{figure*}

Figure \ref{fig:s_prop}, left panels, shows similar comparisons
between both samples for the $r-$band absolute magnitude and Re (in
physical units) as function of the redshift. Like in the case of the
CMD, there is a fair agreement between the distributions described by
both samples, in particular in the redshift regime better covered by
the eCALIFA sample ($z<0.03$, as indicated before). At redshift
$z>0.03$ the eCALIFA sample is clearly incomplete, as illustrated by
the comparison with the NSA$|$Re$_{lim}$ distributions. It is worth
noticing that this was the limiting redshift of the original CALIFA
MS. Despite of these differences in the coverage of the absolute
magnitude and effective radius with redshift, the distribution of one
parameter as a function of the other is pretty similar for both
samples (left panel of Fig. \ref{fig:s_prop}). Indeed, $\sim$90\%\ of the eCALIFA galaxies are encircled by the lowest density contour of the  NSA$|$Re$_{lim}$ sample shown
in this figure, comprising 95\%\ of these objects.

\begin{table*}
\begin{center}
\caption{Observational properties of the galaxy sample$^*$}
\begin{tabular}{rccrrcccrrrrr}\hline\hline
  ID & cubename & galaxy & RA & DEC & z & type & M$_{abs,r}$ & g-r    & Re       & PA &  $\epsilon$ & V$_{cor}$ \\
      &                   &             & deg & deg &   &        &  mag   & mag & $\arcsec$ & deg &             & Mpc$^3$ \\
\hline
  1 & IC5376 & IC5376 & 0.3325 & 34.5265 & 0.01645 & Sb & -19.81 & 0.73 & 9.95 & -86.1 & 0.97 & 3250155.2 \\
2 & UGC00005 & UGC00005 & 0.7733 & -1.9129 & 0.02405 & Sbc & -20.68 & 0.56 & 14.61 & -43.9 & 0.86 & 2008811.2 \\
3 & NGC7819 & NGC7819 & 1.1018 & 31.4725 & 0.01634 & Sc & -19.42 & 0.47 & 19.3 & 4.3 & 0.85 & 1352577.3 \\
4 & UGC00029 & UGC00029 & 1.1403 & 28.3022 & 0.0293 & E1 & -20.98 & 0.7 & 5.57 & 85.1 & 0.53 & 41513.4 \\
5 & IC1528 & IC1528 & 1.2724 & -7.0925 & 0.01261 & Sbc & -18.41 & 0.52 & 23.34 & -16.1 & 0.91 & 2563246.9 \\
6 & NGC7824 & NGC7824 & 1.2754 & 6.9209 & 0.02024 & Sab & -21.06 & 0.62 & 7.29 & 65.4 & 0.74 & 66001.7 \\
7 & UGC00036 & UGC00036 & 1.3078 & 6.7728 & 0.02075 & Sab & -20.88 & 0.67 & 7.03 & -71.8 & 0.87 & 165870.3 \\
8 & NGC0001 & NGC0001 & 1.816 & 27.7085 & 0.01473 & Sbc & -20.38 & 0.66 & 8.08 & 5.1 & 0.74 & 34274.9 \\
9 & NGC0023 & NGC0023 & 2.4726 & 25.9244 & 0.01482 & Sb & -20.98 & 0.6 & 29.66 & 63.2 & 0.77 & 2017008.3 \\
10 & NGC0036 & NGC0036 & 2.8426 & 6.3899 & 0.01978 & Sb & -21.09 & 0.66 & 18.18 & -69.0 & 0.82 & 1565578.4 \\
\hline
\end{tabular}\label{tab:sample} 
\end{center}
Subset of the properties of the galaxy sample: (1)  CALIFA ID \citep{sanchez16}; (2) IFS cube name; (3) galaxy name, (4) right ascension; (5) declination; (6) redshift; (7) Hubble type; (8) $r-$band absolute magnitude; (9) $g-r$ color; (10) effective radius; (11) position angle; (12) eccentricity; and (13) equivalent volume accessible for the considered target.\\ $^*$We present just ten galaxies, the remaining ones are presented electronically.
\end{table*}

\begin{figure}
 \minipage{0.99\textwidth}
 \includegraphics[width=8.5cm]{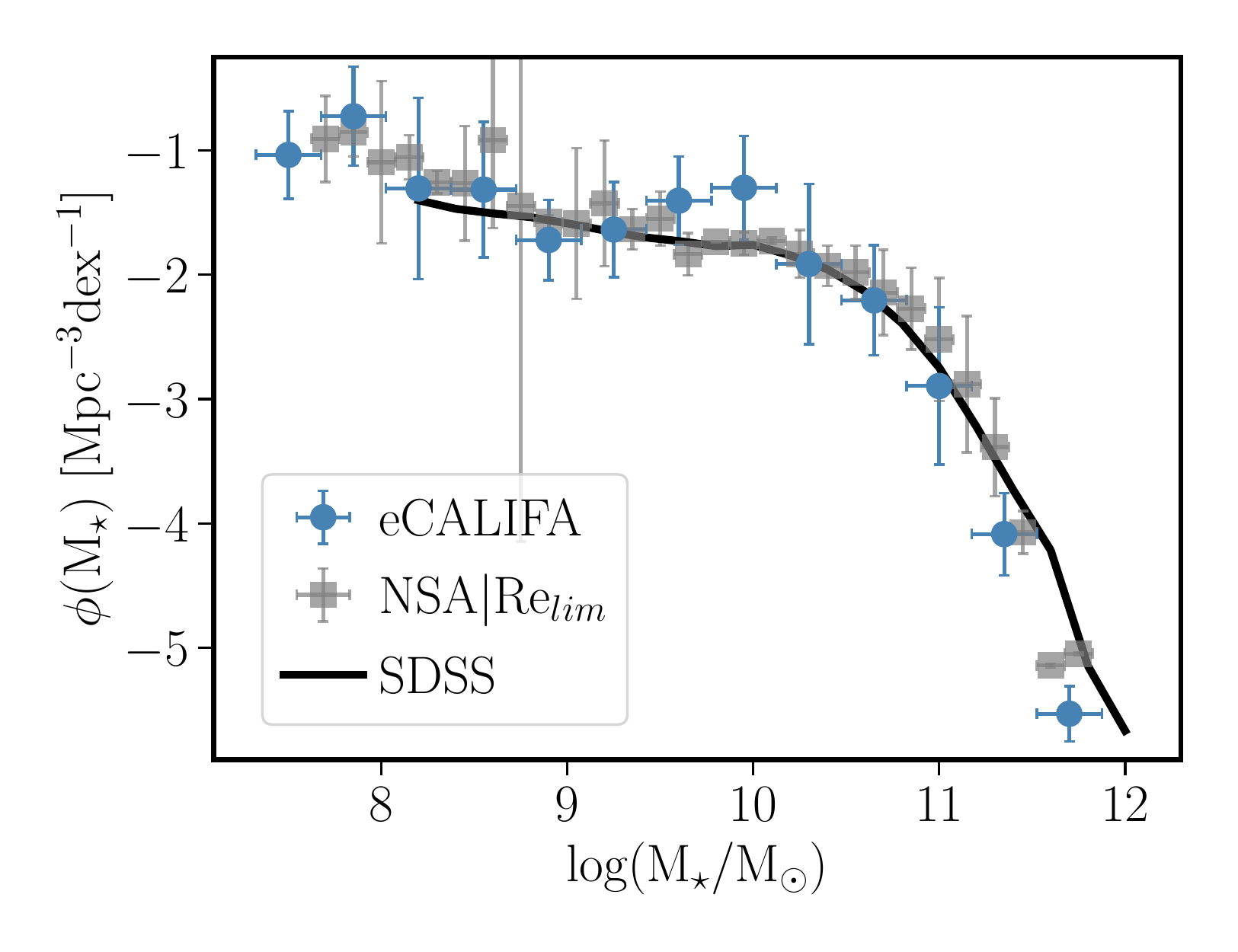}
 \endminipage
 \caption{Mass function estimated for the eCALIFA compilation (blue-solid circles) assuming that it behaves as a diameter selected sample. Error bars illustrate the width of the mass bins (x-axis) and the error propagation based on a Monte-Carlo iteration for all parameters involved in the derivation of Volume correction (see the text). For comparison purposes we include (i) the same estimation for the NSA-SDSS sub-sample selected adopting the same diameter, magnitude and redshift range as the eCALIFA galaxies (grey-solid squares) and (ii) the mass-function derived using the full SDSS-NSA catalog \citep{blanton+2017}, already published in \citet{sanchez22}. There is a considerably good agreement between the three mass functions, despite the differences between the three samples.}
 \label{fig:Vcor}
\end{figure}

\subsection{Volumen correction}
\label{sec:Vcor}

All these comparisons indicate that indeed the eCALIFA behaves, at a
first order, as a diameter-selected sample at the same redshift range
extracted from a luminosity limited sample (i.e., the NSA$|$Re$_{lim}$
sub-sample). If this is the case, this sample could be representative
of the nearby universe when applying the proper volume
correction. Following \citet{walcher14}, we derive the available
volume per galaxy using the V$_{max}$ method described by
\citet{schmidt68}. \citet{dejo94} adopted this method, deriving the
equation of the accessible volume for a galaxy selected within a range
of isophotal sizes (projected diameters). Considering Re as a tracer of the
diameter of a galaxy, the same equation can be adopted to derive the accessible volume
for a given limiting effective radius:
\begin{equation}\label{eq:vol}
\begin{array}{l}
V_{lim} (z,Re) = 4\pi \frac{D_L(z)}{(1+z)^2} \Big(\frac{Re}{Re_{lim}} \Big)^3\\
\end{array}
\end{equation}
where $D_L(z)$ is the luminosity distance at redshift $z$ and
Re$_{lim}$ is the largest limiting effective radius adopted in the
selection criteria (defined by the FoV of the IFU, in our case). To estimate the luminosity distance at a given
redshift we assume a pure Hubble flow distance and the cosmology
assumed along this exploration. This is a good approximation for the
average redshift of the considered samples ($z\sim$0.015). However, it
may introduce considerable errors for the lowest redshift objects,
where the average galaxy velocity with respect to the cosmic web is of
the same order of the bulk velocity due to the cosmic expansion
($z<0.001$).  Fortunately we have just 2 objects below this limit (
NGC 6789 and NGC 0784).

The inverse of $V_{lim}$ is the weight of each galaxy in any
estimated global property (e.g. the star-formation or mass density in
the considered volume) or distribution (e.g., the luminosity or the
mass function). When selecting galaxies within a range of effective
radius, from Re$_{min}$ to Re$_{max}$, the corresponding weight ($\omega$) is
derived as the difference between the inverse of the corresponding V$_{min}$
and V$_{max}$, and estimated using Eq. \ref{eq:vol}. This indeed
defines an equivalent accessible volume that is estimated by the equation:
\begin{equation}\label{eq:vol_corr}
\frac{1}{V_{cor}}= \omega = \frac{1}{V_{max}}-\frac{1}{V_{min}}
\end{equation}
Note that $V_{cor}$, the equivalent volume correction depends not only
on the redshift and Re of the object, but also on the minimum and
maximum Re adopted in the selection of the sample. Finally, a
correction must be applied if the accessible solid angle for the
sample is smaller than the full sky, which involves a global scaling
factor. Further corrections can be applied to take into account other limiting
factors, just as the flux-limit introduced by a limiting magnitude.

We acknowledge that this derivation relies on multiple simplifications
and assumptions that have been broadly discussed in previous studies,
in particular in \citet{walcher14}. However, despite of them, the
approach is valid. To demonstrate it so, we have first applied it to the
NSA$|$Re$_{lim}$ sub-sample, a sample selected imposing a well defined
range of redshifts, magnitudes and in particular effective radius over
a large and well defined catalog of galaxies.  Based on this volume
corrections we estimate the mass-function, using the stellar masses
provided by the SDSS-NSA catalog \citep{blanton+2017}, shifted by a
constant to correct for the difference cosmology and to match the
\citet{salpeter55} initial mass function (IMF). Figure \ref{fig:Vcor}
shows the resulting mass-function compared to the one derived for the
full SDSS-NSA catalog \citep[Fig. 17 of][]{sanchez22}. There
is a good agreement between the two mass-functions in the regime of
masses in common, validating the adopted volume correction. Based on
this result we estimate the volume correction using this approach for
the eCALIFA sample and repeat the estimation of the mass-function. For
doing so we estimate the stellar mass from the $g-$ and $r-$band
photometric data obtained from the segmented PS images described
during the isophotal analysis (Sec. \ref{sec:gal_seg}), implementing
the relation between the mass-to-light ($\Upsilon_\star$) and the $g-r$
colors published by \citep{bell00} for the same IMF. We refer to this
stellar mass as M$_{\star,phot}$ hereafter. The result of this analysis is included in
Fig. \ref{fig:Vcor}. Like in the case of the NSA$|$Re$_{lim}$
sub-sample, the mass function estimated for the eCALIFA sample follows
pretty well the know distribution described by the full NSA-SDSS
catalog, for the range of stellar masses in common. The strongest
deviations are found at (i) $\sim$10$^{10}$M$_\odot$, where the
eCALIFA sample seems to present a slight excess of galaxies, and (ii)
at $>$10$^{11.5}$M$_\odot$, where the eCALIFA sample (and maybe the
NSA$|$Re$_{lim}$ sub-sample), present a slight defect of
galaxies. The excess of galaxies at $\sim$10$^{10}$M$_\odot$ could be due to the local fluctuations in the galaxy density in this mass regime, already noticed in \citet{walcher14}. In addition we note that for the stellar masses below
10$^{8.5}$M$_\odot$, not sampled by the published NSA-SDSS mass
function, the agreement between the two mass functions derived using
the described volume correction is reasonable good. 

Based on all the results described along this section we can fairly
claim that the eCALIFA sample, once corrected by the proper volume
coverage, can be adopted as a representative sample of galaxies in the covered redshift
range (i.e., the nearby Universe). Table \ref{tab:sample} presents the main observational properties of the galaxies discussed along this section.

\begin{figure*}
 \minipage{0.99\textwidth}
 \includegraphics[width=18cm]{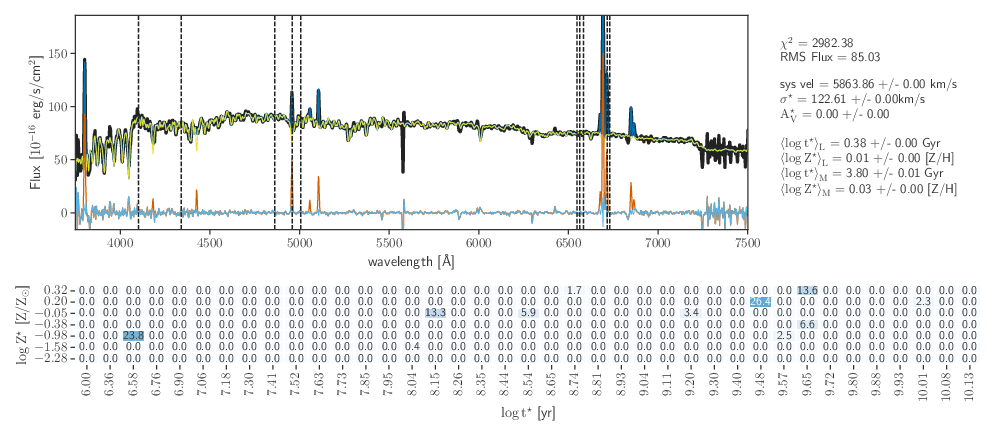}
 \endminipage
 \caption{Example of the analysis performed by {\sc pyFIT3D} for the two aperture spectra extracted from each IFS datacube. The current figure corresponds to the central spectrum of the archetype galaxy NGC\ 5947. The top panel shows: (i) the observed spectrum (black line); (ii) the best fitted stellar population model (yellow line); (iii) the combination of this stellar population model with the model of the analyzed emission lines (steelblue line), i.e., the stellar ; (iv) the residual of the subtraction of the stellar population model to the original spectrum (orange line) and (v) the residual of the subtraction of the combined model (stellar population plus emission lines) to the original spectrum (lightblue line). Dashed-dotted line show the location of the fitted emission lines in the rest-frame, illustrating the redshift. The main properties of the stellar population derived from the fitting are listed in the right panel, including: (i) the systemic velocity ({\it sys vel}); (ii) the velocity dispersion ($\sigma_\star$); (iii) the dust attenuation (${\rm A}^\star_{\rm V}$); (iv) the luminosity-weighted age (\ageLW or $t^\star_{L}$) and metallicity (\metLW or  $Z^\star_{L}$), and (v) the mass-weighted age (\ageMW or $t^\star_{M}$) and metallicity (\metMW or $Z^\star_{M}$). Note that the errors for the {\it sys vel}, {$\sigma_\star$} and ${\rm A}^\star_{\rm V}$ are all zero, as they were derived in the first step of the fitting procedure, as described in the text. For the rest of the parameters the nominal errors are also very low, as they do not included systematics from the procedure. The light-weights of each SSP in the adopted library (i.e., the fraction of light that each SSP contributes to the best fitted model in  the $V$-band) is shown in the bottom panel. Each row corresponds to one of the 7 metallicites and each column to one of the 39 ages included in the SSP library.}
 \label{fig:FIT3D}
\end{figure*}

\section{Analysis}
\label{sec:analysis}

We perform a set of analyses in order to extract
the main spectroscopic properties of the studied galaxies and to evaluate
quality and usability of the new reduced dataset.  In
particular we explore the spectra extracted in two different apertures
for each galaxy: (i) a circular aperture of 1.5$\arcsec$/diameter
located at the center; and (ii) the integrated spectra, obtained by
co-adding all the spectra within the IFS datacube corresponding to the
spaxels with a S/N$>$3 at 5500\AA. The first of those spectra would allow us to
characterize the spectroscopic properties of both the stellar
populations and the ionized gas in the inner regions of the galaxies,
at scales of $\sim$500 pc, making use of the improvement in the
spatial resolution introduced by the new reduction. The inner regions
are particularly interesting as they are the locations where AGN
activity, galactic winds, and the effect of star-formation quenching
are more frequently observed. On the other hand, the integrated
spectra would allow to characterize extensive quantities (such as the
integrated stellar mass and integrated star-formation rate), the
dominant ionizing source and/or the star-formation stage.

\subsection{Spectral fitting}
\label{sec:spec_fit}

We analyze each individual spectrum using {\sc pyFIT3D}
\citep{pipe3d,pypipe3d} to derive the main properties of stellar populations and ionized gas components.  For doing so, the code generates a model of the stellar
population spectrum based on the linear combination of a set of single
stellar populations (SSP) included in a given library. This model is
convolved by a Gaussian function to take into account the velocity
dispersion ($\sigma_\star$) and shifted according to the systemic
velocity ($v_\star$). In addition, the multi-SSP model is attenuated
adopting a given extinction curve \citep[][, in this particular
implementation]{cardelli+1989}, and a certain dust attenuation value
in the $V-$band (A$_{\rm V,\star}$). These three non-linear parameters
are derived by {\sc pyFIT3D} based on a brute exploration of the space
of parameters, adopting a limited version of the SSP-library, prior
to the fit with the full library, that we describe later on.

The treatment of the emission lines is done by a two steps
procedure. First, the wavelength location of the emission lines is
masked out and a preliminar model of the stellar population is
created. This model is then subtracted to the original spectrum.  The
resulting residual spectrum comprises just the emission lines (plus
noise and imperfections associated to the modelling of the stellar
population). Then, each emission line of a defined set (\oii 3727, \Hd, \Hg, \Hb, \oiii 4959,5007, \Ha, \nii 6548,84 and \sii 6717,31, in this particular case)
is modelled using a single
Gaussian function, deriving the flux intensity, velocity and velocity
dispersion. The combined model for all the emission
lines is subtracted to the original spectrum. This final spectrum,
comprising just the stellar population component (plus noise and
imperfections associated to the modelling of the emission lines), is
fitted using the full SSP library (shifted, convolved and dust
attenuated).  Finally, a Monte-Carlo iteration is adopted to estimate
the errors for each derived parameter.  The procedure has demonstrated
to produce reliable results, being contrasted against ad-hoc created
simulations and mock galaxy spectra generated from hydrodynamical
simulations \citep[e.g.,][]{pipe3d,pypipe3d,ibarra19,sarm23}. Indeed,
it has been adopted by previous several studies, in particular focused
on the analysis of IFS data
\citep[e.g.][]{mariana16,sanchez18,sanchez22}.

Figure \ref{fig:FIT3D} shows an example of the fitting procedure,
corresponding to the analysis of the integrated spectrum of the
archetype galaxy NGC\ 5947. It illustrates how well the observed
spectrum is reproduced by the best-fitted model for both the stellar
population and emission line models. The typical level and shape of
the residuals is appreciated, with the location of the strong
night-sky emission lines easily identified \citep[e.g., at 5577\AA\ or
beyond 7200\AA,][]{sanchez07a}. The distribution by age and
metallicity of the individual weight ($w_{ssp,\star,L}$) of each SSP
in the final model (i.e., the fraction of light in the $V$-band), is also
included in the figure. In the current analysis we adopt the {\sc
  MaStar\_sLOG} SSP library, generated by an updated version the
\citet{bc03} stellar population synthesis code \citep[described in
Appendix A of][]{sanchez22}, using the MaStar stellar library
\citet{yan19}. The {\sc MaStar\_sLOG} library comprises 273 SSP
templates, covering 39 ages (from 1 Myr to 13.5 Gyr), in a
pseudo-logarithmic sampling, and 7 metallicities (Z/Z$_\odot$~=~0.006,
0.029, 0.118, 0.471, 1, 1.764, 2.353). A solar $\alpha$-enhancement is
considered for all SSPs in the adopted library (i.e.,
[$\alpha$/Fe]=0).  This library was recently adopted to analyze the
last data-release of the MaNGA \citep{manga} IFS galaxy survey
\citep{sanchez22}, and therefore, by adopting it the results can be
more easily compared with.

\begin{table*}
\begin{center}
\caption{Stellar population properties derived by {\sc pyFIT3D} for both the central and integrated spectra of each galaxy.}
\begin{tabular}{ccrrrrrrrr}\hline\hline
  cubename & aperture & \ageLW & \metLW & \ageMW & \metMW & A$_{\rm V,\star}$ & $\sigma_\star$ & log($\Upsilon_\star$) & log(M$_\star$) \\
                    &                  & Gyr   &   f$_{mass}$    &  Gyr & f$_{mass}$    &  mag & km s$^{-1}$ & M$_\odot$/L$_\odot$ & M$_\odot$ \\
  \hline
IC5376 & int  & 3.2659 $\pm$ 0.4899 & 0.0138 $\pm$ 0.0021 & 5.1993 $\pm$ 0.7799 & 0.0088 $\pm$ 0.0013 & 0.4 $\pm$ 0.1 & 128.4 $\pm$ 19.3 & 0.76 & 10.07\\
 " & cen  & 6.8726 $\pm$ 1.0309 & 0.0172 $\pm$ 0.0026 & 7.6401 $\pm$ 1.146 & 0.0168 $\pm$ 0.0025 & 0.5 $\pm$ 0.1 & 130.0 $\pm$ 19.5 & 1.05 & 9.32\\
UGC00005 & int  & 0.8419 $\pm$ 0.1263 & 0.0161 $\pm$ 0.0024 & 3.0662 $\pm$ 0.4599 & 0.019 $\pm$ 0.0028 & 0.4 $\pm$ 0.1 & 287.8 $\pm$ 43.2 & 0.53 & 10.47\\
 " & cen  & 2.7528 $\pm$ 0.4129 & 0.0169 $\pm$ 0.0025 & 5.2094 $\pm$ 0.7814 & 0.0167 $\pm$ 0.0025 & 0.7 $\pm$ 0.1 & 62.6 $\pm$ 9.4 & 0.83 & 8.99\\
NGC7819 & int  & 0.695 $\pm$ 0.1042 & 0.0093 $\pm$ 0.0014 & 4.9267 $\pm$ 0.739 & 0.0043 $\pm$ 0.0006 & 0.3 $\pm$ 0.0 & 29.9 $\pm$ 4.5 & 0.47 & 9.81\\
 " & cen  & 0.3353 $\pm$ 0.0503 & 0.0037 $\pm$ 0.0006 & 4.2694 $\pm$ 0.6404 & 0.0125 $\pm$ 0.0019 & 0.6 $\pm$ 0.1 & 398.5 $\pm$ 59.8 & 0.57 & 8.56\\
UGC00029 & int  & 7.7201 $\pm$ 1.158 & 0.0191 $\pm$ 0.0029 & 8.7365 $\pm$ 1.3105 & 0.0205 $\pm$ 0.0031 & 0.0 $\pm$ 0.0 & 183.1 $\pm$ 27.5 & 1.12 & 11.01\\
 " & cen  & 5.2782 $\pm$ 0.7917 & 0.029 $\pm$ 0.0044 & 6.146 $\pm$ 0.9219 & 0.0292 $\pm$ 0.0044 & 0.6 $\pm$ 0.1 & 207.4 $\pm$ 31.1 & 1.02 & 9.52\\
IC1528 & int  & 0.9597 $\pm$ 0.144 & 0.0043 $\pm$ 0.0006 & 3.3443 $\pm$ 0.5016 & 0.0112 $\pm$ 0.0017 & 0.2 $\pm$ 0.0 & 242.9 $\pm$ 36.4 & 0.48 & 9.68\\
 " & cen  & 0.8211 $\pm$ 0.1232 & 0.016 $\pm$ 0.0024 & 5.0834 $\pm$ 0.7625 & 0.0139 $\pm$ 0.0021 & 0.6 $\pm$ 0.1 & 102.8 $\pm$ 15.4 & 0.65 & 8.05\\
NGC7824 & int  & 2.7376 $\pm$ 0.4106 & 0.0235 $\pm$ 0.0035 & 2.9638 $\pm$ 0.4446 & 0.029 $\pm$ 0.0044 & 0.0 $\pm$ 0.0 & 157.5 $\pm$ 23.6 & 0.74 & 10.66\\
 " & cen  & 2.4005 $\pm$ 0.3601 & 0.0258 $\pm$ 0.0039 & 2.8525 $\pm$ 0.4279 & 0.0262 $\pm$ 0.0039 & 0.0 $\pm$ 0.0 & 110.0 $\pm$ 16.5 & 0.68 & 8.72\\
UGC00036 & int  & 2.5892 $\pm$ 0.3884 & 0.0254 $\pm$ 0.0038 & 7.646 $\pm$ 1.1469 & 0.0283 $\pm$ 0.0042 & 0.3 $\pm$ 0.0 & 288.8 $\pm$ 43.3 & 0.94 & 10.59\\
 " & cen  & 5.6372 $\pm$ 0.8456 & 0.0293 $\pm$ 0.0044 & 6.5388 $\pm$ 0.9808 & 0.035 $\pm$ 0.0052 & 0.6 $\pm$ 0.1 & 307.2 $\pm$ 46.1 & 1.09 & 9.49\\
NGC0001 & int  & 1.2922 $\pm$ 0.1938 & 0.0159 $\pm$ 0.0024 & 3.4601 $\pm$ 0.519 & 0.0149 $\pm$ 0.0022 & 0.5 $\pm$ 0.1 & 180.4 $\pm$ 27.1 & 0.63 & 10.16\\
 " & cen  & 2.1775 $\pm$ 0.3266 & 0.0288 $\pm$ 0.0043 & 3.6797 $\pm$ 0.552 & 0.0275 $\pm$ 0.0041 & 0.4 $\pm$ 0.1 & 141.6 $\pm$ 21.2 & 0.75 & 9.08\\
NGC0023 & int  & 0.7246 $\pm$ 0.1087 & 0.01 $\pm$ 0.0015 & 4.4113 $\pm$ 0.6617 & 0.0288 $\pm$ 0.0043 & 0.5 $\pm$ 0.1 & 151.1 $\pm$ 22.7 & 0.67 & 10.57\\
 " & cen  & 0.9247 $\pm$ 0.1387 & 0.0202 $\pm$ 0.003 & 4.5153 $\pm$ 0.6773 & 0.0325 $\pm$ 0.0049 & 0.6 $\pm$ 0.1 & 360.0 $\pm$ 54.0 & 0.72 & 9.36\\
NGC0036 & int  & 3.2041 $\pm$ 0.4806 & 0.019 $\pm$ 0.0028 & 10.26 $\pm$ 1.539 & 0.0282 $\pm$ 0.0042 & 0.2 $\pm$ 0.0 & 96.9 $\pm$ 14.5 & 1.02 & 11.05\\
 " & cen  & 3.2059 $\pm$ 0.4809 & 0.0395 $\pm$ 0.0059 & 3.7183 $\pm$ 0.5577 & 0.0394 $\pm$ 0.0059 & 0.3 $\pm$ 0.0 & 178.4 $\pm$ 26.8 & 0.86 & 9.32\\
UGC00139 & int  & 0.1775 $\pm$ 0.0266 & 0.0031 $\pm$ 0.0005 & 4.4885 $\pm$ 0.6733 & 0.0065 $\pm$ 0.001 & 0.6 $\pm$ 0.1 & 37.4 $\pm$ 5.6 & 0.35 & 9.32\\
 " & cen  & 1.5945 $\pm$ 0.2392 & 0.0171 $\pm$ 0.0026 & 2.9027 $\pm$ 0.4354 & 0.0181 $\pm$ 0.0027 & 0.2 $\pm$ 0.0 & 159.5 $\pm$ 23.9 & 0.6 & 7.88\\
MCG-02-02-030 & int  & 3.1502 $\pm$ 0.4725 & 0.0065 $\pm$ 0.001 & 8.8252 $\pm$ 1.3238 & 0.0044 $\pm$ 0.0007 & 0.1 $\pm$ 0.0 & 161.5 $\pm$ 24.2 & 0.85 & 10.13\\
 " & cen  & 0.5685 $\pm$ 0.0853 & 0.0112 $\pm$ 0.0017 & 9.3588 $\pm$ 1.4038 & 0.0158 $\pm$ 0.0024 & 1.3 $\pm$ 0.2 & 309.3 $\pm$ 46.4 & 0.96 & 8.48\\
\hline
\end{tabular}\label{tab:stars} 
We present the values for just ten galaxies. The information for the remaining one is distributed electronically.
\end{center}
\end{table*}


\subsection{Stellar parameters derived by the spectral fitting}
\label{sec:LW}

As discussed in detail in previous studies
\citep[e.g.][]{sanchez21,sanchez22}, the weights or coefficients of
the decomposition can be used (i) to derive the luminosity- and
mass-weighted parameters of the stellar population ($P_{LW}$ and
$P_{MW}$), using the equations:
\begin{equation}
 \begin{array}{l}
   {\rm log} P_{L}  = \Sigma_{ssp} w_{ssp,\star,L} {\rm log} P_{ssp}\\
   \\
{\rm log} P_{M}  = \frac{\Sigma_{ssp} w_{ssp,\star,L} \Upsilon_{ssp,\star} {\rm log}P_{ssp}}{\Sigma_{ssp} w_{ssp,\star,L} \Upsilon_{ssp,\star}}\\
\end{array}
\label{eq:par}    
\end{equation}
\noindent where (a) $w_{ssp,\star,L}$  corresponds to each coefficient of the decomposition, (b) $P_{ssp}$ is the value of the given parameter for a certain SSP (e.g., age, \age, or metallicity, \met), and (c) $\Upsilon_{ssp,\star}$ is the stellar mass-to-light ratio; (ii) to derive integrated properties, such as the stellar mass:
\begin{equation}
\begin{array}{l}
  M_\star= L_V \Sigma_{ssp} w_{ssp,\star,L} \Upsilon_{ssp,\star}\\
\end{array}
  \label{eq:M}  
\end{equation}
where $L_V$ is the dust corrected luminosity in the $V$-band
($L_V = 4\pi DL(z)^2 f_{V} 10^{0.4 A_{\rm V,*}}$),  $f_{V}$ is the observed flux intensity in the $V$-band, $A_{\rm V,*}$ is the dust attenuation affecting the stellar populations (derived by the fitting procedure as described before) and $DL(z)$ is the luminosity distance already described in the previous section; and (iii) to estimate mass- and metallicity-assembly histories of galaxies or regions within galaxies, by applying Equations \ref{eq:M} and \ref{eq:par} to a restricted age range, that corresponds to a particular look-back time \citep[for instance as described in][]{sanchez21,camps22}. From them it is possible to derive the
star-formation and chemical enrichment history, as the derivatives of the previous distributions \citep[e.g.][]{ibarra19}.

\subsection{Ionized-gas dust attenuation}
\label{sec:Av_gas}

The dust attenuation affecting the bulk of the stellar populations
($A_{\rm V,*}$), introduced in the previous section, is in general
different than the one affecting the ISM ($A_{\rm V,gas}$). The reason
for that difference is that the former is embedded within the stars
across a galaxy, not equally affecting all of them, and therefore the
estimated value is the resulting of its net effect. On the contrary,
the later fits better with a screen model, as the dust grains are in
general distributed surrounding the ionized nebulae
\citep[e.g.][]{calz01,wild11,salim20}.

We derive $A_{\rm V,gas}$ for each analyzed spectrum by comparing
the observed \Ha/Hb\ line ratio with the expected value for
photo-ionized nebulae without dust-attenuation (\Ha/\Hb$_0$), and
adopting a MW-like extinction curve \citep[][]{cardelli1989}, with an
extinction factor R$_{\rm V}$=3.1. This is indeed an approximation, as
R$_{\rm V}$ may change galaxy by galaxy and within each galaxy too, as
it is the case in the MW, what adds an uncertainty to the derivation
of A$_{\rm V,gas}$.  The calculation depends on the value of
\Ha/\Hb$_0$.  We use the usual value of 2.86, that corresponds
to a photoionized nebulae with a temperature T$_e$=10$^4$K and an
electron density of $n_e$=10$^2$ cm$^{-3}$ \citep{osterbrock89}. However, we
acknowledge that this value is ill defined, ranging between 2.7 and
3.1 for a range of typical values of photoionized nebulae, being a
topic of study how its affect the estimation of the dust attenuation
and different nebulae properties \citep[e.g.][]{utea21}. Finally, the
derivation of A$_{\rm V,gas}$ depends on how well it is measured \Hb, a line
that it is heavily affected by the modelling of the underlying stellar population.
Being  at least $\sim$3 times weaker than \Ha, in the low-SN regime the derived ratio may present strong asymmetrical deviations from the real value \citep[an usual effect when comparing line ratios of different S/N, e.g.,][]{rosales11}.

We estimate the error associated with A$_{\rm V,gas}$ by propagating
the errors estimated for the emission lines. However, for the reasons
outlined before this error is most probably just a lower-limit to the
real one, that, including systematic effects would be at least
$\sim$0.15 mag in the best case \citep[according to the typical uncertain in the meassurement of emission lines with the adopted technique][]{pypipe3d}. We consider that all values of
A$_{\rm V,gas}$ below this error limit, corresponding to \Ha/\Hb$<$3
or derived from \Ha\ or \Hb\ fluxes below with a S/N$<$2, are ill
defined. All of them corresponds to weak emission lines and low values
of the dust attenuation. We have substituted them by a value of 0.15
mag and set the corresponding errors to zero to indicate that they are
just upper limits. 

\subsection{Star-formation rate}
\label{sec:SFR}

The star-formation rate (SFR) for both apertures is derived from the
dust-corrected \Ha\ luminosity using the formulae:
L$_{H\alpha}=4\pi D_L(z)^2 f_{H\alpha} 10^{0.4 A_{\rm V,gas}}$), where
$f_{H\alpha}$ is the \Ha\ flux intensity derived from the spectral
fitting procedure described before (Sec. \ref{sec:spec_fit}),
A$_{\rm V,gas}$ is the dust attenuation described in the previous
section, and $D_L(z)$ is the luminosity distance corresponding to the
redshift ($z$) of the object. Finally, we use the calibrator derived
by \citet{kennicutt89}, SFR$=8\ 10^{42}$ L$_{H\alpha}$, valid for a
\citep{salpeter55} IMF. Errors of the quantities involved in this
calculation are propagated through the different equations to estimate
the corresponding SFR error.

We have estimated the SFR from \Ha\ irrespectively of the dominant
ionizing source (that we will discuss later on,
Sec. \ref{sec:ion_nat}), following previous studies
\citet{mariana16,sanchez18,sanchez22}. In purity, only in the case
that the gas is ionized by young massive OB-stars is valid this
approximation. However, as noted in those articles, when other ionizing
sources contribute to the \Ha\ luminosity the estimated SFR is just an upper-limit
to the real one, and this is how it should be interpreted.

\subsection{Molecular gas mass}
\label{sec:Mgas}

The molecular gas mass is estimated for the two aperture spectra based
on a dust-to-gas calibrator, following the approach proposed by
\citep{brin14}. The particular calibrator adopted in here is based on
the comparison of the molecular gas mass density obtained from CO
observations \citep{bolatto17} and the ionized-gas dust attenuation
for the EDGE-CALIFA sample presented in \citet{jkbb20}, recently
improved by \citet{jkbb21a}. The estimations based on this approach
have been proved to provide reliable estimations of the molecular gas
densities (and integrated molecular gas) within $\sim$0.15 dex when
comparing with both spatial resolved and single aperture derivations
based on CO observations
\citep[][]{colombo20,jkbb21b,sanchez21,sanchez21b}. The calibrator
provides the molecular gas density ($\Sigma_{\rm gas}$) from the dust
attenuation (A$_{\rm V,gas}$, Sec. \ref{sec:Av_gas}), for each
aperture of each galaxy. From this surface density we estimate the
integrated molecular gas (M$_{\rm gas}$) by multiply by equivalent
area, following \citet{sanchez21,sanchez21b}, i.e., 4$\pi$R$_e^2$,
where R$_e$ is the effective radius (Sec: \ref{sec:gal_seg}). When an
upper-limit was adopted for the dust attenuation, the estimated
molecular gas should be considered as an upper-limit as well. In those
cases the estimated error resulting from the propagation of the
A$_{\rm V,gas}$ value is set to zero.

\subsection{Oxygen abundance}
\label{sec:OH}

The analyzed spectra have not the required spectral resolution and
depth to derive the oxygen abundance (O/H) using the direct method,
even less to estimate the abundances using recombination lines
\citep[see][to gauge the characteristics of the data required to apply
those methods]{peim06,bresolin17}. In the case of CALIFA data, the
direct method was applicable for a small fraction of the galaxies and
at particular star-forming regions \citep[][]{marino13}. Thus, we make
use of strong emission line calibrators that provides with an
estimation to the oxygen valid to explore the global and local trends
in galaxies to the spatial resolution of our data \citep{ARAA}. As we
do not intend to make a detailed study of the systematic differences
between different calibrators depending on their nature
\citep{angel12}, we focus on two calibrators instead of estimating the
oxygen abundance using a set of them \citep[e.g., like in the case
of][]{sanchez22,paola22}. We estimate O/H using the calibrators
proposed by \citet{marino13}, that have derived empirically as a
linear relation between the oxygen abundances estimated using the
direct method for a compiled sample of \HII regions and the logarithm
of \nii/\Ha and (\oiii/\Hb)/(\nii/Ha) line ratios, N2 and O3N2
respectively. Due to their nature these calibrators should be applied
when the dominant ionizing source is due to young/massive OB stars.
Therefore, we derive O/H for those spectra compatible with this kind
of ionization (as selection discussed later one, in
Sec. \ref{sec:ion_nat}). We acknowledge that any contamination by
other ionizing sources may introduce systematic effects in the derivation
of the oxygen abundance that are (somehow) mitigated when using
the spatial resolved information included in our delivered data \citep[e.g.][]{mast14,davies16,lacerda18,vale19}.

\begin{figure}
 \minipage{0.99\textwidth}
 \includegraphics[width=8.5cm]{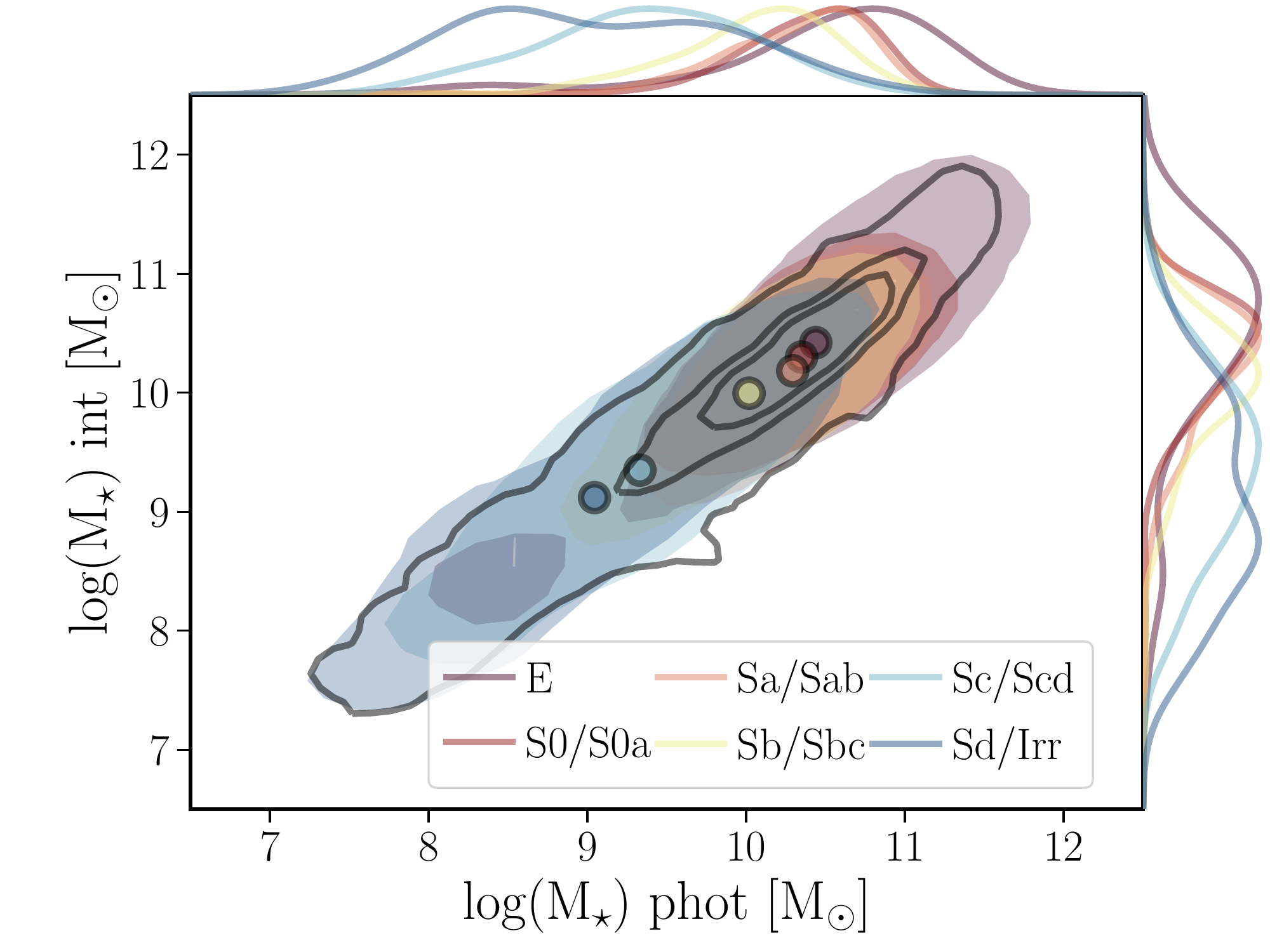}
 \endminipage
 \caption{Comparison between the stellar masses derived using the PS photometry ({\it phot}), described in Sec. \ref{sec:Vcor}, and those obtained using {\sc pyFIT3D} for the integrated spectra ({\it int}). Black-solid contours show the density distribution, with each successive one encircling a 95\%, 65\% and 40\% of the galaxies, respectively. Each shaded region corresponds to the area encircling a 85\%\ of the galaxies of a particular morphological type,
represented by a different color (inset legend). The normalized density distributions of both parameters, also segregated by morphology, are shown in the upper- and right-side panels, with the mean values represented as solid circles, both using the same color code.}
 \label{fig:c_Mass}
\end{figure}

\begin{figure*}
 \minipage{0.99\textwidth}
 \includegraphics[width=6.3cm]{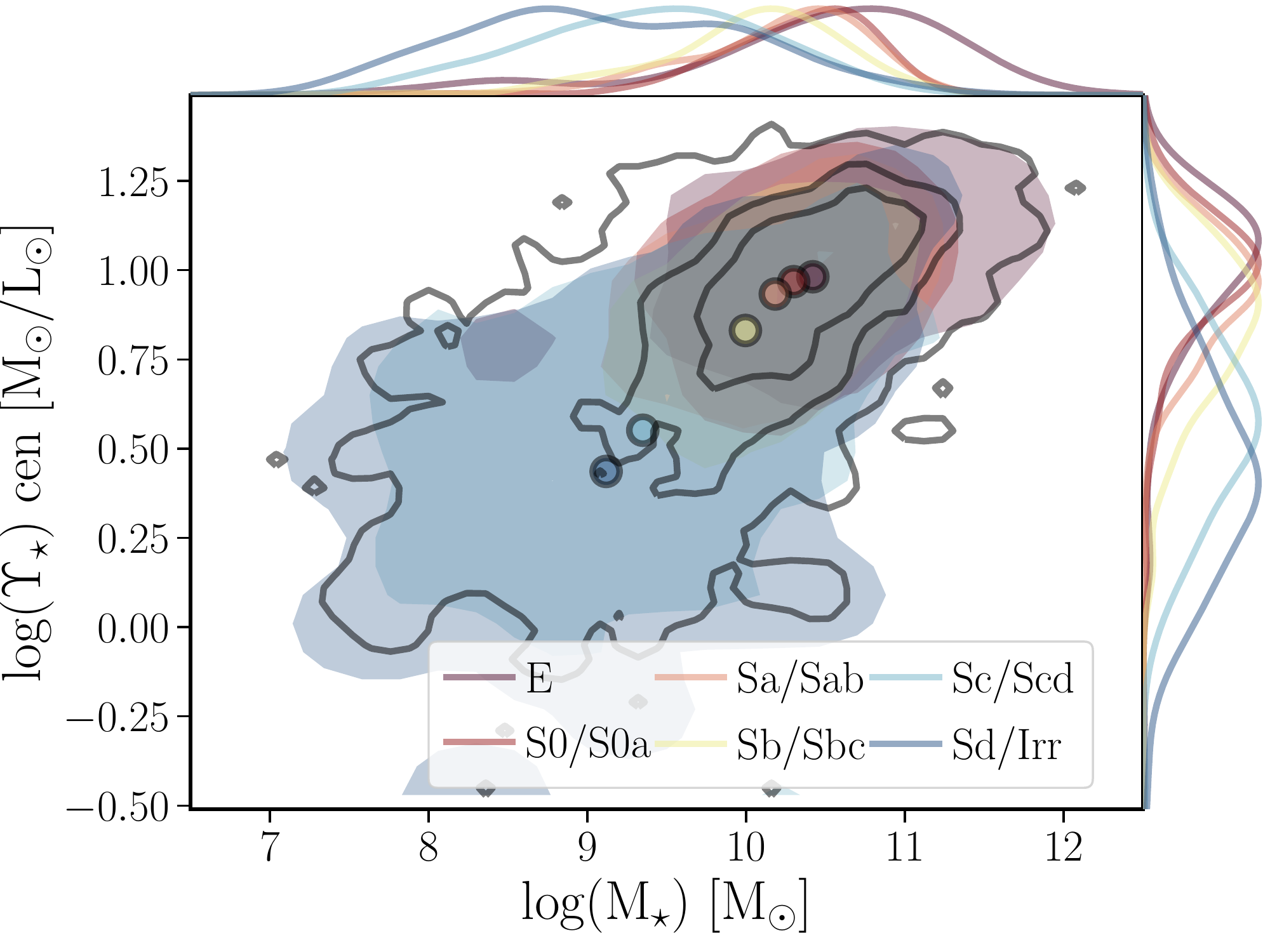}
 \includegraphics[width=6cm]{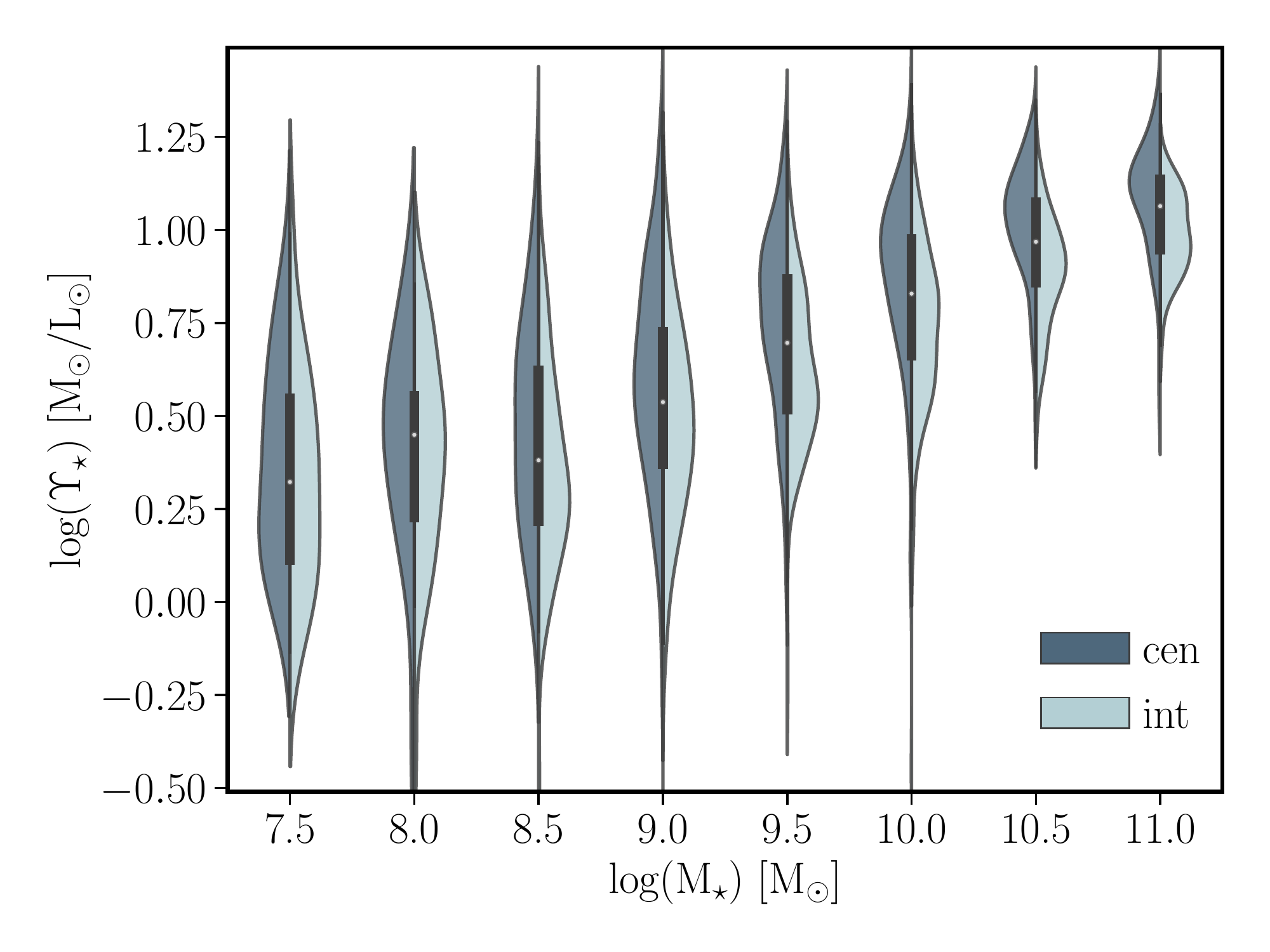}
 \includegraphics[width=6cm]{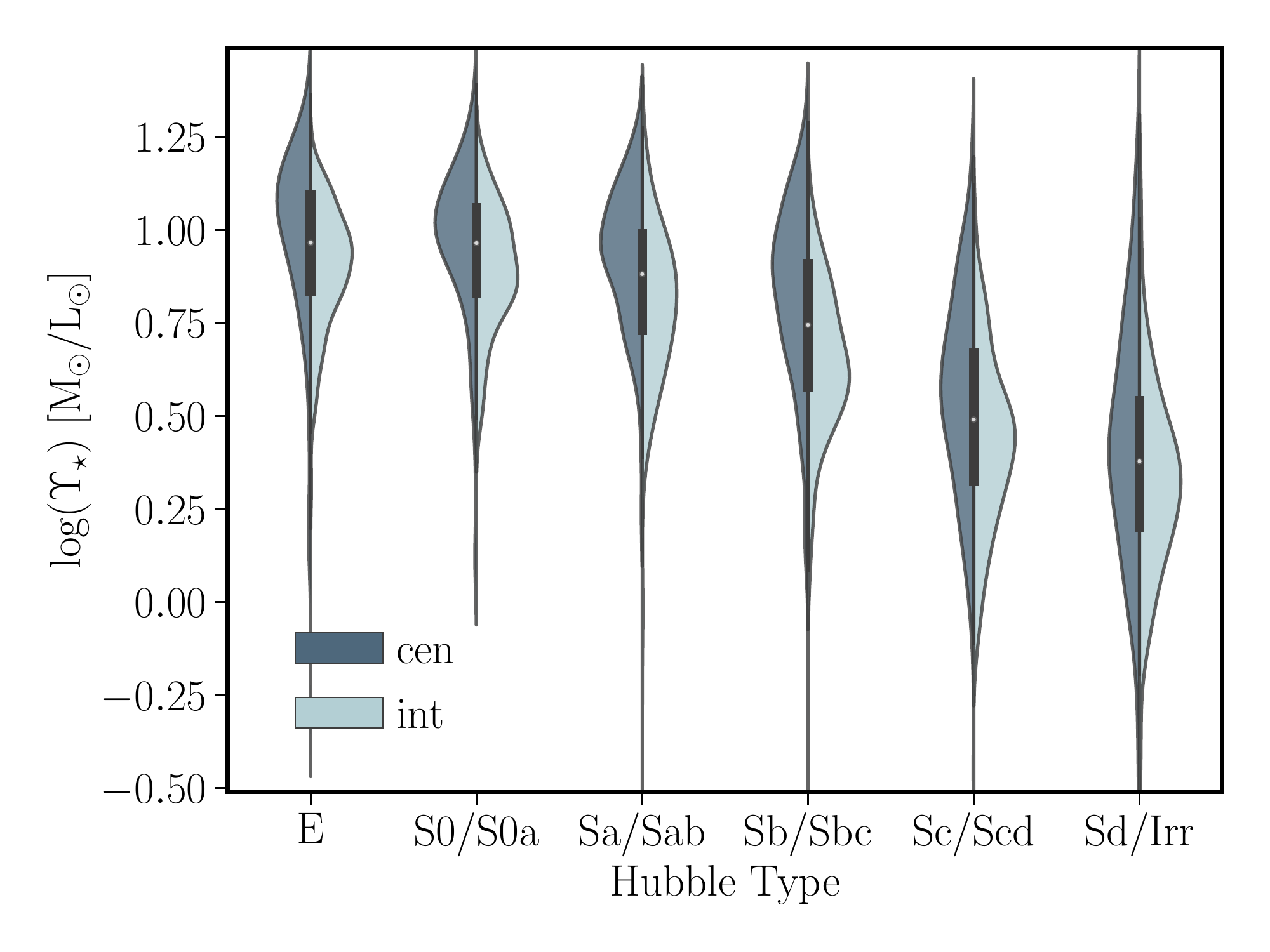}
 \endminipage
 \caption{{\it Left panel:} Distribution of the $g-$band mass-to-light ratio ($\Upsilon_\star$) derived for the central aperture spectra as a function of the stellar mass.  Black-solid contours show the density distribution, with each successive contour encircling a 95\%, 65\% and 40\% of the galaxies, respectively. Each shaded region corresponds to the area encircling a 85\%\ of the galaxies of a particular morphological type, represented by a different color (inset legend). The normalized density distributions of both parameters, segregated by morphology too, are shown in the upper- and right-side panels, with the mean values represented as solid circles, both using the same color code. The violin plots of the $\Upsilon_\star$ parameter derived for the central (cen) and integrated (int) spectra as a function of the stellar mass and the morphology of the galaxies are shown in central and right panels, respectively.}
 \label{fig:ML}
\end{figure*}

\begin{figure}
 \minipage{0.99\textwidth}
 \includegraphics[width=8.5cm]{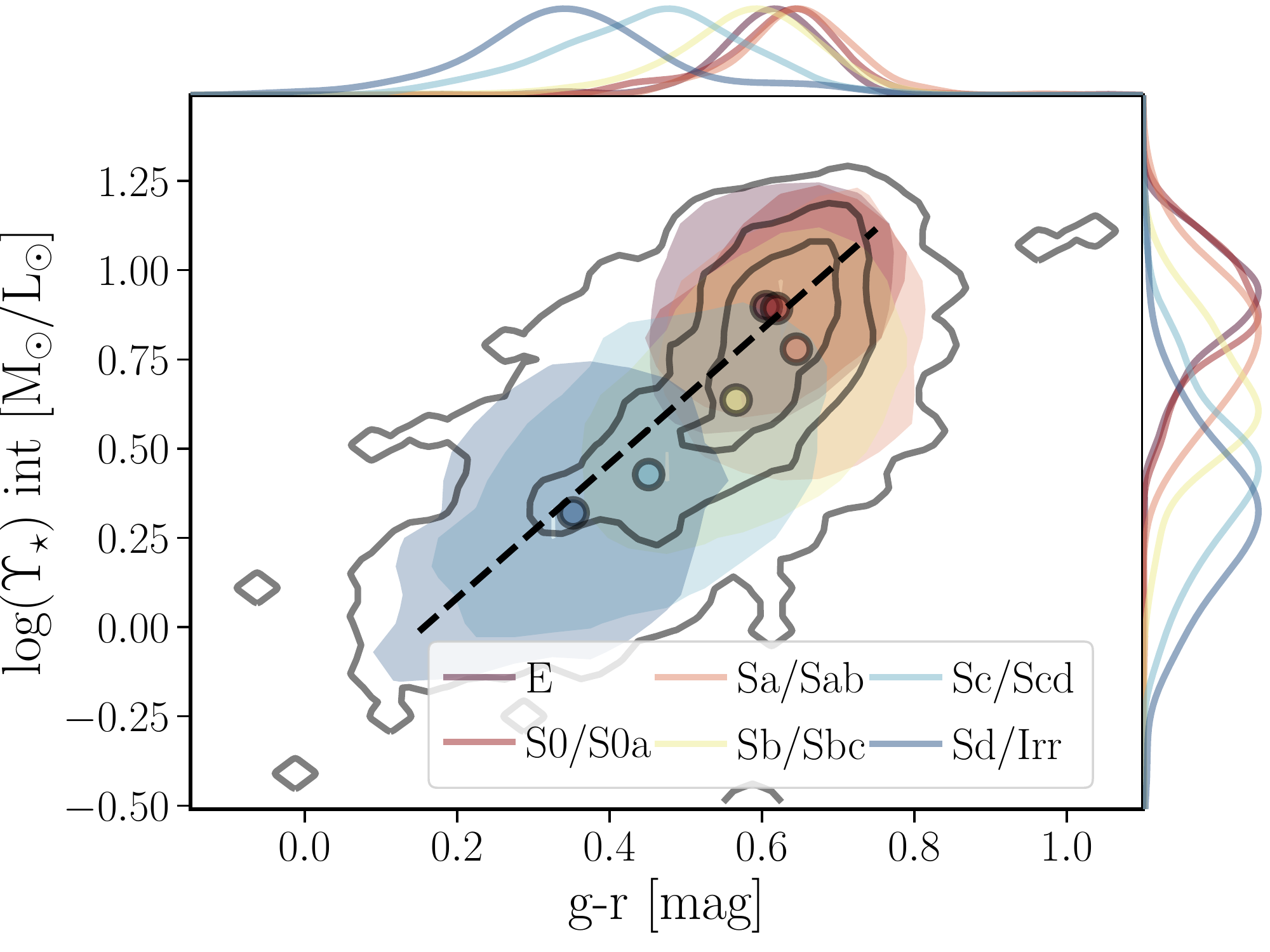}
 \endminipage
 \caption{Distribution of $\Upsilon_\star$ in the $g-$band as a function of the $g-r$ colors both derived for the integrated spectra. All contours, symbols, and colors are similar to the ones adopted in Fig. \ref{fig:c_Mass}. Dashed-line shows the $\Upsilon_\star$ vs. $g-r$ linear relation reported by \citet{rgb18} (Table 1 of that article), shifted to match the IMFs and the photometric systems. The relation match pretty-well with our reported distribution.}
 \label{fig:ML_gr}
\end{figure}

\section{Results}
\label{sec:results}

We present in this section the main results of the analysis described in the previous section, describing the main spectroscopic properties of the eCALIFA sample of galaxies for both the central regions and integrated galaxy wide.

\subsection{Stellar population properties}
\label{sec:stars}

Table \ref{tab:stars} list the main properties of the stellar
populations obtained for the central and integrated spectra ({\it cen}
and {\it int} hereafter), including (i) the LW and MW ages and
metallicities (\ageLW, \ageMW, \metLW and \metMW), (ii) the stellar
velocity dispersion ($\sigma_\star$), (iii) the dust attenuation
(${\rm A}^\star_{\rm V}$), (iv) the mass-to-light ratio
$\Upsilon_\star$ in the $g-$band, and (v) the stellar mass. The
properties derived using the integrated spectra provides with
characteristic properties of the galaxies. On the other hand, those
derived from the central aperture provides information of the most
inner regions. Although the sampled physical region change galaxy by
galaxy, in average corresponds to $\sim$500 pc. Finally, the
comparison between the parameters derived for these central spectra
and those derived using the integrated ones, provides with a rough
estimation of any gradient in the explored property. A possible caveat
is that the central aperture is (obviously) inscribed in the
integrated one. However its typical contribution to the total flux is
$\sim$7\%. Thus, based on Eq. \ref{eq:par}, the parameters derived for
the integrated spectra are contaminated by the value derived for the
central spectra in a similar percentage. In general, this pollution is not
significant in most of the cases.

\subsubsection{Stellar Masses}
\label{sec:mass}

Deriving the mass in galaxies, and in particular, the stellar mass is
a topic of the greatest interest \citep[e.g.][and references
therein]{court13}.  As described before our approach consists of
decomposing the stellar spectrum in a set of SSPs included in a
particular library and from the mass-to-light ratio each of them
($\Upsilon_{\star,SSP}$), their contribution of light to the total
flux, and correcting by the dust attenuation and cosmological
distance, it is derived the stellar mass. However, in previous
sections we already estimated the stellar mass using the photometric
information. Figure \ref{fig:c_Mass} shows a comparison between the
stellar masses derived for the integrated spectra based on the stellar
population analysis (M$_{\star,int}$), and those derived using the PS
photometry (M$_{\star,phot}$, Sec. \ref{sec:Vcor}). This is an usual
sanity check aimed to detect possible problems in the fitting
procedure, or the quality of the data, that are not obvious to a
visual inspection of the spectra themselves
\citep[e.g.][]{sanchez18,sanchez22}. Despite of the evident
differences in the two approaches adopted to estimate both stellar
masses, we find a good agreement, with a clear one-to-one
correspondence ($\Delta$ log(M$_{\star}$)=0.02$\pm$0.22 dex). As a
result of this comparison, a few galaxies were flagged from warning to
bad-quality, as indicated in Sec. \ref{sec:qc}. In addition it is
included the distributions segregated by morphology, showing the
average values too. The loose relation already reported in the
literature \citep[e.g.][]{rgb17} between morphology and stellar mass
is clearly shown, more evident in the average values, with late-type
(early-type) being less (more) massive and covering a wider (narrower)
range of masses.

\begin{figure*}
 \minipage{0.99\textwidth}
 \includegraphics[width=6.3cm]{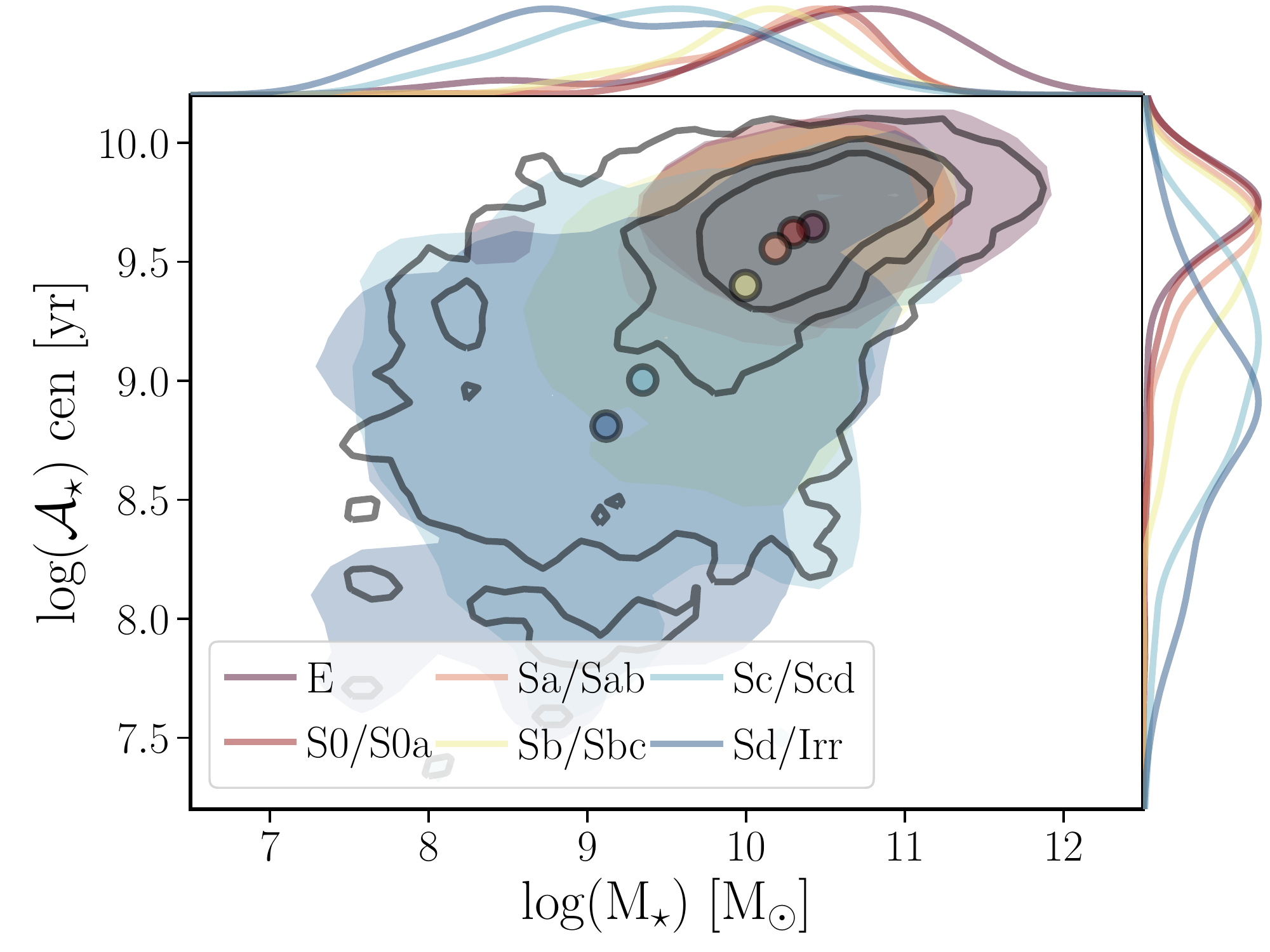}
 \includegraphics[width=6cm]{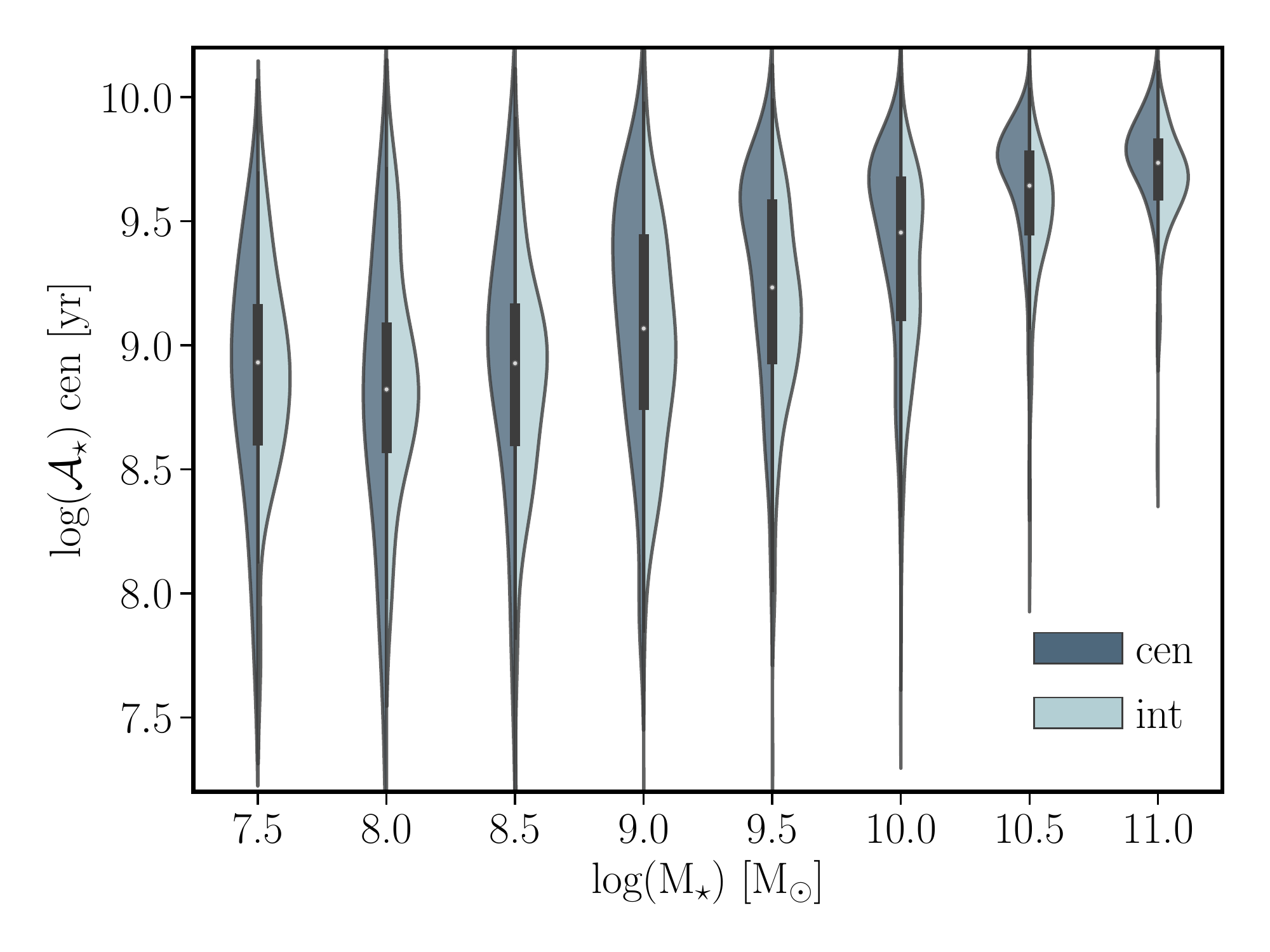}
 \includegraphics[width=6cm]{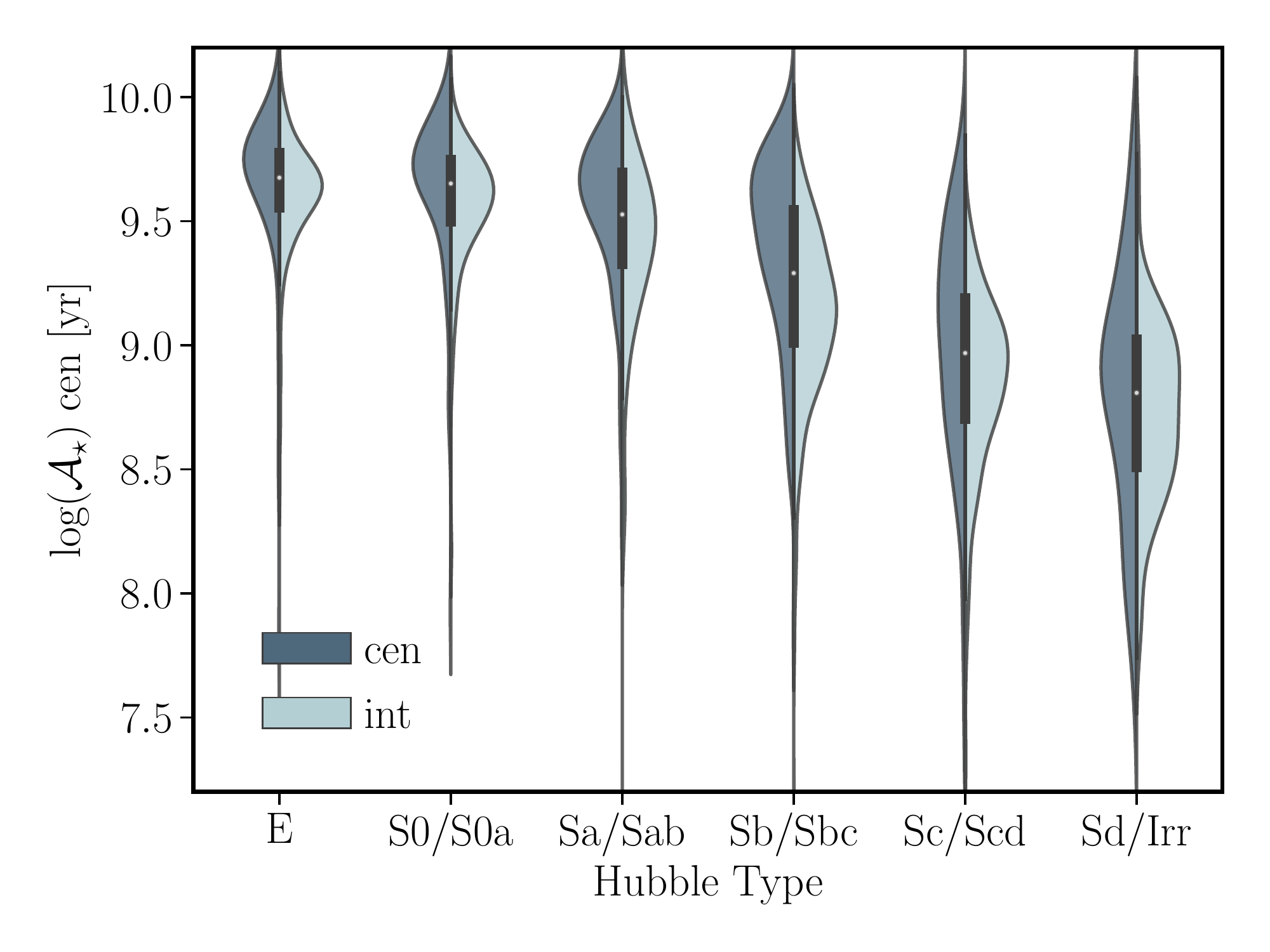}
 \includegraphics[width=6.3cm]{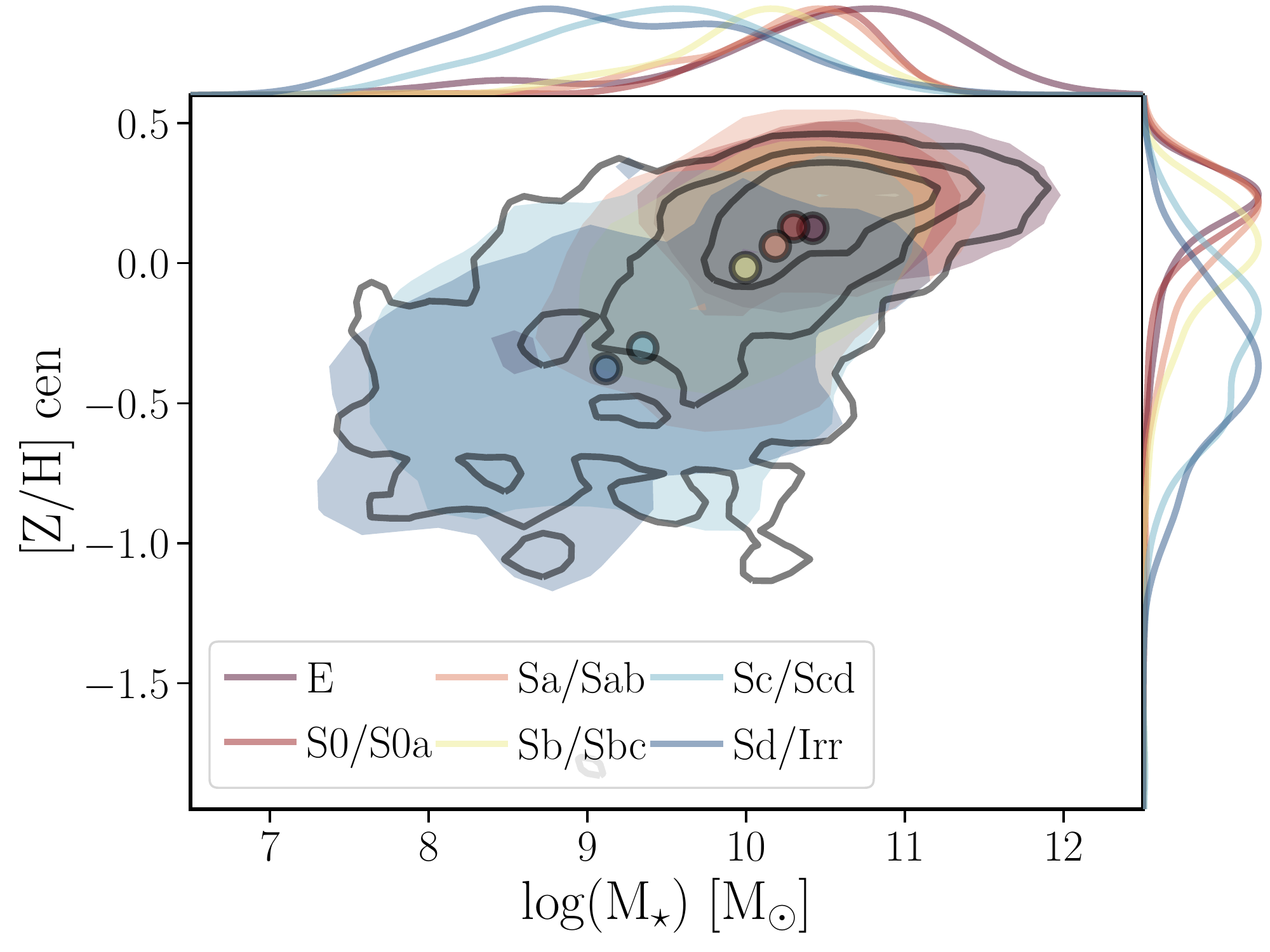}
 \includegraphics[width=6cm]{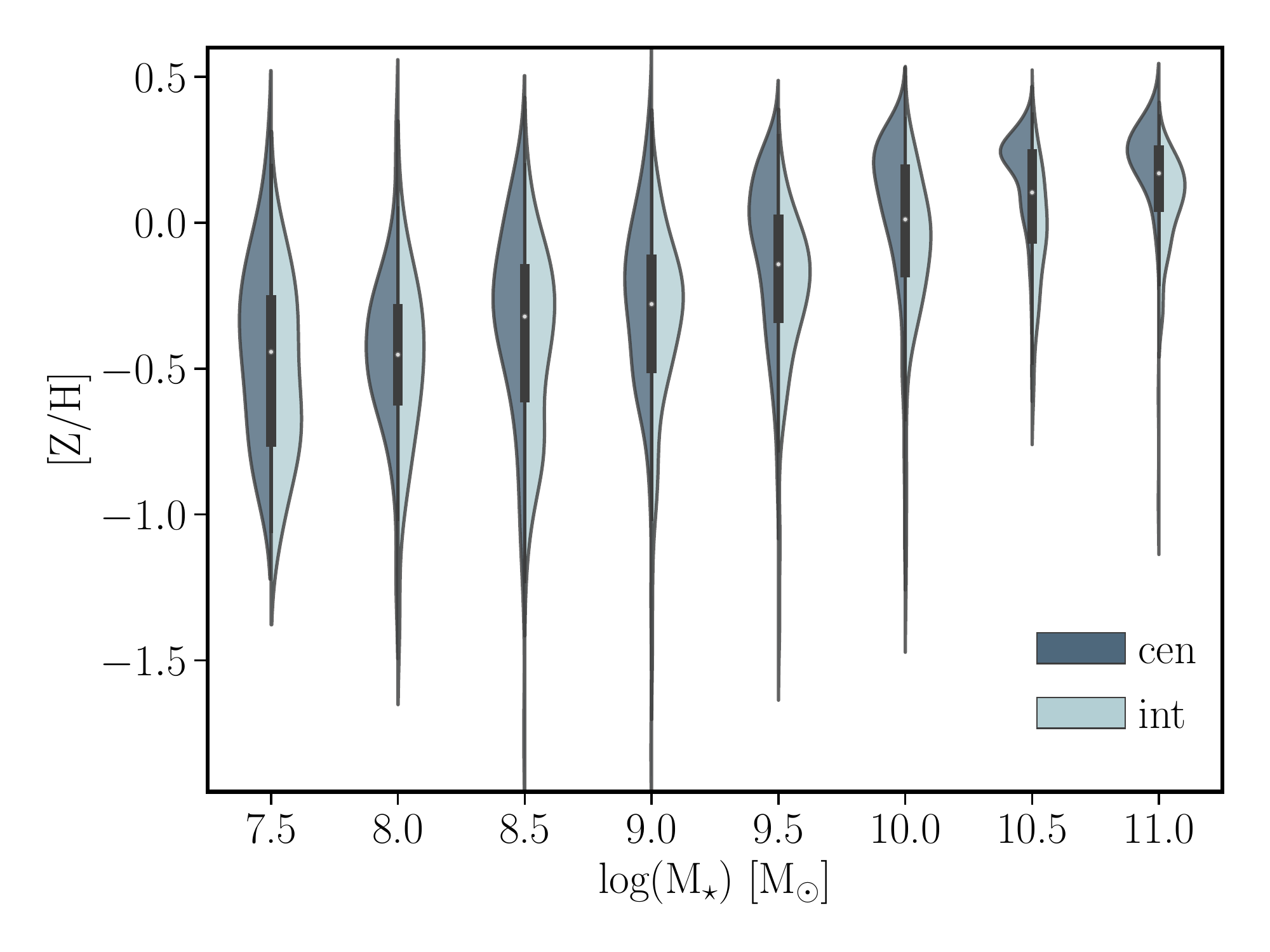}
 \includegraphics[width=6cm]{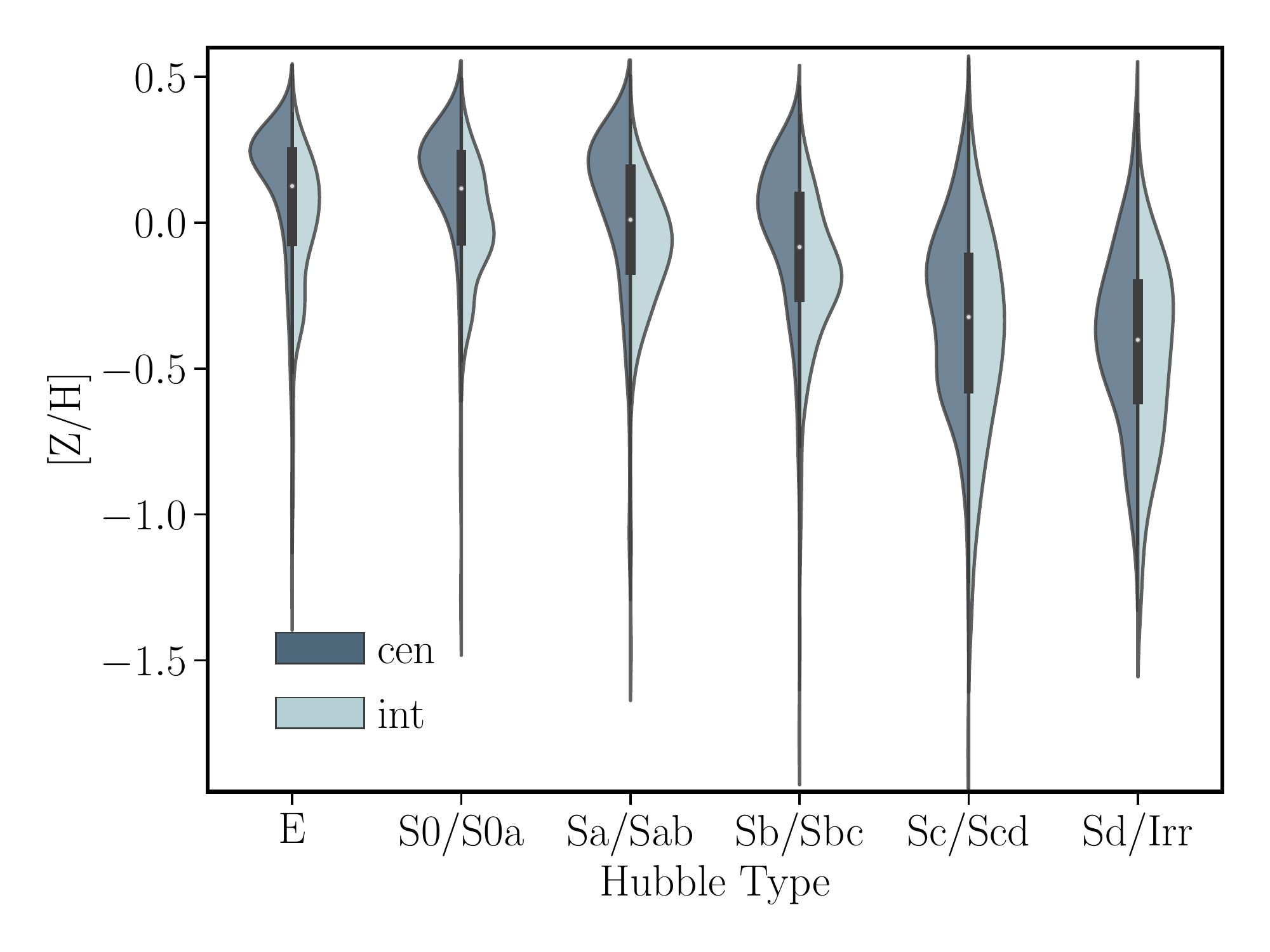}
 \endminipage
 \caption{Distribution of the \ageLW (top panels) and \metLW (bottom panels), derived for the central aperture spectra as a function of the stellar mass ({\it left panels}), together with the violin plots comparing these parameter derived for the central (cen) and integrated (int) spectra as a function of the stellar mass ({\it middle panels}) and the morphology ({\it right panels}). We use the same nomenclature adopted in Fig. \ref{fig:ML} for each panel.}
 \label{fig:AM}
\end{figure*}

\subsubsection{Mass-to-light ratios}
\label{sec:ML}

Figure \ref{fig:ML}, left panel, shows the distribution of
$\Upsilon_\star$ ($g$-band), derived for the central aperture spectra,
as a function of the stellar masses, for all galaxies and segregated
by morphology.  There is a clear trend between both parameters, with
$\Upsilon_\star$ increasing from $\sim$2 M$_\odot$/L$_\odot$ at
low-mass (late-type) galaxies to $\sim$8 M$_\odot$/L$_\odot$ at
high-mass (early-type) ones. This trend it is expected as the
$\Upsilon_\star$ traces the stellar composition in galaxies, having
higher values as older are the stars (being a consequence of young
stars having a much lower $\Upsilon$ than older ones). As more
massive/earlier type galaxies have older stellar populations
\citep[][, and references therein]{kauff03a,blanton09}, they have
larger $\Upsilon_\star$ too.  Since the pioneering studies using
broad-band photometry \citep[e.g.][]{faber1977,pele90}, to the more
recent explorations using IFS data \citep[e.g.][]{rosa14,ARAA,
  sanchez21}, it is well known that galaxies present a radial
gradients in the age of their stellar populations, leading to
variations in the $\Upsilon_\star$ from the inside-out
\citep[e.g.][]{zibetti09,rgb18}, in particular in more massive and
early-type galaxies. Fig. \ref{fig:ML} shows the violin plot of the
$\Upsilon_\star$ derived for the central and integrated apertures as a
function of the stellar mass (middle panel) and the morphology (right
panel). As expected $\Upsilon_{\star}$ has higher values in the
central regions of galaxies than integrated galaxy wide, reflecting
the gradient in the stellar composition. This trend is more clear when
galaxies are segregated by morphology, seeing

The dependence of $\Upsilon_\star$ with galaxy properties, variations from the inside out, and in particular, the analysis of the
correlations with photometric colors, is a topic that has been
addressed by many previous explorations
\citep[e.g.][]{bell00,bell03,zibetti09,gallazzi:2009,taylor:2011,into:2013,roediger:2015,rgb18}. It
is beyond the scope of the current study to make a detailed comparison
with all of them. We will focus on \citet{rgb18}, as it is using a
similar technique (spectral synthesis) for a roughly similar dataset
(the DR3 CALIFA sample), although they performed a fully spatial
resolved analysis, using a different fitting code and a different SSP
library. The main differences between the results are (1) the adopted
IMF, that implies and offset of $\sim$0.29 dex in $\Upsilon_\star$,
constant for all stellar-masses, morphologies, and colors, and (2) the
adopted photometric system for the galaxy colors, $g-r$, Vega in our
case and AB in \citet{rgb18}. This requires an additional offset, that
we took from \citet{fuku07}.

Once applied those offsets we find a very good qualitative and
quantitative agreement between our results.  For instance, the
comparison of the results shown in Fig. \ref{fig:ML} with those
presented in Fig. 2 of \citet{rgb18} (upper-panel), indicates that (i)
the trend between $\Upsilon_\star$ and morphology, (ii) the values
reported for the central regions, and (iii) the average values galaxy
wide for each morphological type agree within a few percentage between
both studies. A more quantitative comparison is included in Figure
\ref{fig:ML_gr}, where it is shown the distribution of
$\Upsilon_\star$ as a function of the $g-r$ color, for the total
sample and segregated by morphology. We include in this figure the
relation reported by \citet{rgb18} for both parameters, once
considered the offsets required to take into account the differences
in the IMF and photometric systems described before. There is a
remarkable agreement between the published relation and the distribution
traced by our measurements.

\subsubsection{Age and metallicity}
\label{sec:AM}

The decomposition of a stellar population in a set of SSPs naturally
provides with an age and metallicity distribution function (ADF and
MDF) in both light and mass, as is illustrated in the bottom panel of
Fig. \ref{fig:FIT3D}.  Since the early studies by \citet{panter03},
these distributions are used to trace the mass-assembly, chemical
enrichment and star-formation histories of galaxies and regions within
them \citep[e.g.][]{asari07,eperez13,ibarra16,camps20,camps22}, as
recently reviewed by \citet{ARAA}. Those distributions can be studied
to compare them with results provided by resolved stellar population,
in particular the MDF \citep[e.g.][]{mejia20}. In some cases, instead
of analyzing the full distributions, it is more convenient to collapse
them into characteristics value such as the luminosity- or
mass-weighted values (i.e., those listed in Table
\ref{tab:stars}). This is particular useful when exploring radial
gradients in those properties or dependencies on global properties
\citep[e.g.][]{gallazzi05,gallazzi06,rosa14,rosa15,godd15,ARAA,sanchez21}.
We should keep in mind that they only provide with a limited
information of the real distribution of those quantities, and that the
information is different depending on (1) how the average is performed
(in a linear or logarithm space), (2) at which wavelength is
normalized the LW, and/or (3) if it is weighted by mass or by
light. In general, LW ages and metallicities, when normalized at
$\sim$5000\AA\ highlights the contribution of young stars with respect
to an underlying content of old stars. This way, \ageLW correlates
pretty well with blue-to-red colors, stellar indices as D4000, the
$\Upsilon_\star$ \citep[e.g.][]{kauff03a,blanton09}, and \metLW
present some trends with the gas-phase oxygen abundance
\citep[e.g.][]{rosa14b,espi22}. On the other hand, MW ages and
metallicities are more representative of the bulk stellar population,
and in general they present higher values \citep[e.g.][]{sanchez21},
being less suitable to explore changes in the stellar populations
sensitive to the presence of young stars.

Figure \ref{fig:AM}, left-panels, shows the distribution of \ageLW
(top panels) and \metLW (bottom panels) derived for the central
aperture as a function of the stellar masses for the full sample of
galaxies, and segregated by morphology. We find a clear trend between
both parameters and both the mass and morphology, with
low-mass/late-type galaxies having a younger/metal-poor stellar
population and high-mass/early-type galaxies having an older/meta-rich
one. In this regards we reproduce well known trends, already described
using aperture limited spectroscopic surveys
\citep[e.g.][]{blanton09,gallazzi06} or integral field-spectroscopy
\citep[e.g.][]{rosa14b,rosa15}. These trends are the consequence of
global downsizing: more massive/early-type galaxies form the bulk of
their stars more early in the cosmological time, having a sharper,
more peaky and shorter SFH than less massive/late-type ones
\citep[e.g.][]{pperez08}. These different SFHs (and mass assembly
history, MAH) produce a different chemical enrichment history (ChEH)
and a final accumulated metallicity for galaxies of different mass and
morphology \citep{panter03}. Recent results have shown that beyond
these average trends, less-massive/later-type galaxies present a wider
range of SFHs (MAHs) and ChEHs than more-massive/ealier-types
\citep{ibarra16,rgb17}. As a consequence, the former present a wider
distribution of \ageLW and \metLW, as indeed it is observed in
Fig. \ref{fig:AM}, and previous results
\citep[e.g.][]{rosa14b}. Furthermore, the shape of the ChEH seem to be
equally to the morphology and to the stellar mass
\citep{camps20,camps22}.

Beside the different patterns in the average evolution of the stellar
populations galaxy-wide for galaxies of different stellar mass and
morphology, there are differences observed at different regions within
galaxies: bulge-dominated/inner-regions, those with higher stellar
mass density ($\Sigma_\star$), present a sharper evolution than
disk-dominated/outer-regions, i.e., those with lower
$\Sigma_\star$. This local downsizing \citep{eperez13}, observed as a
change in the local SFHs (MAHs) and ChEHs from the inside-out, induce
a gradient in the radial distributions of the stellar properties.
Fig. \ref{fig:AM}, shows the violin plots of the distribution of
\ageLW (top panels) and \metLW (bottom panels) for both the central
and integrated apertures, as a function of the stellar masses (central
panels) and morphologies (bottom panels). In the case of the \ageLW,
for all morphologies and for M$_\star>$10$^{8.0}$M$_\odot$, the inner
stellar populations are always older than the outer ones. A somehow
similar pattern is observed in \metLW, although in this case low-mass
($<$10$^{8.5}$M$_\odot$) and later-type galaxies (Sd/Irr), present
similar distributions in both apertures.  These patterns can be
interpreted as an inside-out growth of galaxies earlier than Sd and
more massive than M$_\star>$10$^{8.0-8.5}$M$_\odot$.

\begin{figure*}
 \minipage{0.99\textwidth}
 \includegraphics[width=6.3cm]{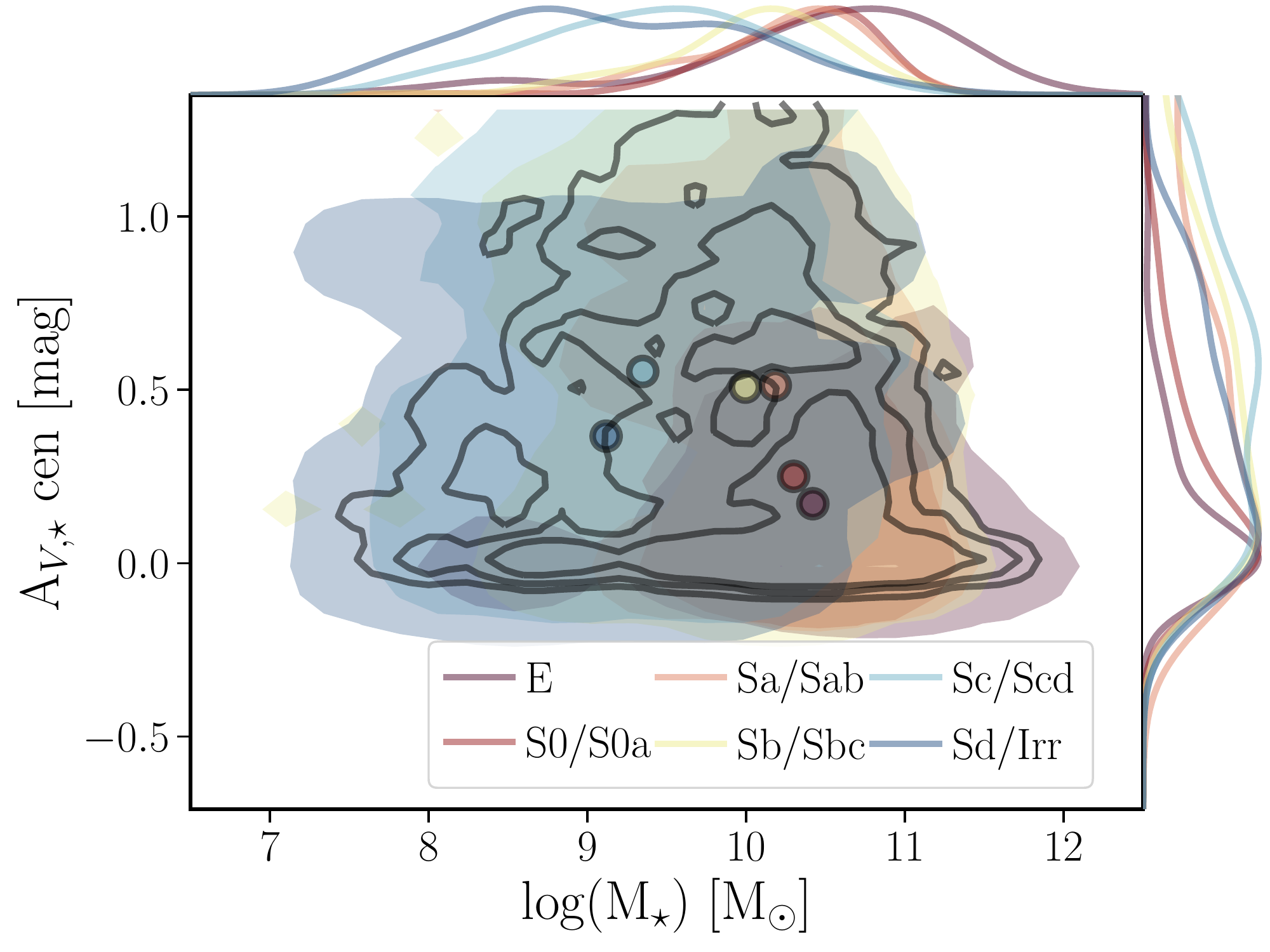}
 \includegraphics[width=6cm]{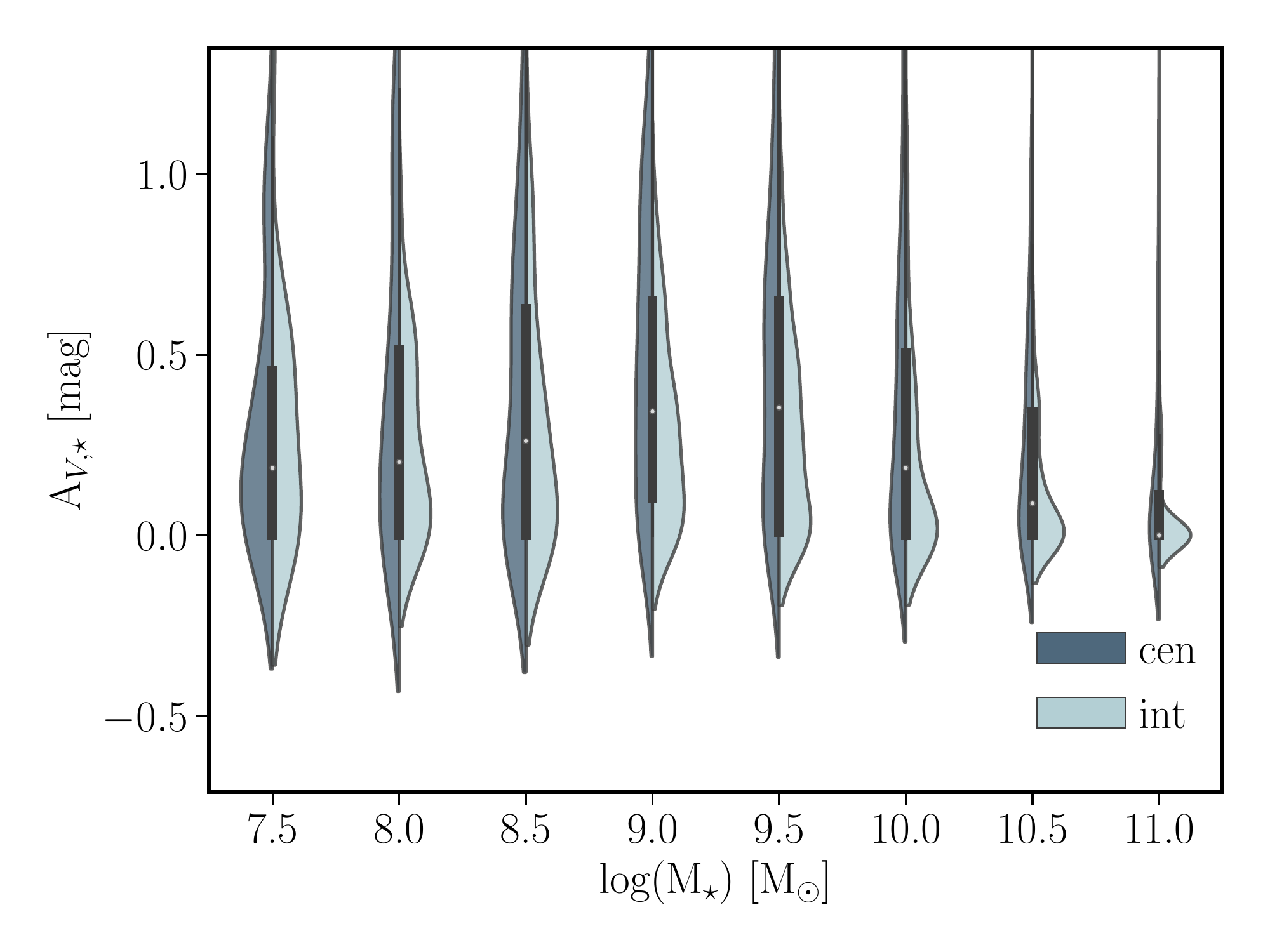}
 \includegraphics[width=6cm]{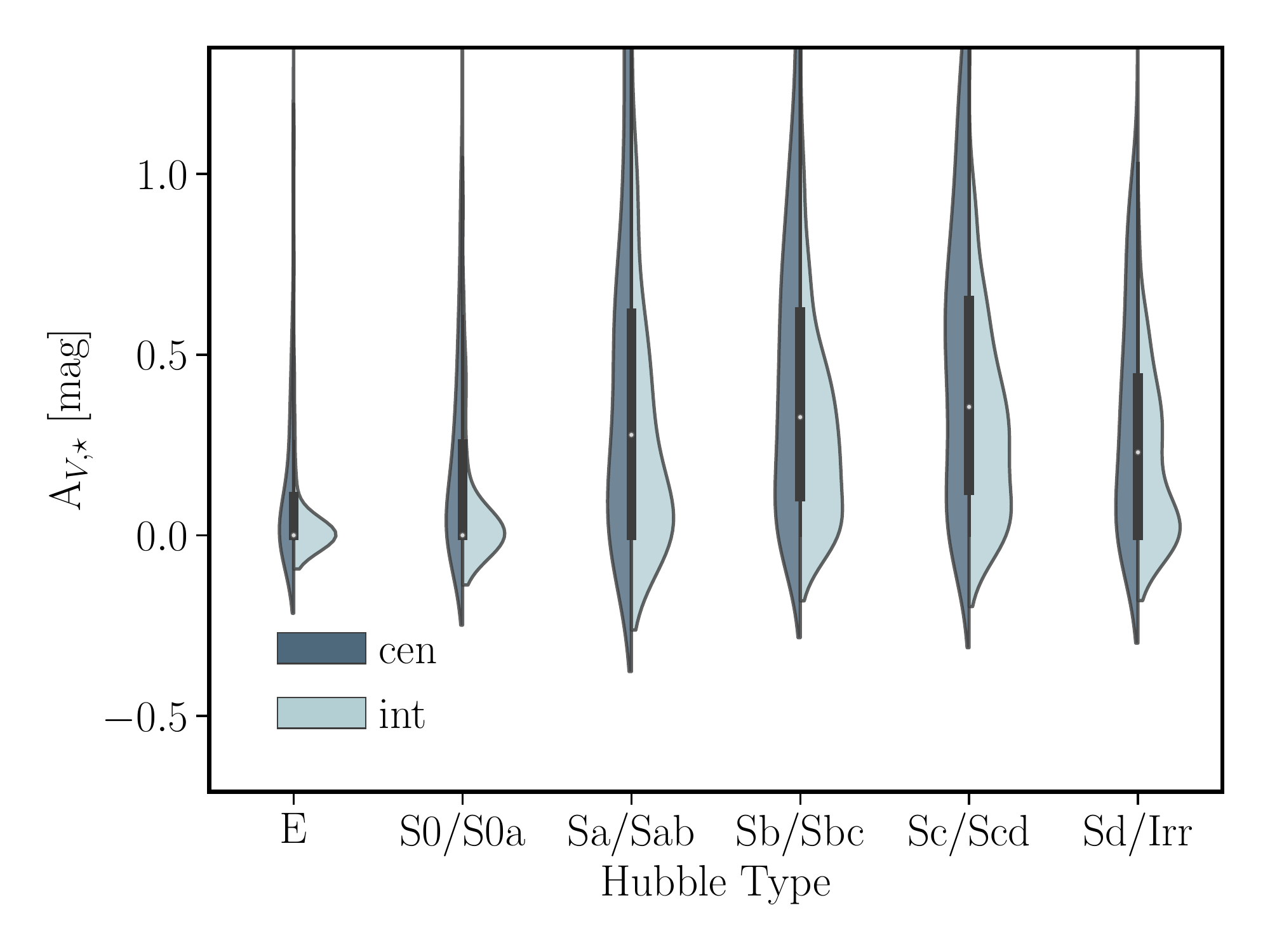}
 \endminipage
 \caption{Distribution of A$_{\star,\rm V}$ derived for the central aperture spectra as a function of the stellar mass ({\it left panel}), together with the violin plots comparing this parameter derived for the central (cen) and integrated (int) spectra as a function of the stellar mass ({\it middle panel}) and the morphology ({\it right panel}). We use the same nomenclature adopted in Fig. \ref{fig:ML} for each panel.}
 \label{fig:Av}
\end{figure*}

The inside-out growth is thought to be connected on how disks acquired
their angular momentum when formed \citep{peebles1969,
  larson1976}. However, the differences in the stellar-populations may
be also connected with an inside-out quenching/aging of the stellar
populations too \citep[e.g.][]{rosa16,sanchez18,ARAA}, which may be
connected to a lack of molecular gas in the inner regions of galaxies,
dynamical stability in bulges, the effect of AGNs, or a combined
effect of all of them \citep[for instance, see the discussion in
][]{bluck19}. For less massive/later type galaxies an outside-in or
homogeneous growth could explain the observed distribution. In general,
it is required to invoke a difference in gas inflow for different
galaxy types and different galactocentric distances to explain this
kind of gradients \citep[e.g.][and references therein]{carigi19}. As
already discussed in the recent review summarizing the results from
different IFS galaxy survey \citep{ARAA}, these scenarios refer to
how fast the star-formation happens in different radial distances with
time, not to the amount of stars. In all galaxies it is found a
negative gradient in the stellar-mass density. Thus, in all of them
there are more stars formed in the central regions along the time than
in the outer ones.

\begin{figure*}
 \minipage{0.99\textwidth}
 \includegraphics[width=6.3cm]{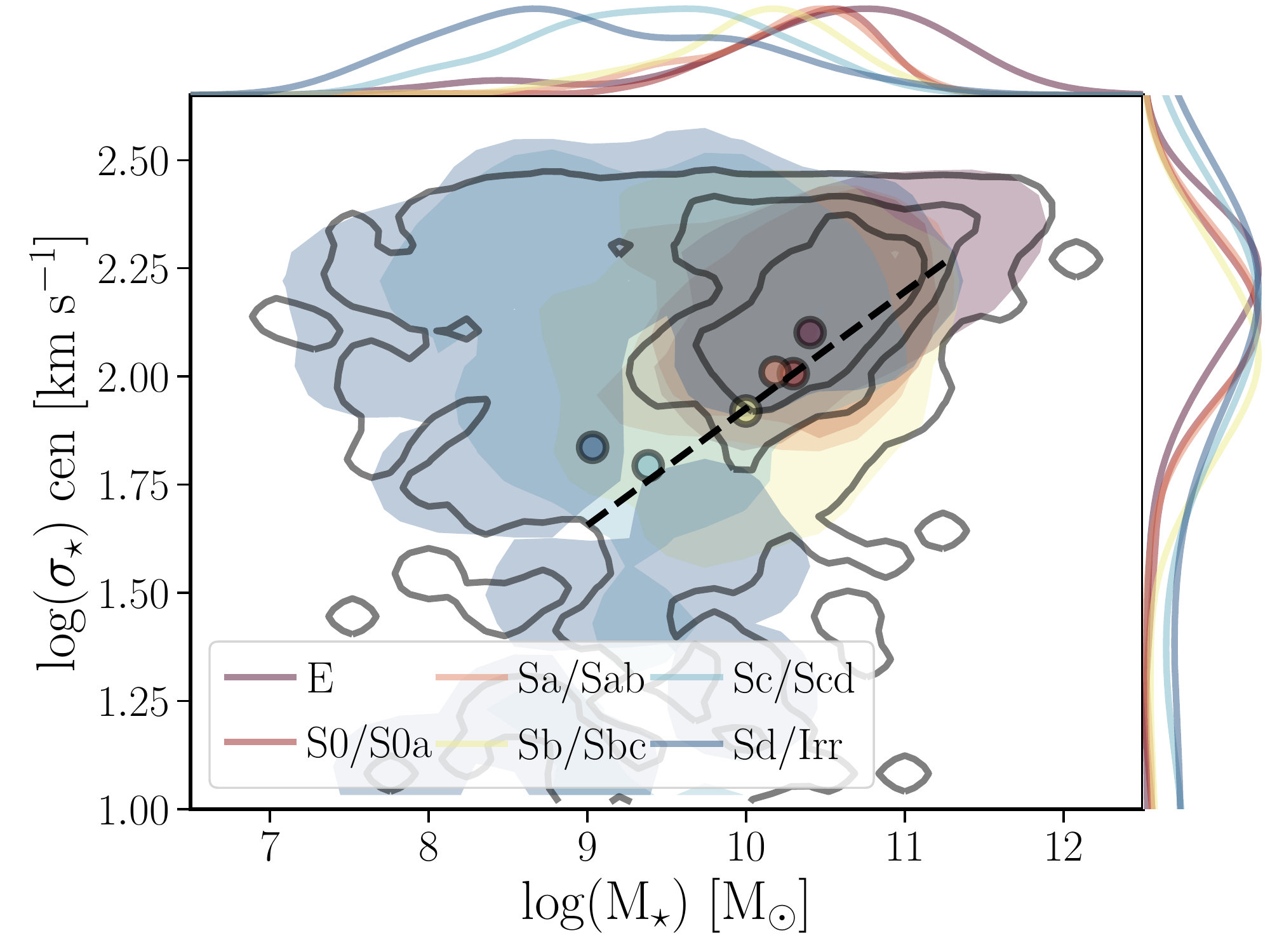}
 \includegraphics[width=6cm]{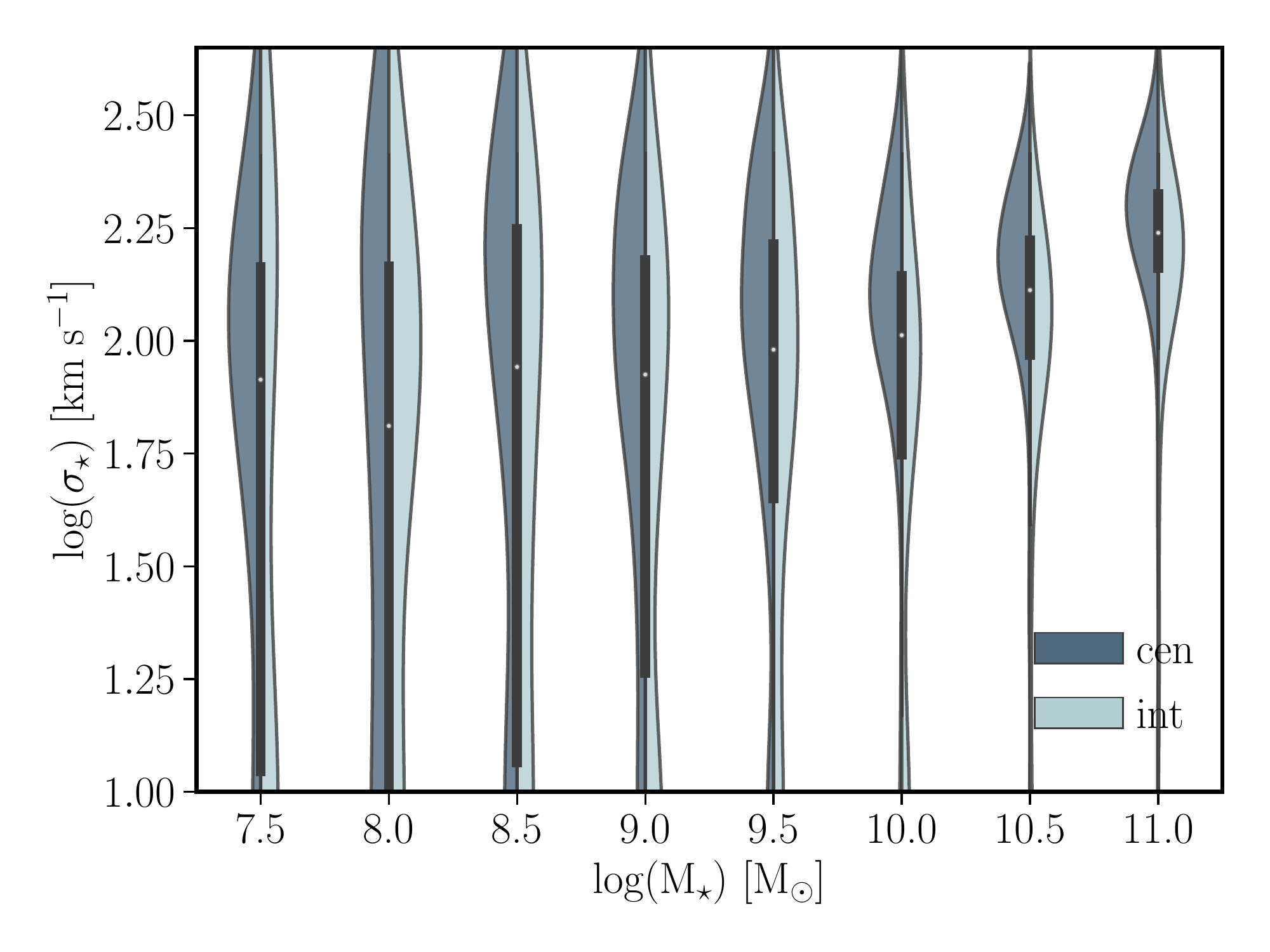}
 \includegraphics[width=6cm]{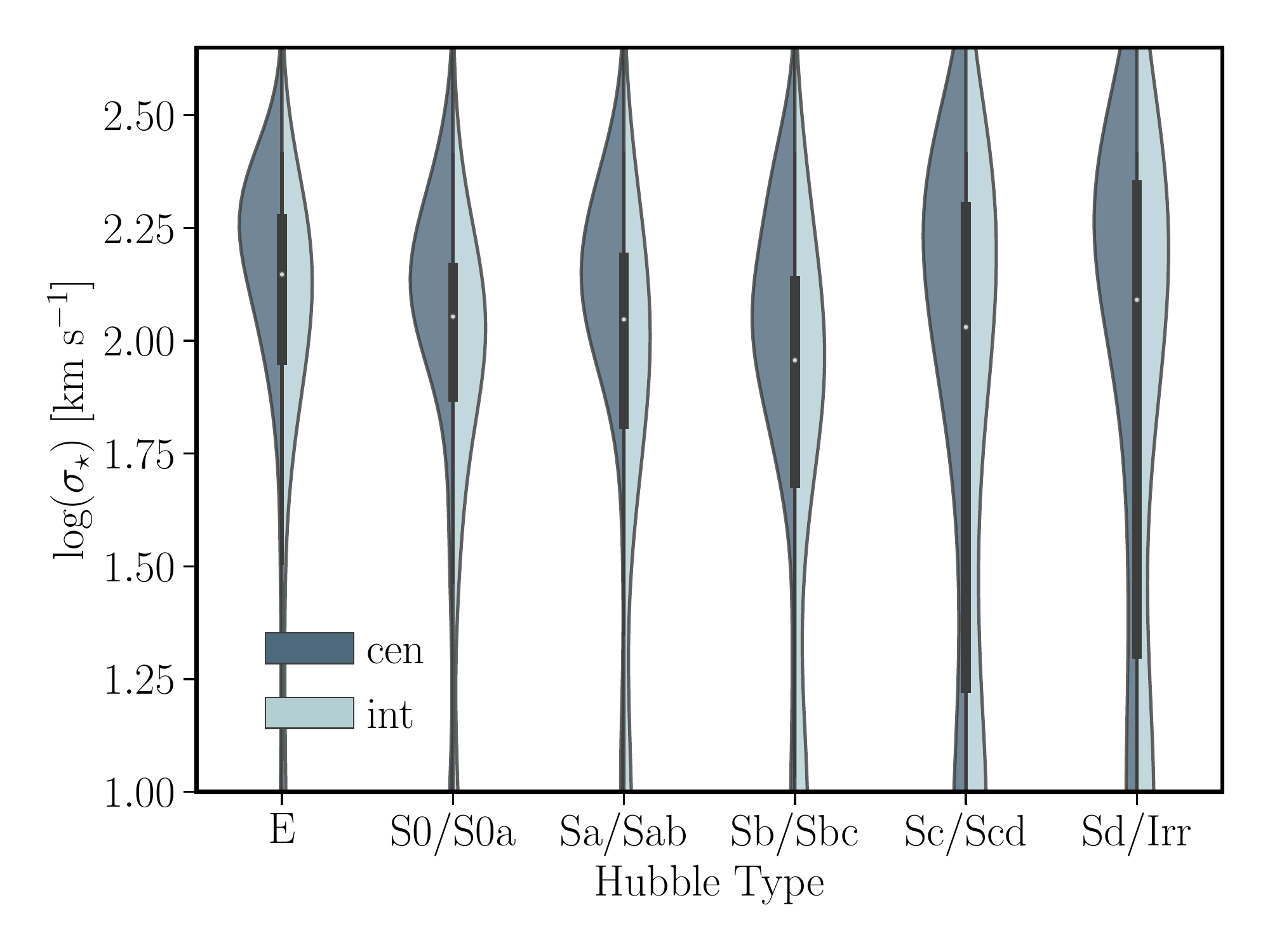}
 \endminipage
 \caption{Distribution of $\sigma_\star$ derived for the central aperture spectra as a function of the stellar mass ({\it left panel}), together with the violin plots comparing this parameter derived for the central (cen) and integrated (int) spectra as a function of the stellar mass ({\it middle panel}) and the morphology ({\it right panel}). We use the same nomenclature adopted in Fig. \ref{fig:ML} for each panel. In addition, the dashed-black line in the left panel corresponds to the Faber-Jackson relation derived for the CALIFA sample \citep{erik18}.}
 \label{fig:disp}
\end{figure*}

\subsubsection{Stellar dust attenuation}
\label{sec:Av_st}

Dust is one of the components of the ISM with the lowest contribution
to its total mass \citep[$<$1\%, e.g.,][]{santini14}. However, it has
an strong impact in the shape of the spectral energy distribution and
in the redistribution of photons from the UV-optical (that are
absorbed and attenuated) to the infrared (due to dust thermal
emission). Dust grains are condensation of metals that are formed in
the atmospheres of evolved stars and supernovae remnants, and
therefore it is a tracer of the evolution of stellar populations
too. Being usually aggregated to molecular clouds, it is as a tracer
of H$_{\rm 2}$ column density, and therefore, of the molecular gas
content \citep[e.g.][]{brin14,jkbb20}. It is known that it affects in
a different way the emission from the ionized gas, in which a pure
screen model provides good estimations of the dust absorption, than to
the integrated stellar populations, where dust is partially embedded
and not fully covering the stellar components
\citep[e.g.][]{calz01,wild11,salim20}. As indicated before {\sc
  pyFIT3D} provides with an estimation of the dust attenuation (A$_{\star,V}$)
affecting the stellar spectrum, assuming that all components of the
stellar population are affected by the same attenuation and adopting a
simple screen model.

Figure \ref{fig:Av}, left-panel, shows the distribution of these dust
attenuation derived for the central aperture spectra as a function of
the stellar mass, for all galaxy types and segregated by
morphology. On the contrary to the parameters explored in the previous
sections there is no simple/clear trend/pattern traced by these
distributions.  For most galaxy types the distribution is centred in
A$_{\star,V}\sim$0 mag, with a narrow range of A$_{\star,V}$ values
around this central one for more massive and early-type galaxies (E
and S0). For later-type, less massive galaxies, the distribution
presents a tail towards larger values. However, the average value is
never higher than $\sim$0.25 mag, and there is no further trend with
the stellar mass or the morphology. The violin plots comparing the
central and integrated values as a function of the stellar mass and
the morphology (Fig. \ref{fig:Av}, middle and right panels,
respectively), show that, in general, the distribution of values is
even more concentrated around zero for the integrated spectra than for
the central ones. This trend is modulated by mass and morphology,
being less clear for E and S0 galaxies, in which both values of
A$_{\star,V}$ are rather low.

If we interpret the dust attenuation as proxy of the molecular gas
content, these results would indicate that more massive/early-type
galaxies, and the outer regions of all galaxies, present a relatively
lower gas content than less massive/late-type galaxies, and the inner
regions of all galaxies. This is indeed has been confirmed by many
previous explorations of the molecular gas content in galaxies
\citep[e.g.][]{saint16,calette18}. However, due to the limitations of
our modelling of the dust attenuation affecting the stellar component, we
should take this result with a certain precaution. We will come back to that
when exploring the dust attenuation derived for the ionized gas.

\subsubsection{Velocity dispersion}
\label{sec:disp}

The analysis provides with the velocity dispersion of the stellar
component ($\sigma_\star$) for the two apertures. The instrumental
resolution of the adopted setup (V500 grating, R$\sim$850),
corresponds to $\sim$120 \kms\ at $\sim$5500\AA. Our simulations
indicate that in case of good S/N the code recovers a reliable
velocity dispersion above $\sim$1/3 of this value \citep[$\sim$40\kms,
for the current data,][]{pypipe3d}. However, it is unable to recover
values below this value.

Figure \ref{fig:disp}, left panel, shows the distribution of
$\sigma_\star$, in logarithm scale, derived for the central aperture
as a function of the stellar mass for all galaxies and segregated by
morphology. Despite of the large scatter, in particular for
low-mass/late-type galaxies, it is observed a clear monotonic increase
of $\sigma_\star$ with both parameters, with a linear trend traced by
the average values (solid circles) for galaxies earlier than
Sd/Irr. \citet{faber76} first shown a tight correlation between the
central velocity dispersion and the integrated luminosity in
early-type galaxies (FJ relation). This relation is supposed to be the
correspondent to the relation between the rotational velocity
and luminosity found for late-type galaxies \citep[TF
relation][]{TF77}. Both of them can be described as a relation between
the corresponding kinematic parameter and the stellar mass (and the
baryonic mass in general), and being unified into a single relation
valid for all morphological types
\citep[e.g.][]{wein06,corte14,erik18}. The usual explanation for both relations
suggest that (i) the central velocity dispersion (rotational velocity) is a good tracer
of the dynamical mass in pressure supported (rotational supported) early-type (late-type) galaxies, and (ii) there is a tight relation between the dynamical mass traced by those parameters and the stellar mass.

When using all morphological types (Fig. \ref{fig:disp}) the relation
between $\sigma_{\star,\rm cen}$ and M$_\star$ is broader, less well
defined than the FJ-relation, showing a larger dispersion for
late-type galaxies \citep[e.g., Fig. 3 of][]{erik18}. However, for
early-type galaxies and early-spirals the average values shown in
Fig. \ref{fig:disp} trace very well the stellar-mass FJ relation
reported in the literature.

Like in the case of other stellar properties, $\sigma_\star$ present
known radial gradients.  This is seen in the violin plots comparing
the central and integrated values for this parameter as a function of
the stellar mass and the morphology (Fig. \ref{fig:disp}, central and
right panel, respectively). All galaxies more massive than
M$_\star>$10$^{9.0}$M$_\odot$, and earlier than Sc (i.e., galaxies
with a massive bulge), have a larger velocity dispersion in the
central aperture than integrated galaxy-wide. However, for
disk-dominated and/or low-mass galaxies there is no evident gradient
velocity dispersion. A possible caveat in here is that in this later case
the reported values are near the minimum $\sigma_\star$ that we
can reliably recover for the current dataset with our code. In any case,
if there is some gradient it is much smoother than the one described
for the more massive and earlier type galaxies.

\begin{table*}
\begin{center}
\caption{Emission line fluxes in units of \fCAL estimated for the central and integrated spectra of all the galaxies.}
\begin{tabular}{ccrrrrrrrrrr}\hline\hline
  cubename & aperture  & \oii 3727 &  \Hb & \oiii 5007 & \Ha & \nii 6584 & \sii 6717 & \sii 6731 \\
  \hline
IC5376 & int  & 399.4 $\pm$ 114.8 & 113.3 $\pm$ 17.0 & 60.9 $\pm$ 9.1 & 531.2 $\pm$ 79.7 & 227.0 $\pm$ 34.0 & 108.6 $\pm$ 16.3 & 72.1 $\pm$ 10.8\\
 " & cen  & 0.9 $\pm$ 18.4 & 1.1 $\pm$ 0.6 & 2.4 $\pm$ 0.6 & 4.5 $\pm$ 0.9 & 6.0 $\pm$ 0.9 & 1.2 $\pm$ 0.9 & 1.0 $\pm$ 0.9\\
UGC00005 & int  & 696.9 $\pm$ 104.5 & 352.4 $\pm$ 52.9 & 129.4 $\pm$ 19.4 & 1527.2 $\pm$ 229.1 & 657.5 $\pm$ 98.6 & 369.2 $\pm$ 55.4 & 264.8 $\pm$ 39.7\\
 " & cen  & 5.3 $\pm$ 3.1 & 6.5 $\pm$ 1.0 & 31.1 $\pm$ 4.7 & 31.2 $\pm$ 4.7 & 37.8 $\pm$ 5.7 & 10.3 $\pm$ 1.5 & 7.9 $\pm$ 1.2\\
NGC7819 & int  & 664.0 $\pm$ 99.6 & 370.7 $\pm$ 55.6 & 125.8 $\pm$ 18.9 & 1379.3 $\pm$ 206.9 & 475.2 $\pm$ 71.3 & 259.7 $\pm$ 39.0 & 177.0 $\pm$ 26.6\\
 " & cen  & 27.3 $\pm$ 4.4 & 46.1 $\pm$ 6.9 & 6.0 $\pm$ 0.9 & 205.1 $\pm$ 30.8 & 74.9 $\pm$ 11.2 & 22.8 $\pm$ 3.4 & 19.6 $\pm$ 2.9\\
UGC00029 & int  & 58.1 $\pm$ 37.3 & 13.5 $\pm$ 16.1 & 16.4 $\pm$ 16.2 & 13.5 $\pm$ 73.8 & 12.4 $\pm$ 73.8 & 59.6 $\pm$ 73.8 & 42.6 $\pm$ 73.8\\
 " & cen  & 0.0 $\pm$ 3.1 & 0.0 $\pm$ 0.6 & 0.0 $\pm$ 0.6 & 1.5 $\pm$ 3.3 & 4.0 $\pm$ 3.3 & 4.4 $\pm$ 3.3 & 1.8 $\pm$ 3.3\\
IC1528 & int  & 451.6 $\pm$ 67.7 & 379.5 $\pm$ 56.9 & 185.7 $\pm$ 27.9 & 1504.6 $\pm$ 225.7 & 534.0 $\pm$ 80.1 & 333.5 $\pm$ 50.0 & 238.9 $\pm$ 35.8\\
 " & cen  & 4.9 $\pm$ 0.7 & 7.0 $\pm$ 1.0 & 1.3 $\pm$ 0.2 & 30.7 $\pm$ 4.6 & 11.7 $\pm$ 1.8 & 4.6 $\pm$ 0.7 & 3.8 $\pm$ 0.6\\
NGC7824 & int  & 491.8 $\pm$ 99.7 & 182.7 $\pm$ 27.4 & 147.7 $\pm$ 23.5 & 412.6 $\pm$ 61.9 & 412.6 $\pm$ 61.9 & 96.9 $\pm$ 31.9 & 133.1 $\pm$ 33.1\\
 " & cen  & -8.4 $\pm$ 2.1 & 3.0 $\pm$ 1.3 & 1.9 $\pm$ 1.2 & 2.9 $\pm$ 0.5 & 2.3 $\pm$ 0.5 & 0.0 $\pm$ 0.6 & 1.0 $\pm$ 0.5\\
UGC00036 & int  & 93.4 $\pm$ 695.2 & 93.8 $\pm$ 14.1 & 45.7 $\pm$ 11.9 & 466.9 $\pm$ 70.0 & 297.4 $\pm$ 44.6 & 53.2 $\pm$ 21.4 & 135.1 $\pm$ 21.4\\
 " & cen  & 14.1 $\pm$ 53.3 & 5.8 $\pm$ 0.9 & 4.6 $\pm$ 0.9 & 39.0 $\pm$ 5.8 & 26.1 $\pm$ 3.9 & 7.5 $\pm$ 1.4 & 6.7 $\pm$ 1.3\\
NGC0001 & int  & 503.6 $\pm$ 75.5 & 454.5 $\pm$ 68.2 & 114.4 $\pm$ 17.2 & 2140.6 $\pm$ 321.1 & 834.4 $\pm$ 125.2 & 323.3 $\pm$ 57.8 & 197.2 $\pm$ 57.8\\
 " & cen  & 3.6 $\pm$ 1.7 & 16.3 $\pm$ 2.4 & 4.6 $\pm$ 0.9 & 78.6 $\pm$ 11.8 & 33.1 $\pm$ 5.0 & 8.2 $\pm$ 1.5 & 5.3 $\pm$ 1.5\\
NGC0023 & int  & 2131.7 $\pm$ 319.8 & 1585.1 $\pm$ 237.8 & 666.2 $\pm$ 99.9 & 7764.9 $\pm$ 1164.7 & 3924.1 $\pm$ 588.6 & 1506.4 $\pm$ 226.0 & 1143.4 $\pm$ 171.5\\
 " & cen  & 59.4 $\pm$ 8.9 & 113.3 $\pm$ 17.0 & 39.8 $\pm$ 6.0 & 548.2 $\pm$ 82.2 & 324.4 $\pm$ 48.7 & 97.1 $\pm$ 14.6 & 80.4 $\pm$ 12.1\\
NGC0036 & int  & 475.6 $\pm$ 99.3 & 367.0 $\pm$ 55.0 & 144.0 $\pm$ 21.6 & 1654.7 $\pm$ 248.2 & 744.6 $\pm$ 111.7 & 302.7 $\pm$ 45.4 & 198.8 $\pm$ 45.1\\
 " & cen  & 2.3 $\pm$ 2.1 & 1.1 $\pm$ 0.6 & 3.1 $\pm$ 0.6 & 5.9 $\pm$ 1.2 & 13.2 $\pm$ 2.0 & 2.9 $\pm$ 1.2 & 2.0 $\pm$ 1.2\\
UGC00139 & int  & 934.6 $\pm$ 140.2 & 352.4 $\pm$ 52.9 & 220.0 $\pm$ 33.0 & 807.4 $\pm$ 121.1 & 236.8 $\pm$ 35.5 & 165.8 $\pm$ 26.2 & 109.4 $\pm$ 25.9\\
 " & cen  & 0.3 $\pm$ 8.1 & 4.0 $\pm$ 0.6 & 0.8 $\pm$ 0.2 & 14.1 $\pm$ 2.1 & 3.2 $\pm$ 1.3 & 2.7 $\pm$ 1.2 & 1.8 $\pm$ 1.2\\
MCG-02-02-030 & int  & 475.8 $\pm$ 141.4 & 388.8 $\pm$ 58.3 & 156.2 $\pm$ 23.4 & 1290.7 $\pm$ 193.6 & 605.2 $\pm$ 90.8 & 281.6 $\pm$ 56.8 & 200.5 $\pm$ 56.1\\
 " & cen  & 3.1 $\pm$ 2.0 & 2.8 $\pm$ 0.6 & 8.0 $\pm$ 1.2 & 24.8 $\pm$ 3.7 & 18.6 $\pm$ 2.8 & 6.9 $\pm$ 2.1 & 6.4 $\pm$ 2.1\\
\hline
\end{tabular}\label{tab:gas} 
We present the values for just ten galaxies. The information for the remaining one is distributed electronically.
\end{center}
\end{table*}

\subsection{Properties of the ionized gas}
\label{sec:ion}

As indicated before, the analysis performed by {\sc pyFIT3D} extracts
the main properties of a set of defined emission lines that were described in Sec. \ref{sec:spec_fit}. Table \ref{tab:gas}
lists the fluxes of the strongest emission lines for each galaxy in
both apertures, with their corresponding errors. No mask have been applied
to the data. Therefore, some reported fluxes may be negative or zero in this table.
We advice the reader to apply the required masks based on the reported errors depending on the requirements of each particular exploration.

\begin{figure*}
 \minipage{0.99\textwidth}
 \includegraphics[width=8.5cm,trim={0 64 0 0},clip]{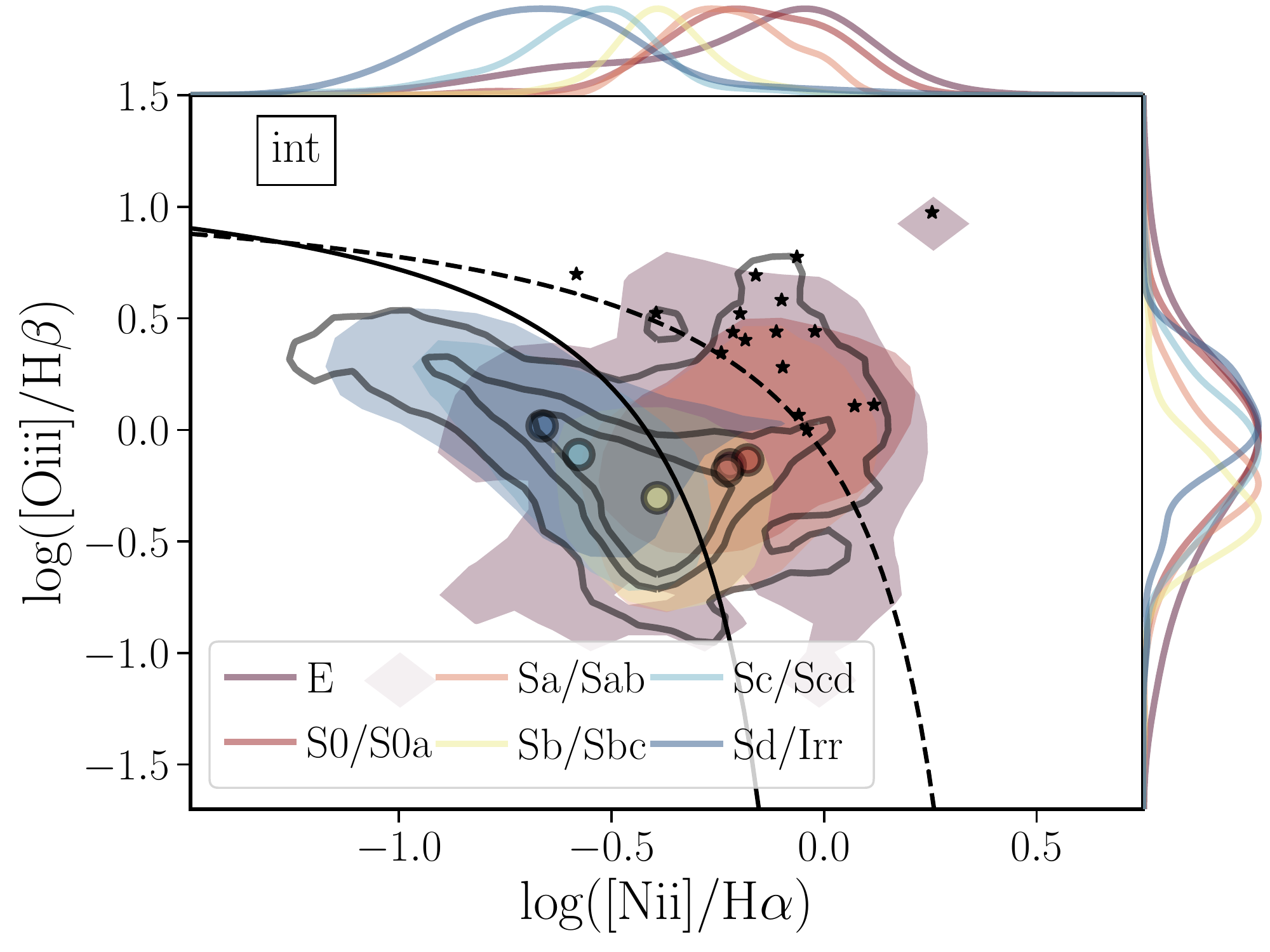}\includegraphics[width=8.5cm,trim={0 64 0 0},clip]{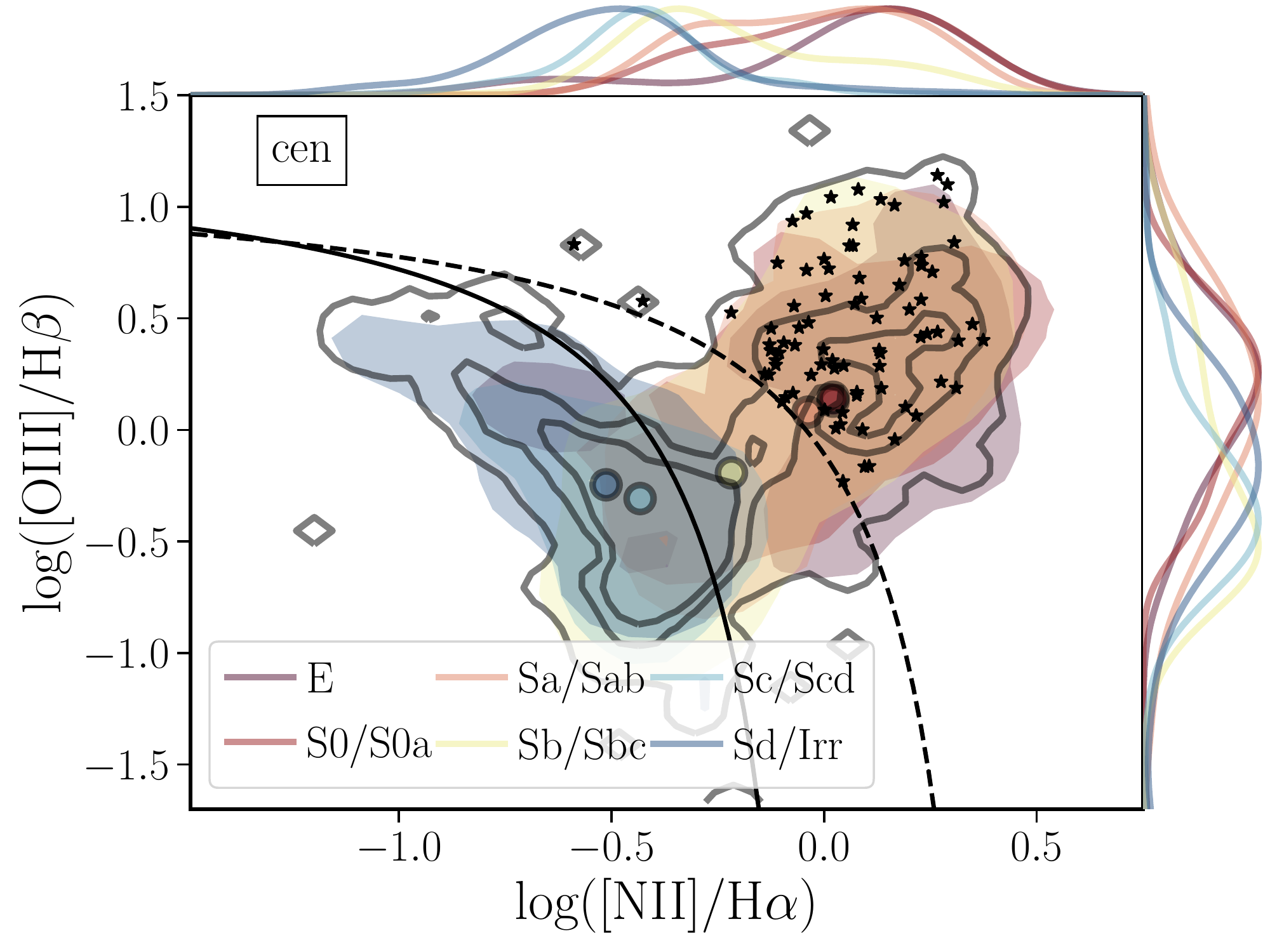}
 \includegraphics[width=8.5cm,trim={0 0 0 43},clip ]{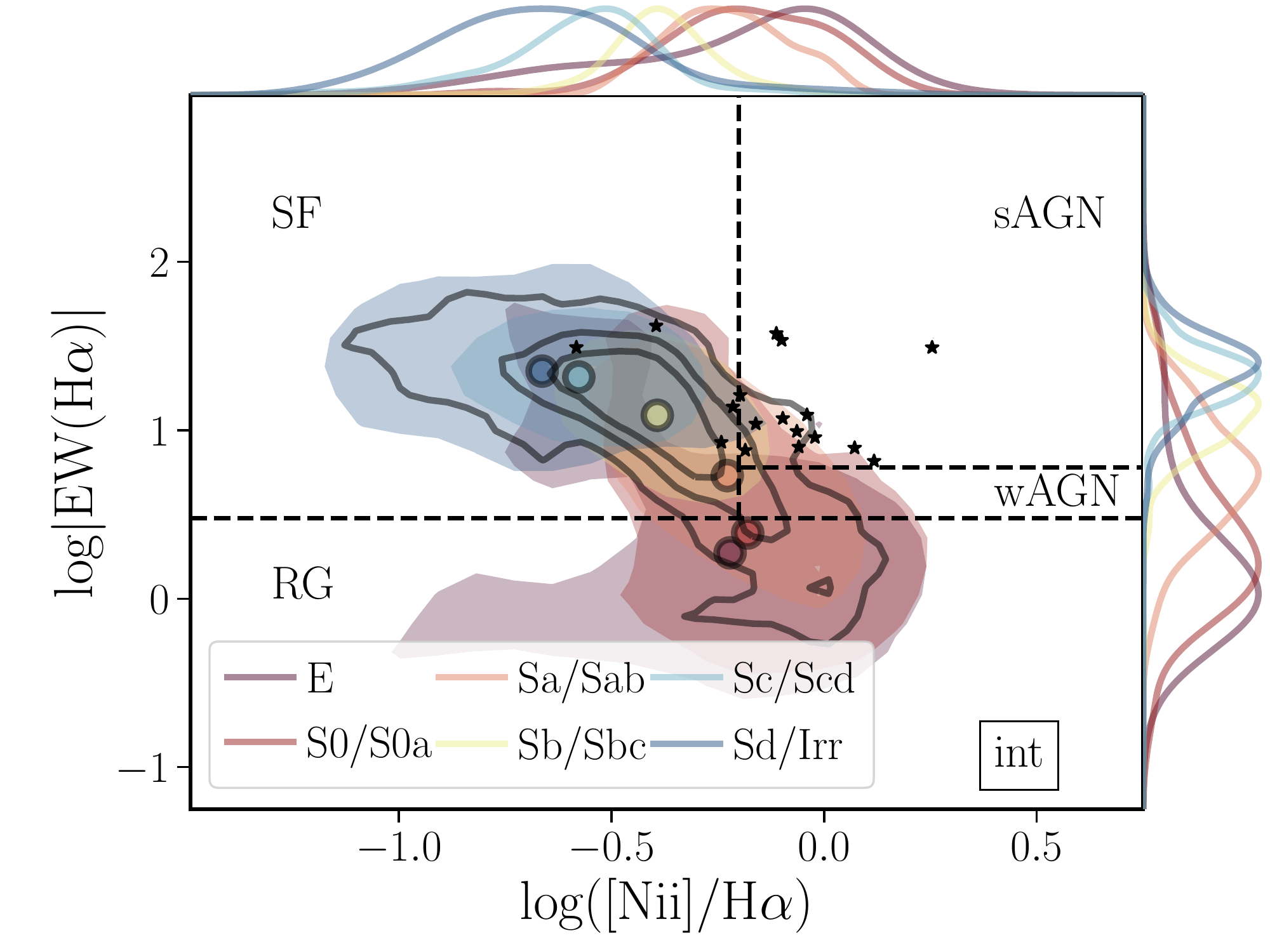}\includegraphics[width=8.5cm,trim={0 0 0 43},clip]{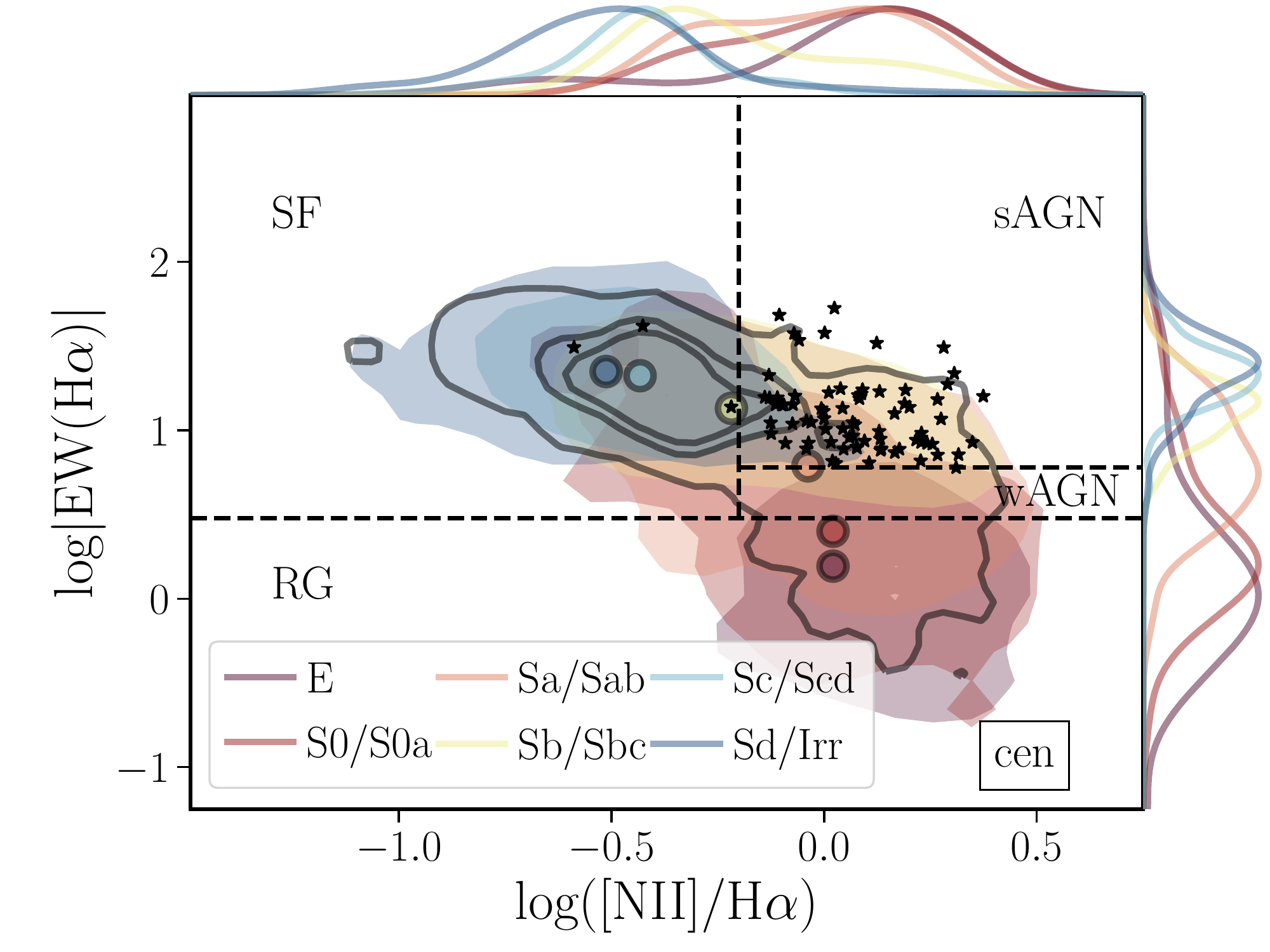}
 \endminipage
 \caption{Distribution of galaxies across the BPT (top panels) and WHAN (bottom panels) based on the emission line properties derived from the integrated (left panels) and central (right panels) spectra. It is adopted the same nomenclature adopted in Fig. \ref{fig:ML} for each panel. In addition, the solid and dashed lines in the BPT diagrams corresponds to the K03 and K01 demarcations lines, respectively. The dashed-lines in the WHAN diagram correspond to the boundaries defined by CF10 between different ionizing sources (as indicated in the legends). Finally, the location of galaxies candidate to host an AGNs, derived based on the central and integrated spectra, are indicated with a back star (thus, they are a different number of possible candidates).}
 \label{fig:BPT}
\end{figure*}

\subsubsection{What ionizes the gas?}
\label{sec:ion_nat}

The segregation between the different ionization sources in galaxies
is a topic of debate \citep[e.g.][]{ARAA,sanchez21}, being
particularly complicated when using aperture limited and/or integrated
spectra, due to the inherent mixing between different ionizing sources
\citep[e.g.][]{mast14,davies16,lacerda18}. For the current exploration
we adopt an heuristic approach, adopting the classification proposed
by \citet{sanchez14} and \citet{sanchez18}. This classification mixes
two frequently used diagnostic diagrams: (i) the classical BPT
diagram, proposed by \citep{baldwin81}, that compares the intensity of
the \oiii/\Hb\ vs. \nii/\Ha\ line ratios; and (ii) the WHAN diagram,
proposed by \citep[][hereafter CF10]{cid-fernandes10}, that introduces
the use of the \EWa\ to distinguish between different ionizing
sources.

Figure \ref{fig:BPT} shows the distribution of galaxies across those
two diagrams for both analyzed apertures. In the case of the BPT
diagrams it is appreciated the classical sea-gulf distribution
frequently described in the literature, that comprises (i) a well
defined branch in the left-side of the diagram, tracing the usual
location of classical \HII\ regions \citep[e.g.][]{osterbrock89}, and
(ii) a more extended and less defined cloud towards the right-side of
the diagram. Despite of the general similarities between the
distributions in this diagram for both the integrated and central
apertures, there are also remarkable differences. For the integrated
(central) spectra the left-side trend is much (less) clearly defined,
comprising a larger (lower) number of galaxies, and the right-side
cloud is more (less) diffuse, covering a more limited (extended) range
towards the upper-right region of the diagram. This clearly indicates
a change in the dominant ionizing source from the central to the outer
regions in galaxies.

When segregating by morphology it is appreciated a somehow similar
trend as the one observed when comparing the central and integrated
apertures: later-type (early-type) galaxies are more frequently
located in the left-side (right-side) branch of the distribution for
both apertures. This trend is observed in both the average values and
the whole distributions. Furthermore, the effect is modulated by the
aperture with (i) a more clear segregation in the integrated aperture
(a larger distinction) than for the central one, and (ii) a general
shift to the right-side for all morphologies in the central
aperture.

These results have been already presented in previous studies using
similar data \citep[e.g.][]{lacerda18,law21,jkbb23}, and discussed in
recent reviews \citep{ARAA,sanchez21}. They clearly indicate that the
dominant ionizing source changes from the inside-out and from earlier
to later type morphologies.  It is known that the left-side branch in
the BPT diagram is the location occupied by classical \HII\ regions
\citep[e.g][]{osterbrock89}, being usually assigned to ionization due
to young massive OB-stars generated in recent SF events. Different
demarcation lines have been proposed to limit the region populated by
this kind of ionization \citep[][,K03 hereafter]{kauff03} and
\citep[][,K01 hereafter]{kewley01}. The nature, actual meaning, and
usability of both boundaries has been extensively discussed in the
literature \citep[e.g.][]{sanchez21}.  On the contrary, the right-side
branch can be populated by ionization due to many different sources,
all of them presenting a harder ionization than OB-stars: (i) active
galactic nuclei \citep[AGNs, e.g.][]{osterbrock89}; (ii) shock
ionization due to high- and low-speed gas winds
\citep[e.g.]{heckman90,bland95,dopita96}; and ionization by hot old
evolved stars \citep[HOLMES][]{floresfajardo11}, beyond their post-AGB
phase, that have lost their external envelop
\citep[e.g.][]{binette94}.

Different approaches have been proposed to distinguish among the
ionization sources that populate the right-branch of the BPT diagram,
including the use of the velocity dispersion
\citep[e.g.][]{law15,dagos19}, explore the shape of the ionized
structures \citep[e.g.][]{carlos20} and/or distribution along the
galactocentric distance \citep[e.g.][]{gomes16a}. Most of them
required the use of the spatially resolved information, as recently
reviewed by \citet{sanchez21}.  However, for aperture limited and
integrated spectra it is frequently adopted the relative strength of
the ionized flux with respect to the underlying continuum,
parametrized by the \EWa. Based on the results by \citet{binette94}
and \citet{sta08}, \citet{cid-fernandes10} proposed the WHAN
diagnostic diagram that compares the distribution of the \EWa as a
function of the \nii/\Ha\ ratio, to distinguish between SF, AGN and
ionization found in retired galaxies \citep[RG, those that are not
forming stars anymore][]{sta08}, i.e., HOLMES/post-AGB
ionization. { It is important to note that low-intensity shock
  ionization \citep[e.g.][]{dopita96} and weak AGNs
  \citep[e.g.][]{ho97a} could be also present in retired galaxies (and
  the center of other galaxies too).  Both ionizating sources would be
  indistinguishable from the proposed mechanism associated to the
  presence of old/evolved stars using the adopted methodology. Their
  frequency is, however, uncertain: (i) \citep{carlos20} found that
  only 12 out of $\sim$260 early-type galaxies of their sample of 635
  galaxies observed with MUSE present clear evidence of shock
  ionization, i.e., filamentary/extended ionized gas structures, with
  relatively high velocity dispersion and \nii/\Ha\ values, and relatively low values of \EWa; (ii)
  \citet{ho97a} found that a non negligible subset of LINERs
  ($\sim$20\%) present broad \Ha\ emission, a clear evidence of the
  presence of an AGN. Considering both numbers together, we could
  establish that in about 25\%\ of the early-type galaxies the
  observed LINER-like ionization could be assigned to an ionizing
  process different than HOLMES/post-AGBs. On the contrary all
  early-type galaxies present the required old-stellar populations to produce
  this kind of ionization, that indeed is ubiquitious observed in the presence
  of old stars and ionized gas \citep[e.g.][]{singh13,belfiore17a}.}. Fig \ref{fig:BPT}, bottom panels, shows the
distribution of our galaxies in this diagram for both considered
apertures. In the case of the integrated spectra we find a well
defined anti-correlation between both parameters, with high (low)
\EWa\ corresponding to low (high) \nii/\Ha\ values. This trend is
followed by the galaxies of different morphologies too, with later
(earlier) type mostly found in the upper-left (bottom-right) regions
of the diagram.  As values of the \EWa $<$3\AA, are assigned to
ionization by old stars { low-intensity shocks and weak AGNs}, this trend clearly indicate that { a combination of those mechanisms produce the observed
  ionization } in early-type galaxies (i.e, E/S0). On the
contrary, the dominant ionization in late-type galaxies, those with a
considerable amount of SF, is indeed compatible with massive
young-stars.

A roughly similar trend is described by the central
apertures. However, in this case, instead of a clear anti-correlation
between the two parameters (\EWa\ and \nii/\Ha), there is less
well-defined trend. The strongest difference is the presence of a bump
in the region of high \EWa\ and high \nii/\Ha~that was assigned to
strong AGN ionization by \citet{cid-fernandes10}. When segregating by
morphology, the negative trend described for the integrated spectra is
appreciated, slightly shifted to higher values of the \nii/\Ha~ratio,
and with the AGN-bump dominated mostly by early-spirals \citep[Sa/Sab
and Sb, in agreement with previous results][]{sanchez18,lacerda20}. This bump is coincident with the cloud towards the upper-right
region described for the same aperture in the BPT diagram. The obvious
conclusion is that in the central aperture the possible contribution
of an AGN or an AGN-like ionization could dominate or significantly
pollute the observed ionization \citep[e.g.][]{davies16}, without
necessarily dominate the ionization galaxy-wide. This result highlights the importance
of defining at which aperture it is classified the ionization of a galaxy,
and demonstrates the importance of introducing the \EWa\ as an extra parameter
to classify the ionization.

%
\begin{table}
\begin{center}
\caption{Distribution of the dominant ionizing sources}
\begin{tabular}{crrrr}\hline\hline
  Ionizing	& \multicolumn{2}{c}{integrated}    &   \multicolumn{2}{c}{central}  \\
  source   &  \# gal. & \% & \# gal. & \% \\
   \hline
  NG      & 164 &  18 &  227 & 25 \\
  UN      &  21  &    2 &   82  &  9\\                                 
  SF       & 576 &  64 &  384 & 43 \\  
  pAGB  & 107  &  12 &   85  &   9\\
  wAGN &   10  &    1 &   37  &   4\\
  sAGN  &   17  &    2 &   80  &   9\\
\hline
\end{tabular}\label{tab:ion_class} 
\end{center}
Number of galaxies segregated by the dominant ionizing source for
the integrated and central apertures: NG (no gas detected), UN (unknown ionizing source),
SF (young massive OB-stars), pAGB (HOLMES/post-AGB old stars), wAGN (weak AGN-like ionization), and sAGN (strong AGN-like ionization). 
\end{table}

Based on all these results, following \citet{sanchez21}, we finally
classified the dominant ionization in each aperture in the following
way: (i) if the \Ha\ flux has a S/N below 3, it is considered that
there is no ionized gas detected (NG); (ii) if \Ha\ is detected above
this threshold, but the S/N of \Hb, \nii\ or \oiii\ is below 1, then we
consider that the source of the ionization is unknown (UN); finally,
if the two detection thresholds are fulfilled, and (iii) the \EWa\ is
below 3\AA, the ionization is classified as HOLMES/post-AGB (pAGB),
irrespectively of its location in the BPT diagram {. We should state clearly
that, as indicated before, we cannot be 100\% sure that the dominant ionizing source in these galaxies is actually related to the presence of old/hot evolved stars, as weak AGNs and low-intensity shocks may also contribute to the observed ionization somehow, or even dominate it \citep{dopita96,ho97a}}; on the other hand
(iv) if the \EWa\ is above 3\AA\ and the line ratios are below the K01
in the BPT diagram, the ionization is considered to be dominated by
young massive stars (SF); finally, (v) if the \EWa\ is above 3\AA\ and
the line ratios are above the K01 in the BPT diagram, the ionization
is considered dominated by an AGN, segregating between a strong AGN
(sAGN) when \EWa$>$6\AA, and a weak AGN (wAGN) when it is below this
value. We should stress that this classification scheme does not distinguish
between AGNs and shock ionization, and therefore when quoting AGN-ionization
we refer to AGN-like ionization (i.e., an ionization not due to an stellar source).

Table \ref{tab:ion_class} show the results of this classification. The
first result to notice is the fraction of galaxies without any
detected or unclassified ionized gas ($\sim$20-30\%), in both
apertures.  This fraction is rather low. However, it is considerably
larger than the one usually reported when using the spatial resolved
IFS with a typical resolution of $\sim$1 kpc \citep[$\sim$10\% or
lower][]{gomes16a,sanchez18}. The reason behind this discrepancy is
the dilution of the emission line signal when using large apertures.
As a consequence, the number of galaxies with an ionization compatible
with HOLMES/post-AGBs, i.e., the weakest one of the considered here,
is very low compared to those previous studies too. This result
highlights the advantage of using spatial resolved spectroscopy even
when deriving integrated or global properties of galaxies.

\begin{figure}
 \minipage{0.99\textwidth}
 \includegraphics[width=8.5cm]{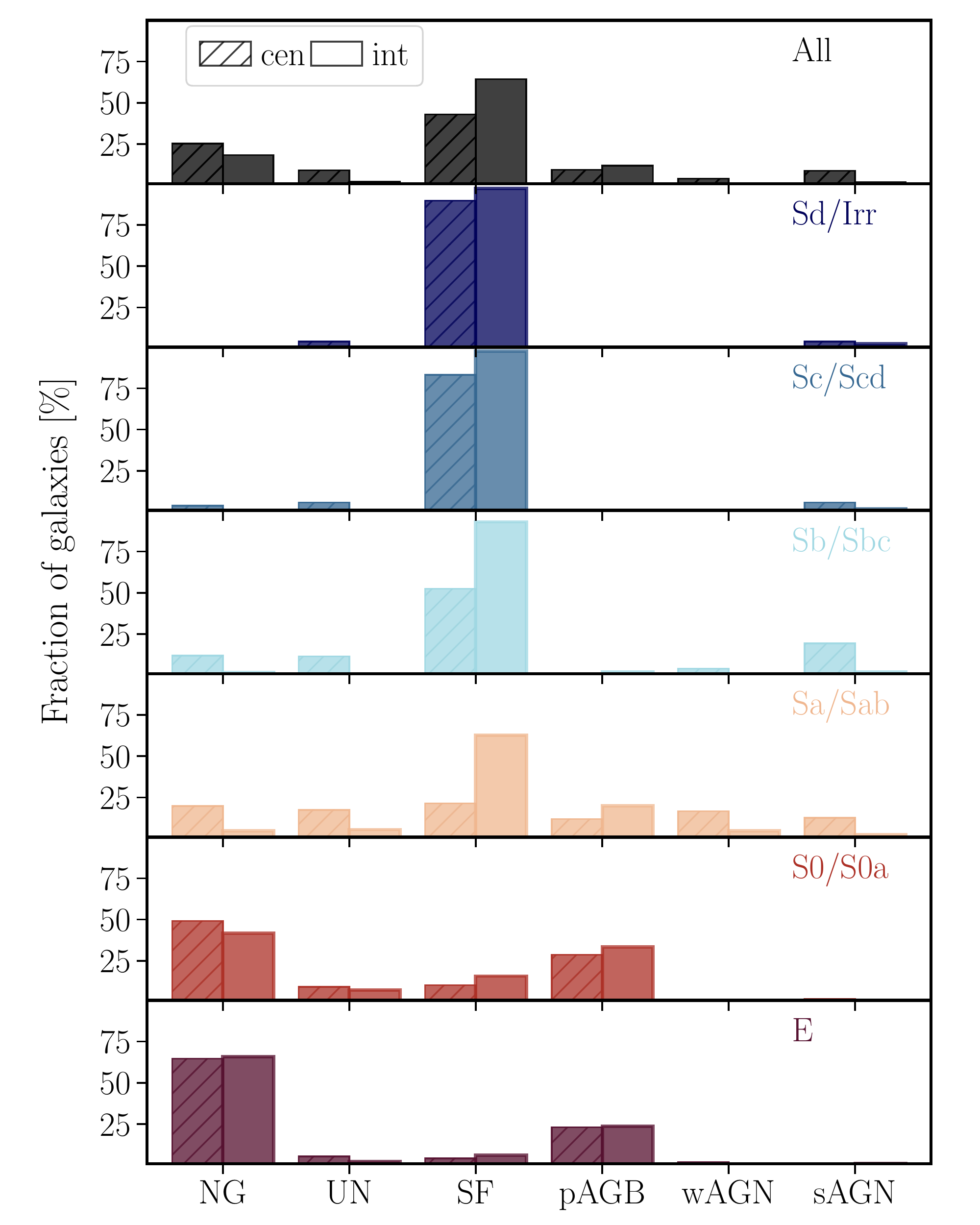}
 \endminipage
 \caption{Distribution of the dominant ionizing sources for the different morphological types for both the central and integrated spectra, labelled as indicated in Tab. \ref{tab:ion_class}.
As described in the text, the dominant ionizing source changes with morphology, with earlier (later) types presenting a larger fraction of (i) galaxies without (with) ionized gas and (i) post-AGB (star-formation) ionization.}
 \label{fig:ion_class}
\end{figure}

The values in Tab. \ref{tab:ion_class} agree with the distributions
shown in Fig \ref{fig:BPT}, for those galaxies with detected ionized
gas: (i) the over-all dominating ionizing source is the presence of
young massive OB stars, in particular for the integrated spectrum
($\sim$64\% of the galaxies); and (ii) the fraction of AGN-like
ionization detected in the integrated spectra is much lower than the
one found in central aperture \citep[$\sim$3\% vs. $\sim$12\%, as expected due to the dilution of the central ionizing source at larger apertures, e.g.,][]{davies16,lacerda18,alban22}, being a
small fraction in both cases. Figure \ref{fig:ion_class} shows the
distribution of the dominant ionizing source for the different
morphological types. Most of the galaxies without a clear detection of
ionized gas, neither in the integrated aperture nor in the central
one, are early type objects (E and S0). In this kind of galaxies, when
detected, the ionization is dominated by HOLMES/post-AGBs, as expected
in retired galaxies \citep{sta08,cid-fernandes10,singh13}. As indicated
before, based on the results using spatial resolved IFS, it is expected
that a substantial fraction of the non-detections present ionization due
to this source that it is diluted in the aperture limited spectra \citep{ARAA}. On the
other hand, the ionization in the latest morphological types (Sc-Sd),
is dominated by young-massive OB stars for both apertures, as expected
in actively star-forming galaxies. They are the objects with the lowest fraction
of non-detections too. Finally, AGN-like ionization, in
particular in the central regions, is more frequently found in
early-spirals (Sa and Sb), although most of them are ionized by
SF-related sources (in particular galaxy wide). Regarding the non-detections they
present a fraction slightly larger than the latest morphological types, in particular
for the central aperture, but significantly lower than that found in the earliest
morphological types.

As a final remark, following \citet{sanchez18}, we consider that a galaxy is clearly hosting an AGN is the dominant ionization corresponds to a strong AGN in the central aperture, irrespectively of which is the dominant ionization in the integrated spectrum \footnote{Note that shock ionization may enter in this category, as indicated before}. On the other hand, we consider that a galaxy is actively forming stars if the dominant ionization is due to young-massive stars in any of both apertures. Therefore, a galaxy could belong to both categories at the same time.

\begin{table*}
\begin{center}
\caption{Main physical properties derived from the emission line intensities.}
\begin{tabular}{cccccccc}\hline\hline
  cubename &  \multicolumn{2}{c}{Dominant Ionization} & \multicolumn{2}{c}{A$_{\rm V,gas}$}& log(SFR)  & 12+log(O/H) & log(M$_{\rm gas}$)\\
                    & int  & cen & int [mag] & cen [mag] & [M$_\odot$yr$^{-1}$] &  &  [M$_\odot$yr$^{-1}$]\\
  \hline
IC5376 & SF & sAGN & 1.55 $\pm$ 0.78 & 0.15 $\pm$ 0.08 & -0.08 $\pm$ 0.01 & 8.54 $\pm$ 4.27 & 9.7 $\pm$ 4.85\\
UGC00005 & SF & sAGN & 1.3 $\pm$ 0.65 & 1.63 $\pm$ 0.82 & 0.64 $\pm$ 0.32 & 8.57 $\pm$ 4.28 & 10.17 $\pm$ 5.08\\
NGC7819 & SF & SF & 0.82 $\pm$ 0.41 & 1.39 $\pm$ 0.7 & 0.1 $\pm$ 0.05 & 8.53 $\pm$ 4.26 & 9.57 $\pm$ 4.78\\
UGC00029 & NG & NG & 0.15 $\pm$ 0.08 & 0.15 $\pm$ 0.08 & -1.66 $\pm$ 2.18 & nan $\pm$ nan & 7.07 $\pm$ 3.54\\
IC1528 & SF & SF & 1.02 $\pm$ 0.51 & 1.32 $\pm$ 0.66 & -0.03 $\pm$ 0.0 & 8.52 $\pm$ 4.26 & 9.75 $\pm$ 4.88\\
NGC7824 & pAGB & pAGB & 0.15 $\pm$ 0.08 & 0.15 $\pm$ 0.08 & -0.5 $\pm$ 0.03 & nan $\pm$ nan & 6.99 $\pm$ 3.5\\
UGC00036 & SF & SF & 1.74 $\pm$ 1.22 & 2.7 $\pm$ 1.58 & 0.13 $\pm$ 0.06 & 8.61 $\pm$ 4.3 & 9.73 $\pm$ 4.86\\
NGC0001 & SF & SF & 1.56 $\pm$ 0.78 & 1.64 $\pm$ 0.82 & 0.44 $\pm$ 0.22 & 8.56 $\pm$ 4.28 & 9.44 $\pm$ 4.72\\
NGC0023 & SF & SF & 1.69 $\pm$ 0.84 & 1.65 $\pm$ 0.82 & 1.04 $\pm$ 0.52 & 8.58 $\pm$ 4.29 & 10.66 $\pm$ 5.33\\
NGC0036 & SF & sAGN & 1.43 $\pm$ 0.72 & 0.15 $\pm$ 0.08 & 0.54 $\pm$ 0.27 & 8.56 $\pm$ 4.28 & 10.29 $\pm$ 5.14\\
UGC00139 & SF & SF & 0.15 $\pm$ 0.08 & 0.63 $\pm$ 1.03 & -0.62 $\pm$ 0.02 & 8.48 $\pm$ 4.24 & 7.19 $\pm$ 3.6\\
MCG-02-02-030 & SF & sAGN & 0.47 $\pm$ 0.65 & 3.54 $\pm$ 3.92 & -0.35 $\pm$ 0.02 & 8.57 $\pm$ 4.28 & 8.43 $\pm$ 4.22\\
\hline
\end{tabular}\label{tab:phy_gas}

We present the values for just ten galaxies. The information for the remaining one is distributed electronically.
\end{center}
\end{table*}

\begin{figure}
 \minipage{0.99\textwidth}
 \includegraphics[width=8.5cm]{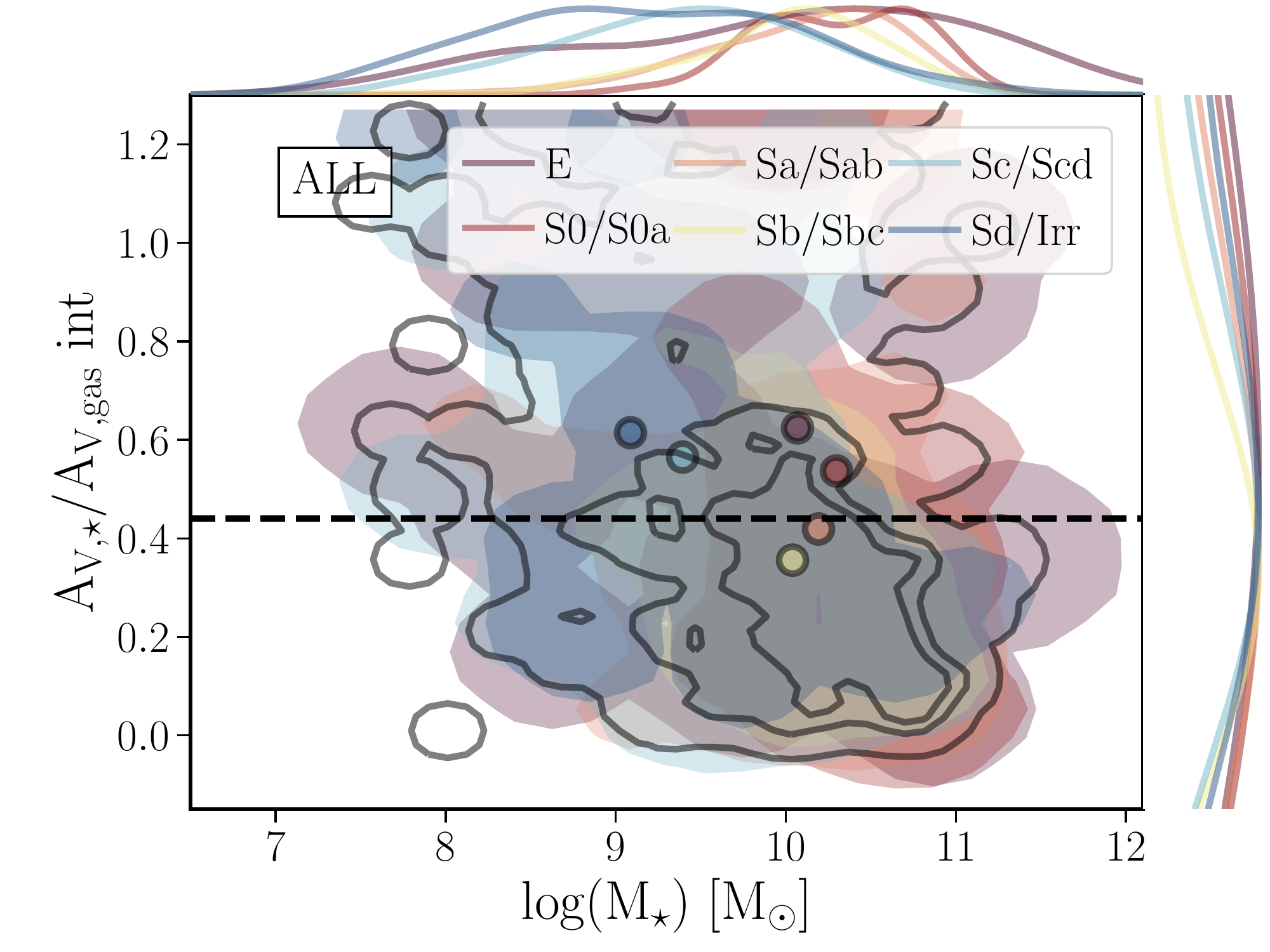}
 
 \includegraphics[width=8.5cm]{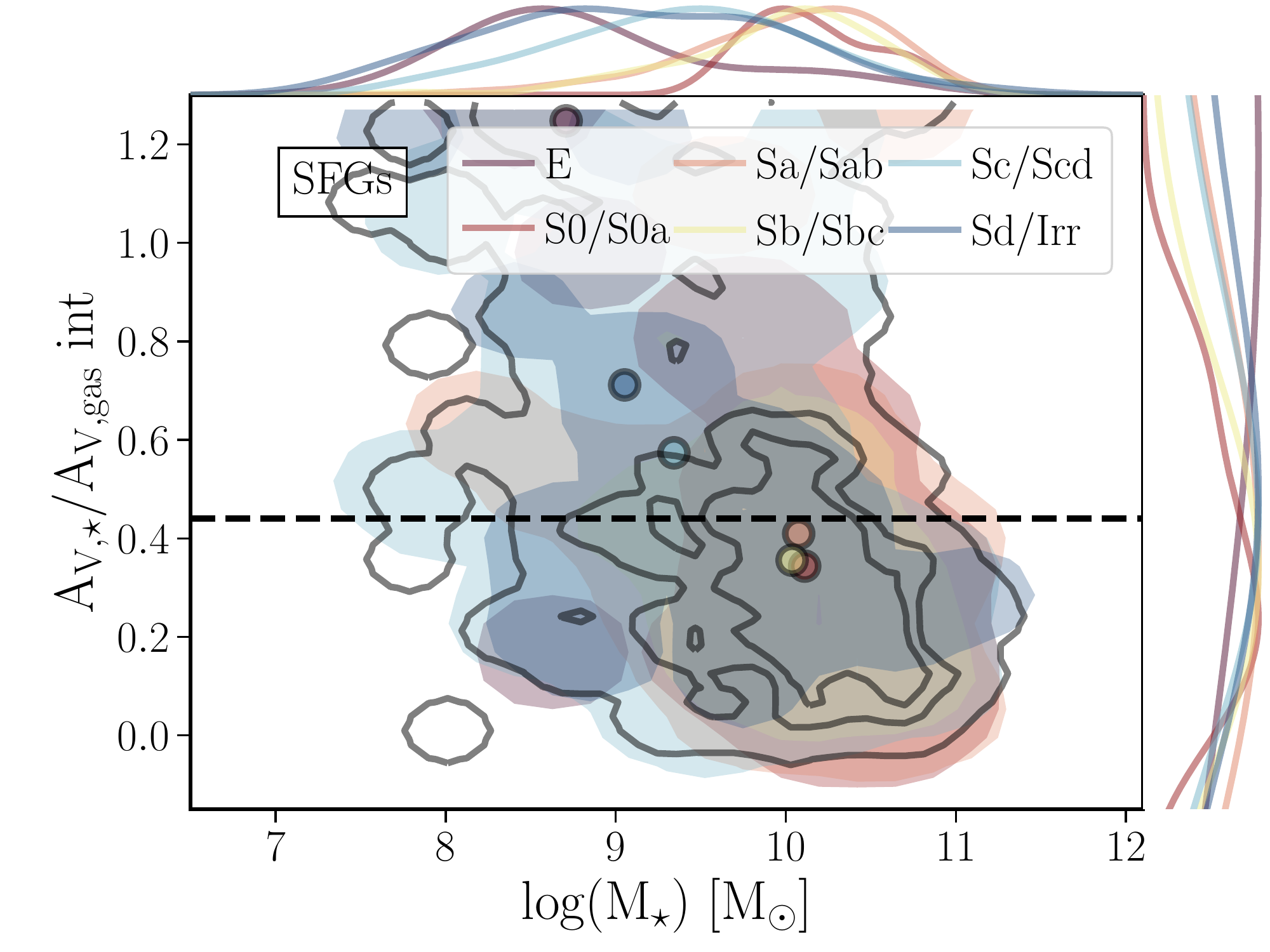}
 \endminipage
 \caption{Distribution of the ratio between the dust attenuation derived from the fitting to the stellar component, A$_{\rm V,\star}$, and the estimated for the ionized gas, A$_{\rm V,gas}$ for the integrated aperture,  as a function of the stellar mass, for the full sample of galaxies (top panel), and the SFGs (bottom panel). The distributions segregated by morphology are shown in both panels too. We use the same nomenclature adopted in left panel of Fig. \ref{fig:ML}. The dashed line corresponds to the ratio relation between both quantities proposed by \citet{calz01}.}
 \label{fig:Av_comp}
\end{figure}

\subsubsection{Ionized gas vs. stellar dust attenuation}
\label{sec:dust_c}

It is well known that the dust affects in a different way the |stellar
continuum than the ionized gas \citep[e.g.][]{calz97,wild11}. There
are many reason of why it is so, the most relevant ones are related to
(i) the different way that dust affects UV radiation that produces the
ionization, and it affects the dominant radiation in the stellar
population (oldish stars), (ii) the effects of the scatter and
redistribution of light in the line of sight and, maybe the most
important (iii) the geometrical distribution of the dust grains with
respect to both components \citep[e.g.][]{calz01}. The effect of the
dust on the observed spectra is well reproduced by a simple screen
model for the ionized gas, that comprises both absorption and scatter
in the line-of-sight (the so-called extinction).  However, the
scenario is more complex for the average stellar population, in which
the dust grains are embedded in the stars, and therefore the relative
geometry and spatial distribution is relevant, including scatter light
redirected into the line-of-sight, an interstellar medium of different
column densities and optical depths, and an incomplete coverage of the
dust for the average stellar population, i.e., a mixing effect of
stars totally and partially obscured and/or not obscured at all
\citep[e.g.][]{salim20}.

The net effect is that dust produces a smoother attenuation of the
stellar spectra than the one expected from a pure screen
model. Consequently, \citet{calz97} and \citet{calz01} estimated that
A$_{\rm V,\star}$$\sim$0.44 A$_{\rm
  V,gas}$, exploring a limited sample of star-forming galaxies (SFGs).
More extensive explorations on larger samples
reported a dependence of this ratio on different galaxy properties, in
particular the specific SFR (sSFR$=\frac{\rm SFR}{\rm
  M_{\star}}$) and the inclination \citep[e.g.][using SDSS data]{wild11}. All these explorations
are limited to SFGs in which  A$_{\rm V,gas}$ is more easily determined
due to strength of the emission lines.

\begin{figure*}
 \minipage{0.99\textwidth}
 \includegraphics[width=6cm]{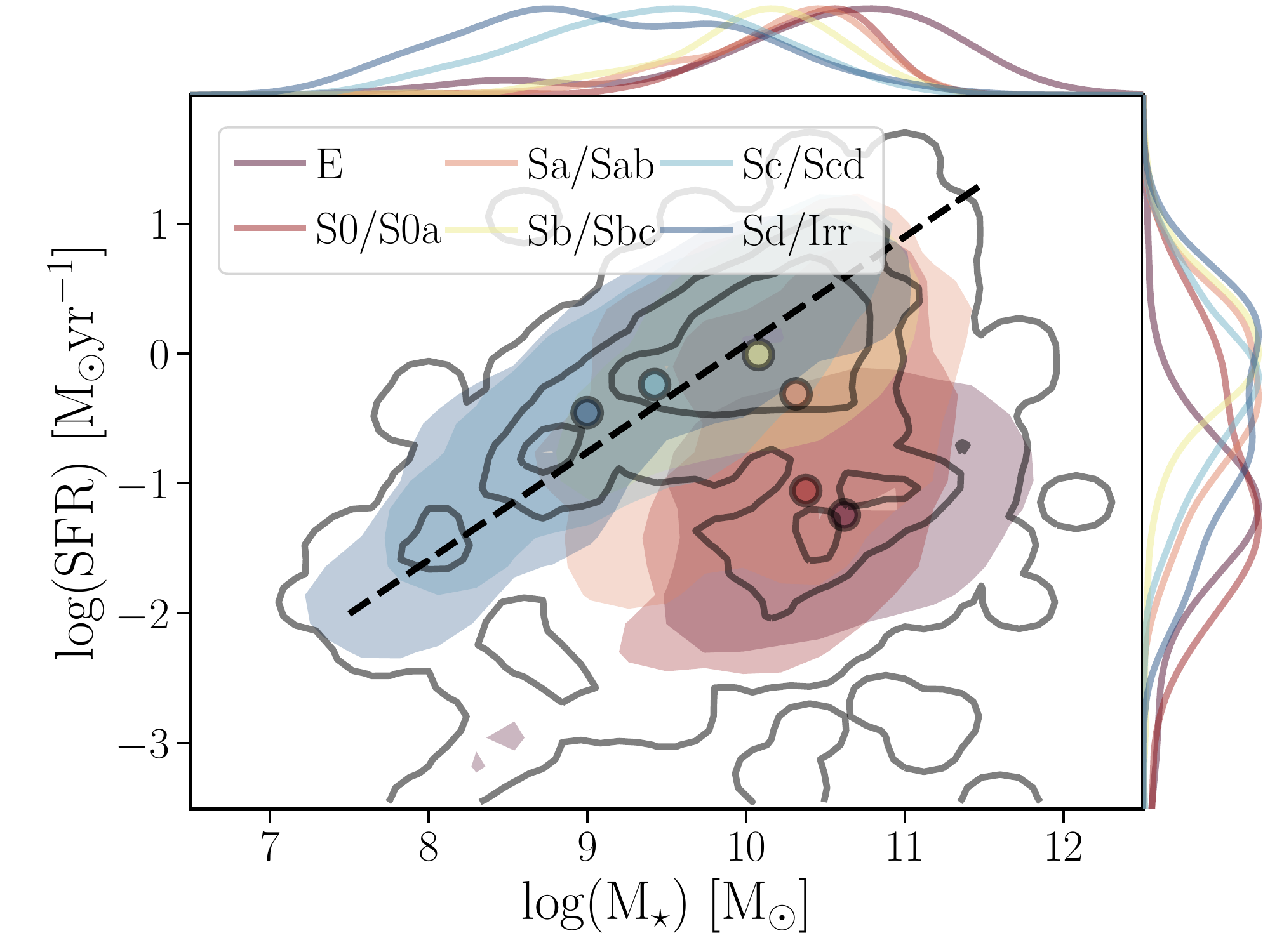}
 \includegraphics[width=6cm]{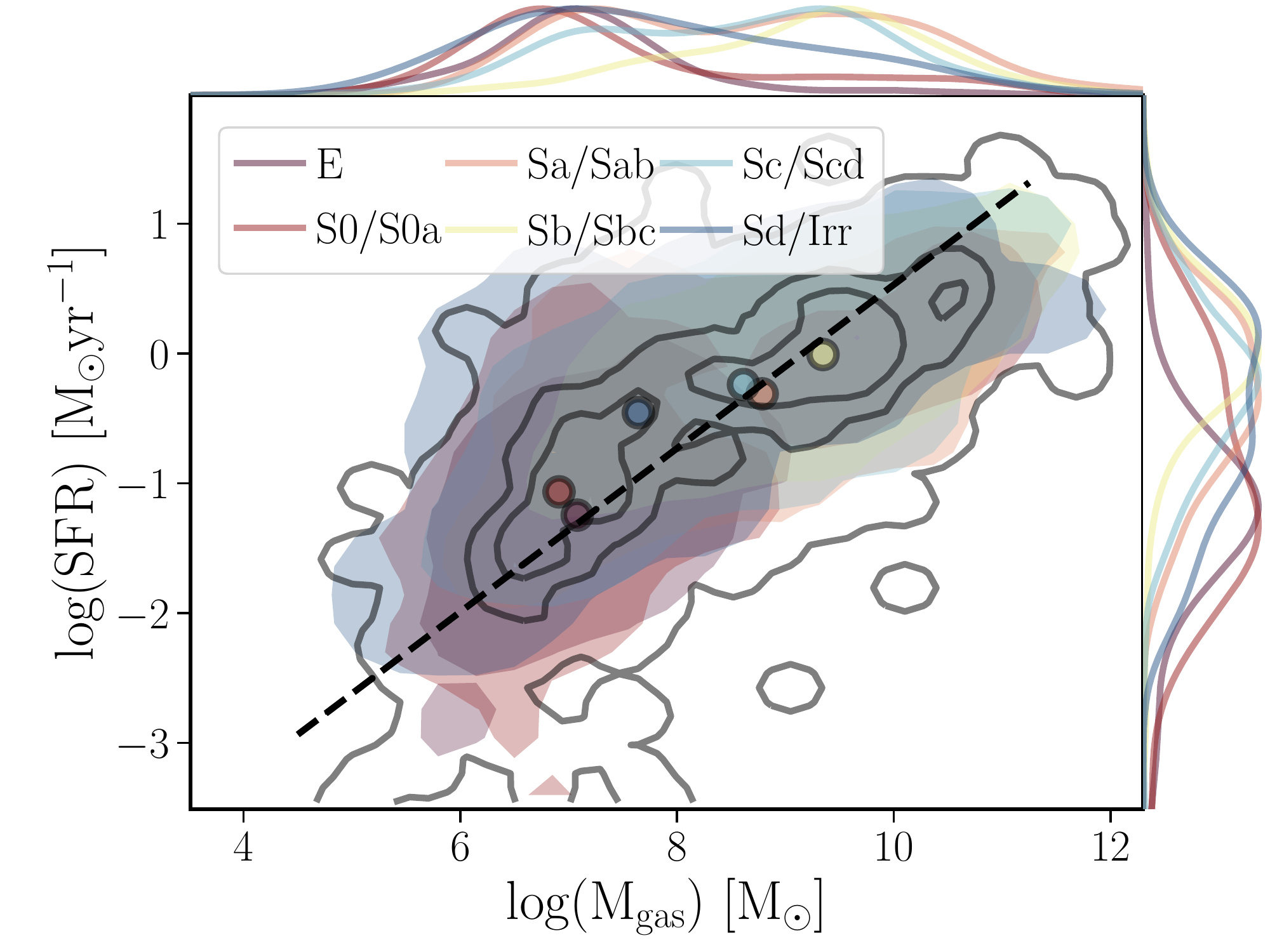}
 \includegraphics[width=6cm]{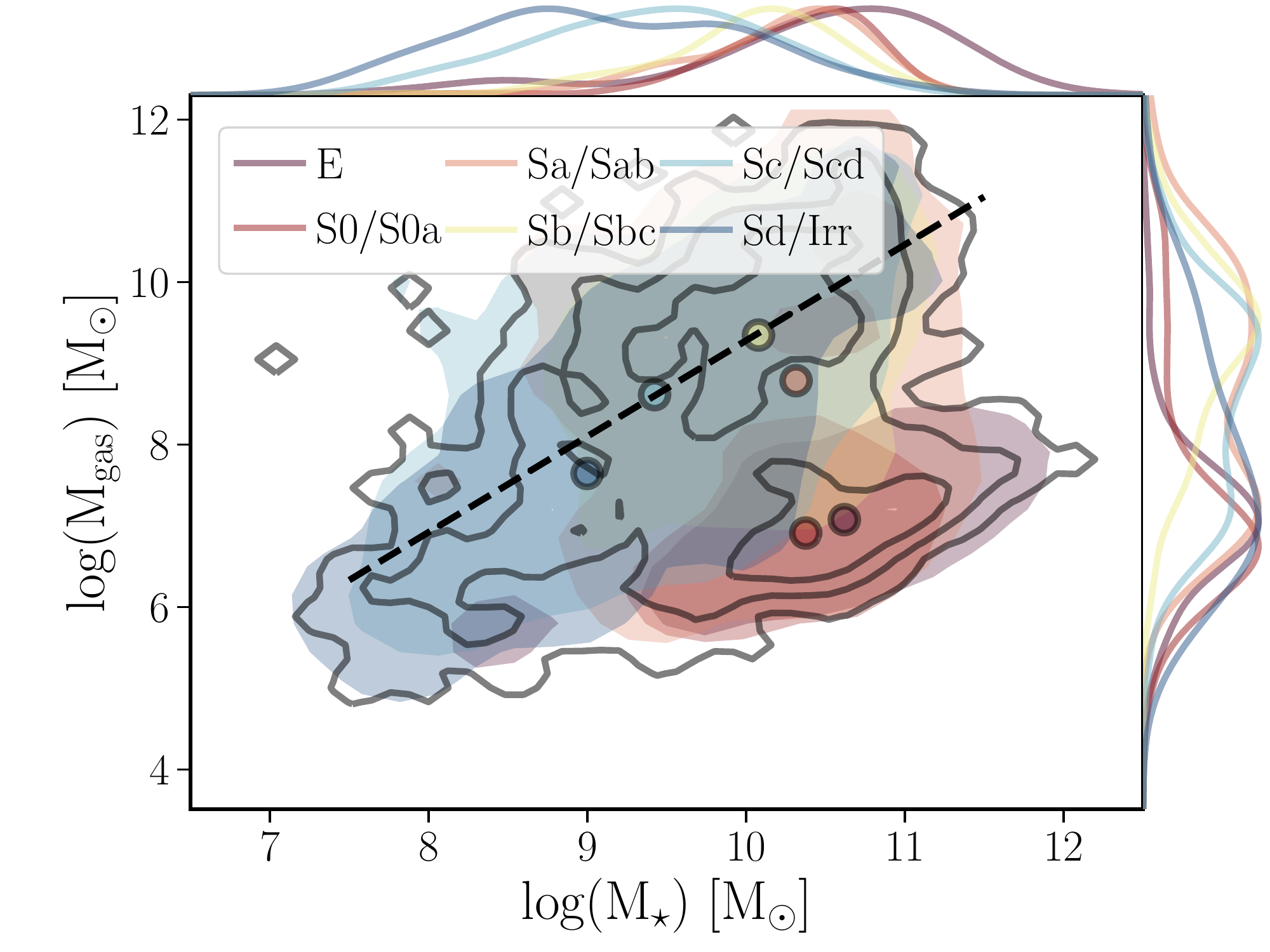}
 \endminipage
 \caption{Distribution of (i) SFR as a function of M$_\star$ (left panel), (ii) SFR as a function of M$_{\rm gas}$ (central panel) and (iii) M$_{\rm gas}$ as a function of M$_\star$  (right panel) for the full sample of galaxies and segregated by morphology. We adopt the same nomenclature adopted in left panel of Fig. \ref{fig:ML}. The dashed-line in each panel corresponds to the extensive form of the SFMS, SK and MGMS relations for SFGs recently reported by \citet{daysi}, using the AMUSING++ compilation.}
 \label{fig:sc_rel}
\end{figure*}

Figure \ref{fig:Av_comp} shows the distribution of the ratio between
both dust attenuations, A$_{\rm V,\star}$/A$_{\rm
  V,gas}$, as a function of
M$_\star$ for the sub-sample of galaxies for which we can estimate
both quantities (420 objects, top-panel) and for those of which
dominant ionization is compatible with SF (344 star-forming galaxies,
SFGs, bottom-panel). First, we notice that the ratio between both dust
attenuations covers a wide range of values, without no clear pattern
with the stellar mass in the case of the full sample of galaxies. If
any, there is a trend with the morphology: (i) the highest values are
found it early-type galaxies
($\sim$0.6, E/S0); (ii) the lowest ones in early-spirals
($\sim$0.2-0.3, Sa-Sbc), and (iii) finally the values rise up again to
a value near to the one reported by \citet{calz01} for late-spirals
($\sim$0.4, Sd/Irr). A more clear pattern with the stellar mass is
observed if we limit the sample to just SFGs, with the ratio
reaching the highest (lowest) values at low (high) masses. We attempt
to explore the distributions reported by \citet{wild11}, in particular
the trend with the sSFR and the inclination, but we did not find clear
patterns.  The only galaxy parameters for which we find some trends
are those that depends directly on the stellar mass, such as the
oxygen abundance and the stellar metallicity.

One possible source of discrepancy between our result and the one
presented in \citet{wild11} is the different aperture at which the
dust attenuation was measured. As indicated before \citet{wild11} used
the 3$\arcsec$ aperture spectra from the SDSS survey, while in here we
are using the galaxy wide (IFU FoV limited) integrated spectra. \citet{jkbb23} already noticed a radial gradient in this dust ratio. For a
better comparison we repeated the exploration using the dust
attenuations estimated for the central aperture. Using these values we
obtain similar average distributions, but with an even larger scatter
in the observed distributions. Therefore, we cannot offer a
satisfactory explanation for the described discrepancy so far, beside
the differences in galaxy samples. The SDSS offers a several order of
magnitudes larger sample than the one explored here. Maybe the
patterns described by \citet{wild11} have a statistical nature and
they only emerge for samples large enough to overcome the large dispersion
introduced by individual objects in more limited samples as the studied here.
We will try to address this issue in a dedicated study using the spatial resolved
information provided by the current data, that could improve the quality of the
derivation of the discussed parameters.

\subsubsection{Star-formation rate scaling relations}
\label{sec:SFMS}

The star-formation (SF) process in galaxies (and regions within them)
generate three relations among the main involved parameters: (i) the
rate at which star-formation happens, SFR, (ii) the accumulated
stellar mass, M$_\star$, and (iii) the ingredient from which stars are
form, i.e., the molecular gas mass, M$_{\rm gas}$. The relation
between SFR and M$_\star$, known as the Star-Formation Main Sequence,
is well described by a linear relation between the logarithm of both
parameters, with a slope $\sim$1 and a standard deviation of
$\sim$0.25 dex \citep[SFMS, e.g.][]{brin04,renzini15}. It has been
observed at a wide range of redshifts, with a strong evolution towards
larger values of SFR and lower values of M$_\star$ in earlier
cosmological times \citep[e.g.][]{speagle14,rodriguez16}, tracing the cosmic evolution
of the SFR rate \citep[e.g.][]{sanchez18b}. The SFMS was first described as a
relation between these two extensive quantities in galaxies. More recently it has been
described as a relation between the surface densities of both quantities, \Ssfr and \Sst (i.e., two extensive quantities), being fulfilled at very different scales in galaxies, from galaxy wide
to $\sim$1kpc scales \citep[][]{ARAA,mariana16,pan18,sanchez21b}.

On the contrary, the relation between the SFR and M$_{\rm gas}$ was
first described as an intensive relation between the star-formation
and the molecular gas surface densities (\Sgas): \citet{kennicutt1998}
shown that the logarithm of both quantities follows a linear relation
with a slope $\sim$1.4, in agreement with the expectations by
\citet{schmidt68}. Like in the case of the SFMS, this relation, known
as the Schmidt-Kennicutt (SK) law, is fulfilled at for a wide range of
galaxy scales \citep[][]{wong02,sanchez21b,daysi}. Although there is
less agreement on its actual slope, being more near to one in the
studies at a kpc-scale \citep[e.g.][]{lin19,sanchez21b}, in all cases
is described as a tight relation, with a scatter of $\sim$0.2 dex
\citep[e.g.][]{bigiel08,leroy13}.

Finally, a third relation, known as the Molecular Gas Main Sequence
(MGMS), has been described between M$_{\rm gas}$ and M$_\star$
\citep[e.g.][]{saint16,calette18}. It follows a similar tight
distribution in the log-log plane of both parameters with a slope near
to one. This relation, like the other two, has a intensive
correspondence that it is fulfilled in a wide range of physical scales
\citep[][]{lin20,sanchez21,sanchez21b,daysi}. The three relations are though
to be strongly inter-connected, and nowadays it is considered that the
self-regulation of the star-formation activity due the feedback produced by
stellar winds is most probably behind them \citep[e.g.][]{ostriker10,sun20,jkbb21b}

The three relations are fulfilled for actively star-forming galaxies,
SFGs (and star-forming areas in galaxies), while galaxies in which the
SF is halted \citep[either quenched or aged][]{corcho23} clearly
deviate from them.  Retired galaxies (and regions within them), i.e.,
those where the star-formation activity has been halted, are found
well below the SFMS, being the distance to this relation
($\Delta$SFMS) a gauge of the quenching/halting stage of those
galaxies (regions) \citep[e.g.][]{colombo20,bluck19,bluck20}. The
comparison between this distance and the offset with respect to the
other two relations is used to determine which is the dominant process
that drives the halting of the SF: (i) the lack of molecular gas (in
$\Delta$SFMS correlates with $\Delta$MGMS) or (ii) a decline in the star-formation
efficiency (if it correlates with $\Delta$SK) \citep[e.g.][]{ellison20,colombo20}.
The universality or not of those relations is a question of debate, being described
considerable variations galaxy to galaxy \citep[e.g.][]{ellison20}, and deviations with the
morphological type \citep[e.g.][]{rosa16,catalan17,mariana19,jairo19}.

Figure \ref{fig:sc_rel} shows the distributions across the
SFR-M$_\star$, SFR-$M_{\rm gas}$ and M$_{\rm gas}$-M$_\star$ diagrams
derived for the integrated spectra of our analyzed sample. The
distributions are shown for all galaxies and segregated by
morphology. For comparison purposes we included one of the most recent
derivations of the three relations described before (SFMS, SK and MGMS
relations), published by \citet{daysi}. In the three diagrams these relations traces regions with clear high density of galaxies. However, in 
the case of the SFR-M$_\star$ and M$_{\rm gas}$-M$_\star$ diagrams the
distribution is bimodal, with late-type galaxies (Sb-Sd) tracing the
SFMS and MGMS relations, early-type galaxies (E/S0) well below both
relations (by $2$ dex in average), and early-spirals (Sa/Sab) located
within both density peaks, slightly below both relations. On the
contrary, the SFR-M$_{\rm gas}$ does not present a clear bimodal
distribution. All galaxies, irrespectively of its morphology seems to be located around
the reported SK-law.

Based on these distributions we confirm that halting/quenching of the
SF strongly depends on the morphology, in agreement with many previous
explorations \citep[e.g.][and references
therein]{blanton+2017,catalan17}. Indeed, it is known that SF activity
happens in the disk of galaxies, being absent of bulges
\citep[e.g.][]{jairo19}. The fact that halting of SF is related with
the presence of bulges (and bulge dominated galaxies), agree with the
scenario in which this process happens from the inside-out in the bulk
population of galaxies \citep[e.g.][]{rosa16,ellison18}. Whether this is
connected with the presence of an AGN or other mechanisms that may
suppress the SF in the presence of a bulge is still a question of
debate \citep[e.g.][]{sanchez18,bluck19,kali22}. Irrespectively of
which is the actual mechanism, the described distributions confirm the
results suggesting that a lack of molecular gas is a necessary process
to halt the SF. Retired galaxies, those below the SFMS more than 1
dex, present a deficit of molecular gas of a similar amount with
respect to the MGMS. In other words, their molecular gas fraction is
considerably lower than the one found for SFGs. On the contrary, they
do not present a similar offset with respect to the SK-law. As this
relation traced the location of nearly constant star-formation
efficiency (SFE=$\frac{SFR}{M_{\rm gas}}$), our results suggest that
the halting of SF, galaxy wide, is not driven by a severe decline in
the SFE.  In this regards we agree with recent results that uses more
direct estimations of the molecular gas based on CO observations
\citep[e.g.][]{colombo20,ellison21b}, although the relevance of the SFE
in the modulation of the SF process is still under debate \citep[e.g.][]{lin19,ellison20,sanchez21b}

\begin{figure}
 \minipage{0.99\textwidth}
 \includegraphics[width=8.5cm]{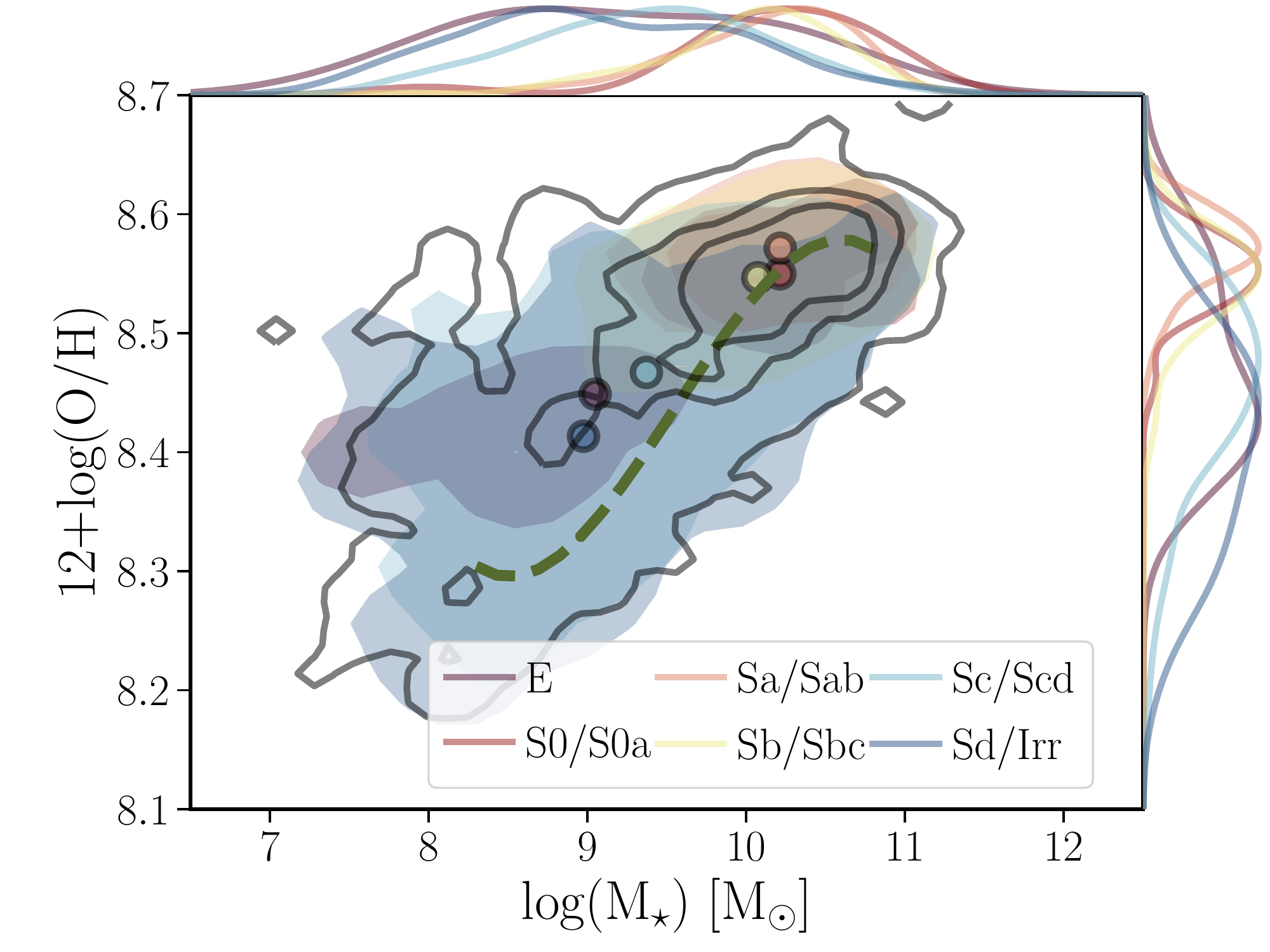}
 \endminipage
 \caption{Distribution of oxygen abundance, 12+log(O/H), as a function of the stellar mass derived for the integrated aperture. We use the same nomenclature adopted in left panel of Fig. \ref{fig:ML}. The green dashed-line corresponds to the MZR reported by \citet{sanchez19} using the oxygen abundance measured at the effective radius for the SAMI survey dataset.}
 \label{fig:MZR}
\end{figure}

\subsubsection{Mass-metallicity relation}
\label{sec:MZR}

One of the main products of the stellar evolution is the generation of
metals (i.e., elements heavier than helium). Those metals can be
expelled during the life-time of stars within the stellar winds, but
primarily are distributed to the ISM in the final stages of the
stellar evolution \citep[e.g.][]{yates12}. The main distributors of
metals are super-novae, with the vast majority of iron-peak elements
being produced in SN-Ia , the result of the collapse of binary
systems.  On the contrary $\alpha$-elements are primarily the result
of core-collapse supernovae, the end-phase of the evolution of very
massive stars \citep[e.g.][]{matt92,woos95,Koba20}. The connection
between the production of $\alpha$-elements and the star-formation
process induces a direct relation between the abundance of those
elements in the ISM and the stellar mass, known as the
mass-metallicity relation \citep[MZR, e.g.][]{pily07}.  This relation
was known for decades as a relation of oxygen abundance with galaxy
luminosity \citep[e.g.][]{vila92}. However, it was not described and
explored in detail until large spectroscopic surveys of galaxies were
available \citep[e.g. SDSS][]{york+2000}. \cite{tremonti04}
demonstrates that the MZR is a tight relation, with a dispersion of
$\sim$0.1 dex \citep[$\sim$0.06 dex in the most recent explorations,
e.g.][]{paola22}, that expands through several orders of magnitude in
stellar mass (from 10$^7$M$_\odot$ to 10$^{13}$M$_\odot$). The MZR
rises from low M$_\star$ values following an almost linear shape down
to M$_\star<$10$^{9.5}$M$_\odot$, bending afterwards and reaching a
plateau at a maximum oxygen abundance value \citep[e.g.][and references
therein]{maio19}. This shape clearly departs from the expectations
from a pure close-box model, with the asymptotic abundance being lower
than the pure expectations from the maximum yield
\citep[e.g.][]{pily07}. Different mechanisms have been proposed to
modulate the shape of the MZR, mostly related with the infall of
pristine (or lower metallicity) gas into the galaxy, outflows of metal
rich gas, and the differential star-formation histories between
galaxies of different final stellar masses
\citep[e.g.][]{SA14,Zhu17,ARAA,maio19}.

\begin{figure*}
 \minipage{0.99\textwidth}
 \includegraphics[width=8.5cm]{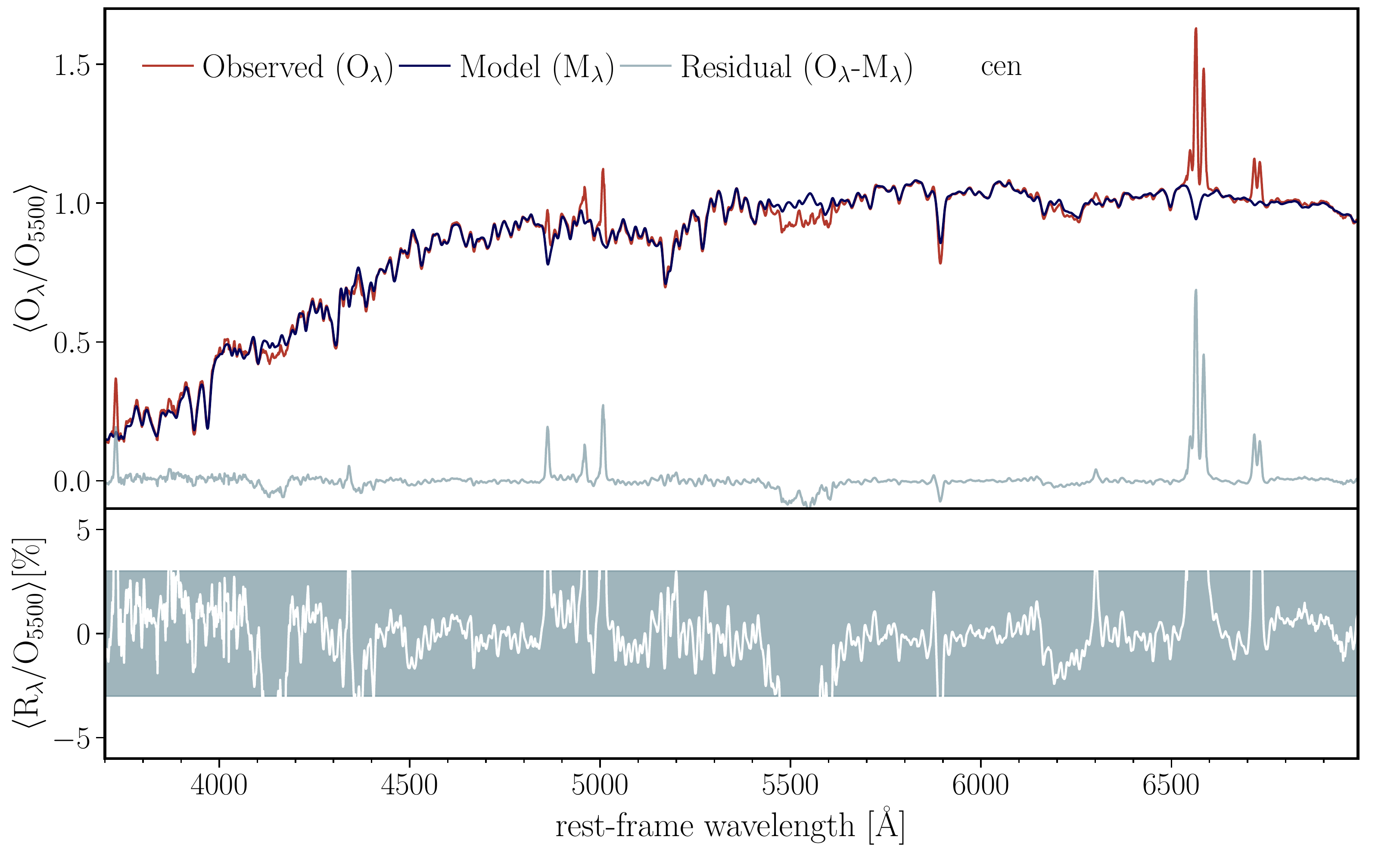}\includegraphics[width=8.5cm]{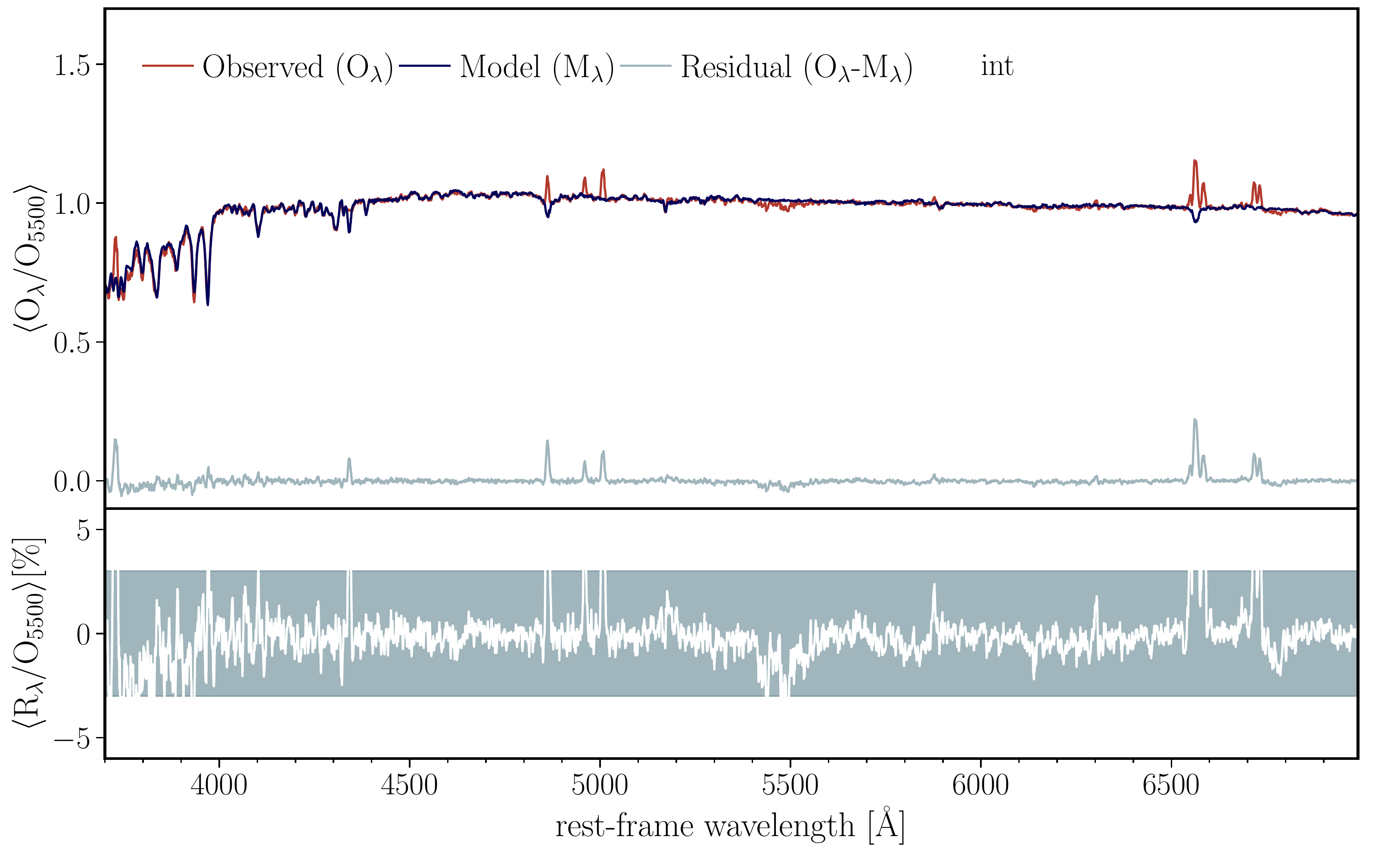}
 \endminipage
 \caption{Statistics of the residual spectra. Top panels shows the mean normalized spectrum extracted from the central 1.5$\arcsec$ aperture (left panels) and integrated across the entire FoV (regions with S/N$>$3, right panels), for the 895 galaxies in the final sample, normalized at $\sim$5500\AA. In addition we include the average of the individual stellar population models generated by {\sc pyFIT3D}, and the corresponding residuals. Bottom panels show a zoom of these residual spectra, in percentage, with a shaded rectangle encompassing the $\pm$3\% area.}
 \label{fig:res_spec}
\end{figure*}

Figure \ref{fig:MZR} shows the distribution of oxygen abundances,
derived from the integrated spectra, as a function of the stellar
masses for the SFGs in our sample (the only ones for which we have a
reliable estimation of the oxygen abundance). In general the
distribution follows the expected shape for the MZR, with a monotonic
increase from low towards high masses. No evident deviation from the
global trend is found when we explore the distributions segregated by
morphology, beside the mass range covered by each morphological type
withing the SFGs subsample. Beside that,  the global distribution present some
differences with the reported trend described for the MZR, at least with the archetypal one \citep{tremonti04}. First, we do not appreciate a clear plateau or
asymptotic oxygen abundance at high M$_\star$. Second, the scatter of
the distribution seems to be larger at low stellar masses than at high
ones. We should note that indeed both differences have been noticed in
previous explorations of the MZR using IFS data similar to the ones
included in this study
\citep{bb17,sanchez17a,sanchez19,cresci19,paola22}. The lack of
plateau at high-mass is most probably due to the poor statistics (low
number) of galaxies in this regime, being more evident when the data
are binned and the average oxygen abundance is estimated in different
intervals of stellar masses \citep[e.g.][]{paola22}. On the other hand,
the larger scatter at lower stellar masses is usually attributed to
the fact that SDSS spectroscopic measurements are restricted to the
central areas of the galaxies, implying an aperture effect on those data.
When using IFS data, and in particular when measuring the oxygen abundance at a particular characteristic radius (e.g., the effective radius) most studies describe a similar increase in the scatter (as indicated before).

An additional difference with previous published results is that the
MZR described by our integrated apertures is shallower than the one
reported using spatial resolved spectroscopic data
\citep[e.g.][]{bb17,cresci19,sanchez19,paola22}. To illustrate this
effect we include in Fig. \ref{fig:MZR} the average relation derived
for the O3N2 and N2 calibrators using the IFS from the SAMI survey
\citep[][]{sami}, applying a global shift in the stellar masses to
take into account the aperture effects due to the limited FoV of the
SAMI IFU \citep[][]{sanchez19}. Despite of this shift there is a
general offset, that increases from high to low masses. We observe
this pattern irrespectively of the adopted parametrization for the MZR
for the considered calibrator from the different publications listed
above (some of them using CALIFA data too). We select this particular one
since \citet{sanchez19} reported a flattening in the distribution at
low stellar masses, what could corresponds to a shallower rise of the
MZR in this regime. However, even in this case, the offset is evident.

As indicated before there is a fundamental difference in the way we
estimate the oxygen abundance for each galaxy in this exploration and
the way it was derived in the quoted explorations using IFS: in here
we first derive the integrated spectrum, and then we estimate the line
fluxes from which we finally estimate the oxygen abundance
\citep[e.g.][, for quoting an example]{sanchez22}. On the previously
quoted articles the spatial resolved spectra are analyzed
individually, deriving the flux intensities spatially resolved. Then,
after selecting the individual spaxels which ionization is compatible
with young-massive OB stars, the spatial distribution of the oxygen
abundance is derived. Finally, the value at the effective radius is
obtained, being considered the representative oxygen abundance of the
bulk galaxy. The reported difference indicate that the oxygen
abundance derived from the integrated spectra is representative of the
bulk distribution only at a first order. Indeed, it is known that the
dilution and the pollution by other ionizing sources alter the line
ratios and the final estimated oxygen abundances
\citep[e.g.][]{mast14,davies16,lacerda18,vale19}, however, to our
knowledge this is the first time that a differential bias depending on
the stellar mass range is reported. We will investigate this effect in
detail in future explorations replicating the analysis that derive the
oxygen abundance at the effective radius using the improved data
reduction, and expanding the comparison using different abundance
calibrators.


\subsection{Statistics of the residual spectra}
\label{sec:res_spec}

The residual spectra, once subtracted the best fitted stellar
population model to each individual spectrum, contains valuable
information of the quality of the data and the accuracy of the data
reduction \citep[e.g.][]{walcher14b,dr2,sanchez16}. Imperfections in
the sky-subtraction, problems in the blue-to-red spectrophotometric,
and even defects in the wavelength calibration are easily identify
when exploring these residual spectra. Following those studies we
perform two different analysis to explore the information included in
the residual spectra. First, we derive the average residual, in order
to enhance all the possible systematic effects/issues. Second, we the
individual residual spectra at the spectral pixel level, to evaluate
its compatibility with the error estimated by the reduction procedure.
{ Despite of its benefits to
  identify all these possible defects and to characterize the quality
  of the spectra, we must remind that this kind of analysis has
  inherent limitations related to the accuracy on how stellar
  population models match real galaxy spectra
  \citep[][]{cidfernandes:2014}. These limitations should be considering
when exploring the results of this analysis.}

To obtain the average residual we first shift each observed, stellar
population model and residual spectrum (as shown in
Fig. \ref{fig:FIT3D}) to a common rest-frame making use of the redshift
(systemic velocity) obtained by the fitting procedure. This shift
involves an inevitable interpolation and re-sampling of the
spectra. Once all spectra are in the common rest-frame, we obtain the
average spectrum for each component (original, model and residual),
for both aperture datasets (central and integrated). We treat them
separately as they correspond to data with two different levels of
S/N, and therefore its separate evaluation would allow to gauge better
the potential impact of any systematic effect.

Figure \ref{fig:res_spec} shows the resulting average spectrum for
each component at each aperture, normalized at $\sim$5500\AA. As
expected there is an evident difference in S/N between the two average
observed spectra, $\sim$100 for the central aperture and $\sim$300 for
the integrated one, reflected in the range of values of the residuals
in the bottom panels. In the case of the integrated spectrum the
residuals present almost no pattern or substructures along the entire
wavelength range shown in this figure. Beside the location of the
strongest emission lines, the residual is restricted to a 1-2\%\ band,
with only two clear deviations: (i) below 3900\AA, where there is a
systematic turn towards values around $\sim-$3-4\%, and (ii) at
$\sim$5500\AA, showing a clear miss-match between the average observed
spectrum and the model, with the former showing a systematic dip. On
the other hand, in the central spectrum the residuals present stronger
patterns and substructures, despite the evident decrease in
signal-to-noise. Although the turn-down at short wavelengths
($<$3900\AA) in the residual spectrum is not appreciated, the dip at
$\sim$5500\AA\ is stronger and more clearly defined. Beside that,
there is a second clear dip observed at $\sim$5100\AA\, and a possible
one at $\sim$6200\AA.

The origin of all these patterns can be explained by a combination of
different reasons. The deviations at the blue end are most probably
due to (i) the difficulty to provide with an accurate
spectrophotometric calibration at the covered wavelength range, (ii)
small but noticeable errors in the estimation of the atmospheric dust
estimation, and (iii) imperfections in the subtraction of the
night-sky spectrum, that is in general bluer than most galaxies
integrated (and central) spectra. { In addition,  there are clear
 miss-matches between the co-added observed spectrum and the
 corresponding co-added model at certain spectral features. The most
 clear one corresponds to NaD ($\sim$5893\AA), although there are
 others like TiO2 ($\lambda$6230\AA) that are also appreciated, in
 particular in the spectrum corresponding to the central
 regions. There is a combination of reasons for those miss-matches:
 (i) a contribution of the ISM to the absorption features (e.g.,
 NaD), (ii) dependence of the absorption features in properties not
 considering when modelling the SSPs (e.g, IMF or [$\alpha$/Fe]
 changes), (iii) intrinsic imperfections in the stellar libraries
 used to generate the SSP and the overall stellar synthesis modelling
 itself.} We should note that despite of those
effects the deviation is of the order of 1-2\% {, when ignoring the particular miss-matches discussed before}. The broad dips at well
defined wavelength regimes are all associated with the location of
strong night-sky and light pollution emission lines at the observatory
\citep{sanchez07a}. Those strong lines usually produce residuals in
the spectra after subtraction (see the well defined residual at
5577\AA\ in the observed spectrum shown in Fig. \ref{fig:FIT3D}). The
blue-shift introduced in the individual spectrum to rearrange all them
into a common rest-frame to produce the described average spectrum
naturally generates a broad feature. The fact that this feature is
negative indicates that somehow there is a systematic over-subtraction
of the strong night-sky and light-pollution emission lines, of the
order of a $\sim$2-3\%. In any case, those wavelength should be
masked in any spectroscopic analysis of the data, and therefore, they have
a limited effect in the quality of any science result.

\begin{figure}
 \minipage{0.99\textwidth}
 \includegraphics[width=8.5cm]{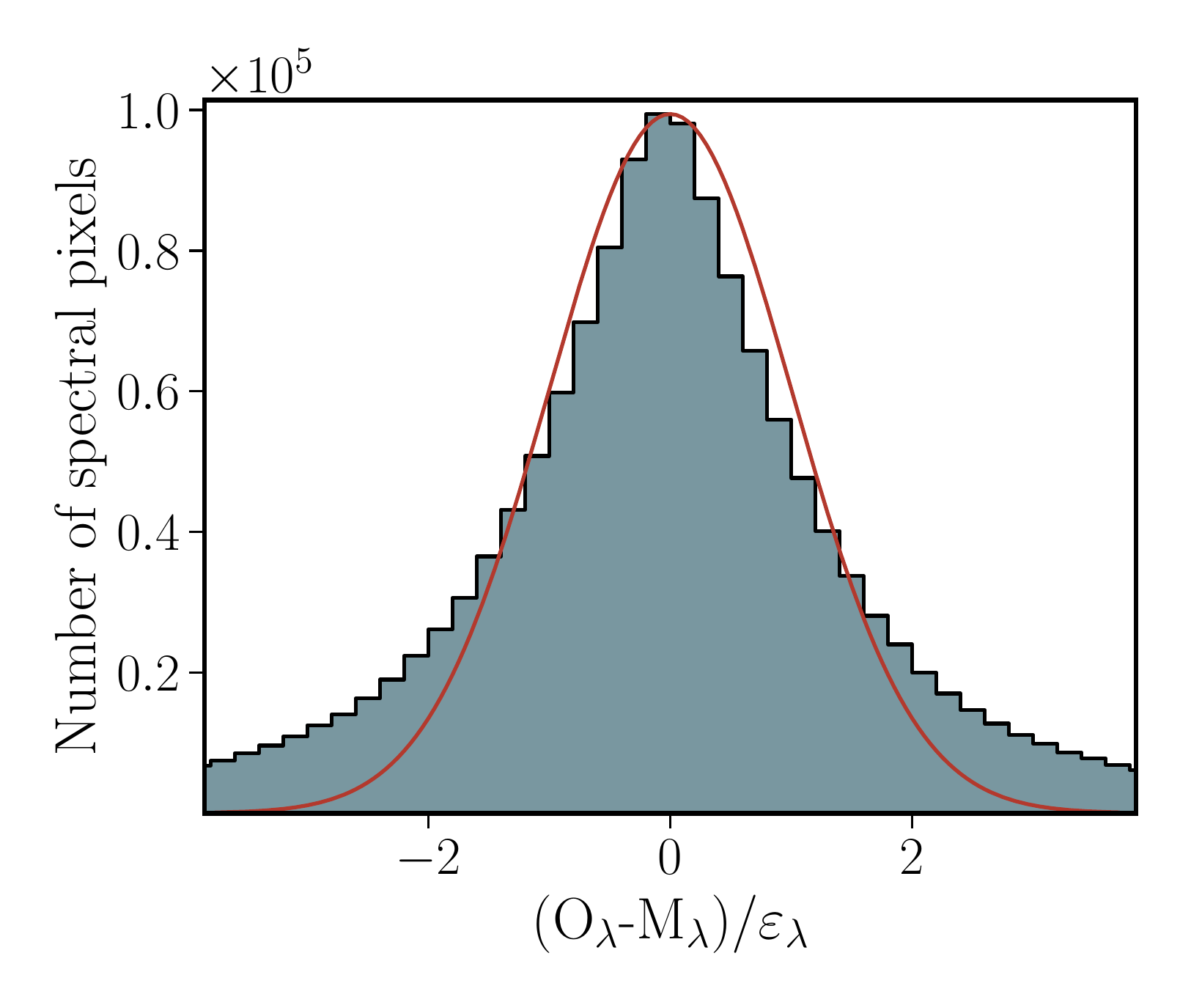}
 \endminipage
 \caption{Histogram of the reduced residuals, (O$_\lambda$-M$_\lambda$)/$\varepsilon_\lambda$,
   where O$_\lambda$ and M$_\lambda$ corresponds to the observed and model spectra for the central aperture of each good quality datacube which average value is shown in Fig. \ref{fig:res_spec}, and $\varepsilon_\lambda$ is the error estimated by the reduction procedure. A total of $\sim$1.5 million spectral pixels are included in the figure. The red line shows a Gaussian function scaled to the peak of the histogram with a width of one}
 \label{fig:noise}
\end{figure}

\subsection{Characterization of the noise}
\label{sec:covar}

Figure \ref{fig:noise} shows the histogram of the individual residuals at each spectral pixel for all the central aperture spectra of all the galaxies in our sample, once removed the best fitted model, relative to the error value estimated by the reduction procedure at the same spectral pixel. In the case of a perfect estimation of the pixel-wide error { and a perfect modelling of the stellar population (that, as we discussed before is not the case)} this distribution should follow a perfect Gaussian distribution with a $\sigma=$1. We include this expected distribution for comparison purposes {, in particular as most of the users would adopt a Poisson distribution for most of their calculations.}
The distribution of residuals follows a well defined symmetrical shape, centred in the zero value, indicating that it is not dominated by any strong systematic deviation, in agreement with the results from the analysis of Fig. \ref{fig:res_spec}. Its shape is clearly sharper/peaky than the one traced by a Gaussian distribution, and present wider wings. { Indeed a Power spectrum would represent better the observed distribution than a Gaussian function.} This indicates that in some spectral pixels the residuals are slightly smaller than the estimated noise, and in other ones they are slightly larger. This is somehow expected, due to the statistical nature of the current comparison: the residuals are, in the best case, a realization of the errors, if the errors are perfectly estimated and all residuals are the result of a pure stochastic fluctuation driven by photon (white) noise. Furthermore, in the presence of systematic effects the residuals should be considerable larger, in absolute value, what may explain the wider tails with respect to the expected distribution. { Indeed, as indicated in the previous section, the adopted stellar spectra modelling introduce several systematic residuals that are not accounted by a pure Poisson statistics.} Despite of those details this comparison shows that statistically speaking the estimated errors are representative (i.e., of the same order) of the observed residuals.


\begin{figure}
 \minipage{0.99\textwidth}
 \includegraphics[width=8.5cm]{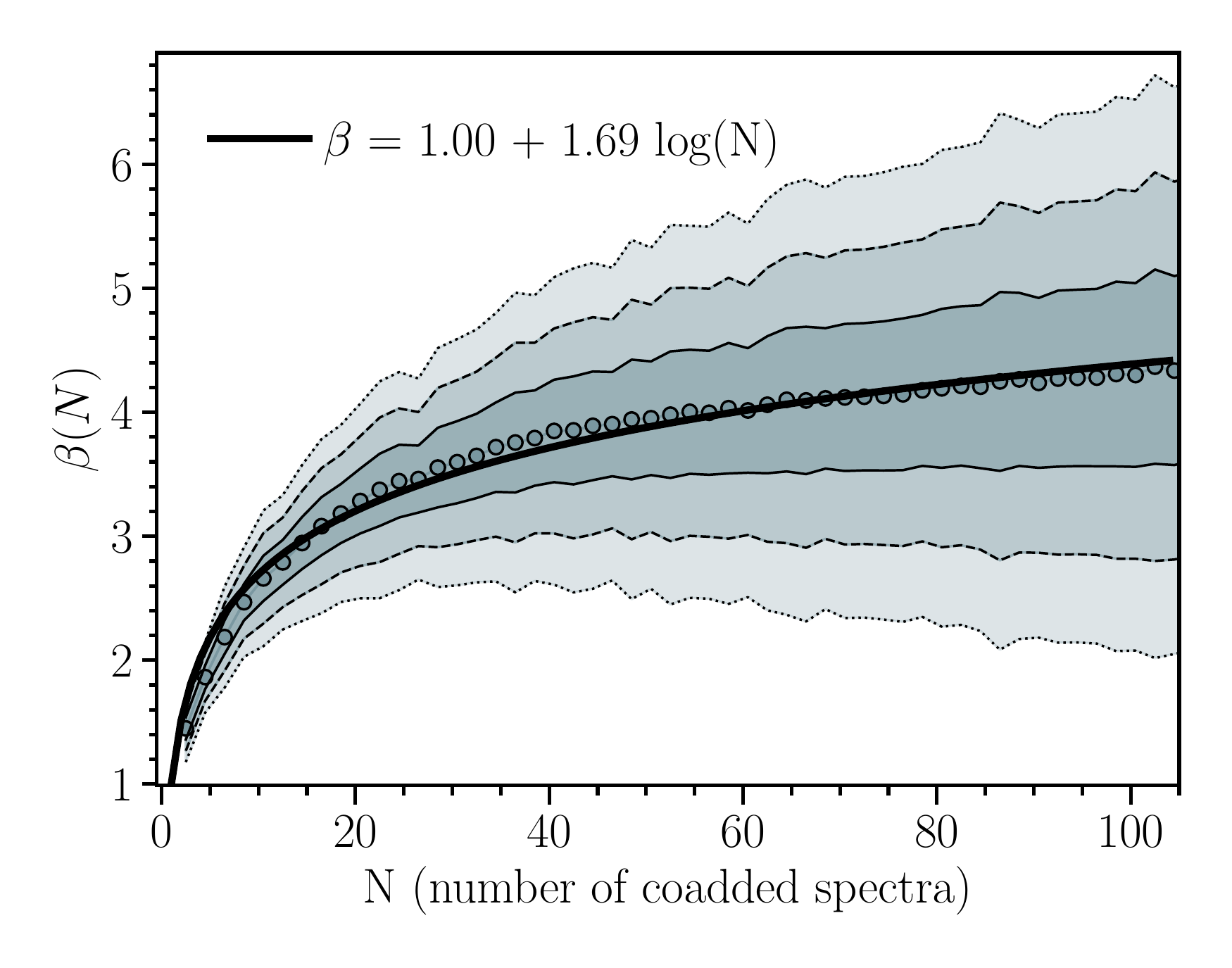}
 \endminipage
 \caption{Correction factor to the noise introduced by the spatial covariance, $\beta(N)$, as a function of the number of co-added adjacent spaxels for the new reduced dataset for a typical target with S/N$>$20. Shaded regions indicate the 1$\sigma$, 2$\sigma$ and 3$\sigma$ width areas around the average value, traced by the grey solid circles. The solid-line shows the best fitted model, following logarithmic parametrization, as shown in the legend.}
   \label{fig:covar}
\end{figure}

Once determined how representative is the error reported by the data reduction of the real one, it is important to illustrate how to propagate it when co-adding spectra.
Any image reconstruction algorithm that combines dithered exposures
with an original aperture (e.g., fibers) larger than the final
selected sampling pixel inevitable generates a noise covariance
between adjacent pixels. This needs to be taken into account for a
correct error propagation when combining/spatial-binning spectra to
increase the signal-to-noise, what it is required in many cases when
analyzing IFS data \citep[e.g.][]{cappellari03}. Some data reduction
procedures provide with the spatial covariance of the individual
datasets to address this issue \citep[e.g.][]{sharp15}. In the
different versions of the data reductions of the CALIFA data we
adopted a simpler and a more practical approach to take into account the
noise correlation, by providing the noise correction factor introduced
by the spatial covariance $\beta(N)$. This parameter is the ratio
between the error estimated on the combined spectrum and the one derived
analytically by propagating the individual errors of each spectra.
By construction $\beta$ depends on the number of co-added
spectra ($N$). To estimate this parameter we randomly generate a set
of circular apertures of different sizes (and number of encircled
spaxels) and extract the corresponding co-added spectra for the final
reduced set of datacubes. We select those spectra with a minimum
S/N$>$20, iterating the procedure until we reach a total of $\sim$1
million spectra.  Then, for each aperture, we estimate (i) the error
empirically as the standard deviation obtained from the detrended
spectra within the 5590-5710$\AA$ range in the rest frame, and (ii)
the expected error, in the absence of co-variance, derived by
propagating the individual errors in each spectra within the considered
aperture, using the same exact procedure. The ratio
between both parameters provides the $\beta$ parameter for the
corresponding spectrum.

Figure \ref{fig:covar} shows the distribution of the resulting $\sim$1
millon estimations of $\beta(N)$ as a function of the number of
co-added spaxels/spectra ($N$). Like in the case of previous data-reductions,
this distribution is well parametrized by a logarithmic function:
\begin{equation}\label{eq:covar}
 \beta(N) = 1 + \alpha log(N)
\end{equation}
with a slope $\alpha=$1.69. This value is considerable larger than the
reported in previous data-releases
\citep[e.g.][]{dr2,sanchez16}. The reason behind that is the finally
selected spaxel size (0.5$\arcsec$), half of the one adopted in
previous data-reductions. As already indicated, the new adopted cube
reconstruction algorithm (Sec. \ref{sec:cube_rec}), improves the
spatial resolution, what forces us to select a smaller spaxel scale
(for a proper sampling of the final PSF). However, it cannot decrease
the level of covariance, that for smaller spaxels is larger as the
original fiber size has not changed. In other words, the same number
of adjacent binned/co-added spectra corresponds to an smaller aperture
(in arcsec) in the current data-set than in the previous ones.
Therefore, they correspond to spaxels covered by a lower number of
originally independent spectra (i.e., different fibers).

We should stress that this derivation of the effect of the co-variance in the
error propagation is completely empirical and does not require the use
of the errors provided by data reduction. 
{ Thus this correction is valid even in the case
  that the original errors are not well represented by a Poisson distribution (see discussion on Fig. \ref{fig:noise} above), as it accounts for the covariance between adjacent spaxels.}
A detail description on how
to use this correction in a practical way when co-coadding spectra was
already included in the articles presenting the second public data-release
of the CALIFA survey \citep[][]{dr2}. We refer the reader to that manuscript
avoiding a repetition of its content.

\section{Summary and conclusions}
\label{sec:summary}

Along this article we have described a new procedure to reconstruct
the datacubes from dithered fiber-feed IFS observations that improves
the final spatial resolution, being now limited by (1) the distance
between adjacent (dithered) fibers and (2) the input/natural
seeing. We apply this procedure to the extended CALIFA dataset, a
compilation of $\sim$1000 galaxy observations in the nearby universe
($z\sim$0.015) using the low-resolution setup adopted along that
survey. This provides us with a remastered version of the datacubes
that we distribute to the community. From this compilation we select a
sub-set of good quality observations, based on a detailed quality
examination, comprising a galaxy sample which main selection criteria
is that its diameter (roughly) match with the FoV of the PPAK
instrument. We characterized the main properties of this sample of 895
galaxies, providing with the location, redshift, morphological type, a
sub-set of the photometric properties. We demonstrate that it is
possible to obtain a rough but reliable volume correction for this
sample, based on the equivalent volume accessible due to the diameter
selection. The information regarding this galaxy sample, including the volume correction, is distributed too.

We extract two spectra for each galaxy, one covering the central
regions and another one covering the full FoV of the IFU, to
demonstrate the quality and usability of the new data. We perform a
spectral fitting on each individual spectrum decoupling the
contribution of the stellar populations and ionized gas, and deriving
a set of central and integrated properties for both components. Among
them we estimate, for the stellar populations, the average ages,
metallicities, and mass-to-light ratios, together with the dust
attenuation affecting them and the velocity dispersion. For the
ionized gas, we estimate the flux intensities of the strongest
emission lines within the wavelength range of the dataset,. From them
we explore the dominant source of the ionization, the dust attenuation
affecting the ionized gas, the star-formation rate and the oxygen
abundance. Using all this dataset we present the main patterns and
distributions of the different parameters and physical properties as a
function of the stellar mass and the morphology of the considered
galaxies. When suitable we compare the behaviour for the inner regions
with those obtain galaxy wide, tracing possible radial trends within
galaxies.  Once more, all these parameters and dataproducts have been
distributed within the current manuscript.

We compare our results with previous ones extracted from the
literature, both using photometric, single aperture and spatial
resolved IFS data. In general we find a considerable agreement with
previous results. A coherent picture emerges, in which (i) the stellar
population content in galaxies, traced by the average ages and
metallicities, (ii) the star-formation stage, (iii) the gas content,
(iv) its metallicity, and (v) the dominant ionizing source strongly
depend on the final stellar mass, morphology and location within each
galaxy (primarily the galactocentric distance). These results agree
with those of the most recent studies on the topic \citep[e.g,][and references therein]{
  blanton+2017,kewley19,maio19,ARAA}.

In a forthcoming article we will present the analysis of the spatial
resolved data distributed here, following \citet{sanchez22}, making a
particular effort in the exploitation of its improved spatial
resolution.

\section*{Acknowledgements}

We acknowledge the comments and suggestions by the anonymous referee
that has helped to improve this manuscript.

We want to thanks Prof. M. Blanton, who motivated us to improve the
datacube reconstruction algorithm with a series of presentations and
discussions on the topic.

S.F.S. thanks the PAPIIT-DGAPA AG100622 project. J.K.B.B. and S.F.S. acknowledge support from the CONACYT grant CF19-39578. R.G.B. acknowledges financial support from the grants CEX2021-001131-S funded by MCIN/AEI/10.13039/501100011033, SEV-2017-0709, and to PID2019-109067-GB100. L.G. acknowledges financial support from the Spanish Ministerio de Ciencia e Innovaci\'on (MCIN), the Agencia Estatal de Investigaci\'on (AEI) 10.13039/501100011033, and the European Social Fund (ESF) "Investing in your future" under the 2019 Ram\'on y Cajal program RYC2019-027683-I and the PID2020-115253GA-I00 HOSTFLOWS project, from Centro Superior de Investigaciones Cient\'ificas (CSIC) under the PIE project 20215AT016, and the program Unidad de Excelencia Mar\'ia de Maeztu CEX2020-001058-M.

This study uses data provided by the Calar Alto Legacy
Integral Field Area (CALIFA) survey (http://califa.caha.es/). Based on
observations collected at the Centro Astron\'omico Hispano Alem\'an
(CAHA) at Calar Alto, operated jointly by the Max-Planck-Institut
f\"ur Astronomie and the Instituto de Astrof\'isica de Andaluc\'ia
(CSIC). 

The Pan-STARRS1 Surveys (PS1) and the PS1 public science archive have been made possible through contributions by the Institute for Astronomy, the University of Hawaii, the Pan-STARRS Project Office, the Max-Planck Society and its participating institutes, the Max Planck Institute for Astronomy, Heidelberg and the Max Planck Institute for Extraterrestrial Physics, Garching, The Johns Hopkins University, Durham University, the University of Edinburgh, the Queen's University Belfast, the Harvard-Smithsonian Center for Astrophysics, the Las Cumbres Observatory Global Telescope Network Incorporated, the National Central University of Taiwan, the Space Telescope Science Institute, the National Aeronautics and Space Administration under Grant No. NNX08AR22G issued through the Planetary Science Division of the NASA Science Mission Directorate, the National Science Foundation Grant No. AST-1238877, the University of Maryland, Eotvos Lorand University (ELTE), the Los Alamos National Laboratory, and the Gordon and Betty Moore Foundation.

This research made use of
Astropy,\footnote{http://www.astropy.org} a community-developed core
Python package for Astronomy \citep{astropy:2013, astropy:2018}.

\section*{Data Availability}
\label{sec:access}

The raw and reduced version of the data explored along this article are accessible through the Calar Alto Observatory Archive\footnote{\url{http://www.caha.es/science-mainmenu-95/public-archives}} and the eCALIFA DR webpage, accessible through the CALIFA webpage\footnote{official: \url{http://califa.caha.es/}, mirror:\url{http://ifs.astroscu.unam.mx/CALIFA_WEB/public_html/}}.

\noindent As part of this data release we distribute:
\begin{itemize}
\item The datacubes reduced using version 2.3 of the data reduction for all the originally compiled observations using the V500 setup:  \url{http://ifs.astroscu.unam.mx/CALIFA/V500/v2.3/reduced/}
\item The datacubes corresponding to the final sample of galaxies described in Sec. \ref{sec:sample}, once masked the possible companion objects and foreground field stars \url{http://ifs.astroscu.unam.mx/CALIFA/V500/v2.3/reduced_masked/}
\item A fitsfile table including all the galaxy properties explored and discussed along this article, comprising the parameters included in Tables \ref{tab:qc}, \ref{tab:sample}, \ref{tab:stars}, \ref{tab:gas} and \ref{tab:phy_gas}: \url{http://ifs.astroscu.unam.mx/CALIFA/V500/v2.2/reduced/}
\item The datacubes reduced using the previous version of the data reduction \citep[ver. 2.2][]{sanchez22}, for the complete compiled observations: \url{http://ifs.astroscu.unam.mx/CALIFA/V500/v2.2/reduced/}
\end{itemize}


\bibliographystyle{mnras}
\bibliography{my_bib} 

\begin{thebibliography}{}
\makeatletter
\relax
\def\mn@urlcharsother{\let\do\@makeother \do\$\do\&\do\#\do\^\do\_\do\%\do\~}
\def\mn@doi{\begingroup\mn@urlcharsother \@ifnextchar [ {\mn@doi@}
  {\mn@doi@[]}}
\def\mn@doi@[#1]#2{\def\@tempa{#1}\ifx\@tempa\@empty \href
  {http://dx.doi.org/#2} {doi:#2}\else \href {http://dx.doi.org/#2} {#1}\fi
  \endgroup}
\def\mn@eprint#1#2{\mn@eprint@#1:#2::\@nil}
\def\mn@eprint@arXiv#1{\href {http://arxiv.org/abs/#1} {{\tt arXiv:#1}}}
\def\mn@eprint@dblp#1{\href {http://dblp.uni-trier.de/rec/bibtex/#1.xml}
  {dblp:#1}}
\def\mn@eprint@#1:#2:#3:#4\@nil{\def\@tempa {#1}\def\@tempb {#2}\def\@tempc
  {#3}\ifx \@tempc \@empty \let \@tempc \@tempb \let \@tempb \@tempa \fi \ifx
  \@tempb \@empty \def\@tempb {arXiv}\fi \@ifundefined
  {mn@eprint@\@tempb}{\@tempb:\@tempc}{\expandafter \expandafter \csname
  mn@eprint@\@tempb\endcsname \expandafter{\@tempc}}}

\bibitem[\protect\citeauthoryear{{Abdurro'uf} et~al.,}{{Abdurro'uf}
  et~al.}{2021}]{DR17}
{Abdurro'uf} et~al., 2021, arXiv e-prints, \href
  {https://ui.adsabs.harvard.edu/abs/2021arXiv211202026A} {p. arXiv:2112.02026}

\bibitem[\protect\citeauthoryear{{Alb{\'a}n} \& {Wylezalek}}{{Alb{\'a}n} \&
  {Wylezalek}}{2023}]{alban22}
{Alb{\'a}n} M.,  {Wylezalek} D.,  2023, \mn@doi [arXiv e-prints]
  {10.48550/arXiv.2302.08519}, \href
  {https://ui.adsabs.harvard.edu/abs/2023arXiv230208519A} {p. arXiv:2302.08519}

\bibitem[\protect\citeauthoryear{{Alvarez-Hurtado}, {Barrera-Ballesteros},
  {S{\'a}nchez}, {Colombo}, {L{\'o}pez-S{\'a}nchez}  \&
  {Aquino-Ort{\'\i}z}}{{Alvarez-Hurtado} et~al.}{2022}]{paola22}
{Alvarez-Hurtado} P.,  {Barrera-Ballesteros} J.~K.,  {S{\'a}nchez} S.~F.,
  {Colombo} D.,  {L{\'o}pez-S{\'a}nchez} A.~R.,   {Aquino-Ort{\'\i}z} E.,
  2022, arXiv e-prints, \href
  {https://ui.adsabs.harvard.edu/abs/2022arXiv220211651A} {p. arXiv:2202.11651}

\bibitem[\protect\citeauthoryear{{Aquino-Ort{\'\i}z}
  et~al.,}{{Aquino-Ort{\'\i}z} et~al.}{2018}]{erik18}
{Aquino-Ort{\'\i}z} E.,  et~al., 2018, \mn@doi [\mnras]
  {10.1093/mnras/sty1522}, \href
  {https://ui.adsabs.harvard.edu/abs/2018MNRAS.479.2133A} {479, 2133}

\bibitem[\protect\citeauthoryear{{Asari}, {Cid Fernandes}, {Stasi{\'n}ska},
  {Torres-Papaqui}, {Mateus}, {Sodr{\'e}}, {Schoenell}  \& {Gomes}}{{Asari}
  et~al.}{2007}]{asari07}
{Asari} N.~V.,  {Cid Fernandes} R.,  {Stasi{\'n}ska} G.,  {Torres-Papaqui}
  J.~P.,  {Mateus} A.,  {Sodr{\'e}} L.,  {Schoenell} W.,   {Gomes} J.~M.,
  2007, \mn@doi [\mnras] {10.1111/j.1365-2966.2007.12255.x}, \href
  {http://adsabs.harvard.edu/abs/2007MNRAS.381..263A} {381, 263}

\bibitem[\protect\citeauthoryear{{Astropy Collaboration} et~al.,}{{Astropy
  Collaboration} et~al.}{2013}]{astropy:2013}
{Astropy Collaboration} et~al., 2013, \mn@doi [\aap]
  {10.1051/0004-6361/201322068}, \href
  {http://adsabs.harvard.edu/abs/2013A%26A...558A..33A} {558, A33}

\bibitem[\protect\citeauthoryear{{Astropy Collaboration} et~al.,}{{Astropy
  Collaboration} et~al.}{2018}]{astropy:2018}
{Astropy Collaboration} et~al., 2018, \mn@doi [aj] {10.3847/1538-3881/aabc4f},
  \href {https://ui.adsabs.harvard.edu/abs/2018AJ....156..123A} {156, 123}

\bibitem[\protect\citeauthoryear{{Bacon} et~al.,}{{Bacon} et~al.}{2010}]{muse}
{Bacon} R.,  et~al., 2010, in SPIE Conf. Series. , \mn@doi{10.1117/12.856027}

\bibitem[\protect\citeauthoryear{{Baldwin}, {Phillips}  \&
  {Terlevich}}{{Baldwin} et~al.}{1981}]{baldwin81}
{Baldwin} J.~A.,  {Phillips} M.~M.,   {Terlevich} R.,  1981, \mn@doi [\pasp]
  {10.1086/130766}, \href {http://adsabs.harvard.edu/abs/1981PASP...93....5B}
  {93, 5}

\bibitem[\protect\citeauthoryear{{Barrera-Ballesteros}, {S{\'a}nchez},
  {Heckman}, {Blanc}  \& {The MaNGA Team}}{{Barrera-Ballesteros}
  et~al.}{2017}]{bb17}
{Barrera-Ballesteros} J.~K.,  {S{\'a}nchez} S.~F.,  {Heckman} T.,  {Blanc}
  G.~A.,   {The MaNGA Team} 2017, \mn@doi [\apj] {10.3847/1538-4357/aa7aa9},
  \href {http://adsabs.harvard.edu/abs/2017ApJ...844...80B} {844, 80}

\bibitem[\protect\citeauthoryear{{Barrera-Ballesteros}
  et~al.,}{{Barrera-Ballesteros} et~al.}{2020}]{jkbb20}
{Barrera-Ballesteros} J.~K.,  et~al., 2020, \mn@doi [\mnras]
  {10.1093/mnras/stz3553}, \href
  {https://ui.adsabs.harvard.edu/abs/2020MNRAS.492.2651B} {492, 2651}

\bibitem[\protect\citeauthoryear{{Barrera-Ballesteros}
  et~al.,}{{Barrera-Ballesteros} et~al.}{2021a}]{jkbb21a}
{Barrera-Ballesteros} J.~K.,  et~al., 2021a, arXiv e-prints, \href
  {https://ui.adsabs.harvard.edu/abs/2021arXiv210102711B} {p. arXiv:2101.02711}

\bibitem[\protect\citeauthoryear{{Barrera-Ballesteros}
  et~al.,}{{Barrera-Ballesteros} et~al.}{2021b}]{jkbb21b}
{Barrera-Ballesteros} J.~K.,  et~al., 2021b, arXiv e-prints, \href
  {https://ui.adsabs.harvard.edu/abs/2021arXiv210104683B} {p. arXiv:2101.04683}

\bibitem[\protect\citeauthoryear{{Barrera-Ballesteros}
  et~al.,}{{Barrera-Ballesteros} et~al.}{2022}]{jkbb23}
{Barrera-Ballesteros} J.~K.,  et~al., 2022, \mn@doi [arXiv e-prints]
  {10.48550/arXiv.2206.07058}, \href
  {https://ui.adsabs.harvard.edu/abs/2022arXiv220607058B} {p. arXiv:2206.07058}

\bibitem[\protect\citeauthoryear{{Belfiore} et~al.,}{{Belfiore}
  et~al.}{2017}]{belfiore17a}
{Belfiore} F.,  et~al., 2017, \mn@doi [\mnras] {10.1093/mnras/stw3211}, \href
  {http://adsabs.harvard.edu/abs/2017MNRAS.466.2570B} {466, 2570}

\bibitem[\protect\citeauthoryear{{Belfiore} et~al.,}{{Belfiore}
  et~al.}{2022}]{belfiore22}
{Belfiore} F.,  et~al., 2022, \mn@doi [\aap] {10.1051/0004-6361/202141859},
  \href {https://ui.adsabs.harvard.edu/abs/2022A&A...659A..26B} {659, A26}

\bibitem[\protect\citeauthoryear{{Bell} \& {de Jong}}{{Bell} \& {de
  Jong}}{2000}]{bell00}
{Bell} E.~F.,  {de Jong} R.~S.,  2000, \mn@doi [\mnras]
  {10.1046/j.1365-8711.2000.03138.x}, \href
  {http://adsabs.harvard.edu/abs/2000MNRAS.312..497B} {312, 497}

\bibitem[\protect\citeauthoryear{{Bell}, {McIntosh}, {Katz}  \&
  {Weinberg}}{{Bell} et~al.}{2003}]{bell03}
{Bell} E.~F.,  {McIntosh} D.~H.,  {Katz} N.,   {Weinberg} M.~D.,  2003, \mn@doi
  [\apjs] {10.1086/378847}, \href
  {http://adsabs.harvard.edu/abs/2003ApJS..149..289B} {149, 289}

\bibitem[\protect\citeauthoryear{{Bigiel}, {Leroy}, {Walter}, {Brinks}, {de
  Blok}, {Madore}  \& {Thornley}}{{Bigiel} et~al.}{2008}]{bigiel08}
{Bigiel} F.,  {Leroy} A.,  {Walter} F.,  {Brinks} E.,  {de Blok} W.~J.~G.,
  {Madore} B.,   {Thornley} M.~D.,  2008, \mn@doi [\aj]
  {10.1088/0004-6256/136/6/2846}, \href
  {http://adsabs.harvard.edu/abs/2008AJ....136.2846B} {136, 2846}

\bibitem[\protect\citeauthoryear{{Binette}, {Magris}, {Stasi{\'n}ska}  \&
  {Bruzual}}{{Binette} et~al.}{1994}]{binette94}
{Binette} L.,  {Magris} C.~G.,  {Stasi{\'n}ska} G.,   {Bruzual} A.~G.,  1994,
  \aap, \href {http://adsabs.harvard.edu/abs/1994A%26A...292...13B} {292, 13}

\bibitem[\protect\citeauthoryear{{Bland-Hawthorn}}{{Bland-Hawthorn}}{1995}]{bland95}
{Bland-Hawthorn} J.,  1995, \pasa, \href
  {http://adsabs.harvard.edu/abs/1995PASA...12..190B} {12, 190}

\bibitem[\protect\citeauthoryear{{Blanton} \& {Moustakas}}{{Blanton} \&
  {Moustakas}}{2009}]{blanton09}
{Blanton} M.~R.,  {Moustakas} J.,  2009, \mn@doi [\araa]
  {10.1146/annurev-astro-082708-101734}, \href
  {http://adsabs.harvard.edu/abs/2009ARA%26A..47..159B} {47, 159}

\bibitem[\protect\citeauthoryear{{Blanton} et~al.,}{{Blanton}
  et~al.}{2017}]{blanton+2017}
{Blanton} M.~R.,  et~al., 2017, \mn@doi [\aj] {10.3847/1538-3881/aa7567}, \href
  {http://adsabs.harvard.edu/abs/2017AJ....154...28B} {154, 28}

\bibitem[\protect\citeauthoryear{{Bluck}, {Maiolino}, {Sanchez}, {Ellison},
  {Thorp}, {Piotrowska}, {Teimoorinia}  \& {Bundy}}{{Bluck}
  et~al.}{2019}]{bluck19}
{Bluck} A. F.~L.,  {Maiolino} R.,  {Sanchez} S.,  {Ellison} S.~L.,  {Thorp}
  M.~D.,  {Piotrowska} J.~M.,  {Teimoorinia} H.,   {Bundy} K.~A.,  2019, arXiv
  e-prints, \href {https://ui.adsabs.harvard.edu/abs/2019arXiv191108857B} {p.
  arXiv:1911.08857}

\bibitem[\protect\citeauthoryear{{Bluck} et~al.,}{{Bluck}
  et~al.}{2020}]{bluck20}
{Bluck} A. F.~L.,  et~al., 2020, \mn@doi [\mnras] {10.1093/mnras/staa2806},
  \href {https://ui.adsabs.harvard.edu/abs/2020MNRAS.499..230B} {499, 230}

\bibitem[\protect\citeauthoryear{{Bolatto} et~al.,}{{Bolatto}
  et~al.}{2017}]{bolatto17}
{Bolatto} A.~D.,  et~al., 2017, \mn@doi [\apj] {10.3847/1538-4357/aa86aa},
  \href {https://ui.adsabs.harvard.edu/abs/2017ApJ...846..159B} {846, 159}

\bibitem[\protect\citeauthoryear{{Bresolin}}{{Bresolin}}{2017}]{bresolin17}
{Bresolin} F.,  2017, in {Knapen} J.~H.,  {Lee} J.~C.,   {Gil de Paz} A.,  eds,
   Astrophysics and Space Science Library Vol. 434, Outskirts of Galaxies.
  p.~145 (\mn@eprint {arXiv} {1612.05278}),
  \mn@doi{10.1007/978-3-319-56570-5_5}

\bibitem[\protect\citeauthoryear{{Brinchmann}, {Charlot}, {White}, {Tremonti},
  {Kauffmann}, {Heckman}  \& {Brinkmann}}{{Brinchmann} et~al.}{2004}]{brin04}
{Brinchmann} J.,  {Charlot} S.,  {White} S.~D.~M.,  {Tremonti} C.,  {Kauffmann}
  G.,  {Heckman} T.,   {Brinkmann} J.,  2004, \mn@doi [\mnras]
  {10.1111/j.1365-2966.2004.07881.x}, \href
  {http://adsabs.harvard.edu/abs/2004MNRAS.351.1151B} {351, 1151}

\bibitem[\protect\citeauthoryear{{Brinchmann}, {Charlot}, {Kauffmann},
  {Heckman}, {White}  \& {Tremonti}}{{Brinchmann} et~al.}{2013}]{brin14}
{Brinchmann} J.,  {Charlot} S.,  {Kauffmann} G.,  {Heckman} T.,  {White}
  S.~D.~M.,   {Tremonti} C.,  2013, \mn@doi [\mnras] {10.1093/mnras/stt551},
  \href {http://adsabs.harvard.edu/abs/2013MNRAS.432.2112B} {432, 2112}

\bibitem[\protect\citeauthoryear{Bruzual \& Charlot}{Bruzual \&
  Charlot}{2003}]{bc03}
Bruzual G.,  Charlot S.,  2003, \mn@doi [Mon. Not. R. Astron. Soc.]
  {10.1046/j.1365-8711.2003.06897.x}, 344, 1000

\bibitem[\protect\citeauthoryear{{Bundy} et~al.,}{{Bundy} et~al.}{2015}]{manga}
{Bundy} K.,  et~al., 2015, \mn@doi [\apj] {10.1088/0004-637X/798/1/7}, \href
  {http://adsabs.harvard.edu/abs/2015ApJ...798....7B} {798, 7}

\bibitem[\protect\citeauthoryear{{Calette}, {Avila-Reese},
  {Rodr{\'\i}guez-Puebla}, {Hern{\'a}ndez-Toledo}  \& {Papastergis}}{{Calette}
  et~al.}{2018}]{calette18}
{Calette} A.~R.,  {Avila-Reese} V.,  {Rodr{\'\i}guez-Puebla} A.,
  {Hern{\'a}ndez-Toledo} H.,   {Papastergis} E.,  2018, \rmxaa, \href
  {https://ui.adsabs.harvard.edu/abs/2018RMxAA..54..443C} {54, 443}

\bibitem[\protect\citeauthoryear{{Calzetti}}{{Calzetti}}{2001}]{calz01}
{Calzetti} D.,  2001, \mn@doi [\pasp] {10.1086/324269}, \href
  {http://adsabs.harvard.edu/abs/2001PASP..113.1449C} {113, 1449}

\bibitem[\protect\citeauthoryear{{Calzetti}, {Meurer}, {Bohlin}, {Garnett},
  {Kinney}, {Leitherer}  \& {Storchi-Bergmann}}{{Calzetti}
  et~al.}{1997}]{calz97}
{Calzetti} D.,  {Meurer} G.~R.,  {Bohlin} R.~C.,  {Garnett} D.~R.,  {Kinney}
  A.~L.,  {Leitherer} C.,   {Storchi-Bergmann} T.,  1997, \mn@doi [\aj]
  {10.1086/118609}, \href
  {https://ui.adsabs.harvard.edu/abs/1997AJ....114.1834C} {114, 1834}

\bibitem[\protect\citeauthoryear{{Camps-Fari{\~n}a}, {Sanchez}, {Lacerda},
  {Carigi}, {Garc{\'\i}a-Benito}, {Mast}  \& {Galbany}}{{Camps-Fari{\~n}a}
  et~al.}{2021}]{camps20}
{Camps-Fari{\~n}a} A.,  {Sanchez} S.~F.,  {Lacerda} E.~A.~D.,  {Carigi} L.,
  {Garc{\'\i}a-Benito} R.,  {Mast} D.,   {Galbany} L.,  2021, \mn@doi [\mnras]
  {10.1093/mnras/stab1018}, \href
  {https://ui.adsabs.harvard.edu/abs/2021MNRAS.504.3478C} {504, 3478}

\bibitem[\protect\citeauthoryear{{Camps-Fari{\~n}a} et~al.,}{{Camps-Fari{\~n}a}
  et~al.}{2022}]{camps22}
{Camps-Fari{\~n}a} A.,  et~al., 2022, arXiv e-prints, \href
  {https://ui.adsabs.harvard.edu/abs/2022arXiv220301159C} {p. arXiv:2203.01159}

\bibitem[\protect\citeauthoryear{{Cano-D{\'{\i}}az} et~al.,}{{Cano-D{\'{\i}}az}
  et~al.}{2016}]{mariana16}
{Cano-D{\'{\i}}az} M.,  et~al., 2016, \mn@doi [\apjl]
  {10.3847/2041-8205/821/2/L26}, \href
  {http://adsabs.harvard.edu/abs/2016ApJ...821L..26C} {821, L26}

\bibitem[\protect\citeauthoryear{{Cano-D{\'\i}az}, {{\'A}vila-Reese},
  {S{\'a}nchez}, {Hern{\'a}ndez-Toledo}, {Rodr{\'\i}guez-Puebla}, {Boquien}  \&
  {Ibarra-Medel}}{{Cano-D{\'\i}az} et~al.}{2019}]{mariana19}
{Cano-D{\'\i}az} M.,  {{\'A}vila-Reese} V.,  {S{\'a}nchez} S.~F.,
  {Hern{\'a}ndez-Toledo} H.~M.,  {Rodr{\'\i}guez-Puebla} A.,  {Boquien} M.,
  {Ibarra-Medel} H.,  2019, \mn@doi [\mnras] {10.1093/mnras/stz1894}, \href
  {https://ui.adsabs.harvard.edu/abs/2019MNRAS.488.3929C} {488, 3929}

\bibitem[\protect\citeauthoryear{{Cappellari}}{{Cappellari}}{2016}]{cappellari16}
{Cappellari} M.,  2016, \mn@doi [\araa] {10.1146/annurev-astro-082214-122432},
  \href {http://adsabs.harvard.edu/abs/2016ARA%26A..54..597C} {54, 597}

\bibitem[\protect\citeauthoryear{{Cappellari} \& {Copin}}{{Cappellari} \&
  {Copin}}{2003}]{cappellari03}
{Cappellari} M.,  {Copin} Y.,  2003, \mn@doi [\mnras]
  {10.1046/j.1365-8711.2003.06541.x}, \href
  {http://adsabs.harvard.edu/abs/2003MNRAS.342..345C} {342, 345}

\bibitem[\protect\citeauthoryear{{Cardelli}, {Clayton}  \& {Mathis}}{{Cardelli}
  et~al.}{1989a}]{cardelli+1989}
{Cardelli} J.~A.,  {Clayton} G.~C.,   {Mathis} J.~S.,  1989a, \mn@doi [\apj]
  {10.1086/167900}, \href {http://adsabs.harvard.edu/abs/1989ApJ...345..245C}
  {345, 245}

\bibitem[\protect\citeauthoryear{{Cardelli}, {Clayton}  \& {Mathis}}{{Cardelli}
  et~al.}{1989b}]{cardelli1989}
{Cardelli} J.~A.,  {Clayton} G.~C.,   {Mathis} J.~S.,  1989b, \mn@doi [\apj]
  {10.1086/167900}, \href {http://adsabs.harvard.edu/abs/1989ApJ...345..245C}
  {345, 245}

\bibitem[\protect\citeauthoryear{{Carigi}, {Peimbert}  \& {Peimbert}}{{Carigi}
  et~al.}{2019}]{carigi19}
{Carigi} L.,  {Peimbert} M.,   {Peimbert} A.,  2019, \mn@doi [\apj]
  {10.3847/1538-4357/aaf28e}, \href
  {https://ui.adsabs.harvard.edu/abs/2019ApJ...873..107C} {873, 107}

\bibitem[\protect\citeauthoryear{{Catal{\'a}n-Torrecilla}
  et~al.,}{{Catal{\'a}n-Torrecilla} et~al.}{2017}]{catalan17}
{Catal{\'a}n-Torrecilla} C.,  et~al., 2017, \mn@doi [\apj]
  {10.3847/1538-4357/aa8a6d}, \href
  {http://adsabs.harvard.edu/abs/2017ApJ...848...87C} {848, 87}

\bibitem[\protect\citeauthoryear{{Chambers} et~al.,}{{Chambers}
  et~al.}{2016}]{PS1}
{Chambers} K.~C.,  et~al., 2016, arXiv e-prints, \href
  {https://ui.adsabs.harvard.edu/abs/2016arXiv161205560C} {p. arXiv:1612.05560}

\bibitem[\protect\citeauthoryear{{Chung}, {Park}  \& {Park}}{{Chung}
  et~al.}{2021}]{chung21}
{Chung} H.,  {Park} C.,   {Park} Y.-S.,  2021, \mn@doi [\apjs]
  {10.3847/1538-4365/ac2828}, \href
  {https://ui.adsabs.harvard.edu/abs/2021ApJS..257...66C} {257, 66}

\bibitem[\protect\citeauthoryear{{Cid Fernandes}, {Stasi{\'n}ska},
  {Schlickmann}, {Mateus}, {Vale Asari}, {Schoenell}  \& {Sodr{\'e}}}{{Cid
  Fernandes} et~al.}{2010}]{cid-fernandes10}
{Cid Fernandes} R.,  {Stasi{\'n}ska} G.,  {Schlickmann} M.~S.,  {Mateus} A.,
  {Vale Asari} N.,  {Schoenell} W.,   {Sodr{\'e}} L.,  2010, \mn@doi [\mnras]
  {10.1111/j.1365-2966.2009.16185.x}, \href
  {http://cdsads.u-strasbg.fr/abs/2010MNRAS.403.1036C} {403, 1036}

\bibitem[\protect\citeauthoryear{{Cid Fernandes} et~al.,}{{Cid Fernandes}
  et~al.}{2014}]{cidfernandes:2014}
{Cid Fernandes} R.,  et~al., 2014, \mn@doi [\aap]
  {10.1051/0004-6361/201321692}, \href
  {http://adsabs.harvard.edu/abs/2014A%26A...561A.130C} {561, A130}

\bibitem[\protect\citeauthoryear{{Colombo} et~al.,}{{Colombo}
  et~al.}{2020}]{colombo20}
{Colombo} D.,  et~al., 2020, arXiv e-prints, \href
  {https://ui.adsabs.harvard.edu/abs/2020arXiv200908383C} {p. arXiv:2009.08383}

\bibitem[\protect\citeauthoryear{{Corcho-Caballero}, {Ascasibar}, {S{\'a}nchez}
   \& {L{\'o}pez-S{\'a}nchez}}{{Corcho-Caballero} et~al.}{2023}]{corcho23}
{Corcho-Caballero} P.,  {Ascasibar} Y.,  {S{\'a}nchez} S.~F.,
  {L{\'o}pez-S{\'a}nchez} {\'A}.~R.,  2023, \mn@doi [\mnras]
  {10.1093/mnras/stad147}, \href
  {https://ui.adsabs.harvard.edu/abs/2023MNRAS.520..193C} {520, 193}

\bibitem[\protect\citeauthoryear{{Cortese} et~al.,}{{Cortese}
  et~al.}{2014}]{corte14}
{Cortese} L.,  et~al., 2014, \mn@doi [\apjl] {10.1088/2041-8205/795/2/L37},
  \href {https://ui.adsabs.harvard.edu/abs/2014ApJ...795L..37C} {795, L37}

\bibitem[\protect\citeauthoryear{{Courteau} et~al.,}{{Courteau}
  et~al.}{2014}]{court13}
{Courteau} S.,  et~al., 2014, \mn@doi [Reviews of Modern Physics]
  {10.1103/RevModPhys.86.47}, \href
  {http://adsabs.harvard.edu/abs/2014RvMP...86...47C} {86, 47}

\bibitem[\protect\citeauthoryear{{Cresci}, {Mannucci}  \& {Curti}}{{Cresci}
  et~al.}{2019}]{cresci19}
{Cresci} G.,  {Mannucci} F.,   {Curti} M.,  2019, \mn@doi [\aap]
  {10.1051/0004-6361/201834637}, \href
  {https://ui.adsabs.harvard.edu/abs/2019A&A...627A..42C} {627, A42}

\bibitem[\protect\citeauthoryear{{Croom} et~al.,}{{Croom} et~al.}{2012}]{sami}
{Croom} S.~M.,  et~al., 2012, \mn@doi [\mnras]
  {10.1111/j.1365-2966.2011.20365.x}, \href
  {http://adsabs.harvard.edu/abs/2012MNRAS.421..872C} {421, 872}

\bibitem[\protect\citeauthoryear{{D'Agostino} et~al.,}{{D'Agostino}
  et~al.}{2019}]{dagos19}
{D'Agostino} J.~J.,  et~al., 2019, \mn@doi [\mnras] {10.1093/mnras/stz1611},
  \href {https://ui.adsabs.harvard.edu/abs/2019MNRAS.487.4153D} {487, 4153}

\bibitem[\protect\citeauthoryear{{Davies} et~al.,}{{Davies}
  et~al.}{2016}]{davies16}
{Davies} R.~L.,  et~al., 2016, \mn@doi [\mnras] {10.1093/mnras/stw1754}, \href
  {http://adsabs.harvard.edu/abs/2016MNRAS.462.1616D} {462, 1616}

\bibitem[\protect\citeauthoryear{{Dopita}, {Koratkar}, {Evans}, {Allen},
  {Bicknell}, {Sutherland}, {Hawley}  \& {Sadler}}{{Dopita}
  et~al.}{1996}]{dopita96}
{Dopita} M.~A.,  {Koratkar} A.~P.,  {Evans} I.~N.,  {Allen} M.,  {Bicknell}
  G.~V.,  {Sutherland} R.~S.,  {Hawley} J.~F.,   {Sadler} E.,  1996, in
  {Eracleous} M.,  {Koratkar} A.,  {Leitherer} C.,   {Ho} L.,  eds,
  Astronomical Society of the Pacific Conference Series Vol. 103, The Physics
  of Liners in View of Recent Observations. p.~44

\bibitem[\protect\citeauthoryear{{Driver} et~al.,}{{Driver}
  et~al.}{2009}]{gamma}
{Driver} S.~P.,  et~al., 2009, \mn@doi [Astronomy and Geophysics]
  {10.1111/j.1468-4004.2009.50512.x}, \href
  {http://adsabs.harvard.edu/abs/2009A%26G....50e..12D} {50, 5.12}

\bibitem[\protect\citeauthoryear{{Ellison}, {S{\'a}nchez}, {Ibarra-Medel},
  {Antonio}, {Mendel}  \& {Barrera-Ballesteros}}{{Ellison}
  et~al.}{2018}]{ellison18}
{Ellison} S.~L.,  {S{\'a}nchez} S.~F.,  {Ibarra-Medel} H.,  {Antonio} B.,
  {Mendel} J.~T.,   {Barrera-Ballesteros} J.,  2018, \mn@doi [\mnras]
  {10.1093/mnras/stx2882}, \href
  {http://adsabs.harvard.edu/abs/2018MNRAS.474.2039E} {474, 2039}

\bibitem[\protect\citeauthoryear{{Ellison}, {Lin}, {Thorp}, {Pan},
  {S{\'a}nchez}, {Bluck}  \& {Belfiore}}{{Ellison} et~al.}{2020a}]{ellison21b}
{Ellison} S.~L.,  {Lin} L.,  {Thorp} M.~D.,  {Pan} H.-A.,  {S{\'a}nchez} S.~F.,
   {Bluck} A. F.~L.,   {Belfiore} F.,  2020a, \mn@doi [\mnras]
  {10.1093/mnrasl/slaa199}, \href
  {https://ui.adsabs.harvard.edu/abs/2020MNRAS.tmpL.238E} {}

\bibitem[\protect\citeauthoryear{{Ellison} et~al.,}{{Ellison}
  et~al.}{2020b}]{ellison20}
{Ellison} S.~L.,  et~al., 2020b, \mn@doi [\mnras] {10.1093/mnrasl/slz179},
  \href {https://ui.adsabs.harvard.edu/abs/2020MNRAS.493L..39E} {493, L39}

\bibitem[\protect\citeauthoryear{{Emsellem} et~al.,}{{Emsellem}
  et~al.}{2022}]{emse22}
{Emsellem} E.,  et~al., 2022, \mn@doi [\aap] {10.1051/0004-6361/202141727},
  \href {https://ui.adsabs.harvard.edu/abs/2022A&A...659A.191E} {659, A191}

\bibitem[\protect\citeauthoryear{{Espinosa-Ponce}, {S{\'a}nchez}, {Morisset},
  {Barrera-Ballesteros}, {Galbany}, {Garc{\'\i}a-Benito}, {Lacerda}  \&
  {Mast}}{{Espinosa-Ponce} et~al.}{2020}]{espi20}
{Espinosa-Ponce} C.,  {S{\'a}nchez} S.~F.,  {Morisset} C.,
  {Barrera-Ballesteros} J.~K.,  {Galbany} L.,  {Garc{\'\i}a-Benito} R.,
  {Lacerda} E.~A.~D.,   {Mast} D.,  2020, \mn@doi [\mnras]
  {10.1093/mnras/staa782}, \href
  {https://ui.adsabs.harvard.edu/abs/2020MNRAS.494.1622E} {494, 1622}

\bibitem[\protect\citeauthoryear{{Espinosa-Ponce}, {S{\'a}nchez}, {Morisset},
  {Barrera-Ballesteros}, {Galbany}, {Garc{\'\i}a-Benito}, {Lacerda}  \&
  {Mast}}{{Espinosa-Ponce} et~al.}{2022}]{espi22}
{Espinosa-Ponce} C.,  {S{\'a}nchez} S.~F.,  {Morisset} C.,
  {Barrera-Ballesteros} J.~K.,  {Galbany} L.,  {Garc{\'\i}a-Benito} R.,
  {Lacerda} E.~A.~D.,   {Mast} D.,  2022, \mn@doi [\mnras]
  {10.1093/mnras/stac456}, \href
  {https://ui.adsabs.harvard.edu/abs/2022MNRAS.tmp..448E} {}

\bibitem[\protect\citeauthoryear{{Faber}}{{Faber}}{1977}]{faber1977}
{Faber} S.,  1977, in Larson B. M. T. . R.~B.,  ed., The Evolution of Galaxies
  and Stellar Populations. p.~157

\bibitem[\protect\citeauthoryear{{Faber} \& {Jackson}}{{Faber} \&
  {Jackson}}{1976}]{faber76}
{Faber} S.~M.,  {Jackson} R.~E.,  1976, \mn@doi [\apj] {10.1086/154215}, \href
  {http://adsabs.harvard.edu/abs/1976ApJ...204..668F} {204, 668}

\bibitem[\protect\citeauthoryear{{Flewelling} et~al.,}{{Flewelling}
  et~al.}{2020}]{PS1_data}
{Flewelling} H.~A.,  et~al., 2020, \mn@doi [\apjs] {10.3847/1538-4365/abb82d},
  \href {https://ui.adsabs.harvard.edu/abs/2020ApJS..251....7F} {251, 7}

\bibitem[\protect\citeauthoryear{{Flores-Fajardo}, {Morisset}, {Stasi{\'n}ska}
  \& {Binette}}{{Flores-Fajardo} et~al.}{2011}]{floresfajardo11}
{Flores-Fajardo} N.,  {Morisset} C.,  {Stasi{\'n}ska} G.,   {Binette} L.,
  2011, \mn@doi [\mnras] {10.1111/j.1365-2966.2011.18848.x}, \href
  {http://adsabs.harvard.edu/abs/2011MNRAS.415.2182F} {415, 2182}

\bibitem[\protect\citeauthoryear{{Fruchter} \& {Hook}}{{Fruchter} \&
  {Hook}}{2002}]{fruchter:2002}
{Fruchter} A.~S.,  {Hook} R.~N.,  2002, \mn@doi [\pasp] {10.1086/338393}, \href
  {http://adsabs.harvard.edu/abs/2002PASP..114..144F} {114, 144}

\bibitem[\protect\citeauthoryear{{Fukugita} et~al.,}{{Fukugita}
  et~al.}{2007}]{fuku07}
{Fukugita} M.,  et~al., 2007, \mn@doi [\aj] {10.1086/518962}, \href
  {http://adsabs.harvard.edu/abs/2007AJ....134..579F} {134, 579}

\bibitem[\protect\citeauthoryear{{Gaia Collaboration} et~al.,}{{Gaia
  Collaboration} et~al.}{2016}]{gaia1}
{Gaia Collaboration} et~al., 2016, \mn@doi [\aap]
  {10.1051/0004-6361/201629272}, \href
  {https://ui.adsabs.harvard.edu/abs/2016A&A...595A...1G} {595, A1}

\bibitem[\protect\citeauthoryear{{Gaia Collaboration} et~al.,}{{Gaia
  Collaboration} et~al.}{2021}]{gaia3}
{Gaia Collaboration} et~al., 2021, \mn@doi [\aap]
  {10.1051/0004-6361/202039657}, \href
  {https://ui.adsabs.harvard.edu/abs/2021A&A...649A...1G} {649, A1}

\bibitem[\protect\citeauthoryear{{Galbany} et~al.,}{{Galbany}
  et~al.}{2018}]{pisco}
{Galbany} L.,  et~al., 2018, \mn@doi [\apj] {10.3847/1538-4357/aaaf20}, \href
  {https://ui.adsabs.harvard.edu/abs/2018ApJ...855..107G} {855, 107}

\bibitem[\protect\citeauthoryear{{Gallazzi} \& {Bell}}{{Gallazzi} \&
  {Bell}}{2009}]{gallazzi:2009}
{Gallazzi} A.,  {Bell} E.~F.,  2009, \mn@doi [\apjs]
  {10.1088/0067-0049/185/2/253}, \href
  {http://adsabs.harvard.edu/abs/2009ApJS..185..253G} {185, 253}

\bibitem[\protect\citeauthoryear{{Gallazzi}, {Charlot}, {Brinchmann}, {White}
  \& {Tremonti}}{{Gallazzi} et~al.}{2005}]{gallazzi05}
{Gallazzi} A.,  {Charlot} S.,  {Brinchmann} J.,  {White} S.~D.~M.,   {Tremonti}
  C.~A.,  2005, \mn@doi [\mnras] {10.1111/j.1365-2966.2005.09321.x}, \href
  {http://adsabs.harvard.edu/abs/2005MNRAS.362...41G} {362, 41}

\bibitem[\protect\citeauthoryear{{Gallazzi}, {Charlot}, {Brinchmann}  \&
  {White}}{{Gallazzi} et~al.}{2006}]{gallazzi06}
{Gallazzi} A.,  {Charlot} S.,  {Brinchmann} J.,   {White} S.~D.~M.,  2006,
  \mn@doi [\mnras] {10.1111/j.1365-2966.2006.10548.x}, \href
  {http://adsabs.harvard.edu/abs/2006MNRAS.370.1106G} {370, 1106}

\bibitem[\protect\citeauthoryear{{Garc{\'{\i}}a-Benito}
  et~al.,}{{Garc{\'{\i}}a-Benito} et~al.}{2015}]{dr2}
{Garc{\'{\i}}a-Benito} R.,  et~al., 2015, \mn@doi [\aap]
  {10.1051/0004-6361/201425080}, \href
  {http://adsabs.harvard.edu/abs/2015A%26A...576A.135G} {576, A135}

\bibitem[\protect\citeauthoryear{{Garc{\'{\i}}a-Benito}
  et~al.,}{{Garc{\'{\i}}a-Benito} et~al.}{2017}]{rgb17}
{Garc{\'{\i}}a-Benito} R.,  et~al., 2017, \mn@doi [\aap]
  {10.1051/0004-6361/201731357}, \href
  {http://adsabs.harvard.edu/abs/2017A%26A...608A..27G} {608, A27}

\bibitem[\protect\citeauthoryear{{Garc{\'\i}a-Benito}, {Gonz{\'a}lez Delgado},
  {P{\'e}rez}, {Cid Fernandes}, {S{\'a}nchez}  \& {de
  Amorim}}{{Garc{\'\i}a-Benito} et~al.}{2019}]{rgb18}
{Garc{\'\i}a-Benito} R.,  {Gonz{\'a}lez Delgado} R.~M.,  {P{\'e}rez} E.,  {Cid
  Fernandes} R.,  {S{\'a}nchez} S.~F.,   {de Amorim} A.~L.,  2019, \mn@doi
  [\aap] {10.1051/0004-6361/201833993}, \href
  {https://ui.adsabs.harvard.edu/abs/2019A&A...621A.120G} {621, A120}

\bibitem[\protect\citeauthoryear{{George} et~al.,}{{George}
  et~al.}{2019}]{george19}
{George} K.,  et~al., 2019, \mn@doi [\mnras] {10.1093/mnras/stz1443}, \href
  {https://ui.adsabs.harvard.edu/abs/2019MNRAS.487.3102G} {487, 3102}

\bibitem[\protect\citeauthoryear{{Goddard} et~al.,}{{Goddard}
  et~al.}{2017}]{godd15}
{Goddard} D.,  et~al., 2017, \mn@doi [\mnras] {10.1093/mnras/stw3371}, \href
  {http://adsabs.harvard.edu/abs/2017MNRAS.466.4731G} {466, 4731}

\bibitem[\protect\citeauthoryear{{Gomes} et~al.,}{{Gomes}
  et~al.}{2016}]{gomes16a}
{Gomes} J.~M.,  et~al., 2016, \mn@doi [\aap] {10.1051/0004-6361/201525976},
  \href {http://adsabs.harvard.edu/abs/2016A%26A...588A..68G} {588, A68}

\bibitem[\protect\citeauthoryear{{Gonz{\'a}lez Delgado}, {P{\'e}rez}, {Cid
  Fernandes}  \& {et~al.}}{{Gonz{\'a}lez Delgado} et~al.}{2014a}]{rosa14}
{Gonz{\'a}lez Delgado} R.~M.,  {P{\'e}rez} E.,  {Cid Fernandes} R.,   {et~al.}
  2014a, \mn@doi [\aap] {10.1051/0004-6361/201322011}, \href
  {http://adsabs.harvard.edu/abs/2014A%26A...562A..47G} {562, A47}

\bibitem[\protect\citeauthoryear{{Gonz{\'a}lez Delgado} et~al.,}{{Gonz{\'a}lez
  Delgado} et~al.}{2014b}]{rosa14b}
{Gonz{\'a}lez Delgado} R.~M.,  et~al., 2014b, \mn@doi [\apjl]
  {10.1088/2041-8205/791/1/L16}, \href
  {http://adsabs.harvard.edu/abs/2014ApJ...791L..16G} {791, L16}

\bibitem[\protect\citeauthoryear{{Gonz{\'a}lez Delgado} et~al.,}{{Gonz{\'a}lez
  Delgado} et~al.}{2015}]{rosa15}
{Gonz{\'a}lez Delgado} R.~M.,  et~al., 2015, \mn@doi [\aap]
  {10.1051/0004-6361/201525938}, \href
  {http://adsabs.harvard.edu/abs/2015A%26A...581A.103G} {581, A103}

\bibitem[\protect\citeauthoryear{{Gonz{\'a}lez Delgado} et~al.,}{{Gonz{\'a}lez
  Delgado} et~al.}{2016}]{rosa16}
{Gonz{\'a}lez Delgado} R.~M.,  et~al., 2016, \mn@doi [\aap]
  {10.1051/0004-6361/201628174}, \href
  {http://adsabs.harvard.edu/abs/2016A%26A...590A..44G} {590, A44}

\bibitem[\protect\citeauthoryear{{Grandmont}, {Drissen}, {Mandar}, {Thibault}
  \& {Baril}}{{Grandmont} et~al.}{2012}]{grand12}
{Grandmont} F.,  {Drissen} L.,  {Mandar} J.,  {Thibault} S.,   {Baril} M.,
  2012, in {McLean} I.~S.,  {Ramsay} S.~K.,   {Takami} H.,  eds,  Society of
  Photo-Optical Instrumentation Engineers (SPIE) Conference Series Vol. 8446,
  Ground-based and Airborne Instrumentation for Astronomy IV. p. 84460U,
  \mn@doi{10.1117/12.926782}

\bibitem[\protect\citeauthoryear{{Green} et~al.,}{{Green}
  et~al.}{2018}]{green18}
{Green} A.~W.,  et~al., 2018, \mn@doi [\mnras] {10.1093/mnras/stx3135}, \href
  {http://adsabs.harvard.edu/abs/2018MNRAS.475..716G} {475, 716}

\bibitem[\protect\citeauthoryear{{Gunn} et~al.,}{{Gunn} et~al.}{1998}]{gunn98}
{Gunn} J.~E.,  et~al., 1998, \mn@doi [\aj] {10.1086/300645}, \href
  {https://ui.adsabs.harvard.edu/abs/1998AJ....116.3040G} {116, 3040}

\bibitem[\protect\citeauthoryear{{Heckman}, {Armus}  \& {Miley}}{{Heckman}
  et~al.}{1990}]{heckman90}
{Heckman} T.~M.,  {Armus} L.,   {Miley} G.~K.,  1990, \mn@doi [\apjs]
  {10.1086/191522}, \href {http://adsabs.harvard.edu/abs/1990ApJS...74..833H}
  {74, 833}

\bibitem[\protect\citeauthoryear{{Ho}, {Filippenko}  \& {Sargent}}{{Ho}
  et~al.}{1997}]{ho97a}
{Ho} L.~C.,  {Filippenko} A.~V.,   {Sargent} W. L.~W.,  1997, \mn@doi [\apj]
  {10.1086/304638}, \href
  {https://ui.adsabs.harvard.edu/abs/1997ApJ...487..568H} {487, 568}

\bibitem[\protect\citeauthoryear{{Husemann} et~al.,}{{Husemann}
  et~al.}{2013}]{dr1}
{Husemann} B.,  et~al., 2013, \mn@doi [\aap] {10.1051/0004-6361/201220582},
  \href {http://adsabs.harvard.edu/abs/2013A%26A...549A..87H} {549, A87}

\bibitem[\protect\citeauthoryear{{Ibarra-Medel} et~al.,}{{Ibarra-Medel}
  et~al.}{2016}]{ibarra16}
{Ibarra-Medel} H.~J.,  et~al., 2016, \mn@doi [\mnras] {10.1093/mnras/stw2126},
  \href {http://adsabs.harvard.edu/abs/2016MNRAS.463.2799I} {463, 2799}

\bibitem[\protect\citeauthoryear{{Ibarra-Medel}, {Avila-Reese}, {S{\'a}nchez},
  {Gonz{\'a}lez-Samaniego}  \& {Rodr{\'{\i}}guez-Puebla}}{{Ibarra-Medel}
  et~al.}{2019}]{ibarra19}
{Ibarra-Medel} H.~J.,  {Avila-Reese} V.,  {S{\'a}nchez} S.~F.,
  {Gonz{\'a}lez-Samaniego} A.,   {Rodr{\'{\i}}guez-Puebla} A.,  2019, \mn@doi
  [\mnras] {10.1093/mnras/sty3256}, \href
  {https://ui.adsabs.harvard.edu/abs/2019MNRAS.483.4525I} {483, 4525}

\bibitem[\protect\citeauthoryear{{Into} \& {Portinari}}{{Into} \&
  {Portinari}}{2013}]{into:2013}
{Into} T.,  {Portinari} L.,  2013, \mn@doi [\mnras] {10.1093/mnras/stt071},
  \href {http://adsabs.harvard.edu/abs/2013MNRAS.430.2715I} {430, 2715}

\bibitem[\protect\citeauthoryear{{Kalinova} et~al.,}{{Kalinova}
  et~al.}{2022}]{kali22}
{Kalinova} V.,  et~al., 2022, \mn@doi [\aap] {10.1051/0004-6361/202243541},
  \href {https://ui.adsabs.harvard.edu/abs/2022A&A...665A..90K} {665, A90}

\bibitem[\protect\citeauthoryear{{Kauffmann} et~al.,}{{Kauffmann}
  et~al.}{2003a}]{kauff03a}
{Kauffmann} G.,  et~al., 2003a, \mn@doi [\mnras]
  {10.1046/j.1365-8711.2003.06291.x}, \href
  {http://adsabs.harvard.edu/abs/2003MNRAS.341...33K} {341, 33}

\bibitem[\protect\citeauthoryear{{Kauffmann} et~al.,}{{Kauffmann}
  et~al.}{2003b}]{kauff03}
{Kauffmann} G.,  et~al., 2003b, \mn@doi [\mnras]
  {10.1111/j.1365-2966.2003.07154.x}, \href
  {http://adsabs.harvard.edu/abs/2003MNRAS.346.1055K} {346, 1055}

\bibitem[\protect\citeauthoryear{{Kelz} et~al.,}{{Kelz} et~al.}{2006}]{kelz06}
{Kelz} A.,  et~al., 2006, \mn@doi [\pasp] {10.1086/497455}, \href
  {http://adsabs.harvard.edu/abs/2006PASP..118..129K} {118, 129}

\bibitem[\protect\citeauthoryear{{Kennicutt}}{{Kennicutt}}{1998}]{kennicutt1998}
{Kennicutt} Jr. R.~C.,  1998, \mn@doi [\apj] {10.1086/305588}, \href
  {http://adsabs.harvard.edu/abs/1998ApJ...498..541K} {498, 541}

\bibitem[\protect\citeauthoryear{{Kennicutt}, {Keel}  \& {Blaha}}{{Kennicutt}
  et~al.}{1989}]{kennicutt89}
{Kennicutt} Jr. R.~C.,  {Keel} W.~C.,   {Blaha} C.~A.,  1989, \mn@doi [\aj]
  {10.1086/115046}, \href {http://adsabs.harvard.edu/abs/1989AJ.....97.1022K}
  {97, 1022}

\bibitem[\protect\citeauthoryear{{Kewley}, {Dopita}, {Sutherland}, {Heisler}
  \& {Trevena}}{{Kewley} et~al.}{2001}]{kewley01}
{Kewley} L.~J.,  {Dopita} M.~A.,  {Sutherland} R.~S.,  {Heisler} C.~A.,
  {Trevena} J.,  2001, \mn@doi [\apj] {10.1086/321545}, \href
  {http://adsabs.harvard.edu/abs/2001ApJ...556..121K} {556, 121}

\bibitem[\protect\citeauthoryear{Kewley, Nicholls  \& Sutherland}{Kewley
  et~al.}{2019}]{kewley19}
Kewley L.~J.,  Nicholls D.~C.,   Sutherland R.~S.,  2019, \mn@doi [Annual
  Review of Astronomy and Astrophysics] {10.1146/annurev-astro-081817-051832},
  57, 511

\bibitem[\protect\citeauthoryear{{Kobayashi}, {Karakas}  \&
  {Lugaro}}{{Kobayashi} et~al.}{2020}]{Koba20}
{Kobayashi} C.,  {Karakas} A.~I.,   {Lugaro} M.,  2020, \mn@doi [\apj]
  {10.3847/1538-4357/abae65}, \href
  {https://ui.adsabs.harvard.edu/abs/2020ApJ...900..179K} {900, 179}

\bibitem[\protect\citeauthoryear{{Lacerda} et~al.,}{{Lacerda}
  et~al.}{2018}]{lacerda18}
{Lacerda} E.~A.~D.,  et~al., 2018, \mn@doi [\mnras] {10.1093/mnras/stx3022},
  \href {http://adsabs.harvard.edu/abs/2018MNRAS.474.3727L} {474, 3727}

\bibitem[\protect\citeauthoryear{{Lacerda}, {S{\'a}nchez}, {Cid Fernandes},
  {L{\'o}pez-Cob{\'a}}, {Espinosa-Ponce}  \& {Galbany}}{{Lacerda}
  et~al.}{2020}]{lacerda20}
{Lacerda} E. A.~D.,  {S{\'a}nchez} S.~F.,  {Cid Fernandes} R.,
  {L{\'o}pez-Cob{\'a}} C.,  {Espinosa-Ponce} C.,   {Galbany} L.,  2020, \mn@doi
  [\mnras] {10.1093/mnras/staa008}, \href
  {https://ui.adsabs.harvard.edu/abs/2020MNRAS.492.3073L} {492, 3073}

\bibitem[\protect\citeauthoryear{{Lacerda}, {S{\'a}nchez},
  {Mej{\'\i}a-Narv{\'a}ez}, {Camps-Fari{\~n}a}, {Espinosa-Ponce},
  {Barrera-Ballesteros}, {Ibarra-Medel}  \& {Lugo-Aranda}}{{Lacerda}
  et~al.}{2022}]{pypipe3d}
{Lacerda} E. A.~D.,  {S{\'a}nchez} S.~F.,  {Mej{\'\i}a-Narv{\'a}ez} A.,
  {Camps-Fari{\~n}a} A.,  {Espinosa-Ponce} C.,  {Barrera-Ballesteros} J.~K.,
  {Ibarra-Medel} H.,   {Lugo-Aranda} A.~Z.,  2022, arXiv e-prints, \href
  {https://ui.adsabs.harvard.edu/abs/2022arXiv220208027L} {p. arXiv:2202.08027}

\bibitem[\protect\citeauthoryear{{Larson}}{{Larson}}{1976}]{larson1976}
{Larson} R.~B.,  1976, \mn@doi [\mnras] {10.1093/mnras/176.1.31}, \href
  {http://adsabs.harvard.edu/abs/1976MNRAS.176...31L} {176, 31}

\bibitem[\protect\citeauthoryear{{Law} et~al.,}{{Law} et~al.}{2015}]{law15}
{Law} D.~R.,  et~al., 2015, \mn@doi [\aj] {10.1088/0004-6256/150/1/19}, \href
  {http://adsabs.harvard.edu/abs/2015AJ....150...19L} {150, 19}

\bibitem[\protect\citeauthoryear{{Law} et~al.,}{{Law} et~al.}{2021}]{law21}
{Law} D.~R.,  et~al., 2021, arXiv e-prints, \href
  {https://ui.adsabs.harvard.edu/abs/2021arXiv211211281L} {p. arXiv:2112.11281}

\bibitem[\protect\citeauthoryear{{Leroy} et~al.,}{{Leroy}
  et~al.}{2013}]{leroy13}
{Leroy} A.~K.,  et~al., 2013, \mn@doi [\aj] {10.1088/0004-6256/146/2/19}, \href
  {http://adsabs.harvard.edu/abs/2013AJ....146...19L} {146, 19}

\bibitem[\protect\citeauthoryear{{Lin} et~al.,}{{Lin} et~al.}{2019}]{lin19}
{Lin} L.,  et~al., 2019, arXiv e-prints, \href
  {https://ui.adsabs.harvard.edu/abs/2019arXiv190911243L} {p. arXiv:1909.11243}

\bibitem[\protect\citeauthoryear{{Lin} et~al.,}{{Lin} et~al.}{2020}]{lin20}
{Lin} L.,  et~al., 2020, arXiv e-prints, \href
  {https://ui.adsabs.harvard.edu/abs/2020arXiv201001751L} {p. arXiv:2010.01751}

\bibitem[\protect\citeauthoryear{{Liu}, {Blanton}  \& {Law}}{{Liu}
  et~al.}{2020}]{liu20}
{Liu} D.,  {Blanton} M.~R.,   {Law} D.~R.,  2020, \mn@doi [\aj]
  {10.3847/1538-3881/ab5aeb}, \href
  {https://ui.adsabs.harvard.edu/abs/2020AJ....159...22L} {159, 22}

\bibitem[\protect\citeauthoryear{{L{\'o}pez-Cob{\'a}}
  et~al.,}{{L{\'o}pez-Cob{\'a}} et~al.}{2020}]{carlos20}
{L{\'o}pez-Cob{\'a}} C.,  et~al., 2020, arXiv e-prints, \href
  {https://ui.adsabs.harvard.edu/abs/2020arXiv200209328L} {p. arXiv:2002.09328}

\bibitem[\protect\citeauthoryear{{L{\'o}pez-S{\'a}nchez}, {Dopita}, {Kewley},
  {Zahid}, {Nicholls}  \& {Scharw{\"a}chter}}{{L{\'o}pez-S{\'a}nchez}
  et~al.}{2012}]{angel12}
{L{\'o}pez-S{\'a}nchez} {\'A}.~R.,  {Dopita} M.~A.,  {Kewley} L.~J.,  {Zahid}
  H.~J.,  {Nicholls} D.~C.,   {Scharw{\"a}chter} J.,  2012, \mn@doi [\mnras]
  {10.1111/j.1365-2966.2012.21145.x}, \href
  {http://adsabs.harvard.edu/abs/2012MNRAS.426.2630L} {426, 2630}

\bibitem[\protect\citeauthoryear{{Maiolino} \& {Mannucci}}{{Maiolino} \&
  {Mannucci}}{2019}]{maio19}
{Maiolino} R.,  {Mannucci} F.,  2019, \mn@doi [\aapr]
  {10.1007/s00159-018-0112-2}, \href
  {https://ui.adsabs.harvard.edu/abs/2019A&ARv..27....3M} {27, 3}

\bibitem[\protect\citeauthoryear{{Marino} et~al.,}{{Marino}
  et~al.}{2013}]{marino13}
{Marino} R.~A.,  et~al., 2013, \mn@doi [\aap] {10.1051/0004-6361/201321956},
  \href {http://adsabs.harvard.edu/abs/2013A%26A...559A.114M} {559, A114}

\bibitem[\protect\citeauthoryear{{M{\'a}rmol-Queralt{\'o}}
  et~al.,}{{M{\'a}rmol-Queralt{\'o}} et~al.}{2011}]{marmol-queralto11}
{M{\'a}rmol-Queralt{\'o}} E.,  et~al., 2011, \mn@doi [\aap]
  {10.1051/0004-6361/201117032}, \href
  {http://adsabs.harvard.edu/abs/2011A%26A...534A...8M} {534, A8}

\bibitem[\protect\citeauthoryear{{Mast} et~al.,}{{Mast} et~al.}{2014}]{mast14}
{Mast} D.,  et~al., 2014, \mn@doi [\aap] {10.1051/0004-6361/201321789}, \href
  {http://adsabs.harvard.edu/abs/2014A%26A...561A.129M} {561, A129}

\bibitem[\protect\citeauthoryear{{Matteucci}}{{Matteucci}}{1992}]{matt92}
{Matteucci} F.,  1992, \memsai, \href
  {https://ui.adsabs.harvard.edu/abs/1992MmSAI..63..301M} {63, 301}

\bibitem[\protect\citeauthoryear{{Mej{\'\i}a-Narv{\'a}ez}, {S{\'a}nchez},
  {Lacerda}, {Carigi}, {Galbany}, {Husemann}  \&
  {Garc{\'\i}a-Benito}}{{Mej{\'\i}a-Narv{\'a}ez} et~al.}{2020}]{mejia20}
{Mej{\'\i}a-Narv{\'a}ez} A.,  {S{\'a}nchez} S.~F.,  {Lacerda} E.~A.~D.,
  {Carigi} L.,  {Galbany} L.,  {Husemann} B.,   {Garc{\'\i}a-Benito} R.,  2020,
  \mn@doi [\mnras] {10.1093/mnras/staa3094}, \href
  {https://ui.adsabs.harvard.edu/abs/2020MNRAS.499.4838M} {499, 4838}

\bibitem[\protect\citeauthoryear{{M{\'e}ndez-Abreu}, {S{\'a}nchez}  \& {de
  Lorenzo-C{\'a}ceres}}{{M{\'e}ndez-Abreu} et~al.}{2019}]{jairo19}
{M{\'e}ndez-Abreu} J.,  {S{\'a}nchez} S.~F.,   {de Lorenzo-C{\'a}ceres} A.,
  2019, \mn@doi [\mnras] {10.1093/mnrasl/slz103}, \href
  {https://ui.adsabs.harvard.edu/abs/2019MNRAS.488L..80M} {488, L80}

\bibitem[\protect\citeauthoryear{{Osterbrock}}{{Osterbrock}}{1989}]{osterbrock89}
{Osterbrock} D.~E.,  1989, {Astrophysics of gaseous nebulae and active galactic
  nuclei}.
University Science Books

\bibitem[\protect\citeauthoryear{{Ostriker}, {McKee}  \& {Leroy}}{{Ostriker}
  et~al.}{2010}]{ostriker10}
{Ostriker} E.~C.,  {McKee} C.~F.,   {Leroy} A.~K.,  2010, \mn@doi [\apj]
  {10.1088/0004-637X/721/2/975}, \href
  {https://ui.adsabs.harvard.edu/abs/2010ApJ...721..975O} {721, 975}

\bibitem[\protect\citeauthoryear{{Pan} et~al.,}{{Pan} et~al.}{2018}]{pan18}
{Pan} H.-A.,  et~al., 2018, \mn@doi [\apj] {10.3847/1538-4357/aaa9bc}, \href
  {http://adsabs.harvard.edu/abs/2018ApJ...854..159P} {854, 159}

\bibitem[\protect\citeauthoryear{{Pan} et~al.,}{{Pan} et~al.}{2022}]{pan22}
{Pan} H.-A.,  et~al., 2022, \mn@doi [\apj] {10.3847/1538-4357/ac474f}, \href
  {https://ui.adsabs.harvard.edu/abs/2022ApJ...927....9P} {927, 9}

\bibitem[\protect\citeauthoryear{{Panter}, {Heavens}  \& {Jimenez}}{{Panter}
  et~al.}{2003}]{panter03}
{Panter} B.,  {Heavens} A.~F.,   {Jimenez} R.,  2003, \mn@doi [\mnras]
  {10.1046/j.1365-8711.2003.06722.x}, \href
  {http://adsabs.harvard.edu/abs/2003MNRAS.343.1145P} {343, 1145}

\bibitem[\protect\citeauthoryear{{Peebles}}{{Peebles}}{1969}]{peebles1969}
{Peebles} P.~J.~E.,  1969, \mn@doi [\apj] {10.1086/149876}, \href
  {http://adsabs.harvard.edu/abs/1969ApJ...155..393P} {155, 393}

\bibitem[\protect\citeauthoryear{{Peimbert} \& {Peimbert}}{{Peimbert} \&
  {Peimbert}}{2006}]{peim06}
{Peimbert} M.,  {Peimbert} A.,  2006, in Revista Mexicana de Astronomia y
  Astrofisica Conference Series. p.~163

\bibitem[\protect\citeauthoryear{{Peletier}, {Davies}, {Illingworth}, {Davis}
  \& {Cawson}}{{Peletier} et~al.}{1990}]{pele90}
{Peletier} R.~F.,  {Davies} R.~L.,  {Illingworth} G.~D.,  {Davis} L.~E.,
  {Cawson} M.,  1990, \mn@doi [\aj] {10.1086/115582}, \href
  {http://adsabs.harvard.edu/abs/1990AJ....100.1091P} {100, 1091}

\bibitem[\protect\citeauthoryear{{P{\'e}rez-Gonz{\'a}lez}, {Trujillo}, {Barro},
  {Gallego}, {Zamorano}  \& {Conselice}}{{P{\'e}rez-Gonz{\'a}lez}
  et~al.}{2008}]{pperez08}
{P{\'e}rez-Gonz{\'a}lez} P.~G.,  {Trujillo} I.,  {Barro} G.,  {Gallego} J.,
  {Zamorano} J.,   {Conselice} C.~J.,  2008, \mn@doi [\apj] {10.1086/591843},
  \href {https://ui.adsabs.harvard.edu/abs/2008ApJ...687...50P} {687, 50}

\bibitem[\protect\citeauthoryear{{P{\'e}rez} et~al.,}{{P{\'e}rez}
  et~al.}{2013}]{eperez13}
{P{\'e}rez} E.,  et~al., 2013, \mn@doi [\apjl] {10.1088/2041-8205/764/1/L1},
  \href {http://adsabs.harvard.edu/abs/2013ApJ...764L...1P} {764, L1}

\bibitem[\protect\citeauthoryear{{Pessa} et~al.,}{{Pessa}
  et~al.}{2022}]{pessa22}
{Pessa} I.,  et~al., 2022, \mn@doi [\aap] {10.1051/0004-6361/202142832}, \href
  {https://ui.adsabs.harvard.edu/abs/2022A&A...663A..61P} {663, A61}

\bibitem[\protect\citeauthoryear{{Pilyugin}, {Thuan}  \&
  {V{\'{\i}}lchez}}{{Pilyugin} et~al.}{2007}]{pily07}
{Pilyugin} L.~S.,  {Thuan} T.~X.,   {V{\'{\i}}lchez} J.~M.,  2007, \mn@doi
  [\mnras] {10.1111/j.1365-2966.2007.11444.x}, \href
  {http://adsabs.harvard.edu/abs/2007MNRAS.376..353P} {376, 353}

\bibitem[\protect\citeauthoryear{{Poggianti} et~al.,}{{Poggianti}
  et~al.}{2017}]{GASP}
{Poggianti} B.~M.,  et~al., 2017, \mn@doi [\apj] {10.3847/1538-4357/aa78ed},
  \href {http://adsabs.harvard.edu/abs/2017ApJ...844...48P} {844, 48}

\bibitem[\protect\citeauthoryear{{Renzini} \& {Peng}}{{Renzini} \&
  {Peng}}{2015}]{renzini15}
{Renzini} A.,  {Peng} Y.-j.,  2015, \mn@doi [\apjl]
  {10.1088/2041-8205/801/2/L29}, \href
  {http://adsabs.harvard.edu/abs/2015ApJ...801L..29R} {801, L29}

\bibitem[\protect\citeauthoryear{Richardson}{Richardson}{1972}]{Richardson:72}
Richardson W.~H.,  1972, \mn@doi [J. Opt. Soc. Am.] {10.1364/JOSA.62.000055},
  62, 55

\bibitem[\protect\citeauthoryear{{Rodr{\'\i}guez-Puebla}, {Primack}, {Behroozi}
   \& {Faber}}{{Rodr{\'\i}guez-Puebla} et~al.}{2016}]{rodriguez16}
{Rodr{\'\i}guez-Puebla} A.,  {Primack} J.~R.,  {Behroozi} P.,   {Faber} S.~M.,
  2016, \mn@doi [\mnras] {10.1093/mnras/stv2513}, \href
  {https://ui.adsabs.harvard.edu/abs/2016MNRAS.455.2592R} {455, 2592}

\bibitem[\protect\citeauthoryear{{Roediger} \& {Courteau}}{{Roediger} \&
  {Courteau}}{2015}]{roediger:2015}
{Roediger} J.~C.,  {Courteau} S.,  2015, \mn@doi [\mnras]
  {10.1093/mnras/stv1499}, \href
  {http://adsabs.harvard.edu/abs/2015MNRAS.452.3209R} {452, 3209}

\bibitem[\protect\citeauthoryear{{Rosales-Ortega}, {D{\'{\i}}az}, {Kennicutt}
  \& {S{\'a}nchez}}{{Rosales-Ortega} et~al.}{2011}]{rosales11}
{Rosales-Ortega} F.~F.,  {D{\'{\i}}az} A.~I.,  {Kennicutt} R.~C.,
  {S{\'a}nchez} S.~F.,  2011, \mn@doi [\mnras]
  {10.1111/j.1365-2966.2011.18870.x}, \href
  {http://adsabs.harvard.edu/abs/2011MNRAS.415.2439R} {415, 2439}

\bibitem[\protect\citeauthoryear{{Rosales-Ortega}, {S{\'a}nchez},
  {Iglesias-P{\'a}ramo}, {D{\'{\i}}az}, {V{\'{\i}}lchez}, {Bland-Hawthorn},
  {Husemann}  \& {Mast}}{{Rosales-Ortega} et~al.}{2012}]{rosales-ortega:2012}
{Rosales-Ortega} F.~F.,  {S{\'a}nchez} S.~F.,  {Iglesias-P{\'a}ramo} J.,
  {D{\'{\i}}az} A.~I.,  {V{\'{\i}}lchez} J.~M.,  {Bland-Hawthorn} J.,
  {Husemann} B.,   {Mast} D.,  2012, \mn@doi [\apjl]
  {10.1088/2041-8205/756/2/L31}, \href
  {http://adsabs.harvard.edu/abs/2012ApJ...756L..31R} {756, L31}

\bibitem[\protect\citeauthoryear{{Roth} et~al.,}{{Roth} et~al.}{2005}]{roth05}
{Roth} M.~M.,  et~al., 2005, \mn@doi [\pasp] {10.1086/429877}, \href
  {http://adsabs.harvard.edu/abs/2005PASP..117..620R} {117, 620}

\bibitem[\protect\citeauthoryear{{Rousseau-Nepton}, {Robert}, {Martin},
  {Drissen}  \& {Martin}}{{Rousseau-Nepton} et~al.}{2018}]{rous18}
{Rousseau-Nepton} L.,  {Robert} C.,  {Martin} R.~P.,  {Drissen} L.,   {Martin}
  T.,  2018, \mnras, 477

\bibitem[\protect\citeauthoryear{{Saintonge} et~al.,}{{Saintonge}
  et~al.}{2016}]{saint16}
{Saintonge} A.,  et~al., 2016, \mn@doi [\mnras] {10.1093/mnras/stw1715}, \href
  {http://adsabs.harvard.edu/abs/2016MNRAS.462.1749S} {462, 1749}

\bibitem[\protect\citeauthoryear{{Salim} \& {Narayanan}}{{Salim} \&
  {Narayanan}}{2020}]{salim20}
{Salim} S.,  {Narayanan} D.,  2020, \mn@doi [\araa]
  {10.1146/annurev-astro-032620-021933}, \href
  {https://ui.adsabs.harvard.edu/abs/2020ARA&A..58..529S} {58, 529}

\bibitem[\protect\citeauthoryear{{Salpeter}}{{Salpeter}}{1955}]{salpeter55}
{Salpeter} E.~E.,  1955, \mn@doi [\apj] {10.1086/145971}, \href
  {http://adsabs.harvard.edu/abs/1955ApJ...121..161S} {121, 161}

\bibitem[\protect\citeauthoryear{{S{\'a}nchez}}{{S{\'a}nchez}}{2006}]{sanchez06a}
{S{\'a}nchez} S.~F.,  2006, \mn@doi [Astronomische Nachrichten]
  {10.1002/asna.200610643}, \href
  {https://ui.adsabs.harvard.edu/abs/2006AN....327..850S} {327, 850}

\bibitem[\protect\citeauthoryear{{S{\'a}nchez}}{{S{\'a}nchez}}{2020}]{ARAA}
{S{\'a}nchez} S.~F.,  2020, \mn@doi [\araa]
  {10.1146/annurev-astro-012120-013326}, \href
  {https://ui.adsabs.harvard.edu/abs/2020ARA&A..5812120S} {58, 99}

\bibitem[\protect\citeauthoryear{{S{\'a}nchez Almeida}, {Elmegreen},
  {Mu{\~n}oz-Tu{\~n}{\'o}n}  \& {Elmegreen}}{{S{\'a}nchez Almeida}
  et~al.}{2014}]{SA14}
{S{\'a}nchez Almeida} J.,  {Elmegreen} B.~G.,  {Mu{\~n}oz-Tu{\~n}{\'o}n} C.,
  {Elmegreen} D.~M.,  2014, \mn@doi [\aapr] {10.1007/s00159-014-0071-1}, \href
  {https://ui.adsabs.harvard.edu/abs/2014A&ARv..22...71S} {22, 71}

\bibitem[\protect\citeauthoryear{{S{\'a}nchez}, {Aceituno}, {Thiele},
  {P{\'e}rez-Ram{\'{\i}}rez}  \& {Alves}}{{S{\'a}nchez}
  et~al.}{2007}]{sanchez07a}
{S{\'a}nchez} S.~F.,  {Aceituno} J.,  {Thiele} U.,  {P{\'e}rez-Ram{\'{\i}}rez}
  D.,   {Alves} J.,  2007, \mn@doi [\pasp] {10.1086/522378}, \href
  {http://adsabs.harvard.edu/abs/2007PASP..119.1186S} {119, 1186}

\bibitem[\protect\citeauthoryear{{S{\'a}nchez}, {Thiele}, {Aceituno},
  {Cristobal}, {Perea}  \& {Alves}}{{S{\'a}nchez} et~al.}{2008}]{sanchez:2008}
{S{\'a}nchez} S.~F.,  {Thiele} U.,  {Aceituno} J.,  {Cristobal} D.,  {Perea}
  J.,   {Alves} J.,  2008, \mn@doi [\pasp] {10.1086/593981}, \href
  {http://adsabs.harvard.edu/abs/2008PASP..120.1244S} {120, 1244}

\bibitem[\protect\citeauthoryear{{S{\'a}nchez} et~al.,}{{S{\'a}nchez}
  et~al.}{2012}]{califa}
{S{\'a}nchez} S.~F.,  et~al., 2012, \mn@doi [\aap]
  {10.1051/0004-6361/201117353}, \href
  {http://adsabs.harvard.edu/abs/2012A%26A...538A...8S} {538, A8}

\bibitem[\protect\citeauthoryear{{S{\'a}nchez} et~al.,}{{S{\'a}nchez}
  et~al.}{2014}]{sanchez14}
{S{\'a}nchez} S.~F.,  et~al., 2014, \mn@doi [\aap]
  {10.1051/0004-6361/201322343}, \href
  {http://adsabs.harvard.edu/abs/2014A%26A...563A..49S} {563, A49}

\bibitem[\protect\citeauthoryear{{S{\'a}nchez} et~al.,}{{S{\'a}nchez}
  et~al.}{2016a}]{pipe3d}
{S{\'a}nchez} S.~F.,  et~al., 2016a, \rmxaa, \href
  {http://adsabs.harvard.edu/abs/2016RMxAA..52...21S} {52, 21}

\bibitem[\protect\citeauthoryear{{S{\'a}nchez} et~al.,}{{S{\'a}nchez}
  et~al.}{2016b}]{sanchez16}
{S{\'a}nchez} S.~F.,  et~al., 2016b, \mn@doi [\aap]
  {10.1051/0004-6361/201628661}, \href
  {http://adsabs.harvard.edu/abs/2016A%26A...594A..36S} {594, A36}

\bibitem[\protect\citeauthoryear{{S{\'a}nchez} et~al.,}{{S{\'a}nchez}
  et~al.}{2017}]{sanchez17a}
{S{\'a}nchez} S.~F.,  et~al., 2017, \mn@doi [\mnras] {10.1093/mnras/stx808},
  \href {http://adsabs.harvard.edu/abs/2017MNRAS.469.2121S} {469, 2121}

\bibitem[\protect\citeauthoryear{{S{\'a}nchez} et~al.,}{{S{\'a}nchez}
  et~al.}{2018}]{sanchez18}
{S{\'a}nchez} S.~F.,  et~al., 2018, \rmxaa, \href
  {http://adsabs.harvard.edu/abs/2018RMxAA..54..217S} {54, 217}

\bibitem[\protect\citeauthoryear{{S{\'a}nchez} et~al.,}{{S{\'a}nchez}
  et~al.}{2019a}]{sanchez18b}
{S{\'a}nchez} S.~F.,  et~al., 2019a, \mn@doi [\mnras] {10.1093/mnras/sty2730},
  \href {https://ui.adsabs.harvard.edu/abs/2019MNRAS.482.1557S} {482, 1557}

\bibitem[\protect\citeauthoryear{{S{\'a}nchez} et~al.,}{{S{\'a}nchez}
  et~al.}{2019b}]{sanchez19}
{S{\'a}nchez} S.~F.,  et~al., 2019b, \mn@doi [\mnras] {10.1093/mnras/stz019},
  \href {http://adsabs.harvard.edu/abs/2019MNRAS.484.3042S} {484, 3042}

\bibitem[\protect\citeauthoryear{{S{\'a}nchez}, {Walcher}, {Lopez-Cob{\'a}},
  {Barrera-Ballesteros}, {Mej{\'\i}a-Narv{\'a}ez}, {Espinosa-Ponce}  \&
  {Camps-Fari{\~n}a}}{{S{\'a}nchez} et~al.}{2021a}]{sanchez21}
{S{\'a}nchez} S.~F.,  {Walcher} C.~J.,  {Lopez-Cob{\'a}} C.,
  {Barrera-Ballesteros} J.~K.,  {Mej{\'\i}a-Narv{\'a}ez} A.,  {Espinosa-Ponce}
  C.,   {Camps-Fari{\~n}a} A.,  2021a, \mn@doi [\rmxaa]
  {10.22201/ia.01851101p.2021.57.01.01}, \href
  {https://ui.adsabs.harvard.edu/abs/2021RMxAA..57....3S} {57, 3}

\bibitem[\protect\citeauthoryear{{S{\'a}nchez} et~al.,}{{S{\'a}nchez}
  et~al.}{2021b}]{sanchez21b}
{S{\'a}nchez} S.~F.,  et~al., 2021b, \mn@doi [\mnras] {10.1093/mnras/stab442},
  \href {https://ui.adsabs.harvard.edu/abs/2021MNRAS.503.1615S} {503, 1615}

\bibitem[\protect\citeauthoryear{{S{\'a}nchez}, {G{\'o}mez Medina},
  {Barrera-Ballesteros}, {Galbany}, {Bolatto}  \& {Wong}}{{S{\'a}nchez}
  et~al.}{2022a}]{daysi}
{S{\'a}nchez} S.~F.,  {G{\'o}mez Medina} D.~C.,  {Barrera-Ballesteros} J.~K.,
  {Galbany} L.,  {Bolatto} A.,   {Wong} T.,  2022a, \mn@doi [arXiv e-prints]
  {10.48550/arXiv.2212.03738}, \href
  {https://ui.adsabs.harvard.edu/abs/2022arXiv221203738S} {p. arXiv:2212.03738}

\bibitem[\protect\citeauthoryear{{S{\'a}nchez} et~al.,}{{S{\'a}nchez}
  et~al.}{2022b}]{sanchez22}
{S{\'a}nchez} S.~F.,  et~al., 2022b, \mn@doi [\apjs]
  {10.3847/1538-4365/ac7b8f}, \href
  {https://ui.adsabs.harvard.edu/abs/2022ApJS..262...36S} {262, 36}

\bibitem[\protect\citeauthoryear{{Santini} et~al.,}{{Santini}
  et~al.}{2014}]{santini14}
{Santini} P.,  et~al., 2014, \mn@doi [\aap] {10.1051/0004-6361/201322835},
  \href {https://ui.adsabs.harvard.edu/abs/2014A&A...562A..30S} {562, A30}

\bibitem[\protect\citeauthoryear{{Sarmiento}, {Huertas-Company}, {Knapen},
  {Ibarra-Medel}, {Pillepich}, {S{\'a}nchez}  \& {Boecker}}{{Sarmiento}
  et~al.}{2022}]{sarm23}
{Sarmiento} R.,  {Huertas-Company} M.,  {Knapen} J.~H.,  {Ibarra-Medel} H.,
  {Pillepich} A.,  {S{\'a}nchez} S.~F.,   {Boecker} A.,  2022, \mn@doi [arXiv
  e-prints] {10.48550/arXiv.2211.11790}, \href
  {https://ui.adsabs.harvard.edu/abs/2022arXiv221111790S} {p. arXiv:2211.11790}

\bibitem[\protect\citeauthoryear{{Schmidt}}{{Schmidt}}{1968}]{schmidt68}
{Schmidt} M.,  1968, \mn@doi [\apj] {10.1086/149446}, \href
  {http://adsabs.harvard.edu/abs/1968ApJ...151..393S} {151, 393}

\bibitem[\protect\citeauthoryear{Shannon}{Shannon}{1948}]{shannon1948}
Shannon C.~E.,  1948, The Bell System Technical Journal, 27, 379

\bibitem[\protect\citeauthoryear{{Sharp} et~al.,}{{Sharp}
  et~al.}{2015}]{sharp15}
{Sharp} R.,  et~al., 2015, \mn@doi [\mnras] {10.1093/mnras/stu2055}, \href
  {https://ui.adsabs.harvard.edu/abs/2015MNRAS.446.1551S} {446, 1551}

\bibitem[\protect\citeauthoryear{{Shepard}}{{Shepard}}{1968}]{shepard1968}
{Shepard} D.,  1968, \mn@doi [Proceedings of the 1968 ACM National Conference]
  {10.1145/800186.810616}, 1, 517

\bibitem[\protect\citeauthoryear{{Singh} et~al.,}{{Singh}
  et~al.}{2013}]{singh13}
{Singh} R.,  et~al., 2013, \mn@doi [\aap] {10.1051/0004-6361/201322062}, \href
  {http://adsabs.harvard.edu/abs/2013A%26A...558A..43S} {558, A43}

\bibitem[\protect\citeauthoryear{{Speagle}, {Steinhardt}, {Capak}  \&
  {Silverman}}{{Speagle} et~al.}{2014}]{speagle14}
{Speagle} J.~S.,  {Steinhardt} C.~L.,  {Capak} P.~L.,   {Silverman} J.~D.,
  2014, \mn@doi [\apjs] {10.1088/0067-0049/214/2/15}, \href
  {http://adsabs.harvard.edu/abs/2014ApJS..214...15S} {214, 15}

\bibitem[\protect\citeauthoryear{{Springel} et~al.,}{{Springel}
  et~al.}{2018}]{springel18}
{Springel} V.,  et~al., 2018, \mn@doi [\mnras] {10.1093/mnras/stx3304}, \href
  {https://ui.adsabs.harvard.edu/abs/2018MNRAS.475..676S} {475, 676}

\bibitem[\protect\citeauthoryear{{Stasi{\'n}ska} et~al.,}{{Stasi{\'n}ska}
  et~al.}{2008}]{sta08}
{Stasi{\'n}ska} G.,  et~al., 2008, \mn@doi [\mnras]
  {10.1111/j.1745-3933.2008.00550.x}, \href
  {http://adsabs.harvard.edu/abs/2008MNRAS.391L..29S} {391, L29}

\bibitem[\protect\citeauthoryear{{Stubbs}, {Doherty}, {Cramer}, {Narayan},
  {Brown}, {Lykke}, {Woodward}  \& {Tonry}}{{Stubbs} et~al.}{2010}]{stub10}
{Stubbs} C.~W.,  {Doherty} P.,  {Cramer} C.,  {Narayan} G.,  {Brown} Y.~J.,
  {Lykke} K.~R.,  {Woodward} J.~T.,   {Tonry} J.~L.,  2010, \mn@doi [\apjs]
  {10.1088/0067-0049/191/2/376}, \href
  {https://ui.adsabs.harvard.edu/abs/2010ApJS..191..376S} {191, 376}

\bibitem[\protect\citeauthoryear{{Sun} et~al.,}{{Sun} et~al.}{2020}]{sun20}
{Sun} J.,  et~al., 2020, \mn@doi [\apjl] {10.3847/2041-8213/abb3be}, \href
  {https://ui.adsabs.harvard.edu/abs/2020ApJ...901L...8S} {901, L8}

\bibitem[\protect\citeauthoryear{{Taylor} et~al.,}{{Taylor}
  et~al.}{2011}]{taylor:2011}
{Taylor} E.~N.,  et~al., 2011, \mn@doi [\mnras]
  {10.1111/j.1365-2966.2011.19536.x}, \href
  {http://adsabs.harvard.edu/abs/2011MNRAS.418.1587T} {418, 1587}

\bibitem[\protect\citeauthoryear{{Tremonti} et~al.,}{{Tremonti}
  et~al.}{2004}]{tremonti04}
{Tremonti} C.~A.,  et~al., 2004, \mn@doi [\apj] {10.1086/423264}, \href
  {http://adsabs.harvard.edu/abs/2004ApJ...613..898T} {613, 898}

\bibitem[\protect\citeauthoryear{{Tully} \& {Fisher}}{{Tully} \&
  {Fisher}}{1977}]{TF77}
{Tully} R.~B.,  {Fisher} J.~R.,  1977, \aap, \href
  {https://ui.adsabs.harvard.edu/abs/1977A&A....54..661T} {54, 661}

\bibitem[\protect\citeauthoryear{{Ueta} \& {Otsuka}}{{Ueta} \&
  {Otsuka}}{2021}]{utea21}
{Ueta} T.,  {Otsuka} M.,  2021, \mn@doi [\pasp] {10.1088/1538-3873/ac20ab},
  \href {https://ui.adsabs.harvard.edu/abs/2021PASP..133i3002U} {133, 093002}

\bibitem[\protect\citeauthoryear{{Vale Asari}, {Couto}, {Cid Fernandes},
  {Stasi{\'n}ska}, {de Amorim}, {Ruschel-Dutra}, {Werle}  \& {Florido}}{{Vale
  Asari} et~al.}{2019}]{vale19}
{Vale Asari} N.,  {Couto} G.~S.,  {Cid Fernandes} R.,  {Stasi{\'n}ska} G.,  {de
  Amorim} A.~L.,  {Ruschel-Dutra} D.,  {Werle} A.,   {Florido} T.~Z.,  2019,
  \mn@doi [\mnras] {10.1093/mnras/stz2470}, \href
  {https://ui.adsabs.harvard.edu/abs/2019MNRAS.489.4721V} {489, 4721}

\bibitem[\protect\citeauthoryear{{Vila-Costas} \& {Edmunds}}{{Vila-Costas} \&
  {Edmunds}}{1992}]{vila92}
{Vila-Costas} M.~B.,  {Edmunds} M.~G.,  1992, \mnras, \href
  {http://adsabs.harvard.edu/abs/1992MNRAS.259..121V} {259, 121}

\bibitem[\protect\citeauthoryear{{Vulcani} et~al.,}{{Vulcani}
  et~al.}{2019}]{vulcani19}
{Vulcani} B.,  et~al., 2019, \mn@doi [\mnras] {10.1093/mnras/stz1829}, \href
  {https://ui.adsabs.harvard.edu/abs/2019MNRAS.488.1597V} {488, 1597}

\bibitem[\protect\citeauthoryear{{Walcher} et~al.,}{{Walcher}
  et~al.}{2014}]{walcher14}
{Walcher} C.~J.,  et~al., 2014, \mn@doi [\aap] {10.1051/0004-6361/201424198},
  \href {http://adsabs.harvard.edu/abs/2014A%26A...569A...1W} {569, A1}

\bibitem[\protect\citeauthoryear{{Walcher}, {Coelho}, {Gallazzi}, {Bruzual},
  {Charlot}  \& {Chiappini}}{{Walcher} et~al.}{2015}]{walcher14b}
{Walcher} C.~J.,  {Coelho} P.~R.~T.,  {Gallazzi} A.,  {Bruzual} G.,  {Charlot}
  S.,   {Chiappini} C.,  2015, \mn@doi [\aap] {10.1051/0004-6361/201525924},
  \href {https://ui.adsabs.harvard.edu/abs/2015A&A...582A..46W} {582, A46}

\bibitem[\protect\citeauthoryear{{Weiner} et~al.,}{{Weiner}
  et~al.}{2006}]{wein06}
{Weiner} B.~J.,  et~al., 2006, \mn@doi [\apj] {10.1086/508922}, \href
  {https://ui.adsabs.harvard.edu/abs/2006ApJ...653.1049W} {653, 1049}

\bibitem[\protect\citeauthoryear{{Wild}, {Charlot}, {Brinchmann}, {Heckman},
  {Vince}, {Pacifici}  \& {Chevallard}}{{Wild} et~al.}{2011}]{wild11}
{Wild} V.,  {Charlot} S.,  {Brinchmann} J.,  {Heckman} T.,  {Vince} O.,
  {Pacifici} C.,   {Chevallard} J.,  2011, \mn@doi [\mnras]
  {10.1111/j.1365-2966.2011.19367.x}, \href
  {https://ui.adsabs.harvard.edu/abs/2011MNRAS.417.1760W} {417, 1760}

\bibitem[\protect\citeauthoryear{{Wong} \& {Blitz}}{{Wong} \&
  {Blitz}}{2002}]{wong02}
{Wong} T.,  {Blitz} L.,  2002, \mn@doi [\apj] {10.1086/339287}, \href
  {http://adsabs.harvard.edu/abs/2002ApJ...569..157W} {569, 157}

\bibitem[\protect\citeauthoryear{{Woosley} \& {Weaver}}{{Woosley} \&
  {Weaver}}{1995}]{woos95}
{Woosley} S.~E.,  {Weaver} T.~A.,  1995, \mn@doi [\apjs] {10.1086/192237},
  \href {http://adsabs.harvard.edu/abs/1995ApJS..101..181W} {101, 181}

\bibitem[\protect\citeauthoryear{{Yan} et~al.,}{{Yan} et~al.}{2016}]{renbin16}
{Yan} R.,  et~al., 2016, \mn@doi [\aj] {10.3847/0004-6256/151/1/8}, \href
  {http://adsabs.harvard.edu/abs/2016AJ....151....8Y} {151, 8}

\bibitem[\protect\citeauthoryear{{Yan} et~al.,}{{Yan} et~al.}{2019}]{yan19}
{Yan} R.,  et~al., 2019, \mn@doi [\apj] {10.3847/1538-4357/ab3ebc}, \href
  {https://ui.adsabs.harvard.edu/abs/2019ApJ...883..175Y} {883, 175}

\bibitem[\protect\citeauthoryear{{Yates}, {Henriques}, {Thomas}, {Kauffmann},
  {Johansson}  \& {White}}{{Yates} et~al.}{2013}]{yates12}
{Yates} R.~M.,  {Henriques} B.,  {Thomas} P.~A.,  {Kauffmann} G.,  {Johansson}
  J.,   {White} S. D.~M.,  2013, \mn@doi [\mnras] {10.1093/mnras/stt1542},
  \href {https://ui.adsabs.harvard.edu/abs/2013MNRAS.435.3500Y} {435, 3500}

\bibitem[\protect\citeauthoryear{{York} et~al.,}{{York}
  et~al.}{2000a}]{york2000}
{York} D.~G.,  et~al., 2000a, \mn@doi [\aj] {10.1086/301513}, \href
  {http://adsabs.harvard.edu/abs/2000AJ....120.1579Y} {120, 1579}

\bibitem[\protect\citeauthoryear{{York}, {Adelman}, {Anderson}, {Anderson}  \&
  {et al.}}{{York} et~al.}{2000b}]{york+2000}
{York} D.~G.,  {Adelman} J.,  {Anderson} Jr. J.~E.,  {Anderson} S.~F.,   {et
  al.} 2000b, \mn@doi [\aj] {10.1086/301513}, \href
  {http://adsabs.harvard.edu/abs/2000AJ....120.1579Y} {120, 1579}

\bibitem[\protect\citeauthoryear{{Zhu}, {Barrera-Ballesteros}, {Heckman},
  {Zakamska}, {S{\'a}nchez}, {Yan}  \& {Brinkmann}}{{Zhu} et~al.}{2017}]{Zhu17}
{Zhu} G.~B.,  {Barrera-Ballesteros} J.~K.,  {Heckman} T.~M.,  {Zakamska} N.~L.,
   {S{\'a}nchez} S.~F.,  {Yan} R.,   {Brinkmann} J.,  2017, \mn@doi [\mnras]
  {10.1093/mnras/stx740}, \href
  {https://ui.adsabs.harvard.edu/abs/2017MNRAS.468.4494Z} {468, 4494}

\bibitem[\protect\citeauthoryear{{Zhu} et~al.,}{{Zhu} et~al.}{2018}]{zhu18b}
{Zhu} L.,  et~al., 2018, \mn@doi [Nature Astronomy]
  {10.1038/s41550-017-0348-1}, \href
  {https://ui.adsabs.harvard.edu/abs/2018NatAs...2..233Z} {2, 233}

\bibitem[\protect\citeauthoryear{{Zibetti}, {Charlot}  \& {Rix}}{{Zibetti}
  et~al.}{2009}]{zibetti09}
{Zibetti} S.,  {Charlot} S.,   {Rix} H.,  2009, \mn@doi [\mnras]
  {10.1111/j.1365-2966.2009.15528.x}, \href
  {http://adsabs.harvard.edu/abs/2009MNRAS.400.1181Z} {400, 1181}

\bibitem[\protect\citeauthoryear{{de Jong} \& {van der Kruit}}{{de Jong} \&
  {van der Kruit}}{1994}]{dejo94}
{de Jong} R.~S.,  {van der Kruit} P.~C.,  1994, \aaps, \href
  {https://ui.adsabs.harvard.edu/abs/1994A&AS..106..451D} {106, 451}

\makeatother
\end{thebibliography}

\end{document}